\documentclass[12pt,twoside]{article}

\usepackage{amsmath}
\usepackage{graphicx}
\usepackage{epsfig}
\usepackage[figuresright]{rotating}
\usepackage{pstcol}
\usepackage{cite}
\usepackage[dvips]{hyperref}                              
\hypersetup{ pdfborder=0 0 0, urlbordercolor=1 0 0 }      
\usepackage{relsize}                                      

\definecolor{hfviolet}{rgb}{0.5,0,0.5}
\definecolor{hflightcyan}{rgb}{0.1,1.0,1.0}

\def\hflmode{Mode}
\def\hflpdg2004{PDG 2004}
\def\hflbabar{\mbox{\slshape B\kern-0.1em{\smaller A}\kern-0.1em B\kern-0.1em{\smaller A\kern-0.2em R}}}
\def\hflbelle{\mbox{Belle}}
\def\hflcdf{\mbox{CDF}}

\def\hfld0{\mbox{D0}}
\def\hflavg{Average}

\def\hfpubhotcolor{red}
\def\hfpubcolor{magenta}
\def\hfpuboldcolor{black}

\def\hfprehotcolor{blue}
\def\hfprecolor{cyan}
\def\hfpreoldcolor{hflightcyan}

\def\hfdefcolor{black}
\def\hferrcolor{yellow}
\def\hfsuperceededcolor{hfviolet}
\def\hfwaitingcolor{green}
\def\hfinactivecolor{hfviolet}
\def\hfnoquocolor{hfviolet}
\def\hfdeftext#1{\textcolor{\hfdefcolor}{#1}}
\def\hflabel#1{\textcolor{\hfdefcolor}{$#1$}}
\def\hfavg#1{\textcolor{\hfdefcolor}{$#1$}}

\def\hfpdg#1{\textcolor{\hfdefcolor}{$#1$}}
\def\hfwaiting#1{\textcolor{\hfwaitingcolor}{$#1$}}
\def\hfpubhot#1{\textcolor{\hfpubhotcolor}{$#1$}}
\def\hfprehot#1{\textcolor{\hfprehotcolor}{$#1$}}
\def\hfwaitingtext#1{\textcolor{\hfwaitingcolor}{#1}}
\def\hfpubhottext#1{\textcolor{\hfpubhotcolor}{#1}}
\def\hfprehottext#1{\textcolor{\hfprehotcolor}{#1}}
\def\hfpub#1{\textcolor{\hfpubcolor}{$#1$}}
\def\hfpre#1{\textcolor{\hfprecolor}{$#1$}}
\def\hfpubold#1{\textcolor{\hfpuboldcolor}{$#1$}}
\def\hfpreold#1{\textcolor{\hfpreoldcolor}{$#1$}}
\def\hfpubtext#1{\textcolor{\hfpubcolor}{#1}}
\def\hfpretext#1{\textcolor{\hfprecolor}{#1}}
\def\hfpuboldtext#1{\textcolor{\hfpuboldcolor}{#1}}
\def\hfpreoldtext#1{\textcolor{\hfpreoldcolor}{#1}}

\def\hfsuperceeded#1{\textcolor{\hfsuperceededcolor}{$#1$}}

\def\hferrortext#1{\textcolor{\hferrcolor}{#1}}
\def\hfsuperceededtext#1{\textcolor{\hfsuperceededcolor}{#1}}
\def\hfinactivetext#1{\textcolor{\hfinactivecolor}{#1}}
\def\hfnoquotext#1{\textcolor{\hfnoquocolor}{#1}}
\def\hfbb#1{#1M $B\bar{B}$ pairs}

\def\hffootnotemark#1{\tiny{$^{#1}$}}
\def\hffootnotetext#1{\tiny{#1}}

\def\hffootitemsep{-0.7}

\def\hfnewp{new particles }
\def\hfdstr{strange D mesons }
\def\hfbary{baryons }
\def\hfjpsi{$J/\psi(1S)$ }
\def\hfochm{charmonium other than $J/\psi(1S)$ }
\def\hfmuld{multiple $D$, $D^{*}$ or $D^{**}$ mesons }
\def\hfsgdx{a single $D^{*}$ or $D^{**}$ meson }
\def\hfsgld{a single D meson }

\def\hfurl#1{\href{http://hfag.phys.ntu.edu.tw#1}{http://hfag.phys.ntu.edu.tw#1}} 
\def\hfhref#1#2{\href{http://hfag.phys.ntu.edu.tw#1}{#2}} 
\def\hftabletype{sidewaystable}
\def\hftableposn{!htbp}
\def\hfaftercapspace{\vspace*{2mm}}

\def\hfnewcaption#1#2#3#4#5#6#7{\caption{ #1 of #2 modes producing #3 #4, #5. The latest version is available at: \hfurl{#6} \label{#7}  }\hfaftercapspace}
\newcommand\hftabletlcell{\rule{0pt}{2.6ex}}  
\newcommand\hftableblcell{\rule[-1.2ex]{0pt}{0pt}}

\def\hfmetadata#1{}      

\def\hfurl#1{http://hfag.phys.ntu.edu.tw#1} 
\newcommand{\mysection}[1]{\section[#1]{\boldmath #1}}
\newcommand{\mysubsection}[1]{\subsection[#1]{\boldmath #1}}
\newcommand{\mysubsubsection}[1]{\subsubsection[#1]{\boldmath #1}}
\newcommand{\mysubsubsubsection}[1]{\subsubsubsection{\boldmath #1}}

\newcommand{\lesssim}{\ensuremath{\raise-.5ex\hbox{$\buildrel<\over\sim$}\,}} 

\def\dof{{\rm dof}}

\newcommand\VCKM{{V}}
\newcommand\etacpf{{\eta_f}}
\newcommand\etacp{{\eta}}

\renewcommand\Im{{\rm Im}} 
\renewcommand\Re{{\rm Re}}

\newcommand\Abar{\kern 0.18em\overline{\kern -0.18em A}{}}
\newcommand\Af{A_f}
\newcommand\Abarf{\Abar_f}
\newcommand\Afbar{A_{\bar f}}
\newcommand\Abarfbar{\Abar_{\bar f}}
\newcommand\Acp{{\cal A}}
\newcommand\Adirnoncp{\ensuremath{\langle{\cal A}_{f\bar f}\rangle}\xspace}
\newcommand\mc{\multicolumn}
\newcommand\ph{\phantom}
\newcommand {\cbf}{\ensuremath{{\cal B}}}

\newcommand {\vcb}{\ensuremath{|V_{cb}|}}
\newcommand {\vub}{\ensuremath{|V_{ub}|}}

\def\Bp      {\ensuremath{B^{+}}}
\def\Bm      {\ensuremath{B^{-}}}
\def\Bz      {\ensuremath{B^{0}}}
\def\Bs      {\ensuremath{B_{s}}}

\newcommand{\BzbDplnu}    {\ensuremath{\bar{B}^{0}\to D^{+}\ell^{-}\nub}}
\newcommand{\BzbDstarlnu} {\ensuremath{\bar{B}^{0}\to D^{*+}\ell^{-}\nub}}

\usepackage{relsize}

\def\beq{\begin{equation}}
\def\eeq#1{\label{#1}\end{equation}}
\def\eeqn{\end{equation}}

\def\beqa{\begin{eqnarray}}
\def\eeqa#1{\label{#1}\end{eqnarray}}
\def\eeqan{\end{eqnarray}}

\let\bar=\overbar

\def\etal{{\it et al.}}
\def\ie{{\it i.e.}}
\def\eg{{\it e.g.}}
\def\etc{{\it etc.}}
\def\cf{{\it cf.}}

\def\Dslash{\ensuremath{\not{\hbox{\kern-4pt $D$}}}\xspace}
\def\dslash{\not{\hbox{\kern-2pt $\del$}}}

\def\BR{\mbox{\rm BR}}
\def\ee{e^+e^-}

\def\msb{{\bar{\ssstyle M \kern -1pt S}}}

\RequirePackage{xspace}

\usepackage{relsize}
\def\babar{\mbox{\slshape B\kern-0.1em{\smaller A}\kern-0.1em
    B\kern-0.1em{\smaller A\kern-0.2em R}}\xspace}
\def\belle{\mbox{\normalfont Belle}\xspace}
\newcommand{\dzero}{D\O\xspace}

\def\ee         {\ensuremath{e^-e^-}\xspace}

\def\nub        {\ensuremath{\overline{\nu}}\xspace}

\def\nub        {\ensuremath{\overline{\nu}}\xspace}

\def\ubar  {\ensuremath{\overline u}\xspace}

\def\sbar  {\ensuremath{\overline s}\xspace}

\def\b  {\ensuremath{b}\xspace}

\def\piz   {\ensuremath{\pi^0}\xspace}

\def\pip   {\ensuremath{\pi^+}\xspace}
\def\pim   {\ensuremath{\pi^-}\xspace}

\def\etapr {\ensuremath{\eta^{\prime}}\xspace}

\def\Kbar  {\kern 0.2em\overline{\kern -0.2em K}{}\xspace}

\def\Kmp   {\ensuremath{K^\mp}\xspace}
\def\Kp    {\ensuremath{K^+}\xspace}
\def\Km    {\ensuremath{K^-}\xspace}
\def\KS    {\ensuremath{K^0_{\scriptscriptstyle S}}\xspace} 
\def\KL    {\ensuremath{K^0_{\scriptscriptstyle L}}\xspace}

\def\Kstar   {\ensuremath{K^*}\xspace}

\def\Kstarmp   {\ensuremath{K^{*\mp}}\xspace}
\def\Kz   {\ensuremath{K^0}\xspace}
\def\Kzb   {\ensuremath{\Kbar^0}\xspace}
\def\KzKzb {\ensuremath{K^0 \kern -0.16em \Kzb}\xspace}

\def\Dz    {\ensuremath{D^0}\xspace}
\def\Dbar  {\kern 0.2em\overline{\kern -0.2em D}{}\xspace}

\def\Dzb   {\ensuremath{\Dbar^0}\xspace}
\def\DzDzb {\ensuremath{D^0 {\kern -0.16em \Dzb}}\xspace}

\def\Dstar   {\ensuremath{D^*}\xspace}

\def\DorDstar   {\ensuremath{D^{(*)}}\xspace}
\def\DorDstarz  {\ensuremath{D^{(*)0}}\xspace}
\def\DorDstarzb {\ensuremath{\Dbar^{(*)0}}\xspace}

\def\Ds    {\ensuremath{D^+_s}\xspace}

\def\Bz    {\ensuremath{B^0}\xspace}
\def\B     {\ensuremath{B}\xspace}
\def\Bbar  {\kern 0.18em\overline{\kern -0.18em B}{}\xspace}
\def\Bb    {\ensuremath{\Bbar}\xspace}
\def\Bzb   {\ensuremath{\Bbar^0}\xspace}
\def\Bu    {\ensuremath{B^+}\xspace}

\def\Bmp   {\ensuremath{B^\mp}\xspace}
\def\Bs    {\ensuremath{B_s}\xspace}
\def\Bsb   {\ensuremath{\Bbar_s}\xspace}
\def\BB    {\ensuremath{B\Bbar}\xspace} 
\def\BzBzb {\ensuremath{B^0 {\kern -0.16em \Bzb}}\xspace}

\def\jpsi  {\ensuremath{{J\mskip -3mu/\mskip -2mu\psi\mskip 2mu}}\xspace}

\mathchardef\Upsilon="7107
\def\Y#1S{\ensuremath{\Upsilon{(#1S)}}\xspace}

\mathchardef\Deltares="7101
\mathchardef\Xi="7104
\mathchardef\Lambda="7103
\mathchardef\Sigma="7106
\mathchardef\Omega="710A
\def\Deltabar   {\kern 0.25em\overline{\kern -0.25em \Deltares}{}\xspace}
\def\Lbar {\kern 0.2em\overline{\kern -0.2em\Lambda\kern 0.05em}\kern-0.05em{}\xspace}
\def\Sigbar{\kern 0.2em\overline{\kern -0.2em \Sigma}{}\xspace}
\def\Xibar{\kern 0.2em\overline{\kern -0.2em \Xi}{}\xspace}
\def\Obar{\kern 0.2em\overline{\kern -0.2em \Omega}{}\xspace}
\def\Nbar{\kern 0.2em\overline{\kern -0.2em N}{}\xspace}
\def\Xb{\kern 0.2em\overline{\kern -0.2em X}{}}

\def\BR{{\ensuremath{\cal B}}}

\newcommand{\tev}{\ensuremath{\mathrm{Te\kern -0.1em V}}\xspace}
\newcommand{\gev}{\ensuremath{\mathrm{Ge\kern -0.1em V}}\xspace}
\newcommand{\mev}{\ensuremath{\mathrm{Me\kern -0.1em V}}\xspace}
\newcommand{\kev}{\ensuremath{\mathrm{ke\kern -0.1em V}}\xspace}
\newcommand{\ev}{\ensuremath{\mathrm{e\kern -0.1em V}}\xspace}
\newcommand{\gevc}{\ensuremath{{\mathrm{Ge\kern -0.1em V\!/}c}}\xspace}
\newcommand{\mevc}{\ensuremath{{\mathrm{Me\kern -0.1em V\!/}c}}\xspace}
\newcommand{\gevcc}{\ensuremath{{\mathrm{Ge\kern -0.1em V\!/}c^2}}\xspace}
\newcommand{\mevcc}{\ensuremath{{\mathrm{Me\kern -0.1em V\!/}c^2}}\xspace}

\def\mus  {\ensuremath{\rm \,\mus}\xspace}

\def\ps   {\ensuremath{\rm \,ps}\xspace}

\def\mus        {\ensuremath{\,\mu{\rm s}}\xspace}    
\def\ps         {\ensuremath{{\rm \,ps}}\xspace}  
\def\gsim{{~\raise.15em\hbox{$>$}\kern-.85em
          \lower.35em\hbox{$\sim$}~}\xspace}
\def\lsim{{~\raise.15em\hbox{$<$}\kern-.85em
          \lower.35em\hbox{$\sim$}~}\xspace}

\def\CP                 {\ensuremath{C\!P}\xspace}
\def\CPT                {\ensuremath{C\!PT}\xspace}

\def\pep2{PEP-II}

\def\rhobar {\ensuremath{\overline{\rho}}\xspace}
\def\etabar {\ensuremath{\overline{\eta}}\xspace}

\def\Vub  {\ensuremath{|V_{ub}|}\xspace}

\def\stwob{\ensuremath{\sin\! 2 \beta   }\xspace}

\def\deltamd{\ensuremath{{\rm \Delta}m_d}\xspace}

\xspace

\newcommand{\epjc}      [1]  {{Eur.\ Phys.\ Jour.\ {\bf C}{\bf #1}}}

\newcommand{\npb}       [1]  {{Nucl.\ Phys.\ {\bf B{\bf #1}}}}

\newcommand{\plb}       [1]  {{Phys.\ Lett.\ B~{\bf #1}}}   

\def\jetset74   {\mbox{\tt Jetset \hspace{-0.5em}7.\hspace{-0.2em}4}}

\newcommand{\aerr}[4]   {\mbox{${{#1}^{+ #2}_{- #3}\pm #4}$}}
\newcommand{\berr}[4]   {\mbox{${{#1}\pm #2^{+ #3}_{- #4}}$}}
\newcommand{\cerr}[3]   {\mbox{${{#1}^{+ #2}_{- #3}}$}}
\newcommand{\aerrsy}[5] {\mbox{${{#1}^{+ #2 + #4}_{- #3 - #5}}$}}

\newcommand{\err}[3]   {\mbox{${{#1}\pm{#2}\pm{#3}}$}}
\newcommand{\nodata}{$$}
\def\etapr{{\eta^{\prime}}}

\def\sgline{\noalign{\vskip 0.10truecm\hrule\vskip 0.10truecm}}
\def\sglinespt{\noalign{\vskip 0.05truecm\hrule}}
\def\sglinespb{\noalign{\hrule\vskip 0.05truecm}}
\newcommand{\prl}      [1]  {{Phys.\ Rev.\ Lett.\ {\bf #1}}}
\newcommand{\prd}      [1]  {{Phys.\ Rev.\ D~{\bf #1}}}

\newcommand{\kz}    {\mbox{$K^0$}}

\newcommand{\RPP}{}

\renewcommand{\mysection}[1]{\section[#1]{#1}} 

\begin{document}

\setcounter{page}{1}

\title{\begin{flushleft}
\end{flushleft}
\vskip 20pt
Averages of $b$-hadron Properties \\
at the End of 2005 }
\author{Heavy Flavor Averaging Group (HFAG)\footnote{
The HFAG members involved in producing the averages for the end of 2005
update are:
   E.~Barberio, 
   I.~Bizjak, 
   S.~Blyth, 
   G.~Cavoto,     
   P.~Chang, 
   J.~Dingfelder, 
   S.~Eidelman, 
   T.~Gershon,    
   R.~Godang,     
   R.~Harr,       
   A.~H\"ocker,     
   T.~Iijima, 
   R.~Kowalewski,   
   F.~Lehner, 
   A.~Limosani, 
   C.-J.~Lin, 
   O.~Long,       
   V.~Luth,    
   M.~Morii,  
   S.~Prell, 
   O.~Schneider, 
   J.~Smith, 
   A.~Stocchi,     
   S.~Tosi, 
   K.~Trabelsi,    
   R.~Van~Kooten, 
   C.~Voena, 
   and C.~Weiser. 
  }
 }
\maketitle
\thispagestyle{empty}
\begin{abstract}
This article reports world averages for measurements on $b$-hadron
properties obtained by the Heavy Flavor Averaging Group (HFAG) using the
available results as of at the end of 2005. In the averaging, the
input parameters used in the various analyses are adjusted (rescaled) to
common values, and all known correlations are taken into account.
The averages include lifetimes, neutral meson mixing parameters,
parameters of semileptonic decays, branching fractions of $B$ decays to
final states with open charm, charmonium and no charm, and measurements
related to \CP asymmetries. 
\end{abstract}

\newpage
\tableofcontents
\newpage


\mysection{Introduction}
\label{sec:intro}

Flavor dynamics is an important element in understanding the nature of
particle physics.  The accurate knowledge of properties of heavy flavor
hadrons, especially $b$ hadrons, plays an essential role for
determination of the Cabbibo-Kobayashi-Maskawa (CKM)
matrix~\cite{ref_ckm}. Since asymmetric-energy $e^+e^-$ $B$ factories
started their operation, the size of available $B$ meson samples has
dramatically increased and the accuracies of measurements have been
improved. Tevatron experiments also started to provide rich results on
$B$ hadron decays with increased Run-II data samples. 
 
The Heavy Flavor Averaging Group (HFAG) has been formed in 2002,
continuing the activities of LEP Heavy Flavor Steering
group~\cite{LEPHFS}, to provide averages for measurements
of $b$-flavor related quantities. The HFAG consists of
representatives and contact persons from the experimental groups:
\babar, \belle, CDF, CLEO, \dzero, and LEP. 

The HFAG is currently organized into five subgroups:
\begin{itemize}
\item the ``Lifetime and Mixing'' group provides averages for $b$-hadron
  lifetimes, $b$-hadron fractions in $\Upsilon(4S)$ decay and high
  energy collisions, and various parameters in $B^0$ and $B_s^0$
  oscillation (mixing);

\item the ``Semileptonic $B$ Decays'' group provides averages
for inclusive and exclusive $B$-decay branching fractions, and
estimates of the CKM matrix elements $|V_{cb}|$ and $|V_{ub}|$;

\item the ``$\CP(t)$ and Unitarity Triangle Angles'' group provides
  averages for time-dependent $\CP$ asymmetry parameters and angles of
  the $B$ unitarity triangle;

\item the ``Rare Decays'' group provides averages of branching fractions
  and their asymmetries between $B$ and $\bar B$ for charmless mesonic, 
  radiative, leptonic, and baryonic $B$ decays;

\item the ``$B$ to Charm Decays'' group provides averages of branching
  fractions for $B$ decays to final states involving open charm mesons
  or charmonium.
\end{itemize}

The first two subgroups continue the activities from LEP working groups
with some reorganization (merging four groups into two groups). The
latter three groups have been newly formed to provide averages for
results which are available from $B$ factory experiments.

This article is an update of the Winter 2005 HFAG
document~\cite{hfag_hepex_winter2005}, and we report the world averages
using the available results as of the end of 2005. All results that are
publicly available, including recent preliminary results, are used in
the averages.  We do not use preliminary results which remain
unpublished for a long time or for which no publication is planned. 
Close contacts have been established between representatives from
the experiments and members of different subgroups in charge of the
averages, to ensure that the data are prepared in a form suitable
for combinations.  

We do not scale the error of an average (as is presently done by the
Particle Data Group~\cite{Eidelman:2004wy}) in case $\chi^2/\dof > 1$,
where $\dof$ is the number of degrees of freedom in the average
calculation. In such a case, we examine the systematics of each
measurement and try to understand them. Unless we find possible
systematic discrepancies between the measurements, we do not make any
special treatment for the calculated error. We provide the confidence
level of the fit as an indicator for the consistency of the measurements
included in the average. We attach a warning message in case that some
special treatment was necessary to calculate the average or the
approximation used in the average calculation may not be good enough 
(\eg, Gaussian error is used in averaging although the likelihood 
indicates non-Gaussian behavior).

Section~\ref{sec:method} describes the methodology for calculating
averages for various quantities used by the HFAG. In the averaging, the
input parameters used in the various analyses are adjusted (rescaled) to
common values, and, where possible, known correlations are taken into
account. The general philosophy and tools for calculations of averages
are presented. 
Sections~\ref{sec:life_mix}--\ref{sec:BtoCharm} describe the averaging of 
the quantities from each of the subgroups mentioned above.
A brief summary of the averages described in this article is given in
Sec.~\ref{sec:summary}.   

 The complete listing of averages and plots described in this article
are also available on the HFAG web page:
 
 {\tt http://www.slac.stanford.edu/xorg/hfag } and 

 {\tt http://belle.kek.jp/mirror/hfag } (KEK mirror site).

\section{Methodology } \label{sec:method} 

The general averaging problem that HFAG faces is to combine the
information provided by different measurements of the same parameter,
to obtain our best estimate of the parameter's value and
uncertainty. The methodology described here focuses on the problems of
combining measurements performed with different systematic assumptions
and with potentially-correlated systematic uncertainties. Our methodology
relies on the close involvement of the people performing the
measurements in the averaging process.

Consider two hypothetical measurements of a parameter $x$, which might
be summarized as
\begin{align*}
x &= x_1 \pm \delta x_1 \pm \Delta x_{1,1} \pm \Delta x_{2,1} \ldots \\
x &= x_2 \pm \delta x_2 \pm \Delta x_{1,2} \pm \Delta x_{2,2} \ldots
\; ,
\end{align*}
where the $\delta x_k$ are statistical uncertainties, and
the $\Delta x_{i,k}$ are contributions to the systematic
uncertainty. One popular approach is to combine statistical and
systematic uncertainties in quadrature
\begin{align*}
x &= x_1 \pm \left(\delta x_1 \oplus \Delta x_{1,1} \oplus \Delta
x_{2,1} \oplus \ldots\right) \\
x &= x_2 \pm \left(\delta x_2 \oplus \Delta x_{1,2} \oplus \Delta
x_{2,2} \oplus \ldots\right)
\end{align*}
and then perform a weighted average of $x_1$ and $x_2$, using their
combined uncertainties, as if they were independent. This approach
suffers from two potential problems that we attempt to address. First,
the values of the $x_k$ may have been obtained using different
systematic assumptions. For example, different values of the \Bz
lifetime may have been assumed in separate measurements of the
oscillation frequency $\deltamd$. The second potential problem is that
some contributions of the systematic uncertainty may be correlated
between experiments. For example, separate measurements of $\deltamd$
may both depend on an assumed Monte-Carlo branching fraction used to
model a common background.

The problems mentioned above are related since, ideally, any quantity $y_i$
that $x_k$ depends on has a corresponding contribution $\Delta x_{i,k}$ to the
systematic error which reflects the uncertainty $\Delta y_i$ on $y_i$
itself. We assume that this is the case, and use the values of $y_i$ and
$\Delta y_i$ assumed by each measurement explicitly in our
averaging (we refer to these values as $y_{i,k}$ and $\Delta y_{i,k}$
below). Furthermore, since we do not lump all the systematics
together,
we require that each measurement used in an average have a consistent
definition of the various contributions to the systematic uncertainty.
Different analyses often use different decompositions of their systematic
uncertainties, so achieving consistent definitions for any potentially
correlated contributions requires close coordination between HFAG and
the experiments. In some cases, a group of
systematic uncertainties must be lumped to obtain a coarser
description that is consistent between measurements. Systematic uncertainties
that are uncorrelated with any other sources of uncertainty appearing
in an average are lumped with the statistical error, so that the only
systematic uncertainties treated explicitly are those that are
correlated with at least one other measurement via a consistently-defined
external parameter $y_i$. When asymmetric statistical or systematic
uncertainties are quoted, we symmetrize them since our combination
method implicitly assumes parabolic likelihoods for each measurement.

\begin{figure}
\begin{center}
\includegraphics[width=6.0in]{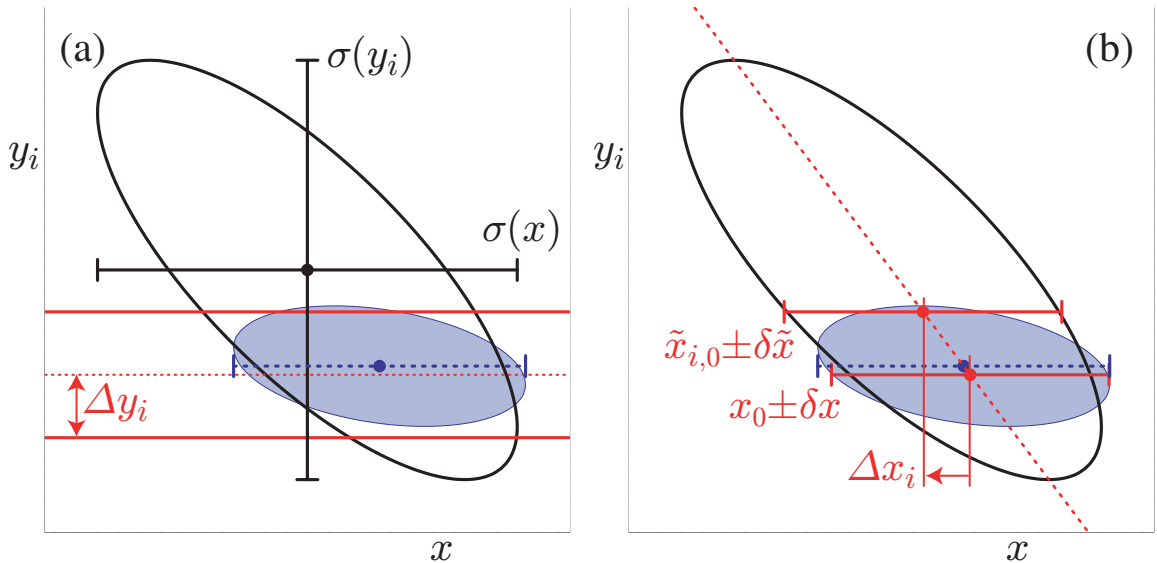}
\end{center}
\caption{The left-hand plot, (a), compares the 68\% confidence-level
  contours of a
  hypothetical measurement's unconstrained (large ellipse) and
  constrained (filled ellipse) likelihoods, using the Gaussian
  constraint on $y_i$ represented by the horizontal band. The solid
  error bars represent the statistical uncertainties, $\sigma(x)$ and
  $\sigma(y_i)$, of the unconstrained likelihood. The dashed
  error bar shows the statistical error on $x$ from a
  constrained simultaneous fit to $x$ and $y_i$. The right-hand plot,
  (b), illustrates the method described in the text of performing fits
  to $x$ only with $y_i$ fixed at different values. The dashed
  diagonal line between these fit results has the slope
  $\rho(x,y_i)\sigma(y_i)/\sigma(x)$ in the limit of a parabolic
  unconstrained likelihood. The result of the constrained simultaneous
  fit from (a) is shown as a dashed error bar on $x$.}
\label{fig:singlefit}
\end{figure}

The fact that a measurement of $x$ is sensitive to the value of $y_i$
indicates that, in principle, the data used to measure $x$ could
equally-well be used for a simultaneous measurement of $x$ and $y_i$, as
illustrated by the large contour in Fig.~\ref{fig:singlefit}(a) for a hypothetical
measurement. However, we often have an external constraint $\Delta
y_i$ on the value of $y_i$ (represented by the horizontal band in
Fig.~\ref{fig:singlefit}(a)) that is more precise than the constraint
$\sigma(y_i)$ from
our data alone. Ideally, in such cases we would perform a simultaneous
fit to $x$ and $y_i$, including the external constraint, obtaining the
filled $(x,y)$ contour and corresponding dashed one-dimensional estimate of
$x$ shown in Fig.~\ref{fig:singlefit}(a). Throughout, we assume that
the external constraint $\Delta y_i$ on $y_i$ is Gaussian.

In practice, the added technical complexity of a constrained fit with
extra free parameters is not justified by the small increase in
sensitivity, as long as the external constraints $\Delta y_i$ are
sufficiently precise when compared with the sensitivities $\sigma(y_i)$
to each $y_i$ of the data alone. Instead, the usual procedure adopted
by the experiments is to perform a baseline fit with all $y_i$ fixed
to nominal values $y_{i,0}$, obtaining $x = x_0 \pm \delta
x$. This baseline fit neglects the uncertainty due to $\Delta y_i$, but
this error can be mostly recovered by repeating the fit separately for
each external parameter $y_i$ with its value fixed at $y_i = y_{i,0} +
\Delta y_i$ to obtain $x = \tilde{x}_{i,0} \pm \delta\tilde{x}$, as
illustrated in Fig.~\ref{fig:singlefit}(b). The absolute shift,
$|\tilde{x}_{i,0} - x_0|$, in the central value of $x$ is what the
experiments usually quote as their systematic uncertainty $\Delta x_i$
on $x$ due to the unknown value of $y_i$. Our procedure requires that
we know not only the magnitude of this shift but also its sign. In the
limit that the unconstrained data is represented by a parabolic
likelihood, the signed shift is given by
\begin{equation}
\Delta x_i = \rho(x,y_i)\frac{\sigma(x)}{\sigma(y_i)}\,\Delta y_i \;,
\end{equation}
where $\sigma(x)$ and $\rho(x,y_i)$ are the statistical uncertainty on
$x$ and the correlation between $x$ and
$y_i$ in the unconstrained data.
While our procedure is not
equivalent to the constrained fit with extra parameters, it yields (in
the limit of a parabolic unconstrained likelihood) a central value
$x_0$ that agrees 
to ${\cal O}(\Delta y_i/\sigma(y_i))^2$ and an uncertainty $\delta x
\oplus \Delta x_i$ that agrees to ${\cal O}(\Delta y_i/\sigma(y_i))^4$.

In order to combine two or more measurements that share systematics
due to the same external parameters $y_i$, we would ideally perform a
constrained simultaneous fit of all data samples to obtain values of
$x$ and each $y_i$, being careful to only apply the constraint on each
$y_i$ once. This is not practical since we generally do not have
sufficient information to reconstruct the unconstrained likelihoods
corresponding to each measurement. Instead, we perform the two-step
approximate procedure described below.

\begin{figure}
\begin{center}
\includegraphics[width=6.0in]{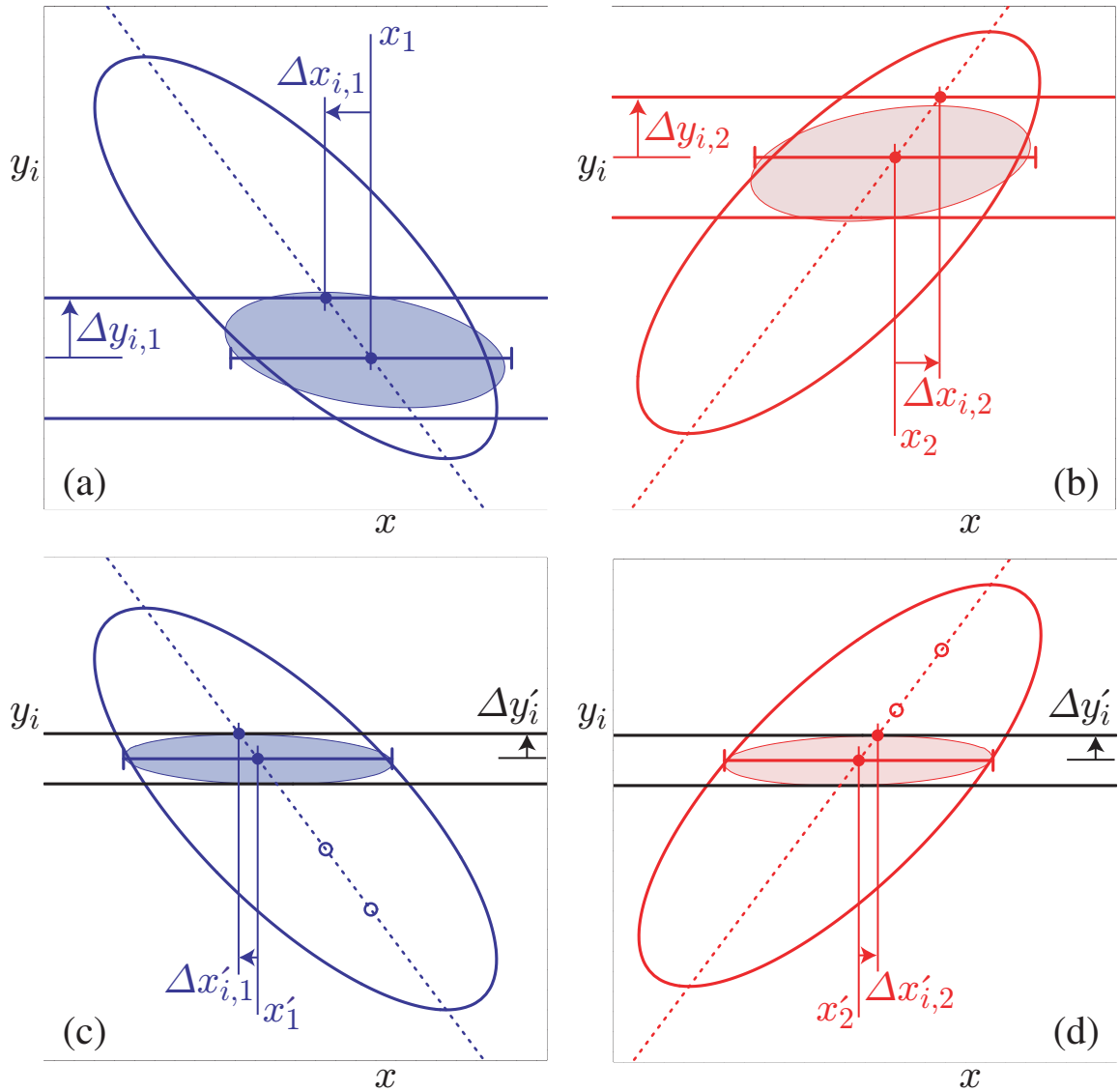}
\end{center}
\caption{The upper plots, (a) and (b), show examples of two individual
  measurements to be combined. The large ellipses represent their
  unconstrained likelihoods, and the filled ellipses represent their
  constrained likelihoods. Horizontal bands indicate the different
  assumptions about the value and uncertainty of $y_i$ used by each
  measurement. The error bars show the results of the approximate
  method described in the text for obtaining $x$ by performing fits
  with $y_i$ fixed to different values. The lower plots, (c) and (d),
  illustrate the adjustments to accommodate updated and consistent
  knowledge of $y_i$ described in the text. Hollow circles mark the
  central values of the unadjusted fits to $x$ with $y$ fixed, which
  determine the dashed line used to obtain the adjusted values. }
\label{fig:multifit}
\end{figure}

Figs.~\ref{fig:multifit}(a,b) illustrate two
statistically-independent measurements, $x_1 \pm (\delta x_1 \oplus
\Delta x_{i,1})$ and $x_2\pm(\delta x_i\oplus \Delta x_{i,2})$, of the same
hypothetical quantity $x$ (for simplicity, we only show the
contribution of a single correlated systematic due to an external
parameter $y_i$). As our knowledge of the external parameters $y_i$
evolves, it is natural that the different measurements of $x$ will
assume different nominal values and ranges for each $y_i$. The first
step of our procedure is to adjust the values of each measurement to
reflect the current best knowledge of the values $y_i'$ and ranges
$\Delta y_i'$ of the external parameters $y_i$, as illustrated in
Figs.~\ref{fig:multifit}(c,b). We adjust the
central values $x_k$ and correlated systematic uncertainties $\Delta
x_{i,k}$ linearly for each measurement (indexed by $k$) and each
external parameter (indexed by $i$):
\begin{align}
x_k' &= x_k + \sum_i\,\frac{\Delta x_{i,k}}{\Delta y_{i,k}}
\left(y_i'-y_{i,k}\right)\\
\Delta x_{i,k}'&= \Delta x_{i,k}\cdot \frac{\Delta y_i'}{\Delta
  y_{i,k}} \; .
\end{align}
This procedure is exact in the limit that the unconstrained
likelihoods of each measurement is parabolic.

\begin{figure}
\begin{center}
\includegraphics[width=3.5in]{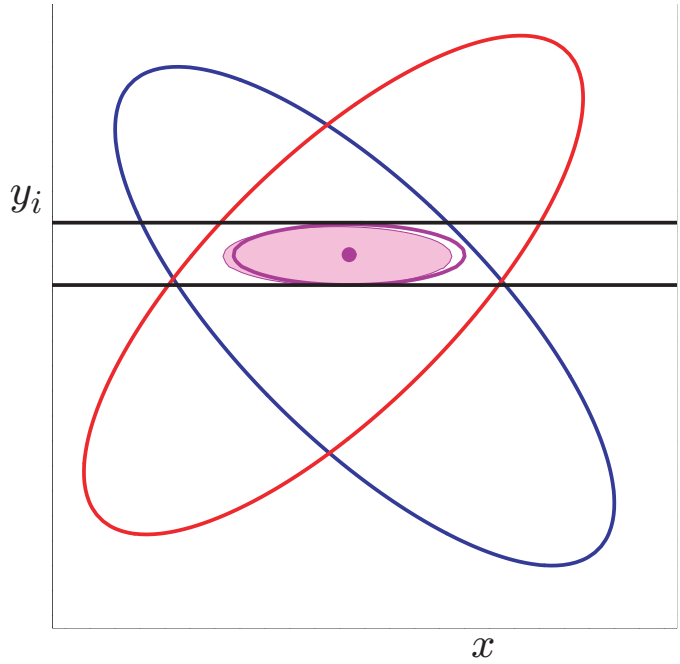}
\end{center}
\caption{An illustration of the combination of two hypothetical
  measurements of $x$ using the method described in the text. The
  ellipses represent the unconstrained likelihoods of each measurement
  and the horizontal band represents the latest knowledge about $y_i$ that
  is used to adjust the individual measurements. The filled small
  ellipse shows the result of the exact method using ${\cal
  L}_{\text{comb}}$ and the hollow small ellipse and dot show the
  result of the approximate method using $\chi^2_{\text{comb}}$.}
\label{fig:fit12}
\end{figure}

The second step of our procedure is to combine the adjusted
measurements, $x_k'\pm (\delta x_k\oplus \Delta x_{k,1}'\oplus \Delta
x_{k,2}'\oplus\ldots)$ using the chi-square 
\begin{equation}
\chi^2_{\text{comb}}(x,y_1,y_2,\ldots) \equiv \sum_k\,
\frac{1}{\delta x_k^2}\left[
x_k' - \left(x + \sum_i\,(y_i-y_i')\frac{\Delta x_{i,k}'}{\Delta y_i'}\right)
\right]^2 + \sum_i\,
\left(\frac{y_i - y_i'}{\Delta y_i'}\right)^2 \; ,
\end{equation}
and then minimize this $\chi^2$ to obtain the best values of $x$ and
$y_i$ and their uncertainties, as illustrated in
Fig.~\ref{fig:fit12}. Although this method determines new values for
the $y_i$, we do not report them since the $\Delta x_{i,k}$ reported
by each experiment are generally not intended for this purpose (for
example, they may represent a conservative upper limit rather than a
true reflection of a 68\% confidence level).

For comparison, the exact method we would
perform if we had the unconstrained likelihoods ${\cal L}_k(x,y_1,y_2,\ldots)$
available for each
measurement is to minimize the simultaneous constrained likelihood
\begin{equation}
{\cal L}_{\text{comb}}(x,y_1,y_2,\ldots) \equiv \prod_k\,{\cal
  L}_k(x,y_1,y_2,\ldots)\,\prod_{i}\,{\cal 
  L}_i(y_i) \; ,
\end{equation}
with an independent Gaussian external constraint on each $y_i$
\begin{equation}
{\cal L}_i(y_i) \equiv \exp\left[-\frac{1}{2}\,\left(\frac{y_i-y_i'}{\Delta
 y_i'}\right)^2\right] \; .
\end{equation}
The results of this exact method are illustrated by the filled ellipses
in Figs.~\ref{fig:fit12}(a,b), and agree with our method in the limit that
each ${\cal L}_k$ is parabolic and that each $\Delta
y_i' \ll \sigma(y_i)$. In the case of a non-parabolic unconstrained
likelihood, experiments would have to provide a description of ${\cal
  L}_k$ itself to allow an improved combination. In the case of some
$\sigma(y_i)\simeq \Delta y_i'$, experiments are advised to perform a
simultaneous measurement of both $x$ and $y$ so that their data will
improve the world knowledge about $y$. 

 The algorithm described above is used as a default in the averages
reported in the following sections.  For some cases, somewhat simplified
or more complex algorithms are used and noted in the corresponding 
sections. 

Following the prescription described above,
the central values and errors are rescaled
to a common set of input parameters in the averaging procedures, 
according to the dependency on
any of these input parameters.
We try to use the most up-to-date values for these common inputs and 
the same values among the HFAG subgroups.
For the parameters whose averages are produced by the HFAG, we use 
the updated values in the current update cycle.  For other external
parameters, we use the most recent PDG values. 

  The parameters and values used in this update cycle are listed in
each subgroup section.

\clearpage
%
%
%
%
%
%
%

%

%
%
%
%

%

\renewcommand{\topfraction}{0.9}

\newcommand{\auth}[1]{#1,}
\newcommand{\coll}[1]{#1 Collaboration,}
\newcommand{\authcoll}[2]{#1 \etal\ (#2 Collaboration),}
\newcommand{\titl}[1]{``#1'',} 
\newcommand{\J}[4]{{#1} {\bf #2}, #3 (#4)}
\newcommand{\subJ}[1]{submitted to #1}
\newcommand{\PRL}[3]{\J{Phys.\ Rev.\ Lett.}{#1}{#2}{#3}}
\newcommand{\subPRL}{\subJ{Phys.\ Rev.\ Lett.}}
\newcommand{\PRD}[3]{\J{Phys.\ Rev.\ D}{#1}{#2}{#3}}
\newcommand{\subPRD}{\subJ{Phys.\ Rev.\ D}}
\newcommand{\ZPC}[3]{\J{Z.\ Phys.\ C}{#1}{#2}{#3}}
\newcommand{\PLB}[3]{\J{Phys.\ Lett.\ B}{#1}{#2}{#3}}
\newcommand{\subPLB}{\subJ{Phys.\ Lett.\ B}}
\newcommand{\EPJC}[3]{\J{Eur.\ Phys.\ J.\ C}{#1}{#2}{#3}}
\newcommand{\NPB}[3]{\J{Nucl.\ Phys.\ B}{#1}{#2}{#3}}
\newcommand{\subNPB}{\subJ{Nucl.\ Phys.\ B}}
\newcommand{\NIMA}[3]{\J{Nucl.\ Instrum.\ Methods A}{#1}{#2}{#3}}
\newcommand{\subNIMA}{\subJ{Nucl.\ Instrum.\ Methods A}}
\newcommand{\JHEP}[3]{\J{J.\ of High Energy Physics }{#1}{#2}{#3}}
\newcommand{\ARNS}[3]{\J{Ann.\ Rev.\ Nucl.\ Sci.}{#1}{#2}{#3}}
\newcommand{\newref}{\\}

\newcommand{\particle}[1]{\ensuremath{#1}\xspace}
\renewcommand{\ee}{\particle{e^+e^-}}
\newcommand{\Ups}{\particle{\Upsilon(4S)}}
\renewcommand{\b}{\particle{b}}
\renewcommand{\B}{\particle{B}}
\newcommand{\Bd}{\particle{B^0}}
\renewcommand{\Bs}{\particle{B^0_s}}
\renewcommand{\Bu}{\particle{B^+}}
\newcommand{\Bc}{\particle{B^+_c}}
\newcommand{\Bdbar}{\particle{\bar{B}^0}}
\newcommand{\Bsbar}{\particle{\bar{B}^0_s}}
\newcommand{\Lb}{\particle{\Lambda_b^0}}
\newcommand{\Xib}{\particle{\Xi_b}}
\newcommand{\Lc}{\particle{\Lambda_c^+}}

\newcommand{\fBs}{\ensuremath{f_{\particle{s}}}\xspace}
\newcommand{\fBd}{\ensuremath{f_{\particle{d}}}\xspace}
\newcommand{\fBu}{\ensuremath{f_{\particle{u}}}\xspace}
\newcommand{\fbb}{\ensuremath{f_{\rm baryon}}\xspace}

\newcommand{\dmd}{\ensuremath{\Delta m_{\particle{d}}}\xspace}
\newcommand{\dms}{\ensuremath{\Delta m_{\particle{s}}}\xspace}
\newcommand{\xd}{\ensuremath{x_{\particle{d}}}\xspace}
\newcommand{\xs}{\ensuremath{x_{\particle{s}}}\xspace}
\newcommand{\yd}{\ensuremath{y_{\particle{d}}}\xspace}
\newcommand{\ys}{\ensuremath{y_{\particle{s}}}\xspace}
\newcommand{\chibar}{\ensuremath{\overline{\chi}}\xspace}
\newcommand{\chid}{\ensuremath{\chi_{\particle{d}}}\xspace}
\newcommand{\chis}{\ensuremath{\chi_{\particle{s}}}\xspace}
\newcommand{\Gd}{\ensuremath{\Gamma_{\particle{d}}}\xspace}
\newcommand{\DGd}{\ensuremath{\Delta\Gd}\xspace}
\newcommand{\DGGd}{\ensuremath{\DGd/\Gd}\xspace}
\newcommand{\Gs}{\ensuremath{\Gamma_{\particle{s}}}\xspace}
\newcommand{\DGs}{\ensuremath{\Delta\Gs}\xspace}
\newcommand{\DGGs}{\ensuremath{\Delta\Gs/\Gs}\xspace}

\renewcommand{\BR}[1]{\particle{{\cal B}(#1)}}
\newcommand{\CL}[1]{#1\%~\mbox{CL}}
\newcommand{\Qjet}{\ensuremath{Q_{\rm jet}}\xspace}

\newcommand{\labe}[1]{\label{equ:#1}}
\newcommand{\labs}[1]{\label{sec:#1}}
\newcommand{\labf}[1]{\label{fig:#1}}
\newcommand{\labt}[1]{\label{tab:#1}}
\newcommand{\refe}[1]{\ref{equ:#1}}
\newcommand{\refs}[1]{\ref{sec:#1}}
\newcommand{\reff}[1]{\ref{fig:#1}}
\newcommand{\reft}[1]{\ref{tab:#1}}
\newcommand{\Ref}[1]{Ref.~\cite{#1}}
\newcommand{\Refs}[1]{Refs.~\cite{#1}}
\newcommand{\Refss}[2]{Refs.~\cite{#1} and \cite{#2}}
\newcommand{\Refsss}[3]{Refs.~\cite{#1}, \cite{#2} and \cite{#3}}
\newcommand{\eq}[1]{(\refe{#1})}
\newcommand{\Eq}[1]{Eq.~(\refe{#1})}
\newcommand{\Eqs}[1]{Eqs.~(\refe{#1})}
\newcommand{\Eqss}[2]{Eqs.~(\refe{#1}) and (\refe{#2})}
\newcommand{\Eqssor}[2]{Eqs.~(\refe{#1}) or (\refe{#2})}
\newcommand{\Eqsss}[3]{Eqs.~(\refe{#1}), (\refe{#2}), and (\refe{#3})}
\newcommand{\Figure}[1]{Figure~\reff{#1}}
\newcommand{\Figuress}[2]{Figures~\reff{#1} and \reff{#2}}
\newcommand{\Fig}[1]{Fig.~\reff{#1}}
\newcommand{\Figs}[1]{Figs.~\reff{#1}}
\newcommand{\Figss}[2]{Figs.~\reff{#1} and \reff{#2}}
\newcommand{\Figsss}[3]{Figs.~\reff{#1}, \reff{#2}, and \reff{#3}}
\newcommand{\Section}[1]{Section~\refs{#1}}
\newcommand{\Sec}[1]{Sec.~\refs{#1}}
\newcommand{\Secs}[1]{Secs.~\refs{#1}}
\newcommand{\Secss}[2]{Secs.~\refs{#1} and \refs{#2}}
\newcommand{\Secsss}[3]{Secs.~\refs{#1}, \refs{#2}, and \refs{#3}}
\newcommand{\Table}[1]{Table~\reft{#1}}
\newcommand{\Tables}[1]{Tables~\reft{#1}}
\newcommand{\Tabless}[2]{Tables~\reft{#1} and \reft{#2}}
\newcommand{\Tablesss}[3]{Tables~\reft{#1}, \reft{#2}, and \reft{#3}}

\newcommand{\subsubsubsection}[1]{\vspace{2ex}\par\noindent {\bf\boldmath\em #1} \vspace{2ex}\par}


\newcommand{\definemath}[2]{\newcommand{#1}{\ensuremath{#2}\xspace}}

\definemath{\hfagCHIBARLEPval}{0.1259}
\definemath{\hfagCHIBARLEPerr}{\pm0.0042}
\definemath{\hfagTAUBDval}{1.527}
\definemath{\hfagTAUBDerr}{\pm0.008}
\definemath{\hfagTAUBUval}{1.643}
\definemath{\hfagTAUBUerr}{\pm0.010}
\definemath{\hfagRTAUBUval}{1.076}
\definemath{\hfagRTAUBUerr}{\pm0.008}
\definemath{\hfagTAUBSval}{1.461}
\definemath{\hfagTAUBSerr}{\pm0.040}
\definemath{\hfagRTAUBSval}{0.957}
\definemath{\hfagRTAUBSerr}{\pm0.027}
\definemath{\hfagTAULBval}{1.288}
\definemath{\hfagTAULBerr}{\pm0.065}
\definemath{\hfagTAUXBval}{1.39}
\definemath{\hfagTAUXBerp}{^{+0.34}}
\definemath{\hfagTAUXBern}{_{-0.28}}
\definemath{\hfagTAUBBval}{1.242}
\definemath{\hfagTAUBBerr}{\pm0.046}
\definemath{\hfagRTAUBBval}{0.813}
\definemath{\hfagRTAUBBerr}{\pm0.030}
\definemath{\hfagTAUBval}{1.568}
\definemath{\hfagTAUBerr}{\pm0.009}
\definemath{\hfagTAUBCval}{0.469}
\definemath{\hfagTAUBCerr}{\pm0.065}
\definemath{\hfagTAUBSSLval}{1.454}
\definemath{\hfagTAUBSSLerr}{\pm0.040}
\definemath{\hfagTAUBSSLXval}{1.457}
\definemath{\hfagTAUBSSLXerr}{\pm0.042}
\definemath{\hfagTAUBSMEANCONval}{1.396}
\definemath{\hfagTAUBSMEANCONerp}{^{+0.044}}
\definemath{\hfagTAUBSMEANCONern}{_{-0.046}}
\definemath{\hfagTAUBSJFval}{1.404}
\definemath{\hfagTAUBSJFerr}{\pm0.066}
\definemath{\hfagRTAUBSSLval}{0.952}
\definemath{\hfagRTAUBSSLerr}{\pm0.027}
\definemath{\hfagRTAUBSMEANCONval}{0.914}
\definemath{\hfagRTAUBSMEANCONerr}{\pm0.030}
\definemath{\hfagRTAUBSMEANCONsig}{2.9}
\definemath{\hfagONEMINUSRTAUBSMEANCONpercent}{(8.6\pm3.0)\%}
\definemath{\hfagRTAULBval}{0.844}
\definemath{\hfagRTAULBerr}{\pm0.043}
\definemath{\hfagTAUBVTXval}{1.572}
\definemath{\hfagTAUBVTXerr}{\pm0.009}
\definemath{\hfagTAUBLEPval}{1.537}
\definemath{\hfagTAUBLEPerr}{\pm0.020}
\definemath{\hfagTAUBJPval}{1.533}
\definemath{\hfagTAUBJPerp}{^{+0.038}}
\definemath{\hfagTAUBJPern}{_{-0.034}}
\definemath{\hfagSDGDGDval}{0.009}
\definemath{\hfagSDGDGDerr}{\pm0.037}
\definemath{\hfagDGSGSval}{+0.35}
\definemath{\hfagDGSGSerp}{^{+0.12}}
\definemath{\hfagDGSGSern}{_{-0.16}}
\definemath{\hfagDGSGSlow}{+0.01}
\definemath{\hfagDGSGSupp}{+0.59}
\definemath{\hfagTAUBSMEANval}{1.42}
\definemath{\hfagTAUBSMEANerp}{^{+0.06}}
\definemath{\hfagTAUBSMEANern}{_{-0.07}}
\definemath{\hfagRHODGSGSTAUBSMEAN}{+0.30}
\definemath{\hfagDGSval}{+0.25}
\definemath{\hfagDGSerp}{^{+0.09}}
\definemath{\hfagDGSern}{_{-0.11}}
\definemath{\hfagDGSlow}{+0.01}
\definemath{\hfagDGSupp}{+0.43}
\definemath{\hfagTAUBSLval}{1.21}
\definemath{\hfagTAUBSLerp}{^{+0.08}}
\definemath{\hfagTAUBSLern}{_{-0.09}}
\definemath{\hfagTAUBSHval}{1.72}
\definemath{\hfagTAUBSHerr}{\pm0.19}
\definemath{\hfagDGSGSCONBDval}{N/A}
\definemath{\hfagDGSGSCONBDerp}{^{N/A}}
\definemath{\hfagDGSGSCONBDern}{_{N/A}}
\definemath{\hfagDGSGSCONval}{+0.31}
\definemath{\hfagDGSGSCONerp}{^{+0.10}}
\definemath{\hfagDGSGSCONern}{_{-0.11}}
\definemath{\hfagDGSGSCONlow}{+0.01}
\definemath{\hfagDGSGSCONupp}{+0.57}
\definemath{\hfagTAUBSMEANCONXval}{1.396}
\definemath{\hfagTAUBSMEANCONXerp}{^{+0.044}}
\definemath{\hfagTAUBSMEANCONXern}{_{-0.046}}
\definemath{\hfagRHODGSGSTAUBSMEANCON}{-0.74}
\definemath{\hfagDGSCONval}{+0.22}
\definemath{\hfagDGSCONerr}{\pm0.08}
\definemath{\hfagDGSCONlow}{+0.05}
\definemath{\hfagDGSCONupp}{+0.32}
\definemath{\hfagTAUBSLCONval}{1.21}
\definemath{\hfagTAUBSLCONerr}{\pm0.08}
\definemath{\hfagTAUBSHCONval}{1.65}
\definemath{\hfagTAUBSHCONerp}{^{+0.07}}
\definemath{\hfagTAUBSHCONern}{_{-0.08}}
\definemath{\hfagDGSGSCONBDCONval}{+0.19}
\definemath{\hfagDGSGSCONBDCONerp}{^{+0.07}}
\definemath{\hfagDGSGSCONBDCONern}{_{-0.08}}
\definemath{\hfagFCWval}{0.507}
\definemath{\hfagFCWerr}{\pm0.007}
\definemath{\hfagFNWval}{0.493}
\definemath{\hfagFNWerr}{\pm0.007}
\definemath{\hfagFFWval}{1.030}
\definemath{\hfagFFWerr}{\pm0.029}
\definemath{\hfagFCNval}{0.513}
\definemath{\hfagFCNerr}{\pm0.013}
\definemath{\hfagFNNval}{0.487}
\definemath{\hfagFNNerr}{\pm0.013}
\definemath{\hfagFFNval}{1.053}
\definemath{\hfagFFNerr}{\pm0.054}
\definemath{\hfagFCval}{0.505}
\definemath{\hfagFCerr}{\pm0.008}
\definemath{\hfagFNval}{0.495}
\definemath{\hfagFNerr}{\pm0.008}
\definemath{\hfagFFval}{1.020}
\definemath{\hfagFFerr}{\pm0.034}
\definemath{\hfagFPRODval}{0.497}
\definemath{\hfagFPRODerr}{\pm0.021}
\definemath{\hfagFSUMval}{0.984}
\definemath{\hfagFSUMerr}{\pm0.031}
\definemath{\hfagFBSNOMIXval}{0.089}
\definemath{\hfagFBSNOMIXerr}{\pm0.021}
\definemath{\hfagFBBNOMIXval}{0.106}
\definemath{\hfagFBBNOMIXerr}{\pm0.018}
\definemath{\hfagFBDNOMIXval}{0.403}
\definemath{\hfagFBDNOMIXerr}{\pm0.011}
\definemath{\hfagCHIBARTEVval}{0.152}
\definemath{\hfagCHIBARTEVerr}{\pm0.013}
\definemath{\hfagCHIBARSFACTOR}{1.9}
\definemath{\hfagCHIBARval}{0.1283}
\definemath{\hfagCHIBARerr}{\pm0.0076}
\definemath{\hfagCHIDUval}{0.182}
\definemath{\hfagCHIDUerr}{\pm0.015}
\definemath{\hfagCHIDWUval}{0.188}
\definemath{\hfagCHIDWUerr}{\pm0.002}
\definemath{\hfagXDWUval}{0.775}
\definemath{\hfagXDWUerr}{\pm0.008}
\definemath{\hfagDMDWval}{0.508}
\definemath{\hfagDMDWsta}{\pm0.003}
\definemath{\hfagDMDWsys}{\pm0.003}
\definemath{\hfagDMDWerr}{\pm0.004}
\definemath{\hfagDMDWUval}{0.507}
\definemath{\hfagDMDWUerr}{\pm0.004}
\definemath{\hfagFBSMIXval}{0.119}
\definemath{\hfagFBSMIXerr}{\pm0.021}
\definemath{\hfagFBSval}{0.104}
\definemath{\hfagFBSerr}{\pm0.014}
\definemath{\hfagFBBval}{0.099}
\definemath{\hfagFBBerr}{\pm0.017}
\definemath{\hfagFBDval}{0.398}
\definemath{\hfagFBDerr}{\pm0.010}
\definemath{\hfagRHOFBBFBS}{-0.154}
\definemath{\hfagRHOFBDFBS}{-0.575}
\definemath{\hfagRHOFBDFBB}{-0.720}
\definemath{\hfagDMDHval}{0.496}
\definemath{\hfagDMDHsta}{\pm0.010}
\definemath{\hfagDMDHsys}{\pm0.009}
\definemath{\hfagDMDHerr}{\pm0.014}
\definemath{\hfagDMDBval}{0.508}
\definemath{\hfagDMDBsta}{\pm0.003}
\definemath{\hfagDMDBsys}{\pm0.003}
\definemath{\hfagDMDBerr}{\pm0.005}
\definemath{\hfagDMDTWODval}{0.509}
\definemath{\hfagDMDTWODsta}{\pm0.005}
\definemath{\hfagDMDTWODsys}{\pm0.003}
\definemath{\hfagDMDTWODerr}{\pm0.006}
\definemath{\hfagTAUBDTWODval}{1.527}
\definemath{\hfagTAUBDTWODsta}{\pm0.007}
\definemath{\hfagTAUBDTWODsys}{\pm0.007}
\definemath{\hfagTAUBDTWODerr}{\pm0.010}
\definemath{\hfagRHOstaDMDTAUBD}{-0.19}
\definemath{\hfagRHOsysDMDTAUBD}{-0.29}
\definemath{\hfagRHODMDTAUBD}{-0.23}
\definemath{\hfagQPval}{1.0015}
\definemath{\hfagQPerr}{\pm0.0039}
\definemath{\hfagASLval}{-0.0030}
\definemath{\hfagASLerr}{\pm0.0078}
\definemath{\hfagREBval}{-0.0007}
\definemath{\hfagREBerr}{\pm0.0020}
\definemath{\hfagDMSWLIMval}{16.6}
\definemath{\hfagDMSWSENSval}{20.0}
\definemath{\hfagDMSXLIMval}{16.6}
\definemath{\hfagDMSXSENSval}{20.1}
\definemath{\hfagDMSDLIMval}{-0.0}
\definemath{\hfagDMSDSENSval}{0.1}
\definemath{\hfagDMSWUPPval}{21.7}
\definemath{\hfagXSWLIMval}{22.4}
\definemath{\hfagCHISWLIMval}{0.49904}

\newcommand{\unit}[1]{~\ensuremath{\rm #1}\xspace}
\renewcommand{\ps}{\unit{ps}}
\newcommand{\invps}{\unit{ps^{-1}}}
\newcommand{\TeV}{\unit{TeV}}
\newcommand{\MeVcc}{\unit{MeV/\mbox{$c$}^2}}
\newcommand{\MeV}{\unit{MeV}}

\definemath{\hfagCHIBARLEP}{\hfagCHIBARLEPval\hfagCHIBARLEPerr}
\definemath{\hfagTAUBD}{\hfagTAUBDval\hfagTAUBDerr\ps}
\definemath{\hfagTAUBDnounit}{\hfagTAUBDval\hfagTAUBDerr}
\definemath{\hfagTAUBU}{\hfagTAUBUval\hfagTAUBUerr\ps}
\definemath{\hfagTAUBUnounit}{\hfagTAUBUval\hfagTAUBUerr}
\definemath{\hfagRTAUBU}{\hfagRTAUBUval\hfagRTAUBUerr}
\definemath{\hfagTAUBS}{\hfagTAUBSval\hfagTAUBSerr\ps}
\definemath{\hfagTAUBSnounit}{\hfagTAUBSval\hfagTAUBSerr}
\definemath{\hfagRTAUBS}{\hfagRTAUBSval\hfagRTAUBSerr}
\definemath{\hfagTAULB}{\hfagTAULBval\hfagTAULBerr\ps}
\definemath{\hfagTAULBnounit}{\hfagTAULBval\hfagTAULBerr}
\definemath{\hfagTAUXBerr}{\hfagTAUXBerp\hfagTAUXBern}
\definemath{\hfagTAUXB}{\hfagTAUXBval\hfagTAUXBerr\ps}
\definemath{\hfagTAUXBnounit}{\hfagTAUXBval\hfagTAUXBerr}
\definemath{\hfagTAUBB}{\hfagTAUBBval\hfagTAUBBerr\ps}
\definemath{\hfagTAUBBnounit}{\hfagTAUBBval\hfagTAUBBerr}
\definemath{\hfagRTAUBB}{\hfagRTAUBBval\hfagRTAUBBerr}
\definemath{\hfagTAUB}{\hfagTAUBval\hfagTAUBerr\ps}
\definemath{\hfagTAUBnounit}{\hfagTAUBval\hfagTAUBerr}
\definemath{\hfagTAUBC}{\hfagTAUBCval\hfagTAUBCerr\ps}
\definemath{\hfagTAUBCnounit}{\hfagTAUBCval\hfagTAUBCerr}
\definemath{\hfagTAUBSSL}{\hfagTAUBSSLval\hfagTAUBSSLerr\ps}
\definemath{\hfagTAUBSSLnounit}{\hfagTAUBSSLval\hfagTAUBSSLerr}
\definemath{\hfagTAUBSSLX}{\hfagTAUBSSLXval\hfagTAUBSSLXerr\ps}
\definemath{\hfagTAUBSSLXnounit}{\hfagTAUBSSLXval\hfagTAUBSSLXerr}
\definemath{\hfagTAUBSMEANCONerr}{\hfagTAUBSMEANCONerp\hfagTAUBSMEANCONern}
\definemath{\hfagTAUBSMEANCON}{\hfagTAUBSMEANCONval\hfagTAUBSMEANCONerr\ps}
\definemath{\hfagTAUBSMEANCONnounit}{\hfagTAUBSMEANCONval\hfagTAUBSMEANCONerr}
\definemath{\hfagTAUBSJF}{\hfagTAUBSJFval\hfagTAUBSJFerr\ps}
\definemath{\hfagTAUBSJFnounit}{\hfagTAUBSJFval\hfagTAUBSJFerr}
\definemath{\hfagRTAUBSSL}{\hfagRTAUBSSLval\hfagRTAUBSSLerr}
\definemath{\hfagRTAUBSMEANCON}{\hfagRTAUBSMEANCONval\hfagRTAUBSMEANCONerr}
\definemath{\hfagRTAULB}{\hfagRTAULBval\hfagRTAULBerr}
\definemath{\hfagTAUBVTX}{\hfagTAUBVTXval\hfagTAUBVTXerr\ps}
\definemath{\hfagTAUBVTXnounit}{\hfagTAUBVTXval\hfagTAUBVTXerr}
\definemath{\hfagTAUBLEP}{\hfagTAUBLEPval\hfagTAUBLEPerr\ps}
\definemath{\hfagTAUBLEPnounit}{\hfagTAUBLEPval\hfagTAUBLEPerr}
\definemath{\hfagTAUBJPerr}{\hfagTAUBJPerp\hfagTAUBJPern}
\definemath{\hfagTAUBJP}{\hfagTAUBJPval\hfagTAUBJPerr\ps}
\definemath{\hfagTAUBJPnounit}{\hfagTAUBJPval\hfagTAUBJPerr}
\definemath{\hfagSDGDGD}{\hfagSDGDGDval\hfagSDGDGDerr}
\definemath{\hfagDGSGSerr}{\hfagDGSGSerp\hfagDGSGSern}
\definemath{\hfagDGSGS}{\hfagDGSGSval\hfagDGSGSerr}
\definemath{\hfagTAUBSMEANerr}{\hfagTAUBSMEANerp\hfagTAUBSMEANern}
\definemath{\hfagTAUBSMEAN}{\hfagTAUBSMEANval\hfagTAUBSMEANerr\ps}
\definemath{\hfagTAUBSMEANnounit}{\hfagTAUBSMEANval\hfagTAUBSMEANerr}
\definemath{\hfagDGSerr}{\hfagDGSerp\hfagDGSern}
\definemath{\hfagDGS}{\hfagDGSval\hfagDGSerr\invps}
\definemath{\hfagDGSnounit}{\hfagDGSval\hfagDGSerr}
\definemath{\hfagTAUBSLerr}{\hfagTAUBSLerp\hfagTAUBSLern}
\definemath{\hfagTAUBSL}{\hfagTAUBSLval\hfagTAUBSLerr\ps}
\definemath{\hfagTAUBSLnounit}{\hfagTAUBSLval\hfagTAUBSLerr}
\definemath{\hfagTAUBSH}{\hfagTAUBSHval\hfagTAUBSHerr\ps}
\definemath{\hfagTAUBSHnounit}{\hfagTAUBSHval\hfagTAUBSHerr}
\definemath{\hfagDGSGSCONBDerr}{\hfagDGSGSCONBDerp\hfagDGSGSCONBDern}
\definemath{\hfagDGSGSCONBD}{\hfagDGSGSCONBDval\hfagDGSGSCONBDerr}
\definemath{\hfagDGSGSCONerr}{\hfagDGSGSCONerp\hfagDGSGSCONern}
\definemath{\hfagDGSGSCON}{\hfagDGSGSCONval\hfagDGSGSCONerr}
\definemath{\hfagTAUBSMEANCONXerr}{\hfagTAUBSMEANCONXerp\hfagTAUBSMEANCONXern}
\definemath{\hfagTAUBSMEANCONX}{\hfagTAUBSMEANCONXval\hfagTAUBSMEANCONXerr\ps}
\definemath{\hfagTAUBSMEANCONXnounit}{\hfagTAUBSMEANCONXval\hfagTAUBSMEANCONXerr}
\definemath{\hfagDGSCON}{\hfagDGSCONval\hfagDGSCONerr\invps}
\definemath{\hfagDGSCONnounit}{\hfagDGSCONval\hfagDGSCONerr}
\definemath{\hfagTAUBSLCON}{\hfagTAUBSLCONval\hfagTAUBSLCONerr\ps}
\definemath{\hfagTAUBSLCONnounit}{\hfagTAUBSLCONval\hfagTAUBSLCONerr}
\definemath{\hfagTAUBSHCONerr}{\hfagTAUBSHCONerp\hfagTAUBSHCONern}
\definemath{\hfagTAUBSHCON}{\hfagTAUBSHCONval\hfagTAUBSHCONerr\ps}
\definemath{\hfagTAUBSHCONnounit}{\hfagTAUBSHCONval\hfagTAUBSHCONerr}
\definemath{\hfagDGSGSCONBDCONerr}{\hfagDGSGSCONBDCONerp\hfagDGSGSCONBDCONern}
\definemath{\hfagDGSGSCONBDCON}{\hfagDGSGSCONBDCONval\hfagDGSGSCONBDCONerr}
\definemath{\hfagFCW}{\hfagFCWval\hfagFCWerr}
\definemath{\hfagFNW}{\hfagFNWval\hfagFNWerr}
\definemath{\hfagFFW}{\hfagFFWval\hfagFFWerr}
\definemath{\hfagFCN}{\hfagFCNval\hfagFCNerr}
\definemath{\hfagFNN}{\hfagFNNval\hfagFNNerr}
\definemath{\hfagFFN}{\hfagFFNval\hfagFFNerr}
\definemath{\hfagFC}{\hfagFCval\hfagFCerr}
\definemath{\hfagFN}{\hfagFNval\hfagFNerr}
\definemath{\hfagFF}{\hfagFFval\hfagFFerr}
\definemath{\hfagFPROD}{\hfagFPRODval\hfagFPRODerr}
\definemath{\hfagFSUM}{\hfagFSUMval\hfagFSUMerr}
\definemath{\hfagFBSNOMIX}{\hfagFBSNOMIXval\hfagFBSNOMIXerr}
\definemath{\hfagFBBNOMIX}{\hfagFBBNOMIXval\hfagFBBNOMIXerr}
\definemath{\hfagFBDNOMIX}{\hfagFBDNOMIXval\hfagFBDNOMIXerr}
\definemath{\hfagCHIBARTEV}{\hfagCHIBARTEVval\hfagCHIBARTEVerr}
\definemath{\hfagCHIBAR}{\hfagCHIBARval\hfagCHIBARerr}
\definemath{\hfagCHIDU}{\hfagCHIDUval\hfagCHIDUerr}
\definemath{\hfagCHIDWU}{\hfagCHIDWUval\hfagCHIDWUerr}
\definemath{\hfagXDWU}{\hfagXDWUval\hfagXDWUerr}
\definemath{\hfagDMDW}{\hfagDMDWval\hfagDMDWerr\invps}
\definemath{\hfagDMDWnounit}{\hfagDMDWval\hfagDMDWerr}
\definemath{\hfagDMDWfull}{\hfagDMDWval\hfagDMDWsta\hfagDMDWsys\invps}
\definemath{\hfagDMDWnounitfull}{\hfagDMDWval\hfagDMDWsta\hfagDMDWsys}
\definemath{\hfagDMDWU}{\hfagDMDWUval\hfagDMDWUerr\invps}
\definemath{\hfagDMDWUnounit}{\hfagDMDWUval\hfagDMDWUerr}
\definemath{\hfagFBSMIX}{\hfagFBSMIXval\hfagFBSMIXerr}
\definemath{\hfagFBS}{\hfagFBSval\hfagFBSerr}
\definemath{\hfagFBB}{\hfagFBBval\hfagFBBerr}
\definemath{\hfagFBD}{\hfagFBDval\hfagFBDerr}
\definemath{\hfagDMDH}{\hfagDMDHval\hfagDMDHerr\invps}
\definemath{\hfagDMDHnounit}{\hfagDMDHval\hfagDMDHerr}
\definemath{\hfagDMDHfull}{\hfagDMDHval\hfagDMDHsta\hfagDMDHsys\invps}
\definemath{\hfagDMDHnounitfull}{\hfagDMDHval\hfagDMDHsta\hfagDMDHsys}
\definemath{\hfagDMDB}{\hfagDMDBval\hfagDMDBerr\invps}
\definemath{\hfagDMDBnounit}{\hfagDMDBval\hfagDMDBerr}
\definemath{\hfagDMDBfull}{\hfagDMDBval\hfagDMDBsta\hfagDMDBsys\invps}
\definemath{\hfagDMDBnounitfull}{\hfagDMDBval\hfagDMDBsta\hfagDMDBsys}
\definemath{\hfagDMDTWOD}{\hfagDMDTWODval\hfagDMDTWODerr\invps}
\definemath{\hfagDMDTWODnounit}{\hfagDMDTWODval\hfagDMDTWODerr}
\definemath{\hfagDMDTWODfull}{\hfagDMDTWODval\hfagDMDTWODsta\hfagDMDTWODsys\invps}
\definemath{\hfagDMDTWODnounitfull}{\hfagDMDTWODval\hfagDMDTWODsta\hfagDMDTWODsys}
\definemath{\hfagTAUBDTWOD}{\hfagTAUBDTWODval\hfagTAUBDTWODerr\ps}
\definemath{\hfagTAUBDTWODnounit}{\hfagTAUBDTWODval\hfagTAUBDTWODerr}
\definemath{\hfagTAUBDTWODfull}{\hfagTAUBDTWODval\hfagTAUBDTWODsta\hfagTAUBDTWODsys\ps}
\definemath{\hfagTAUBDTWODnounitfull}{\hfagTAUBDTWODval\hfagTAUBDTWODsta\hfagTAUBDTWODsys}
\definemath{\hfagQP}{\hfagQPval\hfagQPerr}
\definemath{\hfagASL}{\hfagASLval\hfagASLerr}
\definemath{\hfagREB}{\hfagREBval\hfagREBerr}
\definemath{\hfagDMSWLIM}{\hfagDMSWLIMval\invps}
\definemath{\hfagDMSWSENS}{\hfagDMSWSENSval\invps}
\definemath{\hfagDMSXLIM}{\hfagDMSXLIMval\invps}
\definemath{\hfagDMSXSENS}{\hfagDMSXSENSval\invps}
\definemath{\hfagDMSDLIM}{\hfagDMSDLIMval\invps}
\definemath{\hfagDMSDSENS}{\hfagDMSDSENSval\invps}
\definemath{\hfagDMSWUPP}{\hfagDMSWUPPval\invps}
\definemath{\hfagXSWLIM}{\hfagXSWLIMval}
\definemath{\hfagCHISWLIM}{\hfagCHISWLIMval}


\mysection{\b-hadron production fractions, lifetimes and mixing parameters}
\labs{life_mix}


Quantities such as \b-hadron production fractions, \b-hadron lifetimes, 
and neutral \B-meson oscillation frequencies have been measured
for many years at high-energy colliders, namely at LEP and SLC 
(\ee colliders at $\sqrt{s}=m_{\particle{Z}}$) as well as at the 
first version of the Tevatron
(\particle{p\bar{p}} collider at $\sqrt{s}=1.8\TeV$). More recently, 
precise measurements of the \Bd and \Bu lifetimes, as well as of the 
\Bd oscillation frequency, have also been performed at the 
asymmetric \B factories, KEKB and PEPII
(\ee colliders at $\sqrt{s}=m_{\Ups}$).
In most cases, these basic quantities, although interesting by themselves,
can now be seen as necessary ingredients for the more complicated and 
refined analyses being currently performed at the asymmetric \B factories
and at the upgraded Tevatron ($\sqrt{s}=1.96\TeV$),
in particular the time-dependent \CP asymmetry measurements.
It is therefore important that the best experimental
values of these quantities continue to be kept up-to-date and improved. 

In several cases, the averages presented in this chapter are indeed
needed and used as input for the results given in the subsequent chapters. 
However, within this chapter, some averages need the knowledge of other 
averages in a circular way. This ``coupling'', which appears through the 
\b-hadron fractions whenever inclusive or semi-exclusive measurements 
have to be considered, has been reduced significantly in the last years 
with increasingly precise exclusive measurements becoming available. 
To cope with this circularity,
a rather involved averaging procedure had been developed, in the framework 
of the former LEP Heavy Flavour Steering Group. This is still in use now
(details can be found in~\cite{LEPHFS}), 
although simplifications can be envisaged in the future when even more 
precise exclusive measurements become available. 

\mysubsection{\b-hadron production fractions}
\labs{fractions}
 
We consider here the relative fractions of the different \b-hadron 
species found in an unbiased sample of weakly-decaying \b hadrons 
produced under some specific conditions. The knowledge of these fractions
is useful to characterize the signal composition in inclusive \b-hadron 
analyses, or to predict the background composition in exclusive analyses.
Many analyses in \B physics need these fractions as input. We distinguish 
here the following two conditions: \Ups decays and 
high-energy collisions. 

\mysubsubsection{\b-hadron production fractions in \Ups decays}
\labs{bfraction}

Only pairs of the two lightest (charged and neutral) \B mesons 
can be produced in \Ups decays, 
and it is enough to determine the following branching 
fractions:
\begin{eqnarray}
f^{+-} & = & \Gamma(\Ups \to \particle{B^+B^-})/
             \Gamma_{\rm tot}(\Ups)  \,, \\
f^{00} & = & \Gamma(\Ups \to \particle{B^0\bar{B}^0})/
             \Gamma_{\rm tot}(\Ups) \,.
\end{eqnarray}
In practice, most analyses measure their ratio
\begin{equation}
R^{+-/00} = f^{+-}/f^{00} = \Gamma(\Ups \to \particle{B^+B^-})/
             \Gamma(\Ups \to \particle{B^0\bar{B}^0}) \,,
\end{equation}
which is easier to access experimentally.
Since an inclusive (but separate) reconstruction of 
\Bu and \Bd is difficult, specific exclusive decay modes, 
${\Bu} \to x^+$ and ${\Bd} \to x^0$, are usually considered to perform 
a measurement of $R^{+-/00}$, whenever they can be related by 
isospin symmetry (for example \particle{\Bu \to J/\psi K^+} and 
\particle{\Bd \to J/\psi K^0}).
Under the assumption that $\Gamma(\Bu \to x^+) = \Gamma(\Bd \to x^0)$, 
\ie\ that isospin invariance holds in these \B decays,
the ratio of the number of reconstructed
$\Bu \to x^+$ and $\Bd \to x^0$ mesons is proportional to
\begin{equation}
\frac{f^{+-}\, \BR{\Bu\to x^+}}{f^{00}\, \BR{\Bd\to x^0}}
= \frac{f^{+-}\, \Gamma({\Bu}\to x^+)\, \tau(\Bu)}%
{f^{00}\, \Gamma({\Bd}\to x^0)\,\tau(\Bd)}
= \frac{f^{+-}}{f^{00}} \, \frac{\tau(\Bu)}{\tau(\Bd)}  \,, 
\end{equation} 
where $\tau(\Bu)$ and $\tau(\Bd)$ are the \Bu and \Bd 
lifetimes respectively.
Hence the primary quantity measured in these analyses 
is $R^{+-/00} \, \tau(\Bu)/\tau(\Bd)$, 
and the extraction of $R^{+-/00}$ with this method therefore 
requires the knowledge of the $\tau(\Bu)/\tau(\Bd)$ lifetime ratio. 

\begin{table}
\caption{Published measurements of the $\Bu/\Bd$ production ratio
in \Ups decays, together with their average (see text).
Systematic uncertainties due to the imperfect knowledge of 
$\tau(\Bu)/\tau(\Bd)$ are included.}
\labt{R_data}
\begin{center}
\begin{tabular}{@{}l@{}c@{\,}cll@{}}
\hline
Experiment & Ref. & Decay modes & Published value of & Assumed value \\
and year & & or method & $R^{+-/00}=f^{+-}/f^{00}$ & of $\tau(\Bu)/\tau(\Bd)$ \\
\hline
CLEO,   2001 & \cite{CLEO_R2001}  & \particle{J/\psi K^{(*)}} 
             & $1.04 \pm0.07 \pm0.04$ & $1.066 \pm0.024$ \\
\babar, 2002 & \cite{BABAR_R2002} & \particle{(c\bar{c})K^{(*)}}
             & $1.10 \pm0.06 \pm0.05$ & $1.062 \pm0.029$\\ 
CLEO,   2002 & \cite{CLEO_R2002}  & \particle{D^*\ell\nu}
             & $1.058 \pm0.084 \pm0.136$ & $1.074 \pm0.028$\\
\belle, 2003 & \cite{BELLE_dmd_dilepton} & dilepton events 
             & $1.01 \pm0.03 \pm0.09$ & $1.083 \pm0.017$\\
\babar, 2004 & \cite{BABAR_R2004} & \particle{J/\psi K}
             & $1.006 \pm0.036 \pm0.031$ & $1.083 \pm0.017$ \\
\hline
Average      & & & \hfagFF~(tot) & \hfagRTAUBU \\
\hline
\end{tabular}
\end{center}
\end{table}

The published measurements of $R^{+-/00}$ are listed 
in \Table{R_data} together with the corresponding assumed values of 
$\tau(\Bu)/\tau(\Bd)$.
All measurements are based on the above-mentioned method, 
except the one from \belle, which is a by-product of the 
\Bd mixing frequency analysis using dilepton events
(but note that it also assumes isospin invariance, 
namely $\Gamma(\Bu \to \ell^+{\rm X}) = \Gamma(\Bd \to \ell^+{\rm X})$).
The latter is therefore treated in a slightly different 
manner in the following procedure used to combine 
these measurements:
\begin{itemize} 
\item each published value of $R^{+-/00}$ from CLEO and \babar
      is first converted back to the original measurement of 
      $R^{+-/00} \, \tau(\Bu)/\tau(\Bd)$, using the value of the 
      lifetime ratio assumed in the corresponding analysis;
\item a simple weighted average of these original
      measurements of $R^{+-/00} \, \tau(\Bu)/\tau(\Bd)$ from 
      CLEO and \babar (which do not depend on the assumed value 
      of the lifetime ratio) is then computed, assuming no 
      statistical or systematic correlations between them;


\item the weighted average of $R^{+-/00} \, \tau(\Bu)/\tau(\Bd)$ 
      is converted into a value of $R^{+-/00}$, using the latest 
      average of the lifetime ratios, $\tau(\Bu)/\tau(\Bd)=\hfagRTAUBU$ 
      (see \Sec{lifetime_ratio});
\item the \belle measurement of $R^{+-/00}$ is adjusted to the 
      current values of $\tau(\Bd)=\hfagTAUBD$ and 
      $\tau(\Bu)/\tau(\Bd)=\hfagRTAUBU$ (see \Sec{lifetime_ratio}),
      using the quoted systematic uncertainties due to these parameters;
\item the combined value of $R^{+-/00}$ from CLEO and \babar is averaged 
      with the adjusted value of $R^{+-/00}$ from \belle, assuming a 100\% 
      correlation of the systematic uncertainty due to the limited 
      knowledge on $\tau(\Bu)/\tau(\Bd)$; no other correlation is considered. 
\end{itemize} 
The resulting global average, 
\begin{equation}
R^{+-/00} = \frac{f^{+-}}{f^{00}} =  \hfagFF \,,
\labe{Rplusminus}
\end{equation}
is consistent with an equal production of charged and neutral \B mesons.

On the other hand, the \babar collaboration has 
recently performed a direct measurement of the $f^{00}$ fraction 
using a novel method, which does not rely on isospin symmetry nor requires 
the knowledge of $\tau(\Bu)/\tau(\Bd)$. Its analysis, 
based on a comparison between the number of events where a single 
$B^0 \to D^{*-} \ell^+ \nu$ decay could be reconstructed and the number 
of events where two such decays could be reconstructed, yields~\cite{BABAR_f00}
\begin{equation}
f^{00}= 0.487 \pm 0.010\,\mbox{(stat)} \pm 0.008\,\mbox{(syst)} \,.
\labe{fzerozero}
\end{equation}

The two results of \Eqss{Rplusminus}{fzerozero} are of very different natures 
and completely independent of each other. 
Their product is equal to $f^{+-} = \hfagFPROD$, 
while another combination of them gives $f^{+-} + f^{00}= \hfagFSUM$, 
compatible with unity.
Assuming\footnote{The first non-$\B\bar{B}$
decay mode of the \Ups has now been observed with a branching ratio
of the order of $10^{-4}$~\cite{BELLE_Ups}, corresponding to a partial
width several times larger than that in the \ee channel.
However, this can still be
neglected and the assumption $f^{+-}+f^{00}=1$ remains valid
in the present context of the determination of $f^{+-}$ and $f^{00}$.}
 $f^{+-}+f^{00}= 1$, also consistent with 
CLEO's observation that the fraction of \Ups decays 
to \BB pairs is larger than 0.96 at \CL{95}~\cite{CLEO_frac_limit},
the results of \Eqss{Rplusminus}{fzerozero}
can be averaged (first converting \Eq{Rplusminus} 
into a value of $f^{00}=1/(R^{+-/00}+1)$) 
to yield the following more precise estimates:
\begin{equation}
f^{00} = \hfagFNW  \,,~~~ f^{+-} = 1 -f^{00} =  \hfagFCW \,,~~~
\frac{f^{+-}}{f^{00}} =  \hfagFFW \,.
\end{equation}

\mysubsubsection{\b-hadron production fractions at high energy}
\labs{fractions_high_energy}

At high energy, all species of weakly-decaying \b hadrons 
can be produced, either directly or in strong and electromagnetic 
decays of excited \b hadrons.
We assume here that the fractions of these different species 
are the same in unbiased samples of high-$p_{\rm T}$ \b jets 
originating from \particle{Z^0} decays or from \particle{p\bar{p}} 
collisions at the Tevatron.
This hypothesis is plausible considering that, in both cases, 
the last step of the jet hadronization is a non-perturbative
QCD process occurring at the scale of $\Lambda_{\rm QCD}$.
On the other hand, there is no strong argument to claim that these 
fractions should be strictly equal, so this assumption 
should be checked experimentally.
Although the available data is not sufficient at 
this time to perform a significant check, 
it is expected that the new data from 
Tevatron Run II may improve this situation and 
allow one to confirm or disprove this assumption with reasonable 
confidence. Meanwhile, the attitude adopted here is that these 
fractions are assumed to be equal at all high-energy colliders
until demonstrated otherwise by experiment.\footnote{It is not unlikely
that the \b-hadron fractions in low-$p_{\rm T}$ jets 
at a hadronic machine be different; in particular, beam-remnant effects may
enhance the \b-baryon production.}

Contrary to what happens in the charm sector where the fractions of \particle{D^+} 
and \particle{D^0} are different, the relative amount of \Bu and \Bd is not affected by the 
electromagnetic decays of excited ${\Bu}^*$ and ${\Bd}^*$ states and strong decays of excited
${\Bu}^{**}$ and ${\Bd}^{**}$ states. Decays of the type \particle{{\Bs}^{**} \to B^{(*)}K}
also contribute to the \Bu and \Bd rates, but with the same magnitude if mass effects
can be neglected.
We therefore assume equal production of \Bu and \Bd. We also  
neglect the production of weakly-decaying states
made of several heavy quarks (like \Bc and other heavy baryons) 
which is known to be very small. Hence, for the purpose of determining 
the \b-hadron fractions, we use the constraints
\begin{equation}
\fBu = \fBd ~~~~\mbox{and}~~~ \fBu + \fBd + \fBs + \fbb = 1 \,,
\labe{constraints}
\end{equation}
where \fBu, \fBd, \fBs and \fbb
are the unbiased fractions of \Bu, \Bd, \Bs and \b-baryons, respectively.

The LEP experiments have measured
$\fBs \times \BR{\Bs\to\particle{D_s^-} \ell^+ \nu_\ell \mbox{$X$}}$~\cite{LEP_fs}, 
$\BR{\b\to\Lb} \times \BR{\Lb\to\Lc\ell^-\bar{\nu}_\ell \mbox{$X$}}$~\cite{DELPHI_fla,ALEPH_fla}
and $\BR{\b\to\Xib^-} \times \BR{\Xi_b^- \to \Xi^-\ell^-\overline\nu_\ell 
\mbox{$X$}}$~\cite{DELPHI_fxi,ALEPH_fxi}
from partially reconstructed final states 
including a lepton, \fbb
from protons identified in \b events~\cite{ALEPH-fbar}, and the 
production rate of charged \b hadrons~\cite{DELPHI-fch}. 
The various \b-hadron fractions 
have also been measured at CDF using electron-charm final states~\cite{CDF_f_ec}
and double semileptonic decays with \particle{\phi\ell} and 
\particle{K^*\ell} final states~\cite{CDF_f_phil_Kstl}.
All these published results have been combined 
following the procedure and assumptions described in~\cite{LEPHFS}
to yield $\fBu=\fBd=\hfagFBDNOMIX$, 
$\fBs=\hfagFBSNOMIX$ and $\fbb=\hfagFBBNOMIX$
under the constraints of \Eq{constraints}.
For this combination, other external inputs are used, \eg\ the branching 
ratios of \B mesons to final states with a \particle{D}, \particle{D^*} or 
\particle{D^{**}} in semileptonic decays, which are needed to evaluate the 
fraction of semileptonic \Bs decays with a \particle{D_s^-} in the final state.


Time-integrated mixing analyses performed with lepton pairs 
from \particle{b\bar{b}} 
events produced at high-energy colliders measure the quantity 
\begin{equation}
\chibar = f'_{\particle{d}} \,\chid + f'_{\particle{s}} \,\chis \,,
\end{equation}
where $f'_{\particle{d}}$ and $f'_{\particle{s}}$ are 
the fractions of \Bd and \Bs hadrons 
in a sample of semileptonic \b-hadron decays, and where \chid and \chis 
are the \Bd and \Bs time-integrated mixing probabilities.
Assuming that all \b hadrons have the same semileptonic decay width implies 
$f'_i = f_i R_i$, where $R_i = \tau_i/\tau_{\particle{b}}$ is the ratio of the lifetime 
$\tau_i$ of species $i$ to the average \b-hadron lifetime 
$\tau_{\particle{b}} = \sum_i f_i \tau_i$.
Hence measurements of the mixing probabilities
\chibar, \chid and \chis can be used to improve our 
knowledge of \fBu, \fBd, \fBs and \fbb.
In practice, the above relations yield another determination of 
\fBs obtained from \fbb and mixing information, 
\begin{equation}
\fBs = \frac{1}{R_{\particle{s}}}
\frac{(1+r)\overline{\chi}-(1-\fbb R_{\rm baryon}) \chid}{(1+r)\chis - \chid} \,,
\labe{fBs-mixing}
\end{equation}
where $r=R_{\particle{u}}/R_{\particle{d}} = \tau(\Bu)/\tau(\Bd)$.

\labs{chibar}
The published measurements of \chibar performed by the LEP
experiments have been combined by the LEP Electroweak Working Group to yield 
$\chibar = \hfagCHIBARLEP$~\cite{LEPEWWG}. This can be compared with a 
recent measurement from CDF, $\chibar = \hfagCHIBARTEV$~\cite{CDF-chibar}, 
obtained from an analysis of the Run I data. The two estimates deviate
from each other by $\hfagCHIBARSFACTOR\,\sigma$,
and could be an indication that the production fractions of \b hadrons 
at the \particle{Z} peak or at the Tevatron are not the same. 
Although this discrepancy 
is not very significant it should be carefully monitored in the future. 
We choose to combine these two results in a simple weighted average,
assuming no correlations, and, following the PDG prescription, we 
multiply the combined uncertainty by \hfagCHIBARSFACTOR to account 
for the discrepancy. Our world average is then
\begin{equation}
\chibar = \hfagCHIBAR \,.
\end{equation}

\begin{table}
\caption{Fractions of the different \b-hadron species in an unbiased sample of 
weakly-decaying \b hadrons produced at high energy, obtained from both direct
and mixing measurements.}
\labt{fractions}
\begin{center}
\begin{tabular}{crcc}
\hline
\b-hadron & \multicolumn{1}{c}{Fraction} & \multicolumn{2}{l}{Correlation coefficients} \\
species   &          & with $\fBd=\fBu$ & and \fBs\\
\hline
\Bd, \Bu   & $\fBd=\fBu = \hfagFBD$  & & \\
\Bs        & $\fBs = \hfagFBS$       & \hfagRHOFBDFBS & \\
\b baryons & $\fbb = \hfagFBB$       & \hfagRHOFBDFBB & \hfagRHOFBBFBS \\
\hline
\end{tabular}
\end{center}
\end{table}

Introducing the latter result in \Eq{fBs-mixing}, together with our world average 
$\chid = \hfagCHIDWU$ (see \Eq{chid} of \Sec{dmd}), the assumption $\chis= 1/2$ 
(justified by the large value of \dms, see \Eq{chis} in \Sec{dms}), the 
best knowledge of the lifetimes (see \Sec{lifetimes}) and the estimate of \fbb given above, 
yields $\fBs = \hfagFBSMIX$, an estimate dominated by the mixing information. 
Taking into account all known correlations (including the one introduced by \fbb), 
this result is then combined with the set of fractions obtained from direct measurements 
(given above), to yield the 
improved estimates of \Table{fractions}, 
still under the constraints of \Eq{constraints}. 
As can be seen, our knowledge on the mixing parameters 
substantially reduces the uncertainty on \fBs, despite the rather strong 
deweighting introduced in the computation of the world average of \chibar.
It should be noted that the results 
are correlated, as indicated in \Table{fractions}.


%
%

\mysubsection{\b-hadron lifetimes}
\labs{lifetimes}

In the spectator model the decay of \b-flavored hadrons $H_b$ is
governed entirely by the flavor changing \particle{b\to Wq} transition
($\particle{q}=\particle{c,u}$).  For this very reason, lifetimes of all
\b-flavored hadrons are the same in the spectator approximation
regardless of the (spectator) quark content of the $H_b$.  In the early
1990's experiments became sophisticated enough to start seeing the
differences of the lifetimes among various $H_b$ species.  The first
theoretical calculations of the spectator quark effects on $H_b$
lifetime emerged only few years earlier.

Currently, most of such calculations are performed in the framework of
the Heavy Quark Expansion, HQE.  In the HQE, under certain assumptions
(most important of which is that of quark-hadron duality), the decay
rate of an $H_b$ to an inclusive final state $f$ is expressed as the sum
of a series of expectation values of operators of increasing dimension,
multiplied by the correspondingly higher powers of $\Lambda_{\rm
QCD}/m_b$:
\begin{equation}
\Gamma_{H_b\to f} = |CKM|^2\sum_n c_n^{(f)}
\Bigl(\frac{\Lambda_{\rm QCD}}{m_b}\Bigr)^n\langle H_b|O_n|H_b\rangle,
\labe{hqe}
\end{equation}
where $|CKM|^2$ is the relevant combination of the CKM matrix elements.
Coefficients $c_n^{(f)}$ of this expansion, known as Operator Product
Expansion~\cite{OPE}, can be calculated perturbatively.  Hence, the HQE
predicts $\Gamma_{H_b\to f}$ in the form of an expansion in both
$\Lambda_{\rm QCD}/m_{\b}$ and $\alpha_s(m_{\b})$.  The precision of
current experiments makes it mandatory to go to the next-to-leading
order in QCD, {\em i.e.}\ to include correction of the order of
$\alpha_s(m_{\b})$ to the $c_n^{(f)}$'s.  All non-perturbative physics
is shifted into the expectation values $\langle H_b|O_n|H_b\rangle$ of
operators $O_n$.  These can be calculated using lattice QCD or QCD sum
rules, or can be related to other observables via the
HQE~\cite{Bigi_1995}.  One may reasonably expect that powers of
$\Lambda_{\rm QCD}/m_{\b}$ provide enough suppression that only the
first few terms of the sum in \Eq{hqe} matter.

Theoretical predictions are usually made for the ratios of the lifetimes
(with $\tau(\Bd)$ chosen as the common denominator) rather than for the
individual lifetimes, for this allows several uncertainties to cancel.
The precision of the current HQE calculations (see
\Refs{nlo_lifetimes,tarantino,Gabbiani_et_al} for the latest updates)
is in some instances already surpassed by the measurements,
\eg\ in the case of $\tau(\Bu)/\tau(\Bd)$.  Also, HQE calculations are
not assumption-free.  More accurate predictions are a matter of progress
in the evaluation of the non-perturbative hadronic matrix elements and
verifying the assumptions that the calculations are based upon.
However, the HQE, even in its present shape, draws a number of important
conclusions, which are in agreement with experimental observations:
\begin{itemize}
\item The heavier the mass of the heavy quark the smaller is the
  variation in the lifetimes among different hadrons containing this
  quark, which is to say that as $m_{\b}\to\infty$ we retrieve the
  spectator picture in which the lifetimes of all $H_b$'s are the same.
   This is well illustrated by the fact that lifetimes are rather
   similar in the \b sector, while they differ by large factors
   in the \particle{c} sector ($m_{\particle{c}}<m_{\b}$).
\item The non-perturbative corrections arise only at the order of
  $\Lambda_{\rm QCD}^2/m_{\b}^2$, which translates into 
  differences among $H_b$ lifetimes of only a few percent.
\item It is only the difference between meson and baryon lifetimes that
  appears at the $\Lambda_{\rm QCD}^2/m_{\b}^2$ level.  The splitting of the
  meson lifetimes occurs at the $\Lambda_{\rm QCD}^3/m_{\b}^3$ level, yet it is
  enhanced by a phase space factor $16\pi^2$ with respect to the leading
  free \b decay.
\end{itemize}

To ensure that certain sources of systematic uncertainty cancel, 
lifetime analyses are sometimes designed to measure a 
ratio of lifetimes.  However, because of the differences in decay
topologies, abundance (or lack thereof) of decays of a certain kind,
{\em etc.}, measurements of the individual lifetimes are more 
common.  In the following section we review the most common
types of the lifetime measurements.  This discussion is followed by the
presentation of the averaging of the various lifetime measurements, each
with a brief description of its particularities.



\mysubsubsection{Lifetime measurements, uncertainties and correlations}

In most cases lifetime of an $H_b$ is estimated from a flight distance
and a $\beta\gamma$ factor which is used to convert the geometrical
distance into the proper decay time.  Methods of accessing lifetime
information can roughly be divided in the following five categories:
\begin{enumerate}
\item {\bf\em Inclusive (flavor blind) measurements}.  These
  measurements are aimed at extracting the lifetime from a mixture of
  \b-hadron decays, without distinguishing the decaying species.  Often
  the knowledge of the mixture composition is limited, which makes these
  measurements experiment-specific.  Also, these
  measurements have to rely on Monte Carlo for estimating the
  $\beta\gamma$ factor, because the decaying hadrons are not fully
  reconstructed.  On the bright side, these usually are the largest
  statistics \b-hadron lifetime measurements that are accessible to a
  given experiment, and can, therefore, serve as an important
  performance benchmark.
\item {\it\bf Measurements in semileptonic decays of a specific
  {\boldmath $H_b$\unboldmath}}.  \particle{W}from \particle{\b\to Wc}
  produces $\ell\nu_l$ pair (\particle{\ell=e,\mu}) in about 21\% of the
  cases.  Electron or muon from such decays is usually a well-detected
  signature, which provides for clean and efficient trigger.
  \particle{c} quark from \particle{b\to Wc} transition and the other
  quark(s) making up the decaying $H_b$ combine into a charm hadron,
  which is reconstructed in one or more exclusive decay channels.
  Knowing what this charmed hadron is allows one to separate, at least
  statistically, different $H_b$ species.  The advantage of these
  measurements is in statistics, which usually is superior to that of the
  exclusively reconstructed $H_b$ decays.  Some of the main
  disadvantages are related to the difficulty of estimating lepton+charm
  sample composition and Monte Carlo reliance for the $\beta\gamma$
  factor estimate.
\item {\bf\em Measurements in exclusively reconstructed hadronic decays}.
  These
  have the advantage of complete reconstruction of decaying $H_b$, which
  allows one to infer the decaying species as well as to perform precise
  measurement of the $\beta\gamma$ factor.  Both lead to generally
  smaller systematic uncertainties than in the above two categories.
  The downsides are smaller branching ratios, larger combinatoric
  backgrounds, especially in $H_b\rightarrow H_c\pi(\pi\pi)$ and
  multi-body $H_c$ decays, or in a hadron collider environment with
  non-trivial underlying event.  $H_b\to J/\psi H_s$ are relatively
  clean and easy to trigger on $J/\psi\to \ell^+\ell^-$, but their
  branching fraction is only about 1\%.
\item {\bf\em Measurements at asymmetric B factories}. In 
  the $\Ups\rightarrow B \bar{B}$ decay, the \B mesons (\Bu or \Bd) are
  essentially at rest in the \Ups rest frame.  This makes lifetime
  measurements impossible with experiments, such as CLEO, in which \Ups
  produced at rest.  At asymmetric \B factories \Ups is boosted
  resulting in \B and \particle{\bar{B}} moving nearly parallel to each
  other.  The lifetime is inferred from the distance $\Delta z$
  separating \B and \particle{\bar{B}} decay vertices and \Ups boost
  known from colliding beam energies.  
  In order to determine the charge of the \B mesons in each event, 
  one of the them is
  fully reconstructed in semileptonic or fully hadronic decay modes.
  The other \B is typically not fully reconstructed, only the position
  of its decay vertex is determined from the remaining tracks in the event.
  These measurements benefit from very large statistics, but suffer from
  poor $\Delta z$ resolution.
\item {\bf\em Direct measurement of lifetime ratios}.  This method has
  so far been only applied in the measurement of $\tau(\Bu)/\tau(\Bd)$.
  The ratio of the lifetimes is extracted from the dependence of the
  observed relative number of \Bu and \Bd candidates (both reconstructed
  in semileptonic decays) on the proper decay time.
\end{enumerate}

In some of the latest analyses, measurements of two (\eg\ $\tau(\Bu)$ and
$\tau(\Bu)/\tau(\Bd)$) or three (\eg\ $\tau(\Bu)$,
$\tau(\Bu)/\tau(\Bd)$, and \dmd) quantities are combined.  This
introduces correlations among measurements.  Another source of
correlations among the measurements are the systematic effects, which
could be common to an experiment or to an analysis technique across the
experiments.  When calculating the averages, such correlations are taken
into account per general procedure, described in
\Ref{lifetime_details}.

\mysubsubsection{Inclusive \b-hadron lifetimes}

The inclusive \b hadron lifetime is defined as $\tau_{\b} = \sum_i f_i
\tau_i$ where $\tau_i$ are the individual species lifetimes and $f_i$ are
the fractions of the various species present in an unbiased sample of
weakly-decaying \b hadrons produced at a high-energy
collider.\footnote{In principle such a quantity could be slightly
different in \particle{Z} decays and a the Tevatron, in case the
fractions of \b-hadron species are not exactly the same; see the
discussion in \Sec{fractions_high_energy}.}  This quantity is certainly
less fundamental than the lifetimes of the individual species, the
latter being much more useful in comparisons of the measurements with
the theoretical predictions.  Nonetheless, we perform the averaging of
the inclusive lifetime measurements for completeness as well as for the
reason that they might be of interest as ``technical numbers.''

\begin{table}[tp]
\caption{Measurements of average \b-hadron lifetimes.}
\labt{lifeincl}
\begin{center}
\begin{tabular}{lcccl} \hline
Experiment &Method           &Data set & $\tau_{\b}$ (ps)       &Ref.\\
\hline
ALEPH  &Dipole               &91     &$1.511\pm 0.022\pm 0.078$ &\cite{ALEIN2}\\
DELPHI &All track i.p.\ (2D) &91--92 &$1.542\pm 0.021\pm 0.045$ &\cite{DELIN0}$^a$\\
DELPHI &Sec.\ vtx            &91--93 &$1.582\pm 0.011\pm 0.027$ &\cite{DELIN}$^a$\\
DELPHI &Sec.\ vtx            &94--95 &$1.570\pm 0.005\pm 0.008$ &\cite{DELB04}\\
L3     &Sec.\ vtx + i.p.     &91--94 &$1.556\pm 0.010\pm 0.017$ &\cite{L3IN1}$^b$\\
OPAL   &Sec.\ vtx            &91--94 &$1.611\pm 0.010\pm 0.027$ &\cite{OPAIN2}\\
SLD    &Sec.\ vtx            &93     &$1.564\pm 0.030\pm 0.036$ &\cite{SLDIN}\\ 
\hline
\multicolumn{2}{l}{Average set 1 (\b vertex)} && \hfagTAUBVTXnounit &\\
\hline\hline
ALEPH  &Lepton i.p.\ (3D)    &91--93 &$1.533\pm 0.013\pm 0.022$ &\cite{ALEIN1}\\
L3     &Lepton i.p.\ (2D)    &91--94 &$1.544\pm 0.016\pm 0.021$ &\cite{L3IN1}$^b$\\
OPAL   &Lepton i.p.\ (2D)    &90--91 &$1.523\pm 0.034\pm 0.038$ &\cite{OPAIN1}\\ 
\hline
\multicolumn{2}{l}{Average set 2 ($\b\to\ell$)} && \hfagTAUBLEPnounit &\\
\hline\hline
CDF1   &\particle{J/\psi} vtx&92--95 &$1.533\pm 0.015^{+0.035}_{-0.031}$ &\cite{CDFIN_BS1} \\ 
\hline\hline
\multicolumn{2}{l}{Average of all above} && \hfagTAUBnounit & \\
\hline
\multicolumn{5}{l}{$^a$ \footnotesize The combined DELPHI result quoted in
\cite{DELIN} is 1.575 $\pm$ 0.010 $\pm$ 0.026 ps.} \\[-0.5ex]
\multicolumn{5}{l}{$^b$ \footnotesize The combined L3 result quoted in \cite{L3IN1} 
is 1.549 $\pm$ 0.009 $\pm$ 0.015 ps.}
\end{tabular}
\end{center}
\end{table}

In practice, an unbiased measurement of the inclusive lifetime is
difficult to achieve, because it would imply an efficiency which is
guaranteed to be the same across species.  So most of the measurements
are biased.  In an attempt to group analyses which are expected to
select the same mixture of \b hadrons, the available results (given in
\Table{lifeincl}) are divided into the following three sets:
\begin{enumerate}
\item measurements at LEP and SLD that accept any \b-hadron decay, based 
      on topological reconstruction (secondary vertex or track impact
      parameters);
\item measurements at LEP based on the identification
      of a lepton from a \b decay; and
\item measurements at the Tevatron based on inclusive 
      \particle{H_b\to J/\psi X} reconstruction, where the
      \particle{J/\psi} is fully reconstructed.
\end{enumerate}

The measurements of the first set are generally considered as estimates
of $\tau_{\b}$, although the efficiency to reconstruct a secondary
vertex most probably depends, in an analysis-specific way, on the number
of tracks coming from the vertex, thereby depending on the type of the
$H_b$.  Even though these efficiency variations can in principle be
accounted for using Monte Carlo simulations (which inevitably contain
assumptions on branching fractions), the $H_b$ mixture in that case can
remain somewhat ill-defined and could be slightly different among
analyses in this set.

On the contrary, the mixtures corresponding to the other two sets of
measurements are better defined in the limit where the reconstruction
and selection efficiency of a lepton or a \particle{J/\psi} from an
$H_b$ does not depend on the decaying hadron type.  These mixtures are
given by the production fractions and the inclusive branching fractions
for each $H_b$ species to give a lepton or a \particle{J/\psi}.  In
particular, under the assumption that all \b hadrons have the same
semileptonic decay width, the analyses of the second set should measure
$\tau(\b\to\ell) = (\sum_i f_i \tau_i^2) /(\sum_i f_i \tau_i)$ which is
necessarily larger than $\tau_{\b}$ if lifetime differences exist.
Given the present knowledge on $\tau_i$ and $f_i$,
$\tau(\b\to\ell)-\tau_{\b}$ is expected to be of the order of 0.01\ps.

Measurements by SLC and LEP experiments are subject to a number of
common systematic uncertainties, such as those due to (lack of knowledge
of) \b and \particle{c} fragmentation, \b and \particle{c} decay models,
\BR{B\to\ell}, \BR{B\to c\to\ell}, \BR{c\to\ell}, $\tau_{\particle{c}}$,
and $H_b$ decay multiplicity.  In the averaging, these systematic
uncertainties are assumed to be 100\% correlated.  The averages for the
sets defined above (also given in \Table{lifeincl}) are
\begin{eqnarray}
\tau(\b~\mbox{vertex}) &=& \hfagTAUBVTX \,,\\
\tau(\b\to\ell) &=& \hfagTAUBLEP  \,, \\
\tau(\b\to\particle{J/\psi}) &=& \hfagTAUBJP\,,
\end{eqnarray}
whereas an average of all measurements, ignoring mixture differences, 
yields \hfagTAUB.

\mysubsubsection{\Bd and \Bu lifetimes and their ratio}
\labs{taubd}
\labs{taubu}
\labs{lifetime_ratio}

\begin{table}[tp]
\caption{Measurements of the \Bd lifetime.}
\labt{lifebd}
\begin{center}
\begin{tabular}{lcccl} \hline
Experiment &Method                    &Data set &$\tau(\Bd)$ (ps)                  &Ref.\\
\hline
ALEPH  &\particle{D^{(*)} \ell}       &91--95 &$1.518\pm 0.053\pm 0.034$          &\cite{ALEB01}\\
ALEPH  &Exclusive                     &91--94 &$1.25^{+0.15}_{-0.13}\pm 0.05$     &\cite{ALEB0}\\
ALEPH  &Partial rec.\ $\pi^+\pi^-$    &91--94 &$1.49^{+0.17+0.08}_{-0.15-0.06}$   &\cite{ALEB0}\\
DELPHI &\particle{D^{(*)} \ell}       &91--93 &$1.61^{+0.14}_{-0.13}\pm 0.08$     &\cite{DELB01}\\
DELPHI &Charge sec.\ vtx              &91--93 &$1.63 \pm 0.14 \pm 0.13$           &\cite{DELB02}\\
DELPHI &Inclusive \particle{D^* \ell} &91--93 &$1.532\pm 0.041\pm 0.040$          &\cite{DELB03}\\
DELPHI &Charge sec.\ vtx              &94--95 &$1.531 \pm 0.021\pm0.031$          &\cite{DELB04}\\
L3     &Charge sec.\ vtx              &94--95 &$1.52 \pm 0.06 \pm 0.04$           &\cite{L3B01}\\
OPAL   &\particle{D^{(*)} \ell}       &91--93 &$1.53 \pm 0.12 \pm 0.08$           &\cite{OPAB0}\\
OPAL   &Charge sec.\ vtx              &93--95 &$1.523\pm 0.057\pm 0.053$          &\cite{OPAB1}\\
OPAL   &Inclusive \particle{D^* \ell} &91--00 &$1.541\pm 0.028\pm 0.023$          &\cite{OPAB2}\\
SLD    &Charge sec.\ vtx $\ell$       &93--95 &$1.56^{+0.14}_{-0.13} \pm 0.10$    &\cite{SLDB01}$^a$\\
SLD    &Charge sec.\ vtx              &93--95 &$1.66 \pm 0.08 \pm 0.08$           &\cite{SLDB01}$^a$\\
CDF1   &\particle{D^{(*)} \ell}       &92--95 &$1.474\pm 0.039^{+0.052}_{-0.051}$ &\cite{CDFB1}\\
CDF1  &Excl.\ \particle{J/\psi K^{*0}}&92--95 &$1.497\pm 0.073\pm 0.032$          &\cite{CDFB2}\\
CDF2  &Excl.\ \particle{J/\psi K^{*0}}&02--04 &$1.539\pm 0.051\pm0.008$           &\cite{CDFB3}$^p$\\
CDF2   &Incl.\ \particle{D^{(*)} \ell}&02--04 &$1.473\pm 0.036\pm0.054$           &\cite{CDFB4}$^p$\\
CDF2   &Excl.\ \particle{D^-(3)\pi}   &02--04 &$1.511\pm 0.023\pm0.013$           &\cite{CDFB5}$^p$\\
CDF2   &Excl. \particle {J/\psi K_S}  &02--04 &$1.503^{+0.050}_{-0.048}\pm0.016$  &\cite{CDFLAM2}$^p$\\
\dzero &Excl. \particle{J/\psi K^{*0}}&02--05 &$1.530\pm0.043\pm0.023$ &\cite{D01_DGs,D0BS1}\\ 
\dzero &Excl. \particle {J/\psi K_S}  &02--04 &$1.40^{+0.11}_{-0.10}\pm0.03$  &\cite{D0LAMB}\\
\babar &Exclusive                     &99--00 &$1.546\pm 0.032\pm 0.022$          &\cite{BABAR1}\\
\babar &Inclusive \particle{D^* \ell} &99--01 &$1.529\pm 0.012\pm 0.029$          &\cite{BABAR2}\\
\babar &Exclusive \particle{D^* \ell} &99--02 &$1.523^{+0.024}_{-0.023}\pm 0.022$ &\cite{BABAR3}\\
\babar &Incl.\ \particle{D^*\pi}, \particle{D^*\rho} 
                                      &99--01 &$1.533\pm 0.034 \pm 0.038$         &\cite{BABAR4}\\
\babar &Inclusive \particle{D^* \ell}
&99--04 &$1.504\pm0.013^{+0.018}_{-0.013}$  &\cite{BABAR5} \\ 
\belle & Exclusive                     & 00--03 & $1.534\pm 0.008\pm0.010$        & \cite{BELLE2}\\
\hline
Average&                               &        & \hfagTAUBDnounit & \\
\hline\hline           
\multicolumn{5}{l}{$^a$ \footnotesize The combined SLD result 
quoted in \cite{SLDB01} is 1.64 $\pm$ 0.08 $\pm$ 0.08 ps.}\\[-0.5ex]
\multicolumn{5}{l}{$^p$ {\footnotesize Preliminary.}}
\end{tabular}
\end{center}
\end{table}

After a number of years of dominating these averages the LEP experiments
yielded the scene to the asymmetric \B~factories and
the Tevatron experiments.  The \B~factories have been very successful in
utilizing their potential -- in only a few years of running, \babar and,
to a greater extent, \belle, have struck a balance between the
statistical and the systematic uncertainties, with both being close to
(or even better than) the impressive 1\%.  In the meanwhile, CDF and
\dzero have emerged as significant contributors to the field as the
Tevatron Run~II data flowed in.  Both appear to enjoy relatively small
systematic effects, and while current statistical uncertainties of their
measurements are factors of 2 to 4 larger than those of their \B-factory
counterparts, both Tevatron experiments stand to increase their samples
by an order of magnitude.

\begin{table}[tbp]
\caption{Measurements of the \Bu lifetime.}
\labt{lifebu}
\begin{center}
\begin{tabular}{lcccl} \hline
Experiment &Method                 &Data set &$\tau(\Bu)$ (ps)                 &Ref.\\
\hline
ALEPH  &\particle{D^{(*)} \ell}    &91--95 &$1.648\pm 0.049\pm 0.035$          &\cite{ALEB01}\\
ALEPH  &Exclusive                  &91--94 &$1.58^{+0.21+0.04}_{-0.18-0.03}$   &\cite{ALEB0}\\
DELPHI &\particle{D^{(*)} \ell}    &91--93 &$1.61\pm 0.16\pm 0.12$             &\cite{DELB01}$^a$\\
DELPHI &Charge sec.\ vtx           &91--93 &$1.72\pm 0.08\pm 0.06$             &\cite{DELB02}$^a$\\
DELPHI &Charge sec.\ vtx           &94--95 &$1.624\pm 0.014\pm 0.018$          &\cite{DELB04}\\
L3     &Charge sec.\ vtx           &94--95 &$1.66\pm  0.06\pm 0.03$            &\cite{L3B01}\\
OPAL   &\particle{D^{(*)} \ell}    &91--93 &$1.52 \pm 0.14\pm 0.09$            &\cite{OPAB0}\\
OPAL   &Charge sec.\ vtx           &93--95 &$1.643\pm 0.037\pm 0.025$          &\cite{OPAB1}\\
SLD    &Charge sec.\ vtx $\ell$    &93--95 &$1.61^{+0.13}_{-0.12}\pm 0.07$     &\cite{SLDB01}$^b$\\
SLD    &Charge sec.\ vtx           &93--95 &$1.67\pm 0.07\pm 0.06$             &\cite{SLDB01}$^b$\\
CDF1   &\particle{D^{(*)} \ell}    &92--95 &$1.637\pm 0.058^{+0.045}_{-0.043}$ &\cite{CDFB1}\\
CDF1   &Excl.\ \particle{J/\psi K} &92--95 &$1.636\pm 0.058\pm 0.025$          &\cite{CDFB2}\\
CDF2   &Excl.\ \particle{J/\psi K} &02--04 &$1.662\pm 0.033\pm 0.008$          &\cite{CDFB3}$^p$\\
CDF2   &Incl.\ \particle{D^0 \ell} &02--04 &$1.653\pm 0.029^{+0.033}_{-0.031}$ &\cite{CDFB4}$^p$\\
CDF2   &Excl.\ \particle{D^0 \pi}  &02--04 &$1.661\pm 0.027\pm0.013$           &\cite{CDFB5}$^p$\\
\babar &Exclusive                  &99--00 &$1.673\pm 0.032\pm 0.023$          &\cite{BABAR1}\\
\belle &Exclusive                  &00--03 &$1.635\pm 0.011\pm 0.011$          &\cite{BELLE2}\\
\hline
Average&                           &       &\hfagTAUBUnounit &\\
\hline\hline
\multicolumn{5}{l}{$^a$ \footnotesize The combined DELPHI result quoted 
in~\cite{DELB02} is $1.70 \pm 0.09$ ps.} \\[-0.5ex]
\multicolumn{5}{l}{$^b$ \footnotesize The combined SLD result 
quoted in~\cite{SLDB01} is $1.66 \pm 0.06 \pm 0.05$ ps.}\\[-0.5ex]
\multicolumn{5}{l}{$^p$ {\footnotesize Preliminary.}}
\end{tabular}
\end{center}
\end{table}

At present time we are in an interesting position of having three sets
of measurements (from LEP/SLC, \B factories and the Tevatron) that
originate from different environments, obtained using substantially
different techniques and are precise enough for incisive comparison.


\begin{table}[tb]
\caption{Measurements of the ratio $\tau(\Bu)/\tau(\Bd)$.}
\labt{liferatioBuBd}
\begin{center}
\begin{tabular}{lcccl} 
\hline
Experiment &Method                 &Data set &Ratio $\tau(\Bu)/\tau(\Bd)$      &Ref.\\
\hline
ALEPH  &\particle{D^{(*)} \ell}    &91--95 &$1.085\pm 0.059\pm 0.018$          &\cite{ALEB01}\\
ALEPH  &Exclusive                  &91--94 &$1.27^{+0.23+0.03}_{-0.19-0.02}$   &\cite{ALEB0}\\
DELPHI &\particle{D^{(*)} \ell}    &91--93 &$1.00^{+0.17}_{-0.15}\pm 0.10$     &\cite{DELB01}\\
DELPHI &Charge sec.\ vtx           &91--93 &$1.06^{+0.13}_{-0.11}\pm 0.10$     &\cite{DELB02}\\
DELPHI &Charge sec.\ vtx           &94--95 &$1.060\pm 0.021 \pm 0.024$         &\cite{DELB04}\\
L3     &Charge sec.\ vtx           &94--95 &$1.09\pm 0.07  \pm 0.03$           &\cite{L3B01}\\
OPAL   &\particle{D^{(*)} \ell}    &91--93 &$0.99\pm 0.14^{+0.05}_{-0.04}$     &\cite{OPAB0}\\
OPAL   &Charge sec.\ vtx           &93--95 &$1.079\pm 0.064 \pm 0.041$         &\cite{OPAB1}\\
SLD    &Charge sec.\ vtx $\ell$    &93--95 &$1.03^{+0.16}_{-0.14} \pm 0.09$    &\cite{SLDB01}$^a$\\
SLD    &Charge sec.\ vtx           &93--95 &$1.01^{+0.09}_{-0.08} \pm0.05$     &\cite{SLDB01}$^a$\\
CDF1   &\particle{D^{(*)} \ell}    &92--95 &$1.110\pm 0.056^{+0.033}_{-0.030}$ &\cite{CDFB1}\\
CDF1   &Excl.\ \particle{J/\psi K} &92--95 &$1.093\pm 0.066 \pm 0.028$         &\cite{CDFB2}\\
CDF2   &Excl.\ \particle{J/\psi K} &02--04 &$1.080\pm 0.042$                   &\cite{CDFB3}$^p$\\
CDF2   &Incl.\ \particle{D \ell}   &02--04 &$1.123\pm0.040^{+0.041}_{-0.039}$  &\cite{CDFB4}$^p$\\
CDF2   &Excl.\ \particle{D \pi}    &02--04 &$1.10\pm 0.02\pm 0.01$             &\cite{CDFB5}$^p$\\
\dzero &\particle{D^{*+} \mu} \particle{D^0 \mu} ratio
	                           &02--04 &$1.080\pm 0.016\pm 0.014$          &\cite{D0B01}\\
\babar &Exclusive                  &99--00 &$1.082\pm 0.026\pm 0.012$          &\cite{BABAR1}\\
\belle &Exclusive                  &00--03 &$1.066\pm 0.008\pm 0.008$          &\cite{BELLE2}\\
\hline
Average&                           &       & \hfagRTAUBU & \\   
\hline\hline
\multicolumn{5}{l}{$^a$ \footnotesize The combined SLD result quoted
	   in~\cite{SLDB01} is $1.01 \pm 0.07 \pm 0.06$.} \\[-0.5ex] 
\multicolumn{5}{l}{$^p$ {\footnotesize Preliminary.}}
\end{tabular}
\end{center}
\end{table}

The averaging of $\tau(\Bu)$, $\tau(\Bd)$ and $\tau(\Bu)/\tau(\Bd)$
measurements is summarized in \Tablesss{lifebd}{lifebu}{liferatioBuBd}.
For $\tau(\Bu)/\tau(\Bd)$ we averaged only the measurements of this
quantity provided by experiments rather than using all available
knowledge, which would have included, for example, $\tau(\Bu)$ and
$\tau(\Bd)$ measurements which did not contribute to any of the ratio
measurements.

The following sources of correlated (within experiment/machine)
systematic uncertainties have been considered:
\begin{itemize}
\item for SLC/LEP measurements -- \particle{D^{**}} branching ratio uncertainties~\cite{LEPHFS},
momentum estimation of \b mesons from \particle{Z^0} decays
(\b-quark fragmentation parameter $\langle X_E \rangle = 0.702 \pm 0.008$~\cite{LEPHFS}),
\Bs and \b baryon lifetimes (see \Secss{taubs}{taulb}),
and \b hadron fractions at high energy (see \Table{fractions}).  
\item for \babar measurements -- alignment, $z$ scale, PEP-II boost,
sample composition (where applicable) 
\item for \dzero and CDF Run~II measurements -- alignment (separately
within each experiment)
\end{itemize}
The resultant averages are:
\begin{eqnarray}
\tau(\Bd) & = & \hfagTAUBD \,, \\
\tau(\Bu) & = & \hfagTAUBU \,, \\
\tau(\Bu)/\tau(\Bd) & = & \hfagRTAUBU \,.
\end{eqnarray}
%
%
%

\mysubsubsection{\Bs lifetime}
\labs{taubs}

Similar to the kaon system, neutral \B mesons contain
short- and long-lived components, since the
light (L) and heavy  (H)
eigenstates, $\B_{\rm L}$ and $\B_{\rm H}$, differ not only
in their masses, but also in their widths 
with $\Delta\Gamma = \Gamma_{\rm L} - \Gamma_{\rm H}$. 
In the case of the \Bs system, $\DGs$ can
be particularly large. The current theoretical
prediction in the Standard Model for
the fractional width difference is
$\DGs/\Gs = 0.12 \pm 0.05$~\cite{delta_gams},
where $\Gs = (\Gamma_{\rm L} + \Gamma_{\rm H})/2$.
Specific measurements of \DGs and \Gs are explained
in \Sec{DGs}, but the result for
\Gs is quoted here.

Neglecting \CP violation in $\Bs-\Bsbar$ mixing, 
which is expected to be small~\cite{delta_gams}, the
\Bs mass eigenstates are also \CP eigenstates. In
the Standard Model assuming no \CP violation in
the \Bs system,
$\Gamma_{\rm L}$ is the width of
the \CP-even state and
$\Gamma_{\rm H}$ the width of
the \CP-odd state.
Final states can be decomposed into
\CP-even and \CP-odd components, each with a different
lifetime.

In view of a possibly substantial width difference,
and the fact that various
decay channels will have different proportions of 
the $\B_{\rm L}$ and $\B_{\rm H}$ eigenstates,
the straight average of all available 
\Bs lifetime measurements
is rather ill-defined.  Therefore,
the \Bs lifetime measurements are broken down into
three categories and averaged separately.

\begin{itemize}
\item 
{\bf\em Flavor-specific decays}, such as semileptonic
$\particle{B_s} \to \particle{D_s \ell \nu}$
or $\particle{B_s} \to \particle {D_s \pi}$, will
have equal 
fractions of $\B_{\rm L}$ and $\B_{\rm H}$ at time
zero, where
$\tau_{\rm L} = 1/\Gamma_{\rm L}$ 
is expected to be the shorter-lived component and
$\tau_{\rm H} = 1/\Gamma_{\rm H}$ 
expected
to be the longer-lived component.  A superposition
of two exponentials thus results with decay
widths $\Gs \pm \DGs /2$.
Fitting to a single exponential one obtains a
measure of the flavor-specific 
lifetime~\cite{Hartkorn_Moser}:
\begin{equation}
\tau(\Bs)_{\rm fs} = \frac{1}{\Gs}
\frac{{1+\left(\frac{\DGs}{2\Gs}\right)^2}}{{1-\left(\frac{\DGs}{2\Gs}\right)^2}
}.
\end{equation}
As given in \Table{lifebs}, the flavor-specific 
\Bs lifetime world average is:
\begin{equation}
\tau(\Bs)_{\rm fs} = \hfagTAUBSSL \,.
\end{equation}
This world average will be used later in \Sec{DGs} in combination
with other measurements to find
$\bar{\tau}(\Bs) = 1/\Gs$ and $\DGs$.

The following correlated systematic errors were considered:
average \B lifetime used in backgrounds,
\Bs decay multiplicity, and branching ratios used to determine 
backgrounds (\eg\ \BR{B\to D_s D}).
A knowledge of the multiplicity of \Bs decays is important for
measurements that partially reconstruct the final state such as 
\particle{\B\to D_s \mbox{$X$}} (where $X$ is not a lepton). 
The boost deduced from Monte Carlo simulation depends on the multiplicity used.
Since this is not well known, the multiplicity in the simulation is
varied and this range of values observed is taken to be a systematic.
Similarly not all the branching ratios for the potential background
processes are measured. Where they are available, the PDG values are
used for the error estimate. Where no measurements are available
estimates can usually be made by using measured branching ratios of
related processes and using some reasonable extrapolation.
\end{itemize}



\begin{table}[tb]
\caption{Measurements of the \Bs lifetime.}
\labt{lifebs}
\begin{center}
\begin{tabular}{lcccl} \hline
Experiment & Method           & Data set & $\tau(\Bs)$ (ps)               & Ref. \\
\hline
ALEPH  & \particle{D_s \ell}  & 91--95 & $1.54^{+0.14}_{-0.13}\pm 0.04$   & \cite{ALEBS1}          \\
CDF1   & \particle{D_s \ell}  & 92--96 & $1.36\pm 0.09 ^{+0.06}_{-0.05}$  & \cite{CDFBS}           \\
DELPHI & \particle{D_s \ell}  & 91--95 & $1.42^{+0.14}_{-0.13}\pm 0.03$   & \cite{DELBS0}          \\
OPAL   & \particle{D_s \ell}  & 90--95 & $1.50^{+0.16}_{-0.15}\pm 0.04$   & \cite{OPABS1_OPALAM2}  \\
\dzero & \particle{D_s \mu}  & 02--04 & $1.420 \pm 0.043 \pm 0.057   $   & \cite{D0BS2}$^p$       \\ 
CDF2   & \particle{D_s \pi, D_s \pi \pi \pi} 
                              & 02--04 & $1.60 \pm 0.10 \pm 0.02      $   & \cite{CDFBS2}$^p$      \\
CDF2   & \particle{D_s \ell}  & 02--04 & $1.381 \pm 0.055 ^{+0.052}_{-0.046} $ &
\cite{CDFBS3}$^p$ \\ \hline
\multicolumn{3}{l}{Average of flavor-specific measurements} &  \hfagTAUBSSLnounit & \\  
\hline
ALEPH  & \particle{D_s h}     & 91--95 & $1.47\pm 0.14\pm 0.08$           & \cite{ALEBS2}          \\
DELPHI & \particle{D_s h}     & 91--95 & $1.53^{+0.16}_{-0.15}\pm 0.07$   & \cite{DELBS1_dms_excl} \\
OPAL   & \particle{D_s} incl. & 90--95 & $1.72^{+0.20+0.18}_{-0.19-0.17}$ & \cite{OPABS2}          \\ 
\hline
\multicolumn{3}{l}{Average of all above \particle{D_s} measurements} &  \hfagTAUBSnounit & \\ 
\hline\hline
CDF1     & \particle{J/\psi\phi} & 92--95  & $1.34^{+0.23}_{-0.19}    \pm 0.05$ & \cite{CDFIN_BS1} \\
CDF2     & \particle{J/\psi\phi} & 02--04  & $1.369 \pm 0.100 ^{+0.008}_{-0.010}$ & \cite{CDFB3}$^p$ \\
\dzero   & \particle{J/\psi\phi} & 02--04  & $1.444^{+0.098}_{-0.090} \pm 0.02$ & \cite{D0BS1}  \\ \hline 
\multicolumn{3}{l}{Average of \particle{J/\psi \phi} measurements} &  \hfagTAUBSJFnounit & \\ 
\hline
\multicolumn{5}{l}{$^p$ \footnotesize Preliminary.}
\end{tabular}
\end{center}
\end{table}

\begin{itemize}
\item
{\bf\em \boldmath $\Bs\to\Ds X$ decays}.
Included in \Table{lifebs} are measurements
of lifetimes using samples of \particle{\Bs} decays to
\particle{D_s} plus
hadrons, and hence into a less known mixture
of \CP-states.  A lifetime
weighted this way can still be a useful input
for analyses examining such an inclusive sample.
These are separated in \Table{lifebs} and combined
with the semileptonic lifetime to obtain:
\begin{equation}
\tau(\Bs)_{\particle{D_s {\rm X}}} = \hfagTAUBS \,.
\end{equation}

\item
{\bf\em Fully exclusive 
{\boldmath \Bs $\to J/\psi \phi$ \unboldmath}decays}
are expected to be
dominated by the \CP-even state and its lifetime.
First measurements of the \CP mix for this decay mode
are outlined in \Sec{DGs}.
CDF and \dzero measurements from this particular mode
\particle{\Bs\to J/\psi\phi} are combined into an
average
given in \Table{lifebs}.  There are no correlations
between the measurements for this fully exclusive
channel, and the world average for this 
specific decay is:
\begin{equation}
\tau(\Bs)_{\particle{J/\psi \phi}} = \hfagTAUBSJF \,.
\end{equation}
A caveat is that different experimental acceptances
will likely lead to different admixtures of the 
\CP-even and \CP-odd states, and fits to a single
exponential may result in inherently different 
measurements of these quantities.
\end{itemize}

Finally, as will be shown in \Sec{DGs}, measurements
of $\DGs$, including separation into
\CP-even and \CP-odd components, give
\begin{equation}
\bar{\tau}(\Bs) = 1/\Gs = \hfagTAUBSMEAN \,,
\end{equation}
and when combined with the flavor-specific lifetime
measurements:
\begin{equation}
\bar{\tau}(\Bs) = 1/\Gs = \hfagTAUBSMEANCON \,.
\end{equation}

\mysubsubsection{\Bc lifetime}
\labs{taubc}

There are currently three measurements of the lifetime of the \Bc meson
from CDF~\cite{CDFBC1,CDFBC2} and \dzero~\cite{D0BC1} using the semileptonic decay
mode \particle{\Bc \to J/\psi \ell} and fitting
simultaneously to the mass and lifetime using the vertex formed
with the leptons from the decay of the \particle{J/\psi} and
the third lepton. Correction factors
to estimate the boost due to the missing neutrino are used.
Mass values of
$6.40 \pm 0.39 \pm 0.13$~GeV/$c^2$ for the CDF Run~1 
result~\cite{CDFBC1} and 
$5.95^{+0.14}_{-0.13} \pm 0.34$~GeV/$c^2$ for the \dzero Run 2 
result~\cite{D0BC1} are
found by fitting
to the tri-lepton invariant mass spectrum. These mass measurements
are consistent within uncertainties.
In the CDF Run~2 result~\cite{CDFBC2}, the mass is fixed
to 6.271~GeV/$c^2$, but then varied between 
6.2 and 6.4~GeV/$c^2$ to assess the systematic error on the
lifetime due to the \Bc mass value.
Correlated systematic errors include the impact
of the uncertainty of the \Bc $p_T$ spectrum on the correction
factors, the level of feed-down from $\psi(2S)$, 
MC modeling of the decay model varying from phase space
to the ISGW model, and mass variations.
Values of the \particle{\Bc} lifetime are given
in \Table{lifebc} and the world average is
determined to be:
\begin{equation}
\tau(\Bc) = \hfagTAUBC \,.
\end{equation}

\begin{table}[tb]
\caption{Measurements of the \Bc lifetime.}
\labt{lifebc}
\begin{center}
\begin{tabular}{lcccl} \hline
Experiment & Method                    & Data set  & $\tau(\Bc)$ (ps)
      & Ref.\\   \hline
CDF1       & \particle{J/\psi \ell} & 92--95  & $0.46^{+0.18}_{-0.16} \pm
 0.03$   & \cite{CDFBC1}  \\ 
CDF2       & \particle{J/\psi e} & 02--04  & $0.474^{+0.073}_{-0.066} \pm
 0.033$   & \cite{CDFBC2}$^p$  \\
 \dzero & \particle{J/\psi \mu} & 02--04  & $0.448^{+0.123}_{-0.096} 
\pm  0.121$   & \cite{D0BC1}$^p$  \\ \hline
  \multicolumn{2}{l}{Average} &   &  \hfagTAUBCnounit
                 &    \\   \hline
\multicolumn{5}{l}{$^p$ \footnotesize Preliminary.}
\end{tabular}
\end{center}
\end{table}

\mysubsubsection{\Lb and \b-baryon lifetimes}
\labs{taulb}

The most precise measurements of the \b-baryon lifetime
originate from two classes of partially reconstructed decays.
In the first class, decays with an exclusively 
reconstructed \Lc baryon
and a lepton of opposite charge are used. These products are
more likely to occur in the decay of \Lb baryons.
In the second class, more inclusive final states with a baryon
(\particle{p}, \particle{\bar{p}}, $\Lambda$, or $\bar{\Lambda}$) 
and a lepton have been used, and these final states can generally
arise from any \b baryon.

The following sources of correlated systematic uncertainties have 
been considered:
experimental time resolution within a given experiment, \b-quark
fragmentation distribution into weakly decaying \b baryons,
\Lb polarization, decay model,
and evaluation of the \b-baryon purity in the selected event samples.
In computing the averages
the central values of the masses are scaled to 
$M(\Lb) = 5624 \pm 9\MeVcc$~\cite{PDGmass} and
$M(\mbox{\b-baryon}) = 5670 \pm 100\MeVcc$.

The meaning of decay model and the correlations are not always clear.
Uncertainties related to the decay model are dominated by
assumptions on the fraction of $n$-body decays.
To be conservative it is assumed
that it is correlated whenever given as an error.
DELPHI varies the fraction of 4-body decays from 0.0 to 0.3. 
In computing the average, the DELPHI
result is corrected for $0.2 \pm 0.2$.

Furthermore, in computing the average,
the semileptonic decay results are corrected for a polarization of 
$-0.45^{+0.19}_{-0.17}$~\cite{LEPHFS} and a 
\Lb fragmentation parameter
$\langle X_E \rangle =0.70\pm 0.03$~\cite{LBFRAG}.




Inputs to the averages are given in \Table{lifelb}.
The world average lifetime of \b baryons is then:
\begin{equation}
\langle\tau(\mbox{\b-baryon})\rangle = \hfagTAUBB \,.
\end{equation}
Keeping only \particle{\Lambda^{\pm}_c \ell^{\mp}}, 
$\Lambda \ell^- \ell^+$, and fully exclusive
final states, as representative of
the \Lb baryon, the following lifetime is obtained:
\begin{equation}
\tau(\Lb) = \hfagTAULB \,. 
\end{equation}

Averaging the measurements based on the $\Xi^{\mp} \ell^{\mp}$
final states~\cite{ALEPH_fxi,DELPHI_fxi} gives
a lifetime value for a sample of events
containing $\Xib^0$ and $\Xib^-$ baryons:
\begin{equation}
\langle\tau(\Xib)\rangle = \hfagTAUXB \,.
\end{equation}

\begin{table}[t]
\caption{Measurements of the \b-baryon lifetimes.
}
\labt{lifelb}
\begin{center}
\begin{tabular}{lcccl} 
\hline
Experiment&Method                &Data set& Lifetime (ps) & Ref. \\\hline
ALEPH  &$\Lc\ell$             & 91--95 &$1.18^{+0.13}_{-0.12} \pm 0.03$ & \cite{ALEPH_fla}$^a$\\
ALEPH  &$\Lambda\ell^-\ell^+$ & 91--95 &$1.30^{+0.26}_{-0.21} \pm 0.04$ & \cite{ALEPH_fla}$^a$\\
CDF1   &$\Lc\ell$             & 91--95 &$1.32 \pm 0.15        \pm 0.06$ & \cite{CDFLAM}\\
CDF2   &$J/\psi \Lambda$      & 02--04 &$1.45^{+0.14}_{-0.13} \pm 0.02$ & \cite{CDFLAM2}$^p$ \\
\dzero &$J/\psi \Lambda$      & 02--04 &$1.22^{+0.22}_{-0.18} \pm 0.04$ & \cite{D0LAMB} \\
DELPHI &$\Lc\ell$             & 91--94 &$1.11^{+0.19}_{-0.18} \pm 0.05$ & \cite{DELLAM0}$^b$\\
OPAL   &$\Lc\ell$, $\Lambda\ell^-\ell^+$ 
                                 & 90--95 & $1.29^{+0.24}_{-0.22} \pm 0.06$ & \cite{OPABS1_OPALAM2}\\ 
\hline
\multicolumn{3}{l}{Average of above 7 (\Lb lifetime)} & \hfagTAULBnounit & \\
\hline
ALEPH  &$\Lambda\ell$         & 91--95 &$1.20 \pm 0.08 \pm 0.06$ & \cite{ALEPH_fla}\\
DELPHI &$\Lambda\ell\pi$ vtx  & 91--94 &$1.16 \pm 0.20 \pm 0.08$        & \cite{DELLAM0}$^b$\\
DELPHI &$\Lambda\mu$ i.p.     & 91--94 &$1.10^{+0.19}_{-0.17} \pm 0.09$ & \cite{DELLAM1}$^b$ \\
DELPHI &\particle{p\ell}      & 91--94 &$1.19 \pm 0.14 \pm 0.07$        & \cite{DELLAM0}$^b$\\
OPAL   &$\Lambda\ell$ i.p.    & 90--94 &$1.21^{+0.15}_{-0.13} \pm 0.10$ & \cite{OPALAM1}$^c$  \\
OPAL   &$\Lambda\ell$ vtx     & 90--94 &$1.15 \pm 0.12 \pm 0.06$        & \cite{OPALAM1}$^c$ \\ 
\hline
\multicolumn{3}{l}{Average of above 13 (\b-baryon lifetime)} & \hfagTAUBBnounit & \\  
\hline\hline
ALEPH  &$\Xi\ell$             & 90--95 &$1.35^{+0.37+0.15}_{-0.28-0.17}$ & \cite{ALEPH_fxi}\\
DELPHI &$\Xi\ell$             & 91--93 &$1.5 ^{+0.7}_{-0.4} \pm 0.3$     & \cite{DELPHI_fxi} \\
\hline
\multicolumn{3}{l}{Average of above 2 (\Xib lifetime)} & \hfagTAUXBnounit & \\
\hline
\multicolumn{5}{l}{$^a$ \footnotesize The combined ALEPH result quoted 
in \cite{ALEPH_fla} is $1.21 \pm 0.11$ ps.} \\[-0.5ex]
\multicolumn{5}{l}{$^b$ \footnotesize The combined DELPHI result quoted 
in \cite{DELLAM0} is $1.14 \pm 0.08 \pm 0.04$ ps.} \\[-0.5ex]
\multicolumn{5}{l}{$^c$ \footnotesize The combined OPAL result quoted 
in \cite{OPALAM1} is $1.16 \pm 0.11 \pm 0.06$ ps.} \\[-0.5ex]
\multicolumn{5}{l}{$^p$ \footnotesize Preliminary.}
\end{tabular}
\end{center}
\end{table}

\mysubsubsection{Summary and comparison with theoretical predictions}
\labs{lifesummary}

Averages of lifetimes of specific \b-hadron species are collected
in \Table{sumlife}.
\begin{table}[t]
\caption{Summary of lifetimes of different \b-hadron species.}
\labt{sumlife}
\begin{center}
\begin{tabular}{lc} \hline
\b-hadron species & Measured lifetime \\ \hline
\Bu                         & \hfagTAUBU   \\
\Bd                         & \hfagTAUBD   \\
\Bs ($\to$ flavor specific) & \hfagTAUBSSL \\
\Bs ($\to J/\psi\phi$)      & \hfagTAUBSJF \\
\Bs ($1/\Gs$)               & \hfagTAUBSMEANCON \\
\Bc                         & \hfagTAUBC   \\ 
\Lb                         & \hfagTAULB   \\
\Xib mixture                & \hfagTAUXB   \\
\b-baryon mixture           & \hfagTAUBB   \\
\b-hadron mixture           & \hfagTAUB    \\
\hline
\end{tabular}
\end{center}
\caption{Measured ratios of \b-hadron lifetimes relative to
the \Bd lifetime and ranges predicted
by theory~\cite{tarantino,Gabbiani_et_al}.}
\labt{liferatio}
\begin{center}
\begin{tabular}{lcc} \hline
Lifetime ratio & Measured value & Predicted range \\ \hline
$\tau(\Bu)/\tau(\Bd)$ & \hfagRTAUBU & 1.04 -- 1.08 \\
$\bar{\tau}(\Bs)/\tau(\Bd)^a$ & \hfagRTAUBSMEANCON & 0.99 -- 1.01 \\
$\tau(\Lb)/\tau(\Bd)$ & \hfagRTAULB & 0.86 -- 0.95    \\
$\tau(\mbox{\b-baryon})/\tau(\Bd)$  & \hfagRTAUBB & 0.86 -- 0.95 \\
\hline
\multicolumn{3}{l}{$^a$ \footnotesize 
Using $\bar{\tau}(\Bs) = 1/\Gs = 2/(\Gamma_{\rm L} + \Gamma_{\rm H})$.
}
\end{tabular}
\end{center}
\end{table}
As described in \Sec{lifetimes},
Heavy Quark Effective Theory
can be employed to explain the hierarchy of
$\tau(\Bc) \ll \tau(\Lb) < \bar{\tau}(\Bs) \approx \tau(\Bd) < \tau(\Bu)$,
and used to predict the ratios between lifetimes.
Typical predictions are compared to the measured 
lifetime ratios in \Table{liferatio}.

A recent prediction of the ratio between the \Bu and \Bd lifetimes,
is $1.06 \pm 0.02$~\cite{tarantino}, in good agreement with experiment. 


The total widths of the \Bs and \Bd mesons
are expected to be very close and differ by at most 
1\%~\cite{equal_lifetimes,Gabbiani_et_al}.
However, the experimental ratio $\bar{\tau}(\Bs)/\tau(\Bd)$,
where $\bar{\tau}(\Bs)=1/\Gs$ is obtained from \DGs and 
flavour-specific lifetime measurements, now appears to be 
smaller than 1 by 
\hfagONEMINUSRTAUBSMEANCONpercent, 
at deviation with respect to the prediction. 

The ratio $\tau(\Lb)/\tau(\Bd)$ has particularly
been the source of theoretical
scrutiny since earlier calculations~\cite{OPE,lblife_early}
predicted a value greater than 0.90, almost two sigma higher
than the world average at the time. More recent calculations
of this ratio that include higher-order effects predict a
lower ratio between the
\Lb and \Bd lifetimes~\cite{tarantino,Gabbiani_et_al}
and reduce this difference.
References~\cite{tarantino,Gabbiani_et_al} present probability density functions
of their predictions with variation of theoretical inputs, and the
indicated ranges in \Table{liferatio}
are the RMS of the distributions from the most probable values.



\mysubsection{Neutral \B-meson mixing}
\labs{mixing}

The $\Bd-\Bdbar$ and $\Bs-\Bsbar$ systems
both exhibit the phenomenon of particle-antiparticle mixing. For each of them, 
there are two mass eigenstates which are linear combinations of the two flavour states,
\B and $\bar{\B}$. 
The heaviest (lightest) of the these mass states is denoted
$\B_{\rm H}$ ($\B_{\rm L}$),
with mass $m_{\rm H}$ ($m_{\rm L}$)
and total decay width $\Gamma_{\rm H}$ ($\Gamma_{\rm L}$). We define
\begin{eqnarray}
\Delta m &=& m_{\rm H} - m_{\rm L} \,, \labe{dm} \\
\Delta \Gamma &=& \Gamma_{\rm L} - \Gamma_{\rm H} \,, \labe{dg}
\end{eqnarray}
where $\Delta m$ is positive by definition, and 
$\Delta \Gamma$ is expected to be positive within
the Standard Model.\footnote{For reason of symmetry in 
\Eqss{dm}{dg}, $\Delta \Gamma$ is sometimes defined with 
the opposite sign. The definition adopted here, i.e.\
\Eq{dg}, is the one used by most experimentalists and many
phenomenologists in \B physics.}

There are four different time-dependent probabilities describing the 
case of a neutral \B meson produced 
as a flavour state and decaying to a flavour-specific final state.
If \CPT is conserved (which  
will be assumed throughout), they can be written as 
\begin{equation}
\left\{
\begin{array}{rcl}
{\cal P}(\B\to\B) & = &  \frac{e^{-\Gamma t}}{2} 
\left[ \cosh\!\left(\frac{\Delta\Gamma}{2}t\right) + \cos\!\left(\Delta m t\right)\right]  \\
{\cal P}(\B\to\bar{\B}) & = &  \frac{e^{-\Gamma t}}{2} 
\left[ \cosh\!\left(\frac{\Delta\Gamma}{2}t\right) - \cos\!\left(\Delta m t\right)\right] 
\left|\frac{q}{p}\right|^2 \\
{\cal P}(\bar{\B}\to\B) & = &  \frac{e^{-\Gamma t}}{2} 
\left[ \cosh\!\left(\frac{\Delta\Gamma}{2}t\right) - \cos\!\left(\Delta m t\right)\right] 
\left|\frac{p}{q}\right|^2 \\
{\cal P}(\bar{\B}\to\bar{\B}) & = &  \frac{e^{-\Gamma t}}{2} 
\left[ \cosh\!\left(\frac{\Delta\Gamma}{2}t\right) + \cos\!\left(\Delta m t\right)\right] 
\end{array} \right. \,,
\labe{oscillations}
\end{equation}
where $t$ is the proper time of the system (\ie\ the time interval between the production 
and the decay in the rest frame of the \B meson) and $\Gamma = 
(\Gamma_{\rm H} + \Gamma_{\rm L})/2 =1/\bar{\tau}(\B)$ 
is the average decay width.
At the \B factories, only the proper-time difference $\Delta t$ between the decays
of the two neutral \B mesons from the \Ups can be determined, but, 
because the two \B mesons evolve coherently (keeping opposite flavours as long as
none of them has decayed), the 
above formulae remain valid 
if $t$ is replaced with $\Delta t$ and the production flavour is replaced by the flavour 
at the time of the decay of the accompanying \B meson in a flavour specific state.
As can be seen in the above expressions,
the mixing probabilities 
depend on three mixing observables:
$\Delta m$, $\Delta\Gamma$,
and $|q/p|^2$ which signals \CP violation in the mixing if $|q/p|^2 \ne 1$.

In the next sections we review in turn the experimental knowledge
on these three parameters, separately 
for the \Bd meson (\dmd, \DGd, $|q/p|_{\particle{d}}$) 
and the \Bs meson (\dms, \DGs, $|q/p|_{\particle{s}}$). 

\mysubsubsection{\Bd mixing parameters}

\subsubsubsection{\boldmath \CP violation parameter $|q/p|_{\particle{d}}$}
\labs{qpd}

Evidence for \CP violation in \Bd mixing
has been searched for,
both with flavor-specific and inclusive \Bd decays, 
in samples where the initial 
flavor state is tagged. In the case of semileptonic 
(or other flavor-specific) decays, 
where the final state tag is 
also available, the following asymmetry
\begin{equation} 
 {\cal A}_{\rm SL} = 
\frac{
N(\hbox{\Bdbar}(t) \to \ell^+      \nu_{\ell} X) -
N(\hbox{\Bd}(t)    \to \ell^- \bar{\nu}_{\ell} X) }{
N(\hbox{\Bdbar}(t) \to \ell^+      \nu_{\ell} X) +
N(\hbox{\Bd}(t)    \to \ell^- \bar{\nu}_{\ell} X) } 
= \frac{|p/q|_{\particle{d}}^2 - |q/p|_{\particle{d}}^2}%
{|p/q|_{\particle{d}}^2 + |q/p|_{\particle{d}}^2}
\labe{ASL}
\end{equation} 
has been measured, either in time-integrated analyses at 
CLEO~\cite{CLEO_chid_CP,CLEO_chid_CP_y,CLEO_CP_semi} 
and CDF~\cite{CDF_CP_semi}, or in time-dependent analyses at 
OPAL~\cite{OPAL_CP_semi}, ALEPH~\cite{ALEPH_CP}, 
\babar~\cite{BABAR_DGd_qp,BABAR_CP_semi} and 
\belle~\cite{BELLE_CP_semi}.
In the inclusive case, also investigated and published
at ALEPH~\cite{ALEPH_CP} and OPAL~\cite{OPAL_CP_incl},
no final state tag is used, and the asymmetry~\cite{incl_asym}
\begin{equation} 
\frac{
N(\hbox{\Bd}(t) \to {\rm all}) -
N(\hbox{\Bdbar}(t) \to {\rm all}) }{
N(\hbox{\Bd}(t) \to {\rm all}) +
N(\hbox{\Bdbar}(t) \to {\rm all}) } 
\simeq
{\cal A}_{\rm SL} \left[ \frac{\dmd}{2\Gd} \sin(\dmd \,t) - 
\sin^2\left(\frac{\dmd \,t}{2}\right)\right] 
\labe{ASLincl}
\end{equation} 
must be measured as a function of the proper time to extract information 
on \CP violation.
In all cases asymmetries compatible with zero have been found,  
with a precision limited by the available statistics. A simple 
average of all published 
results for the \Bd 
meson~\cite{CLEO_chid_CP_y,CLEO_CP_semi,OPAL_CP_semi,ALEPH_CP,BABAR_DGd_qp,BABAR_CP_semi,BELLE_CP_semi,OPAL_CP_incl}
yields 
\begin{equation}
{\cal A}_{\rm SL} = \hfagASL 
\end{equation}
or, equivalently through \Eq{ASL},
\begin{equation}
|q/p|_{\particle{d}} = \hfagQP \,.
\end{equation}
This result\footnote{Early analyses and (perhaps hence) the PDG use the complex
parameter $\epsilon_{\B} = (p-q)/(p+q)$; if \CP violation in the mixing in small, 
${\cal A}_{\rm SL} \cong 4 {\rm Re}(\epsilon_{\B})/(1+|\epsilon_{\B}|^2)$ and our 
current world average  
is ${\rm Re}(\epsilon_{\B})/(1+|\epsilon_{\B}|^2)=\hfagREB$.}, 
summarized in \Table{qoverp},
is compatible 
with no \CP violation in the mixing, an assumption we make for the rest 
of this section.

\begin{table}
\caption{Measurements of \CP violation in \Bd mixing and their average in terms of 
both ${\cal A}_{\rm SL}$ and $|q/p|_{\particle{d}}$.
The individual results are listed as quoted in the original publications, 
or converted\addtocounter{footnote}{-1}\protect\footnotemark\
to an ${\cal A}_{\rm SL}$ value.
When two errors are quoted, the first one is statistical and the second one systematic.
}
\labt{qoverp}
\begin{center}
\begin{tabular}{@{}rcl@{$\,\pm$}l@{$\pm$}ll@{$\,\pm$}l@{$\pm$}l@{}}
\hline
Exp.\ \& Ref. & Method & \multicolumn{3}{c}{Measured ${\cal A}_{\rm SL}$} 
                       & \multicolumn{3}{c}{Measured $|q/p|_{\particle{d}}$} \\
\hline
CLEO   \cite{CLEO_chid_CP_y} & partial hadronic rec. 
                             & $+0.017$ & 0.070 & 0.014 
                             & \multicolumn{3}{c}{} \\
CLEO   \cite{CLEO_CP_semi}   & dileptons 
                             & $+0.013$ & 0.050 & 0.005 
                             & \multicolumn{3}{c}{} \\
CLEO   \cite{CLEO_CP_semi}   & average of above two 
                             & $+0.014$ & 0.041 & 0.006 
                             & \multicolumn{3}{c}{} \\
OPAL   \cite{OPAL_CP_semi}   & leptons     
                             & $+0.008$ & 0.028 & 0.012 
                             & \multicolumn{3}{c}{} \\
OPAL   \cite{OPAL_CP_incl}   & inclusive (\Eq{ASLincl}) 
                             & $+0.005$ & 0.055 & 0.013 
                             & \multicolumn{3}{c}{} \\
ALEPH  \cite{ALEPH_CP}       & leptons 
                             & $-0.037$ & 0.032 & 0.007 
                             & \multicolumn{3}{c}{} \\
ALEPH  \cite{ALEPH_CP}       & inclusive (\Eq{ASLincl}) 
                             & $+0.016$ & 0.034 & 0.009 
                             & \multicolumn{3}{c}{} \\
ALEPH  \cite{ALEPH_CP}       & average of above two 
                             & $-0.013$ & \multicolumn{2}{l}{0.026 (tot)} 
                             & \multicolumn{3}{c}{} \\
\babar \cite{BABAR_CP_semi}  & dileptons
                             & $+0.005$ & 0.012 & 0.014 
                             & 0.998 & 0.006 & 0.007 \\ 
\babar \cite{BABAR_DGd_qp}   & full hadronic rec. 
                             & \multicolumn{3}{c}{}  
                             & $1.029$ & 0.013 & 0.011  \\
\belle \cite{BELLE_CP_semi}  & dileptons 
                             & $-0.0011$ & 0.0079 & 0.0070 
                             & 1.0005 & 0.0040 & 0.0035 \\
\hline
& Average of all above       & \multicolumn{3}{l}{\hfagASL\ (tot)} 
                             & \multicolumn{3}{l}{\hfagQP\  (tot)} \\ 
\hline
\end{tabular}
\end{center}
\end{table}

\subsubsubsection{\boldmath Mass and decay width differences \dmd and \DGd}
\labs{dmd}
\labs{DGd}

\begin{table}
\caption{Time-dependent measurements included in the \dmd average.
The results obtained from multi-dimensional fits involving also 
the \Bd (and \Bu) lifetimes
as free parameter(s)~\cite{BABAR3,BABAR5,BELLE2} 
have been converted into one-dimensional measurements of \dmd.
All the measurements have then been adjusted to a common set of physics
parameters before being combined. 
The CDF2 and \dzero results are preliminary.}
\labt{dmd}
\begin{center}
\begin{tabular}{@{}rc@{}cc@{}c@{}cc@{}c@{}c@{}}
\hline
Experiment & \multicolumn{2}{c}{Method} & \multicolumn{3}{l}{\dmd in\invps}   
                                        & \multicolumn{3}{l}{\dmd in\invps}     \\
and Ref.   &  rec. & tag                & \multicolumn{3}{l}{before adjustment} 
                                        & \multicolumn{3}{l}{after adjustment} \\
\hline
 ALEPH~\cite{ALEPH_dmd}  & \particle{ \ell  } & \particle{ \Qjet  } & $  0.404 $ & $ \pm  0.045 $ & $ \pm  0.027 $ & & & \\
 ALEPH~\cite{ALEPH_dmd}  & \particle{ \ell  } & \particle{ \ell  } & $  0.452 $ & $ \pm  0.039 $ & $ \pm  0.044 $ & & & \\
 ALEPH~\cite{ALEPH_dmd}  & \multicolumn{2}{c}{above two combined} & $  0.422 $ & $ \pm  0.032 $ & $ \pm  0.026 $ & $  0.441 $ & $ \pm  0.032 $ & $ ^{+  0.021 }_{-  0.020 } $ \\
 ALEPH~\cite{ALEPH_dmd}  & \particle{ D^*  } & \particle{ \ell,\Qjet  } & $  0.482 $ & $ \pm  0.044 $ & $ \pm  0.024 $ & $  0.482 $ & $ \pm  0.044 $ & $ \pm  0.024 $ \\
 DELPHI~\cite{DELPHI_dmd}  & \particle{ \ell  } & \particle{ \Qjet  } & $  0.493 $ & $ \pm  0.042 $ & $ \pm  0.027 $ & $  0.504 $ & $ \pm  0.042 $ & $ \pm  0.024 $ \\
 DELPHI~\cite{DELPHI_dmd}  & \particle{ \pi^*\ell  } & \particle{ \Qjet  } & $  0.499 $ & $ \pm  0.053 $ & $ \pm  0.015 $ & $  0.501 $ & $ \pm  0.053 $ & $ \pm  0.015 $ \\
 DELPHI~\cite{DELPHI_dmd}  & \particle{ \ell  } & \particle{ \ell  } & $  0.480 $ & $ \pm  0.040 $ & $ \pm  0.051 $ & $  0.489 $ & $ \pm  0.040 $ & $ ^{+  0.049 }_{-  0.048 } $ \\
 DELPHI~\cite{DELPHI_dmd}  & \particle{ D^*  } & \particle{ \Qjet  } & $  0.523 $ & $ \pm  0.072 $ & $ \pm  0.043 $ & $  0.518 $ & $ \pm  0.072 $ & $ \pm  0.043 $ \\
 DELPHI~\cite{DELPHI_dmd_dms_vtx}  & \particle{ \mbox{vtx}  } & \particle{ \mbox{comb}  } & $  0.531 $ & $ \pm  0.025 $ & $ \pm  0.007 $ & $  0.530 $ & $ \pm  0.025 $ & $ \pm  0.006 $ \\
 L3~\cite{L3_dmd}  & \particle{ \ell  } & \particle{ \ell  } & $  0.458 $ & $ \pm  0.046 $ & $ \pm  0.032 $ & $  0.470 $ & $ \pm  0.046 $ & $ \pm  0.029 $ \\
 L3~\cite{L3_dmd}  & \particle{ \ell  } & \particle{ \Qjet  } & $  0.427 $ & $ \pm  0.044 $ & $ \pm  0.044 $ & $  0.436 $ & $ \pm  0.044 $ & $ \pm  0.042 $ \\
 L3~\cite{L3_dmd}  & \particle{ \ell  } & \particle{ \ell\mbox{(IP)}  } & $  0.462 $ & $ \pm  0.063 $ & $ \pm  0.053 $ & $  0.481 $ & $ \pm  0.063 $ & $ \pm  0.047 $ \\
 OPAL~\cite{OPAL_dmd_dilepton}  & \particle{ \ell  } & \particle{ \ell  } & $  0.430 $ & $ \pm  0.043 $ & $ ^{+  0.028 }_{-  0.030 } $ & $  0.462 $ & $ \pm  0.043 $ & $ ^{+  0.018 }_{-  0.017 } $ \\
 OPAL~\cite{OPAL_dmd_lepton}  & \particle{ \ell  } & \particle{ \Qjet  } & $  0.444 $ & $ \pm  0.029 $ & $ ^{+  0.020 }_{-  0.017 } $ & $  0.467 $ & $ \pm  0.029 $ & $ ^{+  0.015 }_{-  0.014 } $ \\
 OPAL~\cite{OPAL_dmd_dstar}  & \particle{ D^*\ell  } & \particle{ \Qjet  } & $  0.539 $ & $ \pm  0.060 $ & $ \pm  0.024 $ & $  0.544 $ & $ \pm  0.060 $ & $ \pm  0.023 $ \\
 OPAL~\cite{OPAL_dmd_dstar}  & \particle{ D^*  } & \particle{ \ell  } & $  0.567 $ & $ \pm  0.089 $ & $ ^{+  0.029 }_{-  0.023 } $ & $  0.571 $ & $ \pm  0.089 $ & $ ^{+  0.028 }_{-  0.022 } $ \\
 OPAL~\cite{OPAL_dmd_slowpion}  & \particle{ \pi^*\ell  } & \particle{ \Qjet  } & $  0.497 $ & $ \pm  0.024 $ & $ \pm  0.025 $ & $  0.496 $ & $ \pm  0.024 $ & $ \pm  0.025 $ \\
 CDF1~\cite{CDF1_dmd_dlepton_SST}  & \particle{ D\ell  } & \particle{ \mbox{SST}  } & $  0.471 $ & $ ^{+  0.078 }_{-  0.068 } $ & $ ^{+  0.033 }_{-  0.034 } $ & $  0.470 $ & $ ^{+  0.078 }_{-  0.068 } $ & $ ^{+  0.033 }_{-  0.034 } $ \\
 CDF1~\cite{CDF1_dmd_dimuon}  & \particle{ \mu  } & \particle{ \mu  } & $  0.503 $ & $ \pm  0.064 $ & $ \pm  0.071 $ & $  0.513 $ & $ \pm  0.064 $ & $ \pm  0.070 $ \\
 CDF1~\cite{CDF1_dmd_lepton}  & \particle{ \ell  } & \particle{ \ell,\Qjet  } & $  0.500 $ & $ \pm  0.052 $ & $ \pm  0.043 $ & $  0.537 $ & $ \pm  0.052 $ & $ \pm  0.036 $ \\
 CDF1~\cite{CDF1_dmd_dstarlepton}  & \particle{ D^*\ell  } & \particle{ \ell  } & $  0.516 $ & $ \pm  0.099 $ & $ ^{+  0.029 }_{-  0.035 } $ & $  0.523 $ & $ \pm  0.099 $ & $ ^{+  0.028 }_{-  0.035 } $ \\
 CDF2~\cite{CDF2_dmd_dlepton_preliminary}  & \particle{ D^{(*)}\ell  } & \particle{ \mbox{OST}  } & $  0.511 $ & $ \pm  0.020 $ & $ \pm  0.014 $ & $  0.511 $ & $ \pm  0.020 $ & $ \pm  0.014 $ \\
 CDF2~\cite{CDF2_dmd_exclusive_preliminary}  & \particle{ \Bd  } & \particle{ \mbox{comb}  } & $  0.536 $ & $ \pm  0.028 $ & $ \pm  0.006 $ & $  0.536 $ & $ \pm  0.028 $ & $ \pm  0.006 $ \\
 \dzero~\cite{D0_dmd_preliminary}  & \particle{ D^{(*)}\mu  } & \particle{ \mbox{OST}  } & $  0.498 $ & $ \pm  0.026 $ & $ \pm  0.016 $ & $  0.498 $ & $ \pm  0.026 $ & $ \pm  0.016 $ \\
 \babar~\cite{BABAR_dmd_full}  & \particle{ \Bd  } & \particle{ \ell,K,\mbox{NN}  } & $  0.516 $ & $ \pm  0.016 $ & $ \pm  0.010 $ & $  0.520 $ & $ \pm  0.016 $ & $ \pm  0.008 $ \\
 \babar~\cite{BABAR_dmd_dilepton}  & \particle{ \ell  } & \particle{ \ell  } & $  0.493 $ & $ \pm  0.012 $ & $ \pm  0.009 $ & $  0.489 $ & $ \pm  0.012 $ & $ \pm  0.006 $ \\
 \babar~\cite{BABAR5}  & \particle{ D^*\ell\nu\mbox{(part)}  } & \particle{ \ell  } & $  0.511 $ & $ \pm  0.007 $ & $ \pm  0.007 $ & $  0.512 $ & $ \pm  0.007 $ & $ \pm  0.007 $ \\
 \babar~\cite{BABAR3}  & \particle{ D^*\ell\nu  } & \particle{ \ell,K,\mbox{NN}  } & $  0.492 $ & $ \pm  0.018 $ & $ \pm  0.014 $ & $  0.491 $ & $ \pm  0.018 $ & $ \pm  0.013 $ \\
 \belle~\cite{BELLE_dmd_dstarpi_partial}  & \particle{ D^*\pi\mbox{(part)}  } & \particle{ \ell  } & $  0.509 $ & $ \pm  0.017 $ & $ \pm  0.020 $ & $  0.512 $ & $ \pm  0.017 $ & $ \pm  0.019 $ \\
 \belle~\cite{BELLE_dmd_dilepton}  & \particle{ \ell  } & \particle{ \ell  } & $  0.503 $ & $ \pm  0.008 $ & $ \pm  0.010 $ & $  0.506 $ & $ \pm  0.008 $ & $ \pm  0.009 $ \\
 \belle~\cite{BELLE2}  & \particle{ \Bd,D^*\ell\nu  } & \particle{ \mbox{comb}  } & $  0.511 $ & $ \pm  0.005 $ & $ \pm  0.006 $ & $  0.512 $ & $ \pm  0.005 $ & $ \pm  0.006 $ \\
 \hline \\[-2.0ex]
 \multicolumn{6}{l}{World average (all above measurements included):} & $  0.508 $ & $ \pm  0.003 $ & $ \pm  0.003 $ \\

\\[-2.0ex]
\multicolumn{6}{l}{~~~ -- ALEPH, DELPHI, L3, OPAL and CDF1 only:}
     & \hfagDMDHval & \hfagDMDHsta & \hfagDMDHsys \\
\multicolumn{6}{l}{~~~ -- Above measurements of \babar and \belle only:}
     & \hfagDMDBval & \hfagDMDBsta & \hfagDMDBsys \\
\hline
\end{tabular}
\end{center}
\end{table}

Many time-dependent \Bd--\Bdbar oscillation analyses have been performed by the 
ALEPH, \babar, \belle, CDF, \dzero, DELPHI, L3 and OPAL collaborations. 
The corresponding measurements of \dmd are summarized in 
\Table{dmd},
where only the most recent results
are listed (\ie\ measurements superseded by more recent ones have been omitted). 
Although a variety of different techniques have been used, the 
individual \dmd
results obtained at high-energy colliders have remarkably similar precision.
Their average is compatible with the recent and more precise measurements 
from the asymmetric \B factories.
The systematic uncertainties are not negligible; 
they are often dominated by sample composition, mistag probability,
or \b-hadron lifetime contributions.
Before being combined, the measurements are adjusted on the basis of a 
common set of input values, including the averages of the 
\b-hadron fractions and lifetimes given in this report 
(see \Secss{fractions}{lifetimes}).
Some measurements are statistically correlated. 
Systematic correlations arise both from common physics sources 
(fractions, lifetimes, branching ratios of \b hadrons), and from purely 
experimental or algorithmic effects (efficiency, resolution, flavour tagging, 
background description). Combining all published measurements
listed in \Table{dmd}
and accounting for all identified correlations
as described in~\cite{LEPHFS} yields $\dmd = \hfagDMDWfull$.

On the other hand, ARGUS and CLEO have published 
measurements of the time-integrated mixing probability 
\chid~\cite{ARGUS_chid,CLEO_chid_CP,CLEO_chid_CP_y}, 
which average to $\chid =\hfagCHIDU$.
Following \Ref{CLEO_chid_CP_y}, 
the width difference \DGd could 
in principle be extracted from the
measured value of $\Gd=1/\tau(\Bd)$ and the above averages for 
\dmd and \chid 
(provided that \DGd has a negligible impact on 
the \dmd $\tau(\Bd)$ analyses that have assumed $\DGd=0$), 
using the relation
\begin{equation}
\chid = \frac{\xd^2+\yd^2}{2(\xd^2+1)} ~~~ \mbox{with} ~~ \xd=\frac{\dmd}{\Gd} 
~~~ \mbox{and} ~~ \yd=\frac{\DGd}{2\Gd} \,.
\labe{chid_definition}
\end{equation}
However, direct time-dependent studies provide much stronger constraints: 
$|\DGd|/\Gd < 18\%$ at \CL{95} from DELPHI~\cite{DELPHI_dmd_dms_vtx},
and $-6.8\% < {\rm sign}({\rm Re} \lambda_{\CP}) \DGGd < 8.4\%$
at \CL{90} from BABAR~\cite{BABAR_DGd_qp},
where $\lambda_{\CP} = (q/p)_{\particle{d}} (\bar{A}_{\CP}/A_{\CP})$
is defined for a \CP-even final state 
(the sensitivity to the overall sign of 
${\rm sign}({\rm Re} \lambda_{\CP}) \DGGd$ comes
from the use of \Bd decays to \CP final states).
Combining these two results after adjustment to 
$1/\Gd=\tau(\Bd)=\hfagTAUBD$ yields
\begin{equation}
{\rm sign}({\rm Re} \lambda_{\CP}) \DGGd  = \hfagSDGDGD \,.
\end{equation}
The sign of ${\rm Re} \lambda_{\CP}$ is not measured,
but expected to be positive from the global fits
of the Unitarity Triangle within the Standard Model.

Assuming $\DGd=0$ 
and using $1/\Gd=\tau(\Bd)=\hfagTAUBD$,
the \dmd and \chid results are combined through \Eq{chid_definition} 
to yield the 
world average
\begin{equation} 
\dmd = \hfagDMDWU \,,
\labe{dmd}
\end{equation} 
or, equivalently,
\begin{equation} 
\xd= \hfagXDWU ~~~ \mbox{and} ~~~ \chid=\hfagCHIDWU \,.  
\labe{chid}
\end{equation}
\Figure{dmd} compares the \dmd values obtained by the different experiments.

\begin{figure}
\begin{center}
\epsfig{figure=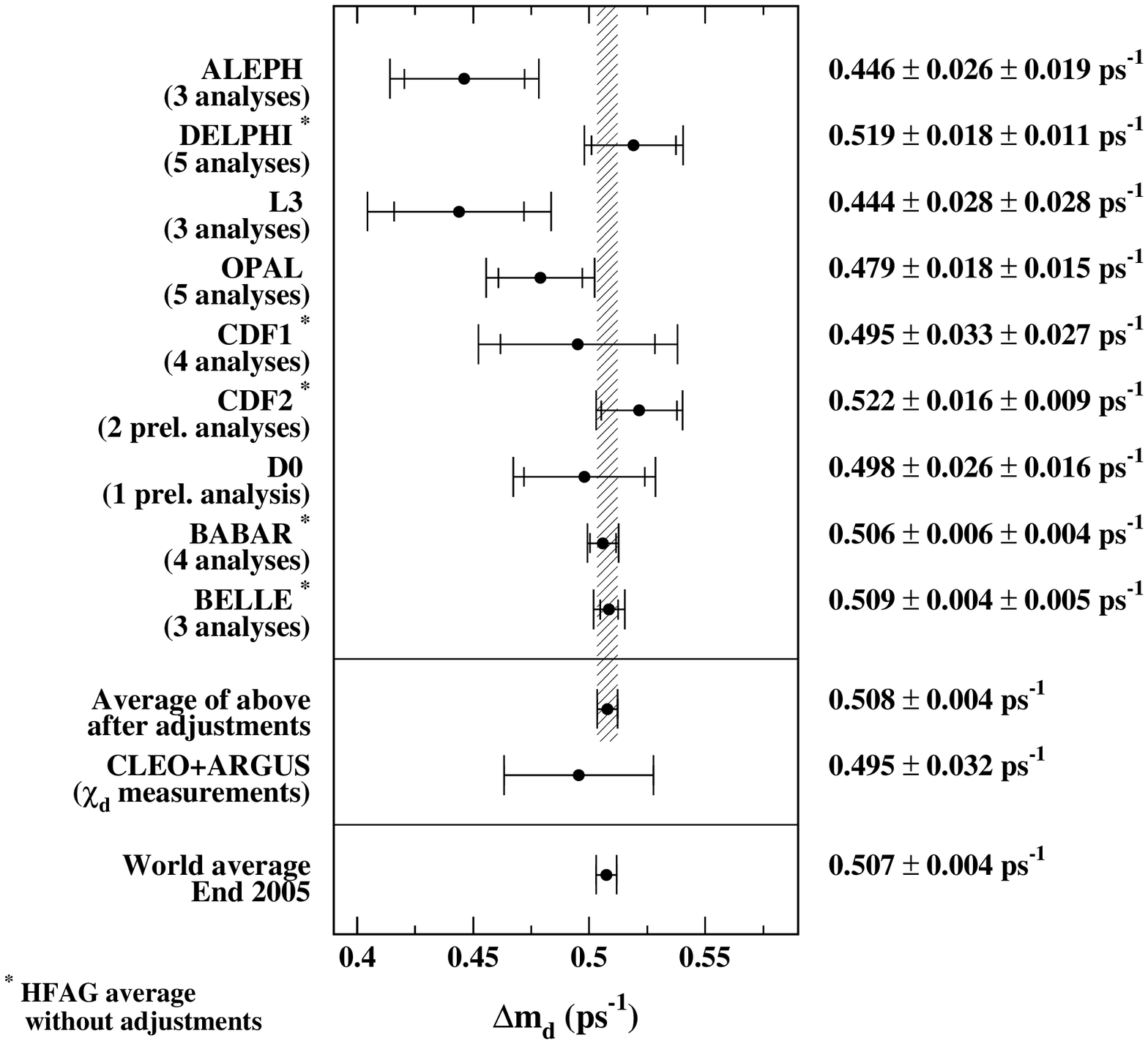,width=\textwidth}
\caption{The \Bd--\Bdbar oscillation frequency \dmd as measured by the different experiments. 
The averages quoted for ALEPH, L3 and OPAL are taken from the original publications, while the 
ones for DELPHI, CDF, \babar, and \belle have been computed from the individual results 
listed in \Table{dmd} without performing any adjustments. The time-integrated measurements 
of \chid from the symmetric \B factory experiments ARGUS and CLEO have been converted 
to a \dmd value using $\tau(\Bd)=\hfagTAUBD$. The two global averages have been obtained 
after adjustments of all the individual \dmd results of \Table{dmd} (see text).}
\labf{dmd}
\end{center}
\end{figure}

The \Bd mixing averages given in \Eqss{dmd}{chid}
and the \b-hadron fractions of \Table{fractions} have been obtained in a fully 
consistent way, taking into account the fact that the fractions are computed using 
the \chid value of \Eq{chid} and that many individual measurements of \dmd
at high energy depend on the assumed values for the \b-hadron fractions.
Furthermore, this set of averages is consistent with the lifetime averages 
of \Sec{lifetimes}.

\begin{table}
\caption{Simultaneous measurements of \dmd and $\tau(\Bd)$, and their average.
The \belle analysis also 
measures $\tau(\Bu)$ at the same time, but it is converted here into a two-dimensional measurement 
of \dmd and $\tau(\Bd)$, for an assumed value of $\tau(\Bu)$. 
The first quoted error on the measurements is statistical
and the second one systematic; in the case of adjusted measurements, the 
latter includes a contribution obtained from the variation of $\tau(\Bu)$ or 
$\tau(\Bu)/\tau(\Bd)$ in the indicated range. Units are\invps\ for \dmd
and\unit{ps} for lifetimes. 
The three different values of $\rho(\dmd,\tau(\Bd))$ correspond 
to the statistical, systematic and total correlation coefficients
between the adjusted measurements of \dmd and $\tau(\Bd)$.}
\labt{dmd2D}
\begin{center}

\begin{tabular}{@{}r@{~}c@{}c@{}c@{~}c@{}c@{}c@{~}c@{}c@{}c@{\hspace{0ex}}c@{}}
\hline
Exp.\ \& Ref.
& \multicolumn{3}{c}{Measured \dmd}   
& \multicolumn{3}{c}{Measured $\tau(\Bd)$}   
& \multicolumn{3}{c}{Measured $\tau(\Bu)$}   
&  Assumed $\tau(\Bu)$ \\
\hline
\babar \cite{BABAR3}  
      & $0.492$ & $\pm 0.018$ & $\pm 0.013$ 
      & $1.523$ & $\pm 0.024$ & $\pm 0.022$ 
      & \multicolumn{3}{c}{---}
      & $(1.083$$\pm 0.017)\tau(\Bd)$ \\  
\babar \cite{BABAR5}  
      & $0.511$ & $\pm 0.007$ & $^{+0.007}_{-0.006}$ 
      & $1.504$ & $\pm 0.013$ & $^{+0.018}_{-0.013}$
      & \multicolumn{3}{c}{---}
      & $1.671$$\pm 0.018$ \\  
\belle \cite{BELLE2}  
      & $0.511$ & $\pm 0.005$ & $\pm 0.006$
      & $1.534$ & $\pm 0.008$ & $\pm 0.010$
      & $1.635$ & $\pm 0.011$ & $\pm 0.011$
      & --- \\  
\cline{2-10}
& \multicolumn{3}{c}{Adjusted \dmd}   
& \multicolumn{3}{c}{Adjusted $\tau(\Bd)$}   
& \multicolumn{3}{c}{$\rho(\dmd,\Bd)$} 
\\
\cline{2-10}
\babar \cite{BABAR3}  
      & $0.492$ & $\pm 0.018$ & $\pm 0.013$ 
      & $1.524$ & $\pm 0.025$ & $\pm 0.022$ 
      & $-0.22$ & $+0.74$ & $+0.16$ 
      & $(\hfagRTAUBUval$$\hfagRTAUBUerr)\tau(\Bd)$ \\  
\babar \cite{BABAR5} 
      & $0.512$ & $\pm 0.007$ & $\pm 0.007$ 
      & $1.506$ & $\pm 0.013$ & $\pm 0.018$
      & $+0.01$ & $-0.85$ & $-0.48$ 
      & $\hfagTAUBUval$$\hfagTAUBUerr$ \\  
\belle \cite{BELLE2}  
      & $0.510$ & $\pm 0.007$ & $\pm 0.005$
      & $1.535$ & $\pm 0.009$ & $\pm 0.009$
      & $-0.27$ & $-0.08$ & $-0.19$ 
      & $\hfagTAUBUval$$\hfagTAUBUerr$ \\  
\hline
\multicolumn{1}{l}{Average} 
      & \hfagDMDTWODval   & \hfagDMDTWODsta   & \hfagDMDTWODsys
      & \hfagTAUBDTWODval & \hfagTAUBDTWODsta & \hfagTAUBDTWODsys
      & \hfagRHOstaDMDTAUBD & \hfagRHOsysDMDTAUBD & \hfagRHODMDTAUBD 
      & $\hfagTAUBUval$$\hfagTAUBUerr$ \\  
\hline 
\end{tabular}
\end{center}
\end{table}

\begin{figure}
\begin{center}
\epsfig{figure=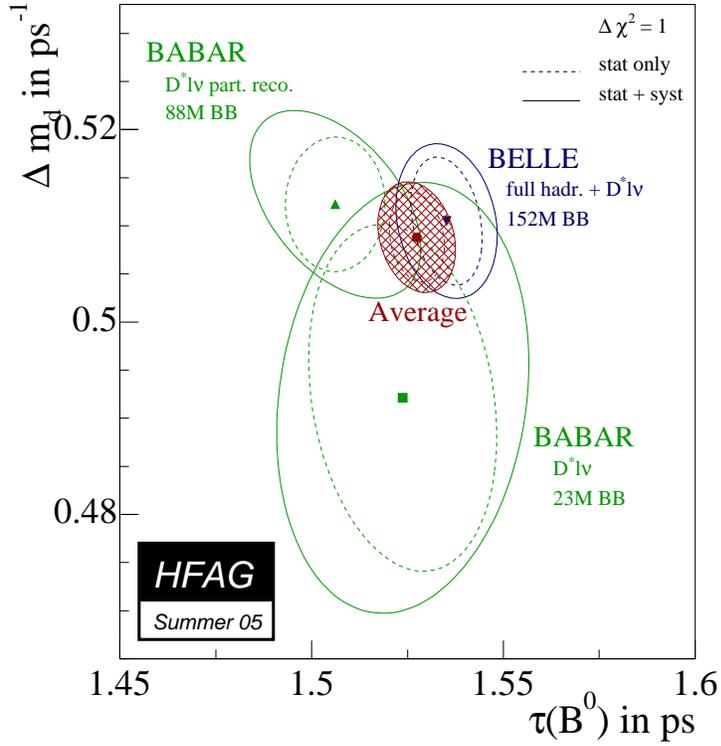,width=0.6\textwidth}
\caption{Simultaneous measurements of
\dmd and $\tau(\Bd)$~\cite{BABAR3,BABAR5,BELLE2}, 
after adjustment to a common set of parameters (see text). 
Statistical and total uncertainties are represented as dashed and
solid contours respectively.
The average of the three measurements
is indicated by a hatched ellipse.}
\labf{dmd2D}
\end{center}
\end{figure}

It should be noted that the most recent (and precise) analyses at the 
asymmetric \B factories measure \dmd
as a result of a multi-dimensional fit. 
Two \babar analyses~\cite{BABAR3,BABAR5},  
based on fully and partially reconstructed $\Bd \to D^*\ell\nu$ decays
respectively, 
extract simultaneously \dmd and $\tau(\Bd)$
while the latest \belle analysis~\cite{BELLE2},  
based on fully reconstructed hadronic \Bd decays and $\Bd \to D^*\ell\nu$ decays, 
extracts simultaneously \dmd, $\tau(\Bd)$ and $\tau(\Bu)$.
The measurements of \dmd and $\tau(\Bd)$ of these three analyses 
are displayed in \Table{dmd2D} and in \Fig{dmd2D}. Their two-dimensional average, 
taking into account all statistical and systematic correlations, and expressed
at $\tau(\Bu)=\hfagTAUBU$, is
\begin{equation}
\left.
\begin{array}{r@{}l}
\dmd = \hfagDMDTWODnounit & \invps \\
\tau(\Bd) = \hfagTAUBDTWODnounit & \ps
\end{array}
\right\}
~\mbox{with a total correlation of \hfagRHODMDTAUBD.}
\end{equation}

\mysubsubsection{\Bs mixing parameters}

\subsubsubsection{\boldmath \CP violation parameter $|q/p|_{\particle{s}}$}
\labs{qps}

No measurement or experimental limit exists on $|q/p|_{\particle{s}}$,
except in the form of a relatively weak constraint from CDF 
on a combination of $|q/p|_{\particle{d}}$ and $|q/p|_{\particle{s}}$,
$f'_{\particle{d}} \,\chid(1-|q/p|^2_{\particle{d}})+
 f'_{\particle{s}} \,\chis(1-|q/p|^2_{\particle{s}})=
0.006\pm 0.017$~\cite{CDF_CP_semi},
using inclusive semileptonic decays of \b hadrons. 
The result is compatible with no \CP violation in the 
mixing, an assumption made in all results described below. 

\subsubsubsection{\boldmath Mass difference \dms}
\labs{dms}

The time-integrated measurements of \chibar (see \Sec{chibar}), when compared to our knowledge
of \chid and the \b-hadron fractions, indicate that \Bs mixing is large, with a value of 
\chis close to its maximal possible value of $1/2$.
However, the time dependence of this mixing (called \Bs oscillations) has not been 
observed yet, mainly because the period of these oscillations turns out to be so small 
that it can't be resolved with the proper-time resolutions achieved so far. 

The statistical significance ${\cal S}$ of a \Bs oscillation signal can be
approximated as~\cite{amplitude}
\begin{equation}
{\cal S} \approx \sqrt{\frac{N}{2}} \,f_{\rm sig}\, (1-2w)\,
\exp{\left(-\left(\dms\sigma_t\right)^2/2\right)}\,,
\labe{significance}
\end{equation}
where $N$ is 
the number of selected and tagged \Bs candidates, 
$f_{\rm sig}$ is the fraction of \Bs signal
in the selected and tagged sample, $w$ is the total mistag probability, 
and $\sigma_t$ is the resolution on proper time.
As can be seen, the quantity ${\cal S}$ decreases very quickly as 
\dms increases: this dependence is controlled by $\sigma_t$, 
which is therefore the most critical parameter for \dms analyses. 
The method widely used for \Bs oscillation searches
consists of measuring a \Bs oscillation amplitude ${\cal A}$
at several different test values of \dms, 
using a maximum likelihood fit based on the functions 
of \Eq{oscillations} where the cosine terms have been multiplied 
by ${\cal A}$.
One expects ${\cal A}=1$ at the true 
value of \dms and to ${\cal A}=0$ at a test value of \dms 
(far) below the true value.
To a good approximation, the statistical uncertainty on ${\cal A}$
is Gaussian and equal to $1/{\cal S}$~\cite{amplitude}.

\begin{figure}
\begin{center}
\epsfig{figure=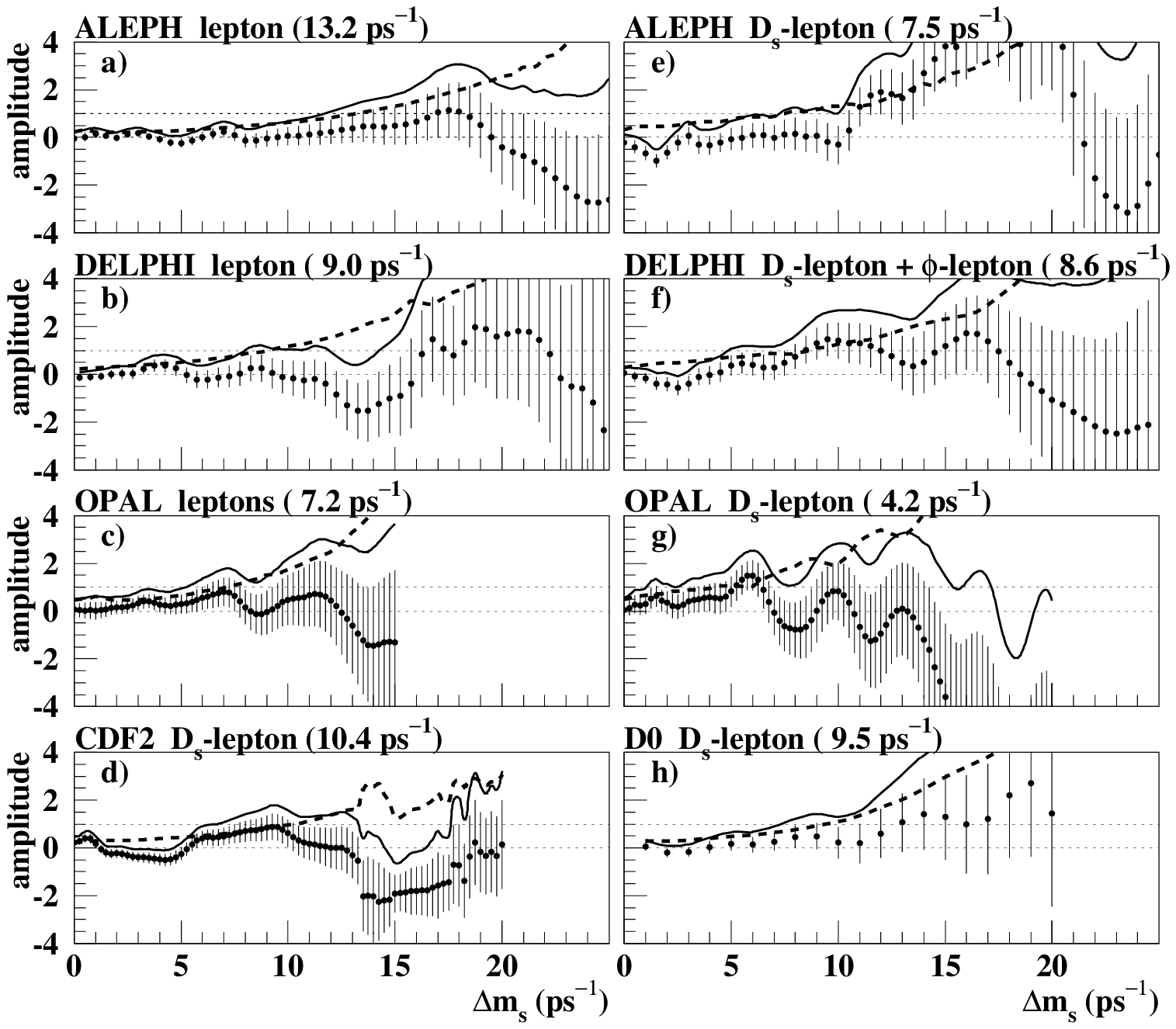,width=\textwidth,%
bbllx=55,bblly=55,bburx=490,bbury=490}
\caption{\Bs-oscillation amplitude spectra, displayed separately for each 
\Bs oscillation analysis. 
The points and error bars represent the measurements of the amplitude ${\cal A}$ and 
their total uncertainties $\sigma_{\cal A}$, adjusted to a set of physics parameters
common to all analyses (including $\fBs=\hfagFBS$).
Values of \dms where the solid curve 
(${\cal A}+1.645\,\sigma_{\cal A}$) is below 1 are excluded at \CL{95}. 
The dashed curve shows $1.645\,\sigma_{\cal A}$; the number in parenthesis indicates where 
this curve is equal to 1, and is a measure of the sensitivity of the analysis. 
a) ALEPH inclusive lepton~\cite{ALEPH_dms},
b) DELPHI inclusive lepton~\cite{DELPHI_dms_last}, 
c) OPAL inclusive lepton and dilepton~\cite{OPAL_dms_l},
d) CDF2 \particle{D_s}-$\ell$ (preliminary)~\cite{CDF2_dms_dslnu_prel},
e) ALEPH \particle{D_s}-$\ell$~\cite{ALEPH_dms}, 
f) DELPHI \particle{D_s}-$\ell$~\cite{DELPHI_dms_last} and $\phi$-$\ell$~\cite{DELPHI_dms_dgs},
g) OPAL \particle{D_s}-$\ell$~\cite{OPAL_dms_dsl},
h) \dzero \particle{D_s}-$\mu$ (preliminary)~\cite{D0_dms_dsmux_prel}.
Continuation on \Fig{individual_amplitudes_2}.}
\labf{individual_amplitudes_1}
\end{center}
\end{figure}

\begin{figure}
\begin{center}
\epsfig{figure=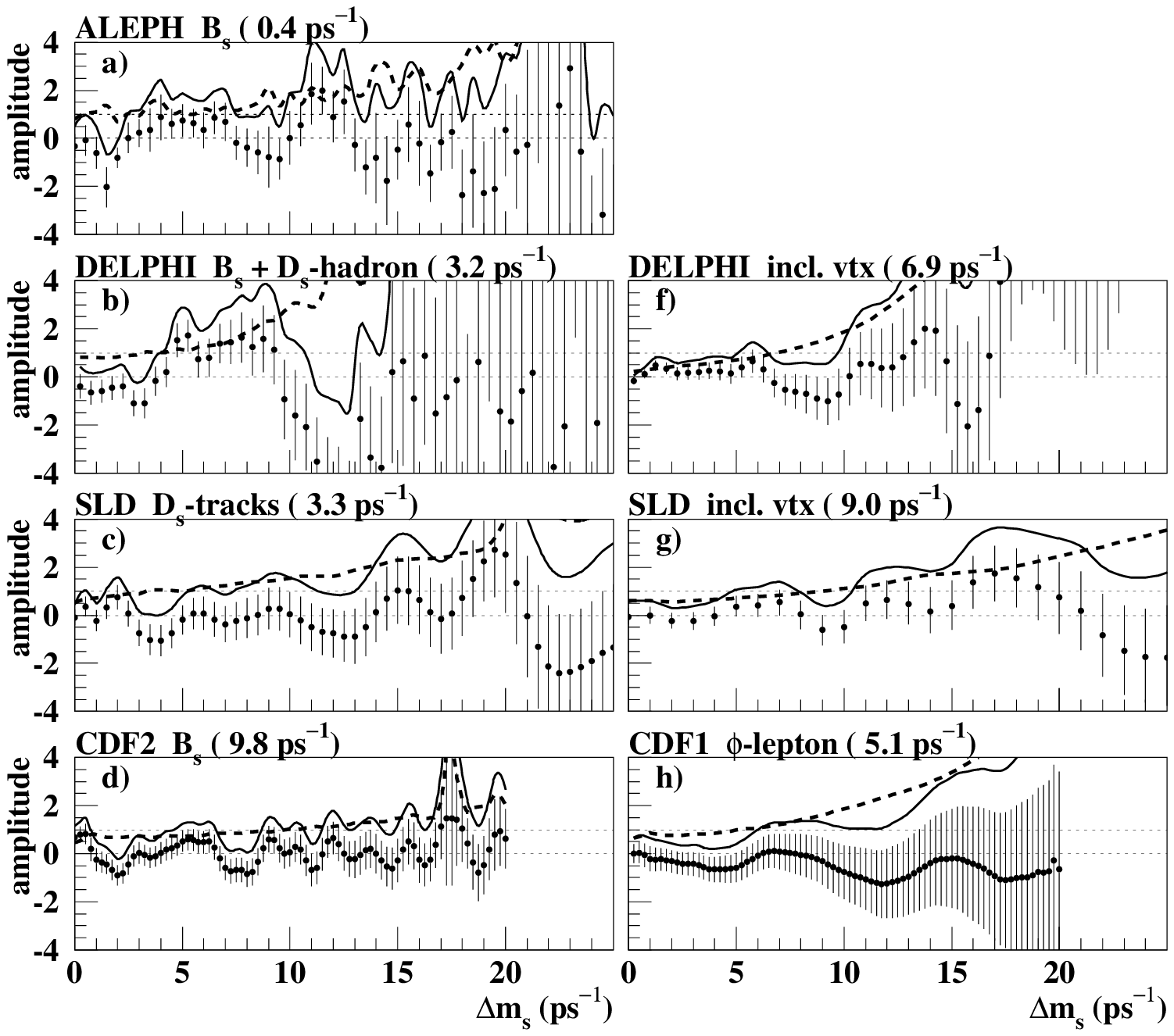,width=\textwidth,%
bbllx=55,bblly=55,bburx=490,bbury=490}
\caption{(continuation of \Fig{individual_amplitudes_1})
\Bs-oscillation amplitude spectra, displayed separately for each 
\Bs oscillation analysis, in the same manner as in \Fig{individual_amplitudes_1}. 
a) ALEPH fully reconstructed \Bs~\cite{ALEPH_dms}, 
b) DELPHI fully reconstructed \Bs and \particle{D_s}-hadron~\cite{DELBS1_dms_excl},
c) SLD \particle{D_s}+tracks~\cite{SLD_dms_ds},
d) CDF2 fully reconstructed \Bs (preliminary)~\cite{CDF2_dms_dspi_prel},
f) DELPHI inclusive vertex~\cite{DELPHI_dmd_dms_vtx}, 
g) SLD inclusive vertex dipole~\cite{SLD_dms_dipole},
h) CDF1 $\phi$-$\ell$~\cite{CDF1_dms}.}
\labf{individual_amplitudes_2}
\end{center}
\end{figure}

\begin{figure}
\begin{center}
\epsfig{figure=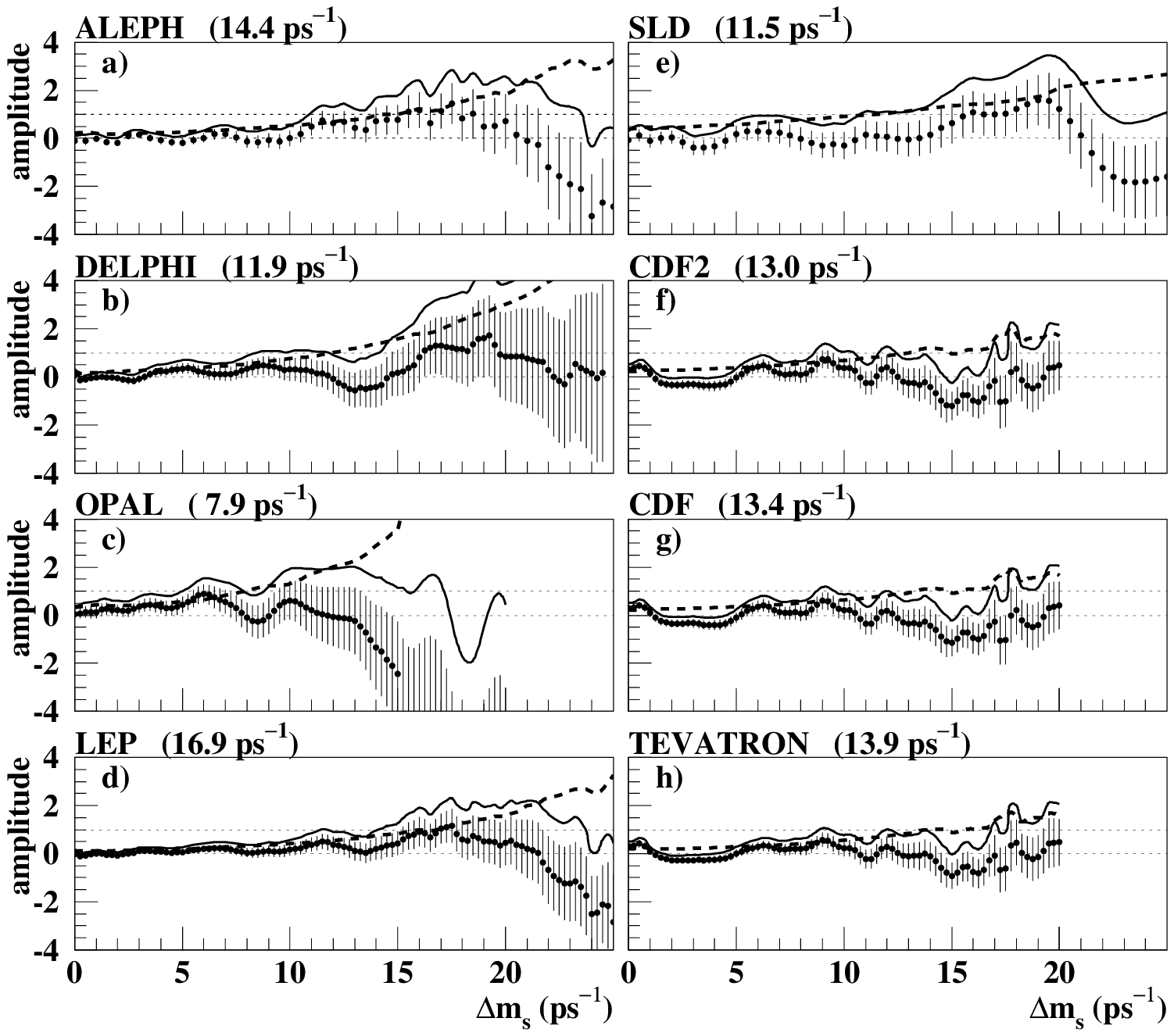,width=\textwidth,%
bbllx=55,bblly=55,bburx=490,bbury=490}
\caption{Combined \Bs-oscillation amplitude spectra, displayed separately for each 
experiment and collider, in the same manner as in \Fig{individual_amplitudes_1}. 
a) ALEPH~\cite{ALEPH_dms}, 
b) DELPHI~\cite{DELBS1_dms_excl,DELPHI_dms_dgs,DELPHI_dmd_dms_vtx,DELPHI_dms_last},
c) OPAL~\cite{OPAL_dms_l,OPAL_dms_dsl}, 
d) LEP~\cite{ALEPH_dms,DELBS1_dms_excl,DELPHI_dms_dgs,DELPHI_dmd_dms_vtx,DELPHI_dms_last,OPAL_dms_l,OPAL_dms_dsl},
e) SLD~\cite{SLD_dms_ds,SLD_dms_dipole},
f) CDF2~\cite{CDF2_dms_dslnu_prel,CDF2_dms_dspi_prel}, 
g) CDF1 and CDF2 together~\cite{CDF1_dms,CDF2_dms_dslnu_prel,CDF2_dms_dspi_prel}, 
h) Tevatron~\cite{CDF1_dms,CDF2_dms_dslnu_prel,CDF2_dms_dspi_prel,D0_dms_dsmux_prel}.
See \Fig{individual_amplitudes_2}h) for CDF1 alone
and \Fig{individual_amplitudes_1}h) for \dzero alone.
}
\labf{individual_amplitudes_3}
\end{center}
\end{figure}

\Figuress{individual_amplitudes_1}{individual_amplitudes_2} show the
amplitude spectra obtained by ALEPH~\cite{ALEPH_dms},
CDF~\cite{CDF1_dms,CDF2_dms_dslnu_prel,CDF2_dms_dspi_prel},
D0~\cite{D0_dms_dsmux_prel}, 
DELPHI~\cite{DELPHI_dmd_dms_vtx,DELPHI_dms_dgs,DELBS1_dms_excl,DELPHI_dms_last},
OPAL~\cite{OPAL_dms_l,OPAL_dms_dsl} and 
SLD~\cite{SLD_dms_dipole,SLD_dms_ds}.\footnote{An unpublished analysis 
from SLD~\cite{SLD_dms_leptDvtx_unpublished}, 
based on an inclusive reconstruction from a 
lepton and a topologically reconstructed \particle{D} meson, 
is not included in the plots or
combined results quoted in this section. However, nothing is known 
to be wrong about this analysis, and including it would increase the 
combined \dms limit of \Eq{dmslimit} by \hfagDMSDLIM and the combined 
sensitivity by \hfagDMSDSENS.}
In each analysis, a particular value of \dms
can be excluded at \CL{95} if ${\cal A}+ 1.645\,\sigma_{\cal A} < 1$, 
where $\sigma_{\cal A}$ is the total uncertainty on ${\cal A}$.
Because of the proper time resolution, the quantity $\sigma_{\cal A}(\dms)$
is an increasing function of \dms (see \Eq{significance} which merely models  
$1/\sigma_{\cal A}(\dms)$ since all results are limited 
by the available statistics). Therefore, 
if the true value of \dms were infinitely large, one 
expects to be able to exclude all values of \dms up to $\dms^{\rm sens}$, 
where $\dms^{\rm sens}$, called here the
sensitivity of the analysis, is defined by
$1.645\,\sigma_{\cal A}(\dms^{\rm sens}) = 1$. 
At LEP times, the most sensitive analyses were the ones based 
on inclusive lepton samples, where reasonable statistics was available. 
Because of their better proper time resolution, the small data samples 
analyzed inclusively at SLD, as well as the few fully reconstructed \Bs decays 
at LEP, turned out to be also very useful to explore the high \dms region.
New preliminary analyses are now available from CDF and \dzero. 
These experiments presently are the only ones active in this area, and 
they are very soon going to reach sufficient statistics so that 
analyses of fully reconstructed hadronic modes can surpass the ones 
based on \particle{D_s \ell \nu \rm X} events.

\begin{figure}
\begin{center}
\epsfig{figure=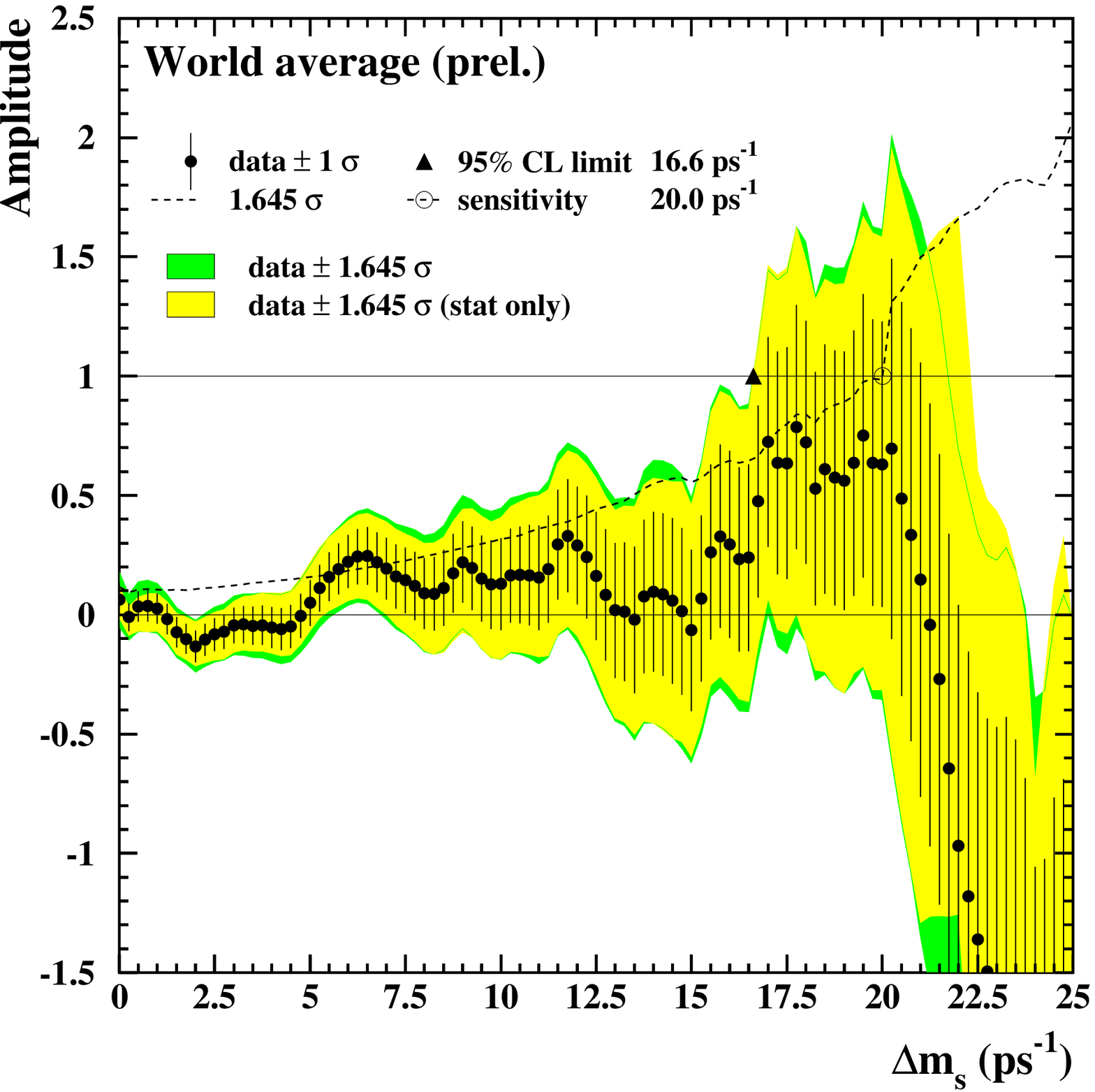,width=\textwidth}
\caption{Combined measurements of the \Bs oscillation amplitude as a 
function of \dms, including all published results and preliminary 
results presented at the Summer and Fall 2005 conferences~%
\cite{ALEPH_dms,CDF1_dms,CDF2_dms_dslnu_prel,CDF2_dms_dspi_prel,D0_dms_dsmux_prel,DELPHI_dmd_dms_vtx,DELPHI_dms_dgs,DELBS1_dms_excl,DELPHI_dms_last,OPAL_dms_l,OPAL_dms_dsl,SLD_dms_dipole,SLD_dms_ds}.
The measurements are dominated by statistical uncertainties. 
Neighboring points are statistically correlated.}
\labf{amplitude}
\end{center}
\end{figure}

These oscillation searches can easily be combined 
by averaging the measured amplitudes ${\cal A}$ at each test value 
of \dms. The combined amplitude spectra for the individual experiments are 
displayed in \Fig{individual_amplitudes_3}, and the world average spectrum is 
displayed in \Fig{amplitude}.
The individual results have been adjusted to common physics inputs, 
and all known correlations have been accounted for; 
in the case of the inclusive analyses, the sensitivities (\ie\ 
the statistical uncertainties on ${\cal A}$), which depend directly 
through \Eq{significance} on the assumed fraction $f_{\rm sig}\sim\fBs$
of \Bs mesons in an unbiased sample of weakly-decaying \b hadrons, 
have also been rescaled to a common average of $\fBs = \hfagFBS$.
The combined sensitivity for \CL{95} exclusion of \dms values is found 
to be\footnote{As can be seen in \Figsss%
{individual_amplitudes_1}{individual_amplitudes_2}{individual_amplitudes_3}, 
as well as from the discontinuity of the dashed curve in \Fig{amplitude},
some experiments did not provide data above 20\invps.
The current combined sensitivity for \CL{95} exclusion
can now only be improved if new amplitude 
measurements are performed above 20\invps.}
\hfagDMSWSENS.
All values of \dms below \hfagDMSWLIM\ are excluded at \CL{95},
which we express as
\begin{equation}
\dms > \rm \hfagDMSWLIM~at~\CL{95} \,.
\labe{dmslimit}
\end{equation}
The values between \hfagDMSWLIM\ and \hfagDMSWUPP\ cannot be excluded, because 
the data is compatible with a signal in this region. However,
no deviation from ${\cal A}=0$ is seen in \Fig{amplitude} that would
indicate the observation of a signal.

It should be noted that most \dms analyses assume no decay-width difference in the \Bs system.
Due to the presence of the $\cosh$ terms in \Eq{oscillations}, a non-zero value of 
\DGs would reduce the oscillation amplitude with a small time-dependent factor that would be 
very difficult to distinguish from time resolution effects.

Convoluting the mean \Bs lifetime that will be obtained
in \Eq{oneoverGs}, $1/\Gs=\hfagTAUBSMEANCON$,
with the limit of \Eq{dmslimit} yields
\begin{equation}
\xs = \frac{\dms}{\Gs} > \rm \hfagXSWLIM~at~\CL{95} \,. 
\labe{xs}
\end{equation}
Using $2\ys = \DGGs=\hfagDGSGSCON$ (see \Eq{DGGs})
and assuming no \CP violation in the mixing, this corresponds to
\begin{equation}
\chis = \frac{\xs^2+\ys^2}{2(\xs^2+1)} > \rm \hfagCHISWLIM~at~\CL{95} \,.
\labe{chis}
\end{equation}


%
%
%

\subsubsubsection{Decay width difference \DGs}
\labs{DGs}



Definitions and an introduction to \DGs can also 
be found in \Sec{taubs}.
Neglecting \CP violation, the mass eigenstates are
also \CP eigenstates, with the short-lived state being
\CP-even and the long-lived one being \CP-odd.
Information on \DGs can be obtained by studying the proper time 
distribution of untagged data samples enriched in 
\Bs mesons~\cite{Hartkorn_Moser}.
In the case of an inclusive \Bs selection~\cite{L3B01} or a semileptonic 
\Bs decay selection~\cite{DELPHI_dms_dgs,CDFBS,D0BS2}, 
both the short- and long-lived
components are present, and the proper time distribution is a superposition 
of two exponentials with decay constants 
$\Gs\pm\DGs/2$.
In principle, this provides sensitivity to both \Gs and 
$(\DGGs)^2$. Ignoring \DGs and fitting for 
a single exponential leads to an estimate of \Gs with a 
relative bias proportional to $(\DGGs)^2$. 
An alternative approach, which is directly sensitive to first order in \DGGs, 
is to determine the lifetime of \Bs candidates decaying to \CP
eigenstates; measurements exist for 
\particle{\Bs\to J/\psi\phi}~\cite{CDFIN_BS1,CDFB3,D0BS1} and 
\particle{\Bs\to D_s^{(*)+} D_s^{(*)-}}~\cite{ALEPH_DGs}, which are 
mostly \CP-even states~\cite{Aleksan}. 
However, more recent
time-dependent angular analyses of \particle{\Bs\to J/\psi\phi} 
allow the simultaneous extraction of \DGGs and the \CP-even and \CP-odd 
amplitudes~\cite{CDF2_DGs,D01_DGs}. 
An estimate of \DGGs
has also been obtained directly from a measurement of the 
\particle{\Bs\to D_s^{(*)+} D_s^{(*)-}} branching ratio~\cite{ALEPH_DGs}, 
under the assumption that 
these decays account for all the \CP-even final states 
(however, no systematic uncertainty due to this assumption is given, so 
the average quoted below will not include this estimate).

\begin{table}
\caption{Experimental constraints on \DGGs. The upper limits,
which have been obtained by the working group, are quoted at the \CL{95}.}
\labt{dgammat}
\begin{center}
\begin{tabular}{l|c|c|c}
\hline
Experiment & Method            & $\Delta \Gs/\Gs$ & Ref.  \\
\hline
L3         & lifetime of inclusive \b-sample              
           & $<0.67$   & \cite{L3B01}      \\
DELPHI     & $\Bsb\to D_s^+\ell^- \overline{\nu_{\ell}} X$, lifetime
	   & $<0.46$   & \cite{DELBS0} \\
ALEPH      & $\Bs\to\phi\phi X$ , 
	     \BR{\Bs \to D_s^{(*)+} D_s^{(*)-}}
	   & $0.26^{+0.30}_{-0.15}$ & \cite{ALEPH_DGs} \\
ALEPH      & $\Bs\to\phi\phi X$, lifetime 
           & $0.45^{+0.80}_{-0.49}$ & \cite{ALEPH_DGs}\\
DELPHI     & $\Bsb \to D_s^+$ hadron, lifetime
           & $<0.69$ & \cite{DELBS0}   \\
CDF1       & $\Bs \to J/\psi\phi$, lifetime
	   & $0.33^{+0.45}_{-0.42}$ & \cite{CDFIN_BS1} \\ \hline
CDF2       & $\Bs \to J/\psi\phi$, time-dependent angular analysis
           & $0.65^{+0.25}_{-0.33} \pm 0.01$ & \cite{CDF2_DGs} \\
\dzero     & $\Bs \to J/\psi\phi$, time-dependent angular analysis
           & $0.24{^{+0.28}_{-0.38}}{^{+0.03}_{-0.04}}$ & \cite{D01_DGs} \\
	 \hline
	 \end{tabular}
	 \end{center}
	 \end{table}

Measurements quoting \DGs results are listed in \Table{dgammat}.
There is significant correlation
between \DGGs and $1/\Gamma_s$. In order to combine these measurements,
the two-dimensional log-likelihood for each measurement
in the $(1/\Gs,\,\DGGs)$ plane is summed and the total
normalized with respect to its minimum.  The one-sigma contour (corresponding
to 0.5 units of log-likelihood greater than the minimum) and
95\% contour are found. 
Inputs as indicated in \Table{dgammat} were used in the combination, 
with the exception of the L3~\cite{L3B01} result since the likelihood
for the results was not available, and the 
ALEPH~\cite{ALEPH_DGs} branching ratio result for the reason
given above.

Results of the combination are shown as the one-sigma contour
labelled ``Direct" in both plots of \Fig{DGs}.  Transformation
of variables from $(1/\Gs,\,\DGGs)$ space to other pairs
of variables such as $(1/\Gs,\,\DGs)$ and 
$(\tau_{\rm L} = 1/\Gamma_{\rm L},\,\tau_{\rm H} = 1/\Gamma_{\rm H})$
are also made.
The resulting one-sigma contour for the latter is shown in
\Fig{DGs}(b).

\begin{figure}
\begin{center}
\epsfig{figure=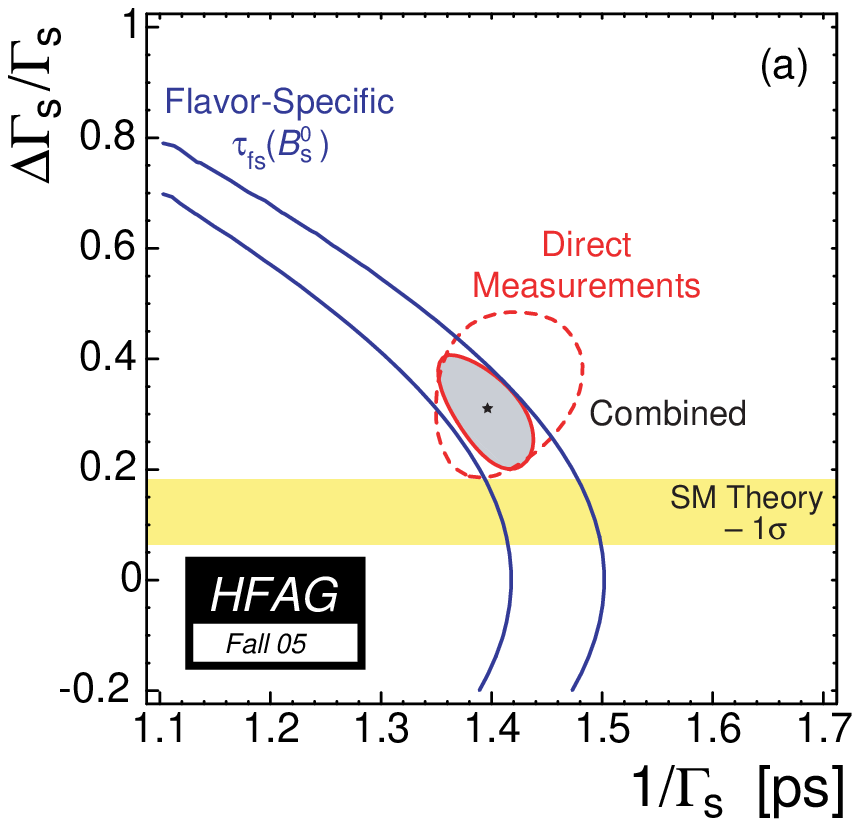,width=0.45\textwidth}
\hfill
\epsfig{figure=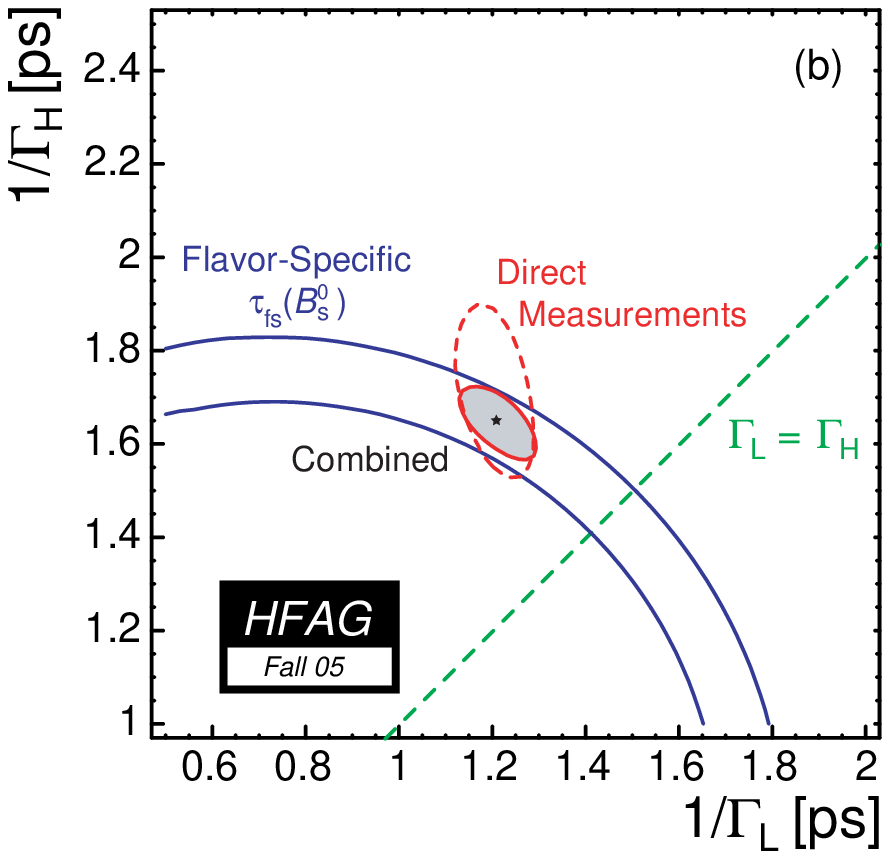,width=0.45\textwidth}
\caption{\DGs combination results with one-sigma contours
($\Delta\log\mathcal{L} = 0.5$) shown for (a) \DGGs versus
$\bar{\tau}(\Bs) = 1/\Gs$  and (b)
$\tau_{\rm H} = 1/\Gamma_{\rm H}$ versus $\tau_{\rm L} = 1/\Gamma_{\rm L}$.
Contours labelled ``Direct" are the result of the combination of
most measurements of \Table{dgammat}, the blue bands are the one-sigma
contours due to the world average of flavor-specific measurements,
and the shaded region the combination of both.
In (b), the diagonal dashed line indicates 
$\Gamma_{\rm L} = \Gamma_{\rm H}$, i.e., where $\DGs = 0$.}
\labf{DGs}
\end{center}
\end{figure}




\newcommand{\comment}[1]{}\comment{

{\em This text below taken by Donatella from previous \DGs notes \ldots}

\newcommand{\Bh}{B^{\rm heavy}_{d,s}}
\newcommand{\Bl}{B^{\rm light}_{d,s}}
\newcommand{\Mh}{m^{\rm heavy}_{d,s}}
\newcommand{\Ml}{m^{\rm light}_{d,s}}
\newcommand{\Gh}{\Gamma^{\rm heavy}_{d,s}}
\newcommand{\Gl}{\Gamma^{\rm light}_{d,s}}
\newcommand{\Gsho}{\Gamma^{\rm short}_{d,s}}
\newcommand{\Glon}{\Gamma^{\rm long}_{d,s}}
\newcommand{\G}{\Gamma_{d,s}}
\newcommand{\tb}{\tau(B^0_{d,s})}
\newcommand{\dg}{\Delta\Gamma_{d,s}}
\newcommand{\tbssemi}{\tau(B_s)_{\rm semi}}
\newcommand{\Bssh}{B^{\rm short}_s}
\newcommand{\tbsshort}{\tau(\Bssh)}

The Standard Model predicts that the \Bs and \Bd can mix before decay. 
The phenomenology of this interaction can be 
described in terms of a $2\times 2$ effective Hamiltonian matrix, 
$M - i\Gamma /2$.
This results in new states called heavy and light, 
$\Bh$  and $\Bl$, for \Bs and \Bd with masses $\Mh$ ,
$\Ml$. Also the  widths  $\Gh$ and $\Gl$ could be different.

Neglecting \CP violation, the mass eigenstates are also \CP eigenstates, the ``long''  state being 
\CP even and the short one being \CP odd.  For convenience of notation, in the following
we therefore substitute 
$\Gl \equiv \Gsho$ and $\Gh \equiv \Glon$, and 
define $\G=1/\tb=(\Glon+\Gsho)/2$ and 
$\Delta \G = \Gsho-\Glon$ which is positive.

$\dg$ is related to the off-diagonal matrix elements, which have been recently 
calculated at the NLO including NLO QCD correction~\cite{Ciuchini2}. 
The theoretical values are:
\begin{equation}
\DGGs = (7.4\pm 2.4 )\times 10^{-2} \,, \hspace{1truecm} \DGGd = (2.42 \pm 0.59 ) \times 10^{-3} \,.
\end{equation}

In the same work the ratio $\DGd/\DGs$ is evaluated since the uncertainties 
coming from higher orders of QCD and $\Lambda_{\rm QCD}/m_{\b}$ corrections cancel out:
\begin{equation}
\DGd/\DGs = (3.2\pm 0.8)\times10^{-2}
\end{equation}

Experimentally \DGs can be measured fitting the lifetime of the 
light and heavy component of the \Bs.
An alternative method is based on the measurement of the  
branching fraction \particle{\Bs\to D_s^{(*)+}D_s^{(*)-}}.
Methods based on lifetime measurements have two different approaches.
Double exponential lifetime fits to samples containing a mixture of \CP eigenstates like
inclusive or semileptonic \Bs decays or $\Bs\to D_s$-hadron have a quadratic sensitivity 
to \DGs.
Whereas the isolation of a single \CP eigenstate as $\Bs\to\phi\phi$ or 
\particle{\Bs\to J/\psi\phi} to extract the lifetime of the \CP-even or odd state have 
a linear dependence on \DGs and it is more sensitive to \DGs but tend 
to suffer from reduced statistics.
The branching fraction method, exploited by ALEPH~\cite{ALEPH-phiphi}, 
is based on several theoretical assumptions~\cite{theoBR}, and allows to have 
information on \DGs only through the branching fraction measurement:
\begin{equation}
\BR{B_s\to D_s^{(*)+}
D_s^{(*)-}} = \frac{\DGs}{\Gs\left(1+\frac{\DGs}{2\Gs}\right)} \,.
\label{eq:dg_ratio}
\end{equation}

The available results are summarized in \Table{dgammat}. 
The values of the limit on \DGGs quoted in the last column of this 
table have been obtained by the working group.

Details on how these measurements are included in the average can be found  
in the previous summaries~\cite{Pcomb}.

\begin{table}
\caption{Experimental constraints on \DGGs. The upper limits,
which have been obtained by the working group, are quoted at the \CL{95}.}
\labt{dgammat}
\begin{center}
\begin{tabular}{|l|c|c|c|} 
\hline
Experiment & Selection        & Measurement            & $\Delta \Gs/\Gs$ \\ 
\hline
L3~\cite{L3B01}         & inclusive \b-sample              &                               & $<0.67$         \\
DELPHI~\cite{DELBS0}     & $\Bsb\to D_s^+\ell^- \overline{\nu_{\ell}} X$ & $\tbssemi=(1.42^{+0.14}_{-0.13}\pm0.03)$~ps  & $<0.46$ \\
others~\cite{ref:others}& $\Bsb\to D_s^+\ell^-  \overline{\nu_{\ell}} X$  & $\tbssemi=(1.46\pm{0.07})$~ps & $<0.30$ \\
ALEPH~\cite{ALEPH-phiphi}      & $\Bs\to\phi\phi X$      & 
$\BR{\Bssh \to D_s^{(*)+} D_s^{(*)-}} =(23\pm10^{+19}_{-~9})\%$       & $0.26^{+0.30}_{-0.15}$ \\
ALEPH~\cite{ALEPH-phiphi}      & $\Bs\to\phi\phi X$      & $\tbsshort=(1.27\pm0.33\pm0.07)$~ps           & 
$0.45^{+0.80}_{-0.49}$ \\ 
DELPHI~\cite{DELBS0}$^a$    & $\Bsb \to D_s^+$ hadron    
&  $\tau_{\rm B^{D_s-had.}_s}=(1.53^{+0.16}_{-0.15}\pm0.07)$~ps                          & $<0.69$         \\
CDF~\cite{CDFB01} & $\Bs \to {\rm J}/\psi\phi$        
& $\tau_{\rm B^{{\rm J}/\psi \phi}_s}=(1.34^{+0.23}_{-0.19}\pm0.05)$~ps & $0.33^{+0.45}_{-0.42}$ \\ 
\hline
\multicolumn{4}{l}{$^a$ \footnotesize 
The value quoted for the measured lifetime differs
slightly from the one quoted in \Table{bs} because it} \\[-1ex]
\multicolumn{4}{l}{~~ \footnotesize 
corresponds to the present status of the analysis in which the information
on \DGs has been obtained.}
\end{tabular}
\end{center}
\end{table}

Here only a short description will be given.

L3 and DELPHI use inclusively reconstructed
\Bs and $\Bs\to \particle{D_s} \ell\nu X$ events respectively.
If those sample are fitted assuming a single exponential lifetime then,
assuming \DGGs is small, the measured lifetime is given by:
\begin{equation}
\tau(\Bs)_{\rm incl.} = \frac{1}{\Gs} \frac{1}{1-\left(\frac{\DGs}{2\Gs}\right)^2}
\quad \quad ; \quad \quad
\tau(\Bs)_{\rm semi.} = \frac{1}{\Gs} 
\frac{{1+\left(\frac{\DGs}{2\Gs}\right)^2}}{{1-\left(\frac{\DGs}{2\Gs}\right)^2}}.     
\end{equation}

The single lifetime fit is thus more sensitive to the effects of 
\DGs in the semileptonic case than in the fully inclusive case.

The same method is used for the \Bs world average lifetime
(recomputed without the DELPHI measurement) obtained
by using only the semileptonic decays and referenced in \Table{dgammat} as {\it others}.

The technique of reconstructing only decays at defined \CP
has been exploited by ALEPH, DELPHI and CDF.

ALEPH reconstructs the decay
$\Bs\to \particle{D_s^{(*)+}D_s^{(*)-}} \to \phi\phi X$
which is predominantly \CP even.
The proper time dependence of the \Bs component is a simple
exponential and the lifetime is related to \DGs via
\begin{equation}
\frac{\Delta \Gs}{\Gs}=2(\frac{1}{\Gs~\tbsshort}-1).  
\end{equation} 
 The same data have been used by ALEPH to exploit the branching fraction method.

DELPHI uses a sample of $\Bs\to D_s$-hadron,
which is expected to have an increased \CP-even component as the contribution
due to \particle{D_s^{(*)+}_s^{(*)-}} events is enhanced by
selection criteria.

CDF reconstructs \particle{\Bs\to J/\psi\phi} with
\particle{J/\psi\to\mu^+\mu^-} and \particle{\phi\to K^+K^-}
where the \CP-even component is equal to $0.84\pm 0.16$ obtained by
combining CLEO~\cite{cleo} measurement of \CP-even fraction in
\particle{\Bd\to J/\psi K^{*0}} and possible SU(3) symmetry
correction.

In order to combine all the measurements~\footnote{L3 is not 
included since the likelihood for the results
was not available} the two-dimensional log-likelihood in the ($1/\Gs$, \DGGs) 
plane is summed and normalized with respect to its minimum.
The 68\%, 95\% and \CL{99} contours of the combined negative 
log-likelihood are shown in \Fig{dgplot} (left)
The corresponding limit on \DGGs is:
\begin{eqnarray}
\DGGs & = & 0.16^{+0.15}_{-0.16}  \,, \\
\DGGs & < & 0.54~\mbox{at \CL{95}} \,. 
\end{eqnarray}

\begin{figure}
\begin{center}
\epsfig{figure=figures/osc/dg_w_notaubd_bw_2d.eps,width=\textwidth}
\epsfig{figure=figures/osc/dg_w_taubd_bw_2d.eps,width=\textwidth}
\end{center}
\caption{Top: 68\%, 95\% and \CL{99} contours of the negative log-likelihood 
distribution in the plane ($1/\Gs$, \DGGs).
Bottom: Same, but with the constraint $1/\Gs \equiv\tau_{\Bd}$} 
\labf{dgplot}
\end{figure}

\begin{figure}
\begin{center}
\epsfig{figure=figures/osc/dg_w_taubd_col_1d,width=\textwidth}
\end{center}
\caption{Probability density distribution for \DGGs after applying the constraint; 
the three shaded regions show the limits at the 68\%, 95\% and \CL{99} respectively.} 
\labf{dgprobplot}
\end{figure}

An improved limit on \DGGs can be obtained by applying the $\tau_{\Bd}=\HFAGtauBd$ constraint.
The world average \Bs lifetime is not used, as its meaning 
is not clear if $\Delta \Gs$ is non-zero.
This is well motivated theoretically, as 
the total widths of the \Bs and \Bd mesons
are expected to be 
equal within less than one percent~\cite{bigilife}, \cite{Beneke}
and \DGd is expected to be small. 
 
The two-dimensional log-likelihood obtained, after including the constraint is shown in 
\Fig{dgplot} (right). The resulting probability density distribution for \DGGs is 
shown in \Fig{dgprobplot}. The corresponding limit on \DGGs is:
\begin{eqnarray}
\DGGs & = & 0.07^{+0.09}_{-0.07} \,, \\
\DGGs & < & 0.29~~\mbox{at \CL{95}} \,.
\end{eqnarray}

} 




Numerical results of the combination of the described inputs
of \Table{dgammat} are:
\begin{eqnarray}
\DGGs &\in& [\hfagDGSGSlow,\hfagDGSGSupp] ~ \mbox{at \CL{95}} \,, \\
\DGGs &=& \hfagDGSGS \,, \\
\DGs &=& \hfagDGS \,, \\
\bar{\tau}(\Bs) = 1/\Gs &=& \hfagTAUBSMEAN \,, \\
\rho(\DGGs, 1/\DGs) &=& \hfagRHODGSGSTAUBSMEAN \,, \\
1/\Gamma_{\rm L} = \tau_{\rm short} &=& \hfagTAUBSL \,, \\
1/\Gamma_{\rm H} = \tau_{\rm long}  &=& \hfagTAUBSH \,. 
\end{eqnarray}

Flavor-specific lifetime measurements are of an equal mix
of \CP-even and \CP-odd states at time zero, and  
if a single exponential function is used in the likelihood
lifetime fit of such a sample~\cite{Hartkorn_Moser}, 
\begin{equation}
\tau(\Bs)_{\rm fs} = \frac{1}{\Gs}
\frac{{1+\left(\frac{\DGs}{2\Gs}\right)^2}}{{1-\left(\frac{\DGs}{2\Gs}\right)^2}
}.
\end{equation}
Using the world average flavor-specific 
lifetime\footnote{The world average of all \Bs lifetime 
measurements using flavour-specific final states is \hfagTAUBSSL; however,
for the purpose of the \DGs extraction, we remove from this average one
DELPHI analysis that is already included in the set of ``direct 
measurements'' and obtain \hfagTAUBSSLX, 
shown as the blue bands on the two plots of \Fig{DGs}.} of \Sec{taubs}
the one-sigma blue bands shown in \Fig{DGs} are obtained. 
Higher-order corrections were checked to be negligible in the
combination.
When these flavor-specific measurements
are combined with the measurements of \Table{dgammat}, the shaded
regions of \Fig{DGs} are obtained, with numerical results:
\begin{eqnarray}
\DGGs &\in& [\hfagDGSGSCONlow,\hfagDGSGSCONupp] ~ \mbox{at \CL{95}} \,, \\
\DGGs &=& \hfagDGSGSCON \,, \labe{DGGs} \\
\DGs &=& \hfagDGSCON \,, \\
\bar{\tau}(\Bs) = 1/\Gs &=& \hfagTAUBSMEANCON \,, \labe{oneoverGs} \\
\rho(\DGGs, 1/\DGs) &=& \hfagRHODGSGSTAUBSMEANCON \,, \\
1/\Gamma_{\rm L} = \tau_{\rm short} &=& \hfagTAUBSLCON \,, \\
1/\Gamma_{\rm H} = \tau_{\rm long}  &=& \hfagTAUBSHCON \,. 
\end{eqnarray}
These results can
be compared with the theoretical prediction of 
$\DGs/\Gs = 0.12 \pm 0.05$~\cite{delta_gams}.


The {\em average} \Bs and \Bd
lifetimes are predicted to be equal within
1\%~\cite{equal_lifetimes,Gabbiani_et_al}
and in the past, an additional constraint applied 
by setting
$\Gs = \Gd$, i.e., $1/\Gs = \tau(\Bd)$,
where $\tau(\Bd) = \hfagTAUBD$ is the world average of experimental results,
including a relative 1\% theoretical uncertainty added
in quadrature with the indicated experimental error.
However, with the increased inconsistency of
the measured values of $1/\Gs= \bar{\tau}(\Bs)$ and $\tau(\Bd)$
at the level of $\hfagRTAUBSMEANCONsig\,\sigma$,
this constraint is no longer applied. 


\clearpage

\mysection{Measurements related to Unitarity Triangle angles
}
\label{sec:cp_uta}

The charge of the ``$\CP(t)$ and Unitarity Triangle angles'' group
is to provide averages of measurements related (mostly) to the 
angles of the Unitarity Triangle (UT).
To date, most of the measurements that can be used to 
obtain model-independent information on the UT angles
come from time-dependent $\CP$ asymmetry analyses.
In cases where considerable theoretical input is required to 
extract the fundamental quantities, no attempt is made to do so at 
this stage. However, straightforward interpretations of the averages 
are given, where possible.

In Sec.~\ref{sec:cp_uta:introduction} 
a brief introduction to the relevant phenomenology is given.
In Sec.~\ref{sec:cp_uta:notations}
an attempt is made to clarify the various different notations in use.
In Sec.~\ref{sec:cp_uta:common_inputs}
the common inputs to which experimental results are rescaled in the
averaging procedure are listed. 
We also briefly introduce the treatment of experimental errors. 
In the remainder of this section,
the experimental results and their averages are given,
divided into subsections based on the underlying quark-level decays.

\mysubsection{Introduction
}
\label{sec:cp_uta:introduction}

The Standard Model Cabibbo-Kobayashi-Maskawa (CKM) quark mixing matrix $\VCKM$ 
must be unitary. A $3 \times 3$ unitary matrix has four free parameters,\footnote{
  In the general case there are nine free parameters,
  but five of these are absorbed into unobservable quark phases.}
and these are conventionally written by the product
of three (complex) rotation matrices~\cite{ref:cp_uta:chau}, where the rotations are 
characterized by the Euler angles $\theta_{12}$, $\theta_{13}$ 
and $\theta_{23}$, which are the mixing angles
between the generations, and one overall phase $\delta$,
\begin{equation}
\label{eq:ckmPdg}
\VCKM =
        \left(
          \begin{array}{ccc}
            V_{ud} & V_{us} & V_{ub} \\
            V_{cd} & V_{cs} & V_{cb} \\
            V_{td} & V_{ts} & V_{tb} \\
          \end{array}
        \right)
        =
        \left(
        \begin{array}{ccc}
        c_{12}c_{13}    
                &    s_{12}c_{13}   
                        &   s_{13}e^{-i\delta}  \\
        -s_{12}c_{23}-c_{12}s_{23}s_{13}e^{i\delta} 
                &  c_{12}c_{23}-s_{12}s_{23}s_{13}e^{i\delta} 
                        & s_{23}c_{13} \\
        s_{12}s_{23}-c_{12}c_{23}s_{13}e^{i\delta}  
                &  -c_{12}s_{23}-s_{12}c_{23}s_{13}e^{i\delta} 
                        & c_{23}c_{13} 
        \end{array}
        \right)
\end{equation}
where $c_{ij}=\cos\theta_{ij}$, $s_{ij}=\sin\theta_{ij}$ for 
$i<j=1,2,3$. 

Following the observation of a hierarchy between the different
matrix elements, the Wolfenstein parameterization~\cite{ref:cp_uta:wolfenstein}
is an expansion of $\VCKM$ in terms of the four real parameters $\lambda$
(the expansion parameter), $A$, $\rho$ and $\eta$. Defining to 
all orders in $\lambda$~\cite{ref:cp_uta:buras}
\begin{eqnarray}
  \label{eq:burasdef}
  s_{12}             &\equiv& \lambda,\nonumber \\ 
  s_{23}             &\equiv& A\lambda^2, \\
  s_{13}e^{-i\delta} &\equiv& A\lambda^3(\rho -i\eta),\nonumber
\end{eqnarray}
and inserting these into the representation of Eq.~(\ref{eq:ckmPdg}), 
unitarity of the CKM matrix is achieved to all orders.
A Taylor expansion of $\VCKM$ leads to the familiar approximation
\begin{equation}
  \label{eq:cp_uta:ckm}
  \VCKM
  = 
  \left(
    \begin{array}{ccc}
      1 - \frac{1}{2}\lambda^2 & \lambda & A \lambda^3 ( \rho - i \eta ) \\
      - \lambda & 1 - \frac{1}{2}\lambda^2 & A \lambda^2 \\
      A \lambda^3 ( 1 - \rho - i \eta ) & - A \lambda^2 & 1 \\
    \end{array}
  \right) + {\cal O}\left( \lambda^4 \right),
\end{equation}
or, at order $\lambda^{5}$
{\small
  \begin{equation}
    \label{eq:cp_uta:ckm_lambda5}
    \VCKM
    =
    \left(
      \begin{array}{ccc}
        1 - \frac{1}{2}\lambda^{2} - \frac{1}{8}\lambda^4 &
        \lambda &
        A \lambda^{3} (\rho - i \eta) \\
        - \lambda + \frac{1}{2} A^2 \lambda^5 \left[ 1 - 2 (\rho + i \eta) \right] &
        1 - \frac{1}{2}\lambda^{2} - \frac{1}{8}\lambda^4 (1+4A^2) &
        A \lambda^{2} \\
        A \lambda^{3} \left[ 1 - (1-\frac{1}{2}\lambda^2)(\rho + i \eta) \right] &
        -A \lambda^{2} + \frac{1}{2}A\lambda^4 \left[ 1 - 2(\rho + i \eta) \right] &
        1 - \frac{1}{2}A^2 \lambda^4
      \end{array} 
    \right) + {\cal O}\left( \lambda^{6} \right).
  \end{equation}
}
The non-zero imaginary part of the CKM matrix,
which is the origin of $\CP$ violation in the Standard Model,
is encapsulated in a non-zero value of $\eta$.



The unitarity relation $\VCKM^\dagger\VCKM = {\mathit 1}$
results in a total of nine expressions,
that can be written as
$\sum_{i=u,c,t} V^*_{ij}V_{ik} = \delta_{jk}$,
where $\delta_{jk}$ is the Kronecker symbol.
Of the off-diagonal expressions ($j \neq k$),
three can be trivially transformed into the other three 
(under $j \leftrightarrow k$),
leaving six relations, in which three complex numbers sum to zero,
which therefore can be expressed as triangles in the complex plane.

\begin{figure}[t]
  \begin{center}
    \resizebox{0.55\textwidth}{!}{\includegraphics{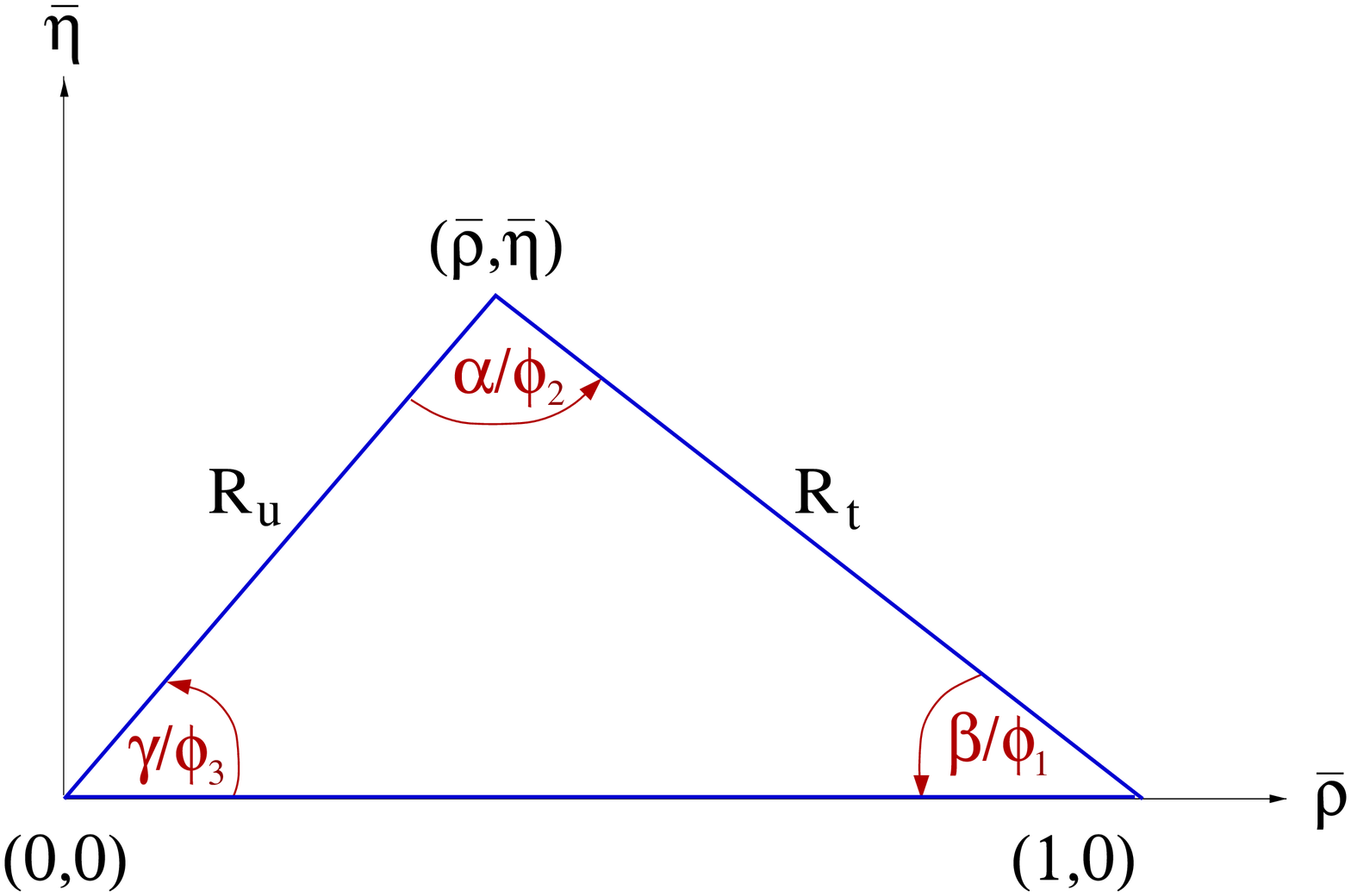}}
    \caption{The Unitarity Triangle.}
    \label{fig:cp_uta:ut}
  \end{center}
\end{figure}

One of these,
\begin{equation}
  \label{eq:cp_uta:ut}
  V_{ud}V^*_{ub} + V_{cd}V^*_{cb} + V_{td}V^*_{tb} = 0,
\end{equation}
is specifically related to $\B$ decays.
The three terms in Eq.~(\ref{eq:cp_uta:ut}) are of the same order 
(${\cal O}\left( \lambda^3 \right)$),
and this relation is commonly known as the Unitarity Triangle.
For presentational purposes,
it is convenient to rescale the triangle by $(V_{cd}V^*_{cb})^{-1}$,
as shown in Fig.~\ref{fig:cp_uta:ut}.

Two popular naming conventions for the UT angles exist in the literature:
\begin{equation}
  \label{eq:cp_uta:abc}
  \alpha  \equiv  \phi_2  = 
  \arg\left[ - \frac{V_{td}V_{tb}^*}{V_{ud}V_{ub}^*} \right],
  \hspace{0.5cm}
  \beta   \equiv   \phi_1 =  
  \arg\left[ - \frac{V_{cd}V_{cb}^*}{V_{td}V_{tb}^*} \right],
  \hspace{0.5cm}
  \gamma  \equiv   \phi_3  =  
  \arg\left[ - \frac{V_{ud}V_{ub}^*}{V_{cd}V_{cb}^*} \right].
  \nonumber
\end{equation}
In this document the $\left( \alpha, \beta, \gamma \right)$ set is used.

The apex of the Unitarity Triangle is written in terms of the 
parameters $\left( \rhobar, \etabar \right)$~\cite{ref:cp_uta:buras}
\begin{equation}
  \label{eq:rhoetabar}
  \rhobar + i\etabar \;\equiv\;
  - \frac{ V_{ud}V_{ub}^* }{ V_{cd}V_{cb}^* }
  = (\rho + i\eta) (1 - \frac{1}{2}\lambda^{2}) + {\cal O}(\lambda^4).
\end{equation}
The exact (to all orders) relation between $\left( \rho, \eta \right)$ and 
$\left( \rhobar, \etabar \right)$ is
\begin{equation}
  \label{eq:rhoetabarinv}
  \rho + i\eta \;=\; 
  \frac{ 
    \sqrt{ 1-A^2\lambda^4 }(\rhobar+i\etabar) 
  }{
    \sqrt{ 1-\lambda^2 } \left[ 1-A^2\lambda^4(\rhobar+i\etabar) \right]
  }.
\end{equation}
The sides $R_u$ and $R_t$ of the Unitarity Triangle 
(the third side being normalized to unity) 
are given by
\begin{eqnarray}
  \label{eq:ru}
  R_u &=& 
  \left|\frac{V_{ud}V_{ub}^*}{V_{cd}V_{cb}^*} \right|
  \;=\; \sqrt{\rhobar^2+\etabar^2}, \\
  \label{eq:rt}
  R_t &=& 
  \left|\frac{V_{td}V_{tb}^*}{V_{cd}V_{cb}^*}\right| 
  \;=\; \sqrt{(1-\rhobar)^2+\etabar^2}.
\end{eqnarray} 

\mysubsection{Notations
}
\label{sec:cp_uta:notations}

Several different notations for $\CP$ violation parameters
are commonly used.
This section reviews those found in the experimental literature,
in the hope of reducing the potential for confusion, 
and to define the frame that is used for the averages.

In some cases, when $\B$ mesons decay into 
multibody final states via broad resonances ($\rho$, $\Kstar$, \etc),
the experimental analyses ignore the effects of interference 
between the overlapping structures.
This is referred to as the quasi-two-body (Q2B) approximation
in the following.

\mysubsubsection{$\CP$ asymmetries
}
\label{sec:cp_uta:notations:pra}

The $\CP$ asymmetry is defined as the difference between the rate 
involving a $b$ quark and that involving a $\bar b$ quark, divided 
by the sum. For example, the partial rate (or charge) asymmetry for 
a charged $\B$ decay would be given as 
\begin{equation}
  \label{eq:cp_uta:pra}
  \Acp_{f} \;\equiv\; 
  \frac{\Gamma(\Bm \to f)-\Gamma(\Bp \to \bar{f})}{\Gamma(\Bm \to f)+\Gamma(\Bp \to \bar{f})}.
\end{equation}

\mysubsubsection{Time-dependent \CP asymmetries in decays to $\CP$ eigenstates
}
\label{sec:cp_uta:notations:cp_eigenstate}

If the amplitudes for $\Bz$ and $\Bzb$ to decay to a final state $f$, 
which is a $\CP$ eigenstate with eigenvalue $\etacpf$,
are given by $\Af$ and $\Abarf$, respectively, 
then the decay distributions for neutral $\B$ mesons, 
with known flavour at time $\Delta t =0$,
are given by
\begin{eqnarray}
  \Gamma_{\Bzb \to f} (\Delta t) & = &
  \frac{e^{-| \Delta t | / \tau(\Bz)}}{4\tau(\Bz)}
  \left[ 
    1 +
      \frac{2\, \Im(\lambda_f)}{1 + |\lambda_f|^2} \sin(\Delta m \Delta t) -
      \frac{1 - |\lambda_f|^2}{1 + |\lambda_f|^2} \cos(\Delta m \Delta t)
  \right], \\
  \Gamma_{\Bz \to f} (\Delta t) & = &
  \frac{e^{-| \Delta t | / \tau(\Bz)}}{4\tau(\Bz)}
  \left[ 
    1 -
      \frac{2\, \Im(\lambda_f)}{1 + |\lambda_f|^2} \sin(\Delta m \Delta t) +
      \frac{1 - |\lambda_f|^2}{1 + |\lambda_f|^2} \cos(\Delta m \Delta t)
  \right].
\end{eqnarray}
Here $\lambda_f = \frac{q}{p} \frac{\Abarf}{\Af}$ 
contains terms related to $\Bz$\textendash$\Bzb$ mixing and to the decay amplitude
(the eigenstates of the effective Hamiltonian in the $\BzBzb$ system 
are $\left| B_\pm \right> = p \left| \Bz \right> \pm q \left| \Bzb \right>$).
This formulation assumes $\CPT$ invariance, 
and neglects possible lifetime differences 
(between the eigenstates of the effective Hamiltonian;
see Section~\ref{sec:mixing} where the mass difference $\Delta m$ is also defined)
in the neutral $\B$ meson system.
The time-dependent $\CP$ asymmetry,
again defined as the difference between the rate 
involving a $b$ quark and that involving a $\bar b$ quark,
is then given by
\begin{equation}
  \label{eq:cp_uta:td_cp_asp}
  \Acp_{f} \left(\Delta t\right) \; \equiv \;
  \frac{
    \Gamma_{\Bzb \to f} (\Delta t) - \Gamma_{\Bz \to f} (\Delta t)
  }{
    \Gamma_{\Bzb \to f} (\Delta t) + \Gamma_{\Bz \to f} (\Delta t)
  } \; = \;
  \frac{2\, \Im(\lambda_f)}{1 + |\lambda_f|^2} \sin(\Delta m \Delta t) -
  \frac{1 - |\lambda_f|^2}{1 + |\lambda_f|^2} \cos(\Delta m \Delta t).
\end{equation}

While the coefficient of the $\sin(\Delta m \Delta t)$ term in 
Eq.~(\ref{eq:cp_uta:td_cp_asp}) is everywhere\footnote
{
  Occasionally one also finds Eq.~(\ref{eq:cp_uta:td_cp_asp}) written as
  $\Acp_{f} \left(\Delta t\right) = 
  {\cal A}^{\rm mix}_f \sin(\Delta m \Delta t) + {\cal A}^{\rm dir}_f \cos(\Delta m \Delta t)$,
  or similar.
} denoted $S_f$:
\begin{equation}
  \label{eq:cp_uta:s_def}
  S_f \;\equiv\; \frac{2\, \Im(\lambda_f)}{1 + \left|\lambda_f\right|^2},
\end{equation}
different notations are in use for the
coefficient of the $\cos(\Delta m \Delta t)$ term:
\begin{equation}
  \label{eq:cp_uta:c_def}
  C_f \;\equiv\; - A_f \;\equiv\; \frac{1 - \left|\lambda_f\right|^2}{1 + \left|\lambda_f\right|^2}.
\end{equation}
The $C$ notation is used by the \babar\  collaboration 
(see \eg~\cite{ref:cp_uta:ccs:babar}), 
and also in this document.
The $A$ notation is used by the \belle\ collaboration
(see \eg~\cite{BELLE2}).

Neglecting effects due to $\CP$ violation in mixing 
(by taking $|q/p| = 1$),
if the decay amplitude contains terms with 
a single weak (\ie, $\CP$ violating) phase
then $\left|\lambda_f\right| = 1$ and one finds
$S_f = -\etacpf \sin(\phi_{\rm mix} + \phi_{\rm dec})$, $C_f = 0$,
where $\phi_{\rm mix}=\arg(q/p)$ and $\phi_{\rm dec}=\arg(\Abarf/\Af)$.
Note that $\phi_{\rm mix}\approx2\beta $ 
in the Standard Model (in the usual phase convention). 
If amplitudes with different weak phases contribute to the decay, 
no clean interpretation of $S_f$ is possible. If the decay amplitudes
have in addition different $\CP$ conserving strong phases,
then $\left| \lambda_f \right| \neq 1$ and no clean interpretation is possible.
The coefficient of the cosine term becomes non-zero,
indicating direct $\CP$ violation.
The sign of $A_f$ as defined above is consistent with that of $\Acp_{f}$ in 
Eq.~(\ref{eq:cp_uta:pra}).

Frequently, we are interested in combining measurements 
governed by similar or identical short-distance physics,
but with different final states
(\eg, $\Bz \to \jpsi \KS$ and $\Bz \to \jpsi \KL$).
In this case, we remove the dependence on the $\CP$ eigenvalue 
of the final state by quoting $-\etacp S_f$.
In cases where the final state is not a $\CP$ eigenstate but has
an effective $\CP$ (see below),
the reported $-\etacp S$ is corrected by the effective $\CP$.

\mysubsubsection{Time-dependent \CP asymmetries in decays to vector-vector final states
}
\label{sec:cp_uta:notations:vv}

Consider \B decays to states consisting of two vector particles,
such as $\jpsi K^{*0}(\to\KS\piz)$, $D^{*+}D^{*-}$ and $\rho^+\rho^-$,
which are eigenstates of charge conjugation but not of parity.\footnote{
  \noindent
  This is not true of all vector-vector final states,
  \eg, $D^{*\pm}\rho^{\mp}$ is clearly not an eigenstate of 
  charge conjugation.
}
In fact, for such a system, there are three possible final states;
in the helicity basis these can be written $h_{-1}, h_0, h_{+1}$.
The $h_0$ state is an eigenstate of parity, and hence of $\CP$;
however, $\CP$ transforms $h_{+1} \leftrightarrow h_{-1}$ (up to 
an unobservable phase). In the transversity basis, these states 
are transformed into  $h_\parallel =  (h_{+1} + h_{-1})/2$ and 
$h_\perp = (h_{+1} - h_{-1})/2$.
In this basis all three states are $\CP$ eigenstates, 
and $h_\perp$ has the opposite $\CP$ to the others.

The amplitudes to these states are usually given by $A_{0,\perp,\parallel}$
(here we use a normalization such that 
$| A_0 |^2 + | A_\perp |^2 + | A_\parallel |^2 = 1$).
Then the effective $\CP$ of the vector-vector state is known if 
$| A_\perp |^2$ is measured.
An alternative strategy is to measure just the longitudinally polarized 
component,  $| A_0 |^2$
(sometimes denoted by $f_{\rm long}$), 
which allows a limit to be set on the effective $\CP$ since
$| A_\perp |^2 \leq | A_\perp |^2 + | A_\parallel |^2 = 1 - | A_0 |^2$.
The most complete treatment for 
neutral $\B$ decays to vector-vector final states
is time-dependent angular analysis 
(also known as time-dependent transversity analysis).
In such an analysis, 
the interference between the $\CP$ even and $\CP$ odd states 
provides additional sensitivity to the weak and strong phases involved.

\subsubsection[Time-dependent asymmetries in decays to multiparticle
final states]{Time-dependent asymmetries in decays to self-conjugate 
  multiparticle final states}
\label{sec:cp_uta:notations:dalitz}

Amplitudes for neutral \B decays into 
self-conjugate multiparticle final states
such as $\pi^+\pi^-\pi^0$, $\jpsi \pi^+\pi^-$ or $D\pi^0$ with $D \to \KS\pi^+\pi^-$
may be written in terms of \CP-even and \CP-odd amplitudes.
As above, the interference between these terms 
provides additional sensitivity to the weak and strong phases
involved in the decay,
and the time-dependence depends on both the sine and cosine
of the weak phase difference.
In order to perform unbinned maximum likelihood fits,
and thereby extract as much information as possible from the distributions,
it is necessary to select a model for the multiparticle decay,
and therefore the results acquire some model dependence
(binned, model independent methods are also possible,
though are not as statistically powerful).
The number of observables depends on the final state (and on the model used);
the key feature is that as long as there are regions where both
\CP-even and \CP-odd amplitudes contribute,
the interference terms will be sensitive to the cosine 
of the weak phase difference.
Therefore, these measurements allow distinction between multiple solutions
for, \eg, the four values of $\beta$ from the measurement of $\sin(2\beta)$.

\mysubsubsection{Time-dependent \CP asymmetries in decays to non-$\CP$ eigenstates
}
\label{sec:cp_uta:notations:non_cp}

Consider a non-$\CP$ eigenstate $f$, and its conjugate $\bar{f}$. 
For neutral $\B$ decays to these final states,
there are four amplitudes to consider:
those for $\Bz$ to decay to $f$ and $\bar{f}$
($\Af$ and $\Afbar$, respectively),
and the equivalents for $\Bzb$
($\Abarf$ and $\Abarfbar$).
If $\CP$ is conserved in the decay, then
$\Af = \Abarfbar$ and $\Afbar = \Abarf$.


The time-dependent decay distributions can be written in many different ways.
Here, we follow Sec.~\ref{sec:cp_uta:notations:cp_eigenstate}
and define $\lambda_f = \frac{q}{p}\frac{\Abarf}{\Af}$ and
$\lambda_{\bar f} = \frac{q}{p}\frac{\Abarfbar}{\Afbar}$.
The time-dependent \CP asymmetries then follow Eq.~(\ref{eq:cp_uta:td_cp_asp}):
\begin{eqnarray}
\label{eq:cp_uta:non-cp-obs}
  {\cal A}_f (\Delta t) \; \equiv \;
  \frac{
    \Gamma_{\Bzb \to f} (\Delta t) - \Gamma_{\Bz \to f} (\Delta t)
  }{
    \Gamma_{\Bzb \to f} (\Delta t) + \Gamma_{\Bz \to f} (\Delta t)
  } & = & S_f \sin(\Delta m \Delta t) - C_f \cos(\Delta m \Delta t), \\
  {\cal A}_{\bar{f}} (\Delta t) \; \equiv \;
  \frac{
    \Gamma_{\Bzb \to \bar{f}} (\Delta t) - \Gamma_{\Bz \to \bar{f}} (\Delta t)
  }{
    \Gamma_{\Bzb \to \bar{f}} (\Delta t) + \Gamma_{\Bz \to \bar{f}} (\Delta t)
  } & = & S_{\bar{f}} \sin(\Delta m \Delta t) - C_{\bar{f}} \cos(\Delta m \Delta t),
\end{eqnarray}
with the definitions of the parameters 
$C_f$, $S_f$, $C_{\bar{f}}$ and $S_{\bar{f}}$,
following Eqs.~(\ref{eq:cp_uta:s_def}) and~(\ref{eq:cp_uta:c_def}).

The time-dependent decay rates are given by
\begin{eqnarray}
  \Gamma_{\Bzb \to f} (\Delta t) & = &
  \frac{e^{-\left| \Delta t \right| / \tau(\Bz)}}{8\tau(\Bz)} 
  ( 1 + \Adirnoncp ) 
  \left\{ 
    1 + S_f \sin(\Delta m \Delta t) - C_f \cos(\Delta m \Delta t) 
  \right\},
  \\
  \Gamma_{\Bz \to f} (\Delta t) & = &
  \frac{e^{-\left| \Delta t \right| / \tau(\Bz)}}{8\tau(\Bz)} 
  ( 1 + \Adirnoncp ) 
  \left\{ 
    1 - S_f \sin(\Delta m \Delta t) + C_f \cos(\Delta m \Delta t) 
  \right\},
  \\
  \Gamma_{\Bzb \to \bar{f}} (\Delta t) & = &
  \frac{e^{-\left| \Delta t \right| / \tau(\Bz)}}{8\tau(\Bz)} 
  ( 1 - \Adirnoncp ) 
  \left\{ 
    1 + S_{\bar{f}} \sin(\Delta m \Delta t) - C_{\bar{f}} \cos(\Delta m \Delta t) 
  \right\},
  \\
  \Gamma_{\Bz \to \bar{f}} (\Delta t) & = &
    \frac{e^{-\left| \Delta t \right| / \tau(\Bz)}}{8\tau(\Bz)} 
  ( 1 - \Adirnoncp ) 
  \left\{ 
    1 - S_{\bar{f}} \sin(\Delta m \Delta t) + C_{\bar{f}} \cos(\Delta m \Delta t) 
  \right\},
\end{eqnarray}
where the time-independent parameter \Adirnoncp
represents an overall asymmetry in the production of the 
$f$ and $\bar{f}$ final states,\footnote{
  This parameter is often denoted ${\cal A}_f$ (or ${\cal A}_{\CP}$),
  but here we avoid this notation to prevent confusion with the
  time-dependent $\CP$ asymmetry.
}
\begin{equation}
  \Adirnoncp = 
  \frac{
    \left( 
      \left| \Af \right|^2 + \left| \Abarf \right|^2
    \right) - 
    \left( 
      \left| \Afbar \right|^2 + \left| \Abarfbar \right|^2
    \right)
  }{
    \left( 
      \left| \Af \right|^2 + \left| \Abarf \right|^2
    \right) +
    \left( 
      \left| \Afbar \right|^2 + \left| \Abarfbar \right|^2
    \right)
  }.
\end{equation}
Assuming $|q/p| = 1$,
the parameters $C_f$ and $C_{\bar{f}}$
can also be written in terms of the decay amplitudes as follows:
\begin{equation}
  C_f = 
  \frac{
    \left| \Af \right|^2 - \left| \Abarf \right|^2 
  }{
    \left| \Af \right|^2 + \left| \Abarf \right|^2
  }
  \hspace{5mm}
  {\rm and}
  \hspace{5mm}
  C_{\bar{f}} = 
  \frac{
    \left| \Afbar \right|^2 - \left| \Abarfbar \right|^2
  }{
    \left| \Afbar \right|^2 + \left| \Abarfbar \right|^2
  },
\end{equation}
giving asymmetries in the decay amplitudes of $\Bz$ and $\Bzb$
to the final states $f$ and $\bar{f}$ respectively.
In this notation, the direct $\CP$ invariance conditions are
$\Adirnoncp = 0$ and $C_f = - C_{\bar{f}}$.
Note that $C_f$ and $C_{\bar{f}}$ are typically non-zero;
\eg, for a flavour-specific final state, 
$\Abarf = \Afbar = 0$ ($\Af = \Abarfbar = 0$), they take the values
$C_f = - C_{\bar{f}} = 1$ ($C_f = - C_{\bar{f}} = -1$).

The coefficients of the sine terms
contain information about the weak phase. 
In the case that each decay amplitude contains only a single weak phase
(\ie, no direct $\CP$ violation),
these terms can be written
\begin{equation}
  S_f = 
  \frac{ 
    - 2 \left| \Af \right| \left| \Abarf \right| 
    \sin( \phi_{\rm mix} + \phi_{\rm dec} - \delta_f )
  }{
    \left| \Af \right|^2 + \left| \Abarf \right|^2
  } 
  \hspace{5mm}
  {\rm and}
  \hspace{5mm}
  S_{\bar{f}} = 
  \frac{
    - 2 \left| \Afbar \right| \left| \Abarfbar \right| 
    \sin( \phi_{\rm mix} + \phi_{\rm dec} + \delta_f )
  }{
    \left| \Afbar \right|^2 + \left| \Abarfbar \right|^2
  },
\end{equation}
where $\delta_f$ is the strong phase difference between the decay amplitudes.
If there is no $\CP$ violation, the condition $S_f = - S_{\bar{f}}$ holds.
If decay amplitudes with different weak and strong phases contribute,
no clean interpretation of $S_f$ and $S_{\bar{f}}$ is possible.

Since two of the $\CP$ invariance conditions are 
$C_f = - C_{\bar{f}}$ and $S_f = - S_{\bar{f}}$,
there is motivation for a rotation of the parameters:
\begin{equation}
\label{eq:cp_uta:non-cp-s_and_deltas}
  S_{f\bar{f}} = \frac{S_{f} + S_{\bar{f}}}{2},
  \hspace{4mm}
  \Delta S_{f\bar{f}} = \frac{S_{f} - S_{\bar{f}}}{2},
  \hspace{4mm}
  C_{f\bar{f}} = \frac{C_{f} + C_{\bar{f}}}{2},
  \hspace{4mm}
  \Delta C_{f\bar{f}} = \frac{C_{f} - C_{\bar{f}}}{2}.
\end{equation}
With these parameters, the $\CP$ invariance conditions become
$S_{f\bar{f}} = 0$ and $C_{f\bar{f}} = 0$. 
The parameter $\Delta C_{f\bar{f}}$ gives a measure of the ``flavour-specificity''
of the decay:
$\Delta C_{f\bar{f}}=\pm1$ corresponds to a completely flavour-specific decay,
in which no interference between decays with and without mixing can occur,
while $\Delta C_{f\bar{f}} = 0$ results in 
maximum sensitivity to mixing-induced $\CP$ violation.
The parameter $\Delta S_{f\bar{f}}$ is related to the strong phase difference 
between the decay amplitudes of $\Bz$ to $f$ and to $\bar f$. 
We note that the observables of Eq.~(\ref{eq:cp_uta:non-cp-s_and_deltas})
exhibit experimental correlations 
(typically of $\sim 20\%$, depending on the tagging purity, and other effects)
between $S_{f\bar{f}}$ and  $\Delta S_{f\bar{f}}$, 
and between $C_{f\bar{f}}$ and $\Delta C_{f\bar{f}}$. 
On the other hand, 
the final state specific observables of Eq.~(\ref{eq:cp_uta:non-cp-obs})
tend to have low correlations.

Alternatively, if we recall that the $\CP$ invariance
conditions at the decay amplitude level are
$\Af = \Abarfbar$ and $\Afbar = \Abarf$,
we are led to consider the parameters~\cite{ref:cp_uta:ckmfitter}
\begin{equation}
  \label{eq:cp_uta:non-cp-directcp}
  {\cal A}_{f\bar{f}} = 
  \frac{
    \left| \Abarfbar \right|^2 - \left| \Af \right|^2 
  }{
    \left| \Abarfbar \right|^2 + \left| \Af \right|^2
  }
  \hspace{5mm}
  {\rm and}
  \hspace{5mm}
  {\cal A}_{\bar{f}f} = 
  \frac{
    \left| \Abarf \right|^2 - \left| \Afbar \right|^2
  }{
    \left| \Abarf \right|^2 + \left| \Afbar \right|^2
  }.
\end{equation}
These are sometimes considered more physically intuitive parameters
since they characterize direct $\CP$ violation 
in decays with particular topologies.
For example, in the case of $\Bz \to \rho^\pm\pi^\mp$
(choosing $f =  \rho^+\pi^-$ and $\bar{f} = \rho^-\pi^+$),
${\cal A}_{f\bar{f}}$ (also denoted ${\cal A}^{+-}_{\rho\pi}$)
parameterizes direct $\CP$ violation
in decays in which the produced $\rho$ meson does not contain the 
spectator quark,
while ${\cal A}_{\bar{f}f}$ (also denoted ${\cal A}^{-+}_{\rho\pi}$)
parameterizes direct $\CP$ violation 
in decays in which it does.
Note that we have again followed the sign convention that the asymmetry 
is the difference between the rate involving a $b$ quark and that
involving a $\bar{b}$ quark, \cf\ Eq.~(\ref{eq:cp_uta:pra}). 
Of course, these parameters are not independent of the 
other sets of parameters given above, and can be written
\begin{equation}
  {\cal A}_{f\bar{f}} =
  - \frac{
    \Adirnoncp + C_{f\bar{f}} + \Adirnoncp \Delta C_{f\bar{f}} 
  }{
    1 + \Delta C_{f\bar{f}} + \Adirnoncp C_{f\bar{f}} 
  }
  \hspace{5mm}
  {\rm and}
  \hspace{5mm}
  {\cal A}_{\bar{f}f} =
  \frac{
    - \Adirnoncp + C_{f\bar{f}} + \Adirnoncp \Delta C_{f\bar{f}} 
  }{
    - 1 + \Delta C_{f\bar{f}} + \Adirnoncp C_{f\bar{f}}  
  }.
\end{equation}
They usually exhibit strong correlations.

We now consider the various notations which have been used 
in experimental studies of
time-dependent $\CP$ asymmetries in decays to non-$\CP$ eigenstates.

\mysubsubsubsection{$\Bz \to D^{*\pm}D^\mp$
}
\label{sec:cp_uta:notations:non_cp:dstard}

The above set of parameters 
($\Adirnoncp$, $C_f$, $S_f$, $C_{\bar{f}}$, $S_{\bar{f}}$),
has been used by both
\babar~\cite{ref:cp_uta:ccd:babar:dd_dstard} and
\belle~\cite{ref:cp_uta:ccd:belle:dstard} in the $D^{*\pm}D^{\mp}$ system
($f = D^{*+}D^-$, $\bar{f} = D^{*-}D^+$).
However, slightly different names for the parameters are used:
\babar\ uses 
(${\cal A}$, $C_{+-}$, $S_{+-}$, $C_{-+}$, $S_{-+}$);
\belle\ uses
(${\cal A}$, $C_{+}$,  $S_{+}$,  $C_{-}$,  $S_{-}$).
In this document, we follow the notation used by \babar.

\mysubsubsubsection{$\Bz \to \rho^{\pm}\pi^\mp$
}
\label{sec:cp_uta:notations:non_cp:rhopi}

In the $\rho^\pm\pi^\mp$ system, the 
($\Adirnoncp$, $C_{f\bar{f}}$, $S_{f\bar{f}}$, $\Delta C_{f\bar{f}}$, 
$\Delta S_{f\bar{f}}$)
set of parameters has been used 
originally by 
\babar~\cite{ref:cp_uta:uud:babar:rhopi_old}, and more recently by
\belle~\cite{ref:cp_uta:uud:belle:rhopi}, in the Q2B approximation;
the exact names\footnote{
  \babar\ has used the notations
  $A_{\CP}^{\rho\pi}$~\cite{ref:cp_uta:uud:babar:rhopi_old} and 
  ${\cal A}_{\rho\pi}$~\cite{ref:cp_uta:uud:babar:rhopi}
  in place of ${\cal A}_{\CP}^{\rho\pi}$.
}
used in this case are
$\left( 
  {\cal A}_{\CP}^{\rho\pi}, C_{\rho\pi}, S_{\rho\pi}, \Delta C_{\rho\pi}, \Delta S_{\rho\pi}
\right)$,
and these names are also used in this document.

Since $\rho^\pm\pi^\mp$ is reconstructed in the final state $\pi^+\pi^-\pi^0$,
the interference between the $\rho$ resonances
can provide additional information about the phases 
(see Sec.~\ref{sec:cp_uta:notations:dalitz}).
\babar~\cite{ref:cp_uta:uud:babar:rhopi} has performed 
a time-dependent Dalitz plot analysis, 
from which the weak phase $\alpha$ is directly extracted.
In such an analysis, the measured Q2B parameters are 
also naturally corrected for interference effects.

\mysubsubsubsection{$\Bz \to D^{\pm}\pi^{\mp}, D^{*\pm}\pi^{\mp}, D^{\pm}\rho^{\mp}$
}
\label{sec:cp_uta:notations:non_cp:dstarpi}

Time-dependent $\CP$ analyses have also been performed for the
final states $D^{\pm}\pi^{\mp}$, $D^{*\pm}\pi^{\mp}$ and $D^{\pm}\rho^{\mp}$.
In these theoretically clean cases, no penguin contributions are possible,
so there is no direct $\CP$ violation.
Furthermore, due to the smallness of the ratio of the magnitudes of the 
suppressed ($b \to u$) and favoured ($b \to c$) amplitudes (denoted $R_f$),
to a very good approximation, $C_f = - C_{\bar{f}} = 1$
(using $f = D^{(*)-}h^+$, $\bar{f} = D^{(*)+}h^-$ $h = \pi,\rho$),
and the coefficients of the sine terms are given by
\begin{equation}
  S_f = - 2 R_f \sin( \phi_{\rm mix} + \phi_{\rm dec} - \delta_f )
  \hspace{5mm}
  {\rm and}
  \hspace{5mm}
  S_{\bar{f}} = - 2 R_f \sin( \phi_{\rm mix} + \phi_{\rm dec} + \delta_f ).
\end{equation}
Thus weak phase information can be cleanly obtained from measurements
of $S_f$ and $S_{\bar{f}}$, 
although external information on at least one of $R_f$ or $\delta_f$ is necessary.
(Note that $\phi_{\rm mix} + \phi_{\rm dec} = 2\beta + \gamma$ for all the decay modes 
in question, while $R_f$ and $\delta_f$ depend on the decay mode.)

Again, different notations have been used in the literature.
\babar\xspace\cite{ref:cp_uta:cud:babar:full,ref:cp_uta:cud:babar:partial}
defines the time-dependent probability function by
\begin{equation}
  f^\pm (\eta, \Delta t) = \frac{e^{-|\Delta t|/\tau}}{4\tau} 
  \left[  
    1 \mp S_\zeta \sin (\Delta m \Delta t) \mp \eta C_\zeta \cos(\Delta m \Delta t) 
  \right],
\end{equation} 
where the upper (lower) sign corresponds to 
the tagging meson being a $\Bz$ ($\Bzb$). 
[Note here that a tagging $\Bz$ ($\Bzb$) corresponds to $-S_\xi$ ($+S_\xi$).]
The parameters $\eta$ and $\zeta$ take the values $+1$ and $+$ ($-1$ and $-$) 
when the final state is, \eg, $D^-\pi^+$ ($D^+\pi^-$). 
However, in the fit, the substitutions $C_\zeta = 1$ and 
$S_\zeta = a \mp \eta b_i - \eta c_i$ are made.\footnote{
  The subscript $i$ denotes tagging category.
}
[Note that, neglecting $b$ terms, $S_+ = a - c$ and $S_- = a + c$, 
so that $a = (S_+ + S_-)/2$, $c = (S_- - S_+)/2$, in analogy to 
the parameters of Eq.~(\ref{eq:cp_uta:non-cp-s_and_deltas}).] 
The subscript $i$ denotes the tagging category. 
These are motivated by the possibility of 
$\CP$ violation on the tag side~\cite{ref:cp_uta:cud:tagside}, 
which is absent for semileptonic $\B$ decays (mostly lepton tags). 
The parameter $a$ is not affected by tag side $\CP$ violation. 
The parameter $b$ only depends on tag side $\CP$ violation parameters 
and is not directly useful for determining UT angles.
A clean interpretation of the $c$ parameter is only possible for 
lepton-tagged events,
so the \babar\ measurements report $c$ measured with those events only.

The parameters used by \belle\ in the analysis using 
partially reconstructed $\B$ decays~\cite{ref:cp_uta:cud:belle:partial}, 
are similar to the $S_\zeta$ parameters defined above. 
However, in the \belle\ convention, 
a tagging $\Bz$ corresponds to a $+$ sign in front of the sine coefficient; 
furthermore the correspondence between the super/subscript 
and the final state is opposite, so that $S_\pm$ (\babar) = $- S^\mp$ (\belle). 
In this analysis, only lepton tags are used, 
so there is no effect from tag side $\CP$ violation. 
In the \belle\ analysis using 
fully reconstructed $\B$ decays~\cite{ref:cp_uta:cud:belle:full}, 
this effect is measured and taken into account using $\Dstar l \nu$ decays; 
in neither \belle\ analysis are the $a$, $b$ and $c$ parameters used. 
In the latter case, the measured parameters are 
$2 R_{D^{(*)}\pi} \sin( 2\phi_1 + \phi_3 \pm \delta_{D^{(*)}\pi} )$; 
the definition is such that 
$S^\pm$ (\belle) = $- 2 R_{\Dstar \pi} \sin( 2\phi_1 + \phi_3 \pm \delta_{\Dstar \pi} )$. 
However, the definition includes an 
angular momentum factor $(-1)^L$~\cite{ref:cp_uta:cud:fleischer}, 
and so for the results in the $D\pi$ system, 
there is an additional factor of $-1$ in the conversion.

Explicitly, the conversion then reads as given in 
Table~\ref{tab:cp_uta:notations:non_cp:dstarpi}, 
where we have neglected the $b_i$ terms used by \babar
(which are zero in the absence of tag side $\CP$ violation).
For the averages in this document,
we use the $a$ and $c$ parameters,
and give the explicit translations used in 
Table~\ref{tab:cp_uta:notations:non_cp:dstarpi2}.
It is to be fervently hoped that the experiments will
converge on a common notation in future.

\begin{table}
  \begin{center} 
    \caption{
      Conversion between the various notations used for 
      $\CP$ violation parameters in the 
      $D^{\pm}\pi^{\mp}$, $D^{*\pm}\pi^{\mp}$ and $D^{\pm}\rho^{\mp}$ systems.
      The $b_i$ terms used by \babar\ have been neglected.
      Recall that $\left( \alpha, \beta, \gamma \right) = \left( \phi_2, \phi_1, \phi_3 \right)$.
    }
    \vspace{0.2cm}
    \setlength{\tabcolsep}{0.0pc}
    \begin{tabular*}{\textwidth}{@{\extracolsep{\fill}}cccc} \hline 
      & \babar\ & \belle\ partial rec. & \belle\ full rec. \\
      \hline
      $S_{D^+\pi^-}$    & $- S_- = - (a + c_i)$ &  N/A  &
      $2 R_{D\pi} \sin( 2\phi_1 + \phi_3 + \delta_{D\pi} )$ \\
      $S_{D^-\pi^+}$    & $- S_+ = - (a - c_i)$ &  N/A  &
      $2 R_{D\pi} \sin( 2\phi_1 + \phi_3 - \delta_{D\pi} )$ \\
      $S_{D^{*+}\pi^-}$ & $- S_- = - (a + c_i)$ & $S^+$ &   
      $- 2 R_{\Dstar \pi} \sin( 2\phi_1 + \phi_3 + \delta_{\Dstar \pi} )$ \\
      $S_{D^{*-}\pi^+}$ & $- S_+ = - (a - c_i)$ & $S^-$ &
      $- 2 R_{\Dstar \pi} \sin( 2\phi_1 + \phi_3 - \delta_{\Dstar \pi} )$ \\
      $S_{D^+\rho^-}$    & $- S_- = - (a + c_i)$ &  N/A  &  N/A  \\
      $S_{D^-\rho^+}$    & $- S_+ = - (a - c_i)$ &  N/A  &  N/A  \\
      \hline 
    \end{tabular*}
    \label{tab:cp_uta:notations:non_cp:dstarpi}
  \end{center}
\end{table}
   
\begin{table}
  \begin{center} 
    \caption{
      Translations used to convert the parameters measured by \belle
      to the parameters used for averaging in this document.
      The angular momentum factor $L$ is $-1$ for $\Dstar\pi$ and $+1$ for $D\pi$.
      Recall that $\left( \alpha, \beta, \gamma \right) = \left( \phi_2, \phi_1, \phi_3 \right)$.
    }
    \vspace{0.2cm}
    \setlength{\tabcolsep}{0.0pc}
    \begin{tabular*}{\textwidth}{@{\extracolsep{\fill}}ccc} \hline 
        & $\Dstar\pi$ partial rec. & $D^{(*)}\pi$ full rec. \\
        \hline
        $a$ & $- (S^+ + S^-)$ &
        $\frac{1}{2} (-1)^{L+1}
        \left(
          2 R_{D^{(*)}\pi} \sin( 2\phi_1 + \phi_3 + \delta_{D^{(*)}\pi} ) + 
          2 R_{D^{(*)}\pi} \sin( 2\phi_1 + \phi_3 - \delta_{D^{(*)}\pi} )
        \right)$ \\
        $c$ & $- (S^+ - S^-)$ & 
        $\frac{1}{2} (-1)^{L+1}
        \left(
          2 R_{D^{(*)}\pi} \sin( 2\phi_1 + \phi_3 + \delta_{D^{(*)}\pi} ) -
          2 R_{D^{(*)}\pi} \sin( 2\phi_1 + \phi_3 - \delta_{D^{(*)}\pi} )
        \right)$ \\
        \hline 
      \end{tabular*}
    \label{tab:cp_uta:notations:non_cp:dstarpi2}
  \end{center}
\end{table}

\mysubsubsubsection{Time-dependent asymmetries in radiative $\B$ decays
}
\label{sec:cp_uta:notations:non_cp:radiative}

As a special case of decays to non-$\CP$ eigenstates,
let us consider radiative $\B$ decays.
Here, the emitted photon has a distinct helicity,
which is in principle observable, but in practice is not usually measured.
Thus the measured time-dependent decay rates 
are given by~\cite{ref:cp_uta:bsg:ags,ref:cp_uta:bsg:aghs}
\begin{eqnarray}
  \Gamma_{\Bzb \to X \gamma} (\Delta t) & = &
  \Gamma_{\Bzb \to X \gamma_L} (\Delta t) + \Gamma_{\Bzb \to X \gamma_R} (\Delta t) \\ \nonumber
  & = &
  \frac{e^{-\left| \Delta t \right| / \tau(\Bz)}}{4\tau(\Bz)} 
  \left\{ 
    1 + 
    \left( S_L + S_R \right) \sin(\Delta m \Delta t) - 
    \left( C_L + C_R \right) \cos(\Delta m \Delta t) 
  \right\},
  \\
  \Gamma_{\Bz \to X \gamma} (\Delta t) & = & 
  \Gamma_{\Bz \to X \gamma_L} (\Delta t) + \Gamma_{\Bz \to X \gamma_R} (\Delta t) \\ \nonumber 
  & = &
  \frac{e^{-\left| \Delta t \right| / \tau(\Bz)}}{4\tau(\Bz)} 
  \left\{ 
    1 - 
    \left( S_L + S_R \right) \sin(\Delta m \Delta t) + 
    \left( C_L + C_R \right) \cos(\Delta m \Delta t) 
  \right\},
\end{eqnarray}
where in place of the subscripts $f$ and $\bar{f}$ we have used $L$ and $R$
to indicate the photon helicity.
In order for interference between decays with and without $\Bz$-$\Bzb$ mixing
to occur, the $X$ system must not be flavour-specific,
\eg, in case of $\Bz \to K^{*0}\gamma$, the final state must be $\KS \pi^0 \gamma$.
The sign of the sine term depends on the $C$ eigenvalue of the $X$ system.
The photons from $b \to q \gamma$ ($\bar{b} \to \bar{q} \gamma$) are predominantly
left (right) polarized, with corrections of order of $m_q/m_b$,
thus interference effects are suppressed.
The predicted smallness of the $S$ terms in the Standard Model
results in sensitivity to new physics contributions.

\mysubsubsection{Asymmetries in $\B \to \DorDstar K^{(*)}$ decays
}
\label{sec:cp_uta:notations:cus}

$\CP$ asymmetries in $\B \to \DorDstar K^{(*)}$ decays are sensitive to $\gamma$.
The neutral $D^{(*)}$ meson produced 
is an admixture of $\DorDstarz$ (produced by a $b \to c$ transition) and 
$\DorDstarzb$ (produced by a colour-suppressed $b \to u$ transition) states.
If the final state is chosen so that both $\DorDstarz$ and $\DorDstarzb$ 
can contribute, the two amplitudes interfere,
and the resulting observables are sensitive to $\gamma$, 
the relative weak phase between 
the two $\B$ decay amplitudes~\cite{ref:cp_uta:cus:bs}.
Various methods have been proposed to exploit this interference,
including those where the neutral $D$ meson is reconstructed 
as a $\CP$ eigenstate (GLW)~\cite{ref:cp_uta:cus:glw},
in a suppressed final state (ADS)~\cite{ref:cp_uta:cus:ads},
or in a self-conjugate three-body final state, 
such as $\KS \pi^+\pi^-$ (Dalitz)~\cite{ref:cp_uta:cus:dalitz}.
It should be emphasized that while each method 
differs in the choice of $D$ decay,
they are all sensitive to the same parameters of the $B$ decay,
and can be considered as variations of the same technique.

Consider the case of $\Bmp \to D \Kmp$,
with $D$ decaying to a final state $f$,
which is accessible to both $\Dz$ and $\Dzb$.
We can write the decay rates for $\Bm$ and $\Bp$ ($\Gamma_\mp$), 
the charge averaged rate ($\Gamma = (\Gamma_- + \Gamma_+)/2$)
and the charge asymmetry 
(${\cal A} = (\Gamma_- - \Gamma_+)/(\Gamma_- + \Gamma_+)$, see Eq.~(\ref{eq:cp_uta:pra})) as 
\begin{eqnarray}
  \label{eq:cp_uta:dk:rate_def}
  \Gamma_\mp  & \propto & 
  r_B^2 + r_D^2 + 2 r_B r_D \cos \left( \delta_B + \delta_D \mp \gamma \right), \\
  \label{eq:cp_uta:dk:av_rate_def}
  \Gamma & \propto &  
  r_B^2 + r_D^2 + 2 r_B r_D \cos \left( \delta_B + \delta_D \right) \cos \left( \gamma \right), \\
  \label{eq:cp_uta:dk:acp_def}
  {\cal A} & = & 
  \frac{
    2 r_B r_D \sin \left( \delta_B + \delta_D \right) \sin \left( \gamma \right)
  }{
    r_B^2 + r_D^2 + 2 r_B r_D \cos \left( \delta_B + \delta_D \right) \cos \left( \gamma \right),  
  }
\end{eqnarray}
where the ratio of $\B$ decay amplitudes\footnote{
  Note that here we use the notation $r_B$ to denote the ratio
  of $\B$ decay amplitudes, 
  whereas in Sec.~\ref{sec:cp_uta:notations:non_cp:dstarpi} 
  we used, \eg, $R_{D\pi}$, for a rather similar quantity.
  The reason is that here we need to be concerned also with 
  $D$ decay amplitudes,
  and so it is convenient to use the subscript to denote the decaying particle.
  Hopefully, using $r$ in place of $R$ will help reduce potential confusion.
} 
is usually defined to be less than one,
\begin{equation}
  \label{eq:cp_uta:dk:rb_def}
  r_B = 
  \frac{
    \left| A\left( \Bm \to \Dzb K^- \right) \right|
  }{
    \left| A\left( \Bm \to \Dz  K^- \right) \right|
  },
\end{equation}
and the ratio of $D$ decay amplitudes is correspondingly defined by
\begin{equation}
  \label{eq:cp_uta:dk:rd_def}
  r_D = 
  \frac{
    \left| A\left( \Dz  \to f \right) \right|
  }{
    \left| A\left( \Dzb \to f \right) \right|
  }.
\end{equation}
The strong phase differences between the $\B$ and $D$ decay amplitudes 
are given by $\delta_B$ and $\delta_D$, respectively.
The values of $r_D$ and $\delta_D$ depend on the final state $f$:
for the GLW analysis, $r_D = 1$ and $\delta_D$ is trivial (either zero or $\pi$),
in the Dalitz plot analysis $r_D$ and $\delta_D$ vary across the Dalitz plot,
and depend on the $D$ decay model used,
for the ADS analysis, the values of $r_D$ and $\delta_D$ are not trivial.

Note that, for given values of $r_B$ and $r_D$, 
the maximum size of ${\cal A}$ (at $\sin \left( \delta_B + \delta_D \right) = 1$)
is $2 r_B r_D \sin \left( \gamma \right) / \left( r_B^2 + r_D^2 \right)$.
Thus even for $D$ decay modes with small $r_D$, 
large asymmetries, and hence sensitivity to $\gamma$, 
may occur for $B$ decay modes with similar values of $r_B$.
For this reason, the ADS analysis of the decay $B^\mp \to D \pi^\mp$ 
is also of interest.

In the GLW analysis, the measured quantities are the 
partial rate asymmetry, and the charge averaged rate,
which are measured both for $\CP$ even and $\CP$ odd $D$ decays.
For the latter, it is experimentally convenient to measure a double ratio,
\begin{equation}
  \label{eq:cp_uta:dk:double_ratio}
  R_{\CP} = 
  \frac{
    \Gamma\left( \Bm \to D_{\CP} \Km  \right) \, / \, \Gamma\left( \Bm \to \Dz \Km \right)
  }{
    \Gamma\left( \Bm \to D_{\CP} \pim \right) \, / \, \Gamma\left( \Bm \to \Dz \pim \right)
  }
\end{equation}
that is normalized both to the rate for the favoured $\Dz \to \Km\pip$ decay, 
and to the equivalent quantities for $\Bm \to D\pim$ decays
(charge conjugate modes are implicitly 
included in Eq.~(\ref{eq:cp_uta:dk:double_ratio})).
In this way the constant of proportionality drops out of 
Eq.~(\ref{eq:cp_uta:dk:av_rate_def}).

For the ADS analysis, using a suppressed $D \to f$ decay,
the measured quantities are again the partial rate asymmetry, 
and the charge averaged rate.
In this case it is sufficient to measure the rate in a single ratio
(normalized to the favoured $D \to \bar{f}$ decay)
since detection systematics cancel naturally;
the observed quantity is then
\begin{equation}
  \label{eq:cp_uta:dk:r_ads}
  R_{\rm ADS} = 
  \frac{
    \Gamma \left( \Bm \to \left[ f \right]_D \Km \right)
  }{
    \Gamma \left( \Bm \to \left[ \bar{f} \right]_D \Km \right)
  }.
\end{equation}
In the ADS analysis, there are an additional two unknowns ($r_D$ and $\delta_D$)
compared to the GLW case.  
However, the value of $r_D$ can be measured using 
decays of $D$ mesons of known flavour.

In the Dalitz plot analysis,
once a model is assumed for the $D$ decay, 
which gives the values of $r_D$ and $\delta_D$ across the Dalitz plot,
it is possible to perform a simultaneous fit to the $B^+$ and $B^-$ samples 
and directly extract $\gamma$, $r_B$ and $\delta_B$.
However, the uncertainties on the phases depend inversely on $r_B$.
Furthermore, $r_B$ is positive definite (and small), 
and therefore tends to be overestimated,
which can lead to an underestimation of the uncertainty.
Some statistical treatment is necessary to correct for this bias.
An alternative approach is to extract from the data the ``Cartesian''
variables
\begin{equation}
  \left( x_\pm, y_\pm \right) = 
  \left( \Re(r_B e^{i(\delta_B\pm\gamma)}), \Im(r_B e^{i(\delta_B\pm\gamma)}) \right) = 
  \left( r_B \cos(\delta_B\pm\gamma), r_B \sin(\delta_B\pm\gamma) \right).
\end{equation}
These are (a) approximately statistically uncorrelated 
and (b) almost Gaussian.
Use of these variables makes the combination of results much simpler.

The relations between the measured quantities and the
underlying parameters are summarized in Table~\ref{tab:cp_uta:notations:dk}.
Note carefully that the hadronic factors $r_B$ and $\delta_B$ 
are different, in general, for each $\B$ decay mode.

\begin{table}
  \begin{center} 
    \caption{
      Summary of relations between measured and physical parameters 
      in GLW and ADS analyses of $\B \to \DorDstar K^{(*)}$.
    }
    \vspace{0.2cm}
    \setlength{\tabcolsep}{1.0pc}
    \begin{tabular}{cc} \hline 
      \mc{2}{c}{GLW analysis} \\
      $R_{\CP\pm}$ & $1 + r_B^2 \pm 2 r_B \cos \left( \delta_B \right) \cos \left( \gamma \right)$ \\
      $A_{\CP\pm}$ & $\pm 2 r_B \sin \left( \delta_B \right) \sin \left( \gamma \right) / R_{\CP\pm}$ \\
      \hline
      \mc{2}{c}{ADS analysis} \\
      $R_{\rm ADS}$ & $r_B^2 + r_D^2 + 2 r_B r_D \cos \left( \delta_B + \delta_D \right) \cos \left( \gamma \right)$ \\
      $A_{\rm ADS}$ & $2 r_B r_D \sin \left( \delta_B + \delta_D \right) \sin \left( \gamma \right) / R_{\rm ADS}$ \\
      \hline
      \mc{2}{c}{Dalitz analysis} \\
      $x_\pm$ & $r_B \cos(\delta_B\pm\gamma)$ \\
      $y_\pm$ & $r_B \sin(\delta_B\pm\gamma)$ \\
      \hline
    \end{tabular}
    \label{tab:cp_uta:notations:dk}
  \end{center}
\end{table}

\mysubsection{Common inputs and error treatment
}
\label{sec:cp_uta:common_inputs}

The common inputs used for rescaling are listed in 
Table~\ref{tab:cp_uta:common_inputs}.
The $\Bz$ lifetime ($\tau(\Bz)$) and mixing parameter ($\Delta m_d$)
averages are provided by the HFAG Lifetimes and Oscillations 
subgroup (Sec.~\ref{sec:life_mix}).
The fraction of the perpendicularly polarized component 
($\left| A_{\perp} \right|^2$) in $\B \to \jpsi \Kstar(892)$ decays,
which determines the $\CP$ composition, 
is averaged from results by 
\babar~\cite{ref:cp_uta:ccs:babar:psi_kstar} and
\belle~\cite{ref:cp_uta:ccs:belle:psi_kstar}.

At present, we only rescale to a common set of input parameters
for modes with reasonably small statistical errors
($b \to c\bar{c}s$ transitions).
Correlated systematic errors are taken into account
in these modes as well.
For all other modes, the effect of such a procedure is 
currently negligible.

\begin{table}
  \begin{center}
    \caption{
      Common inputs used in calculating the averages.
    }
    \vspace{0.2cm}
    \setlength{\tabcolsep}{1.0pc}
    \begin{tabular}{cc} \hline 
      $\tau(\Bz)$ $({\rm ps})$  & $1.528 \pm 0.009$  \\
      $\Delta m_d$ $({\rm ps}^{-1})$ & $0.509 \pm 0.004$ \\
      $\left| A_{\perp} \right|^2 (\jpsi \Kstar)$ & $0.217 \pm 0.010$ \\
      \hline
    \end{tabular}
    \label{tab:cp_uta:common_inputs}
  \end{center}
\end{table}

As explained in Sec.~\ref{sec:intro},
we do not apply a rescaling factor on the error of an average
that has $\chi^2/\dof > 1$ 
(unlike the procedure currently used by the PDG~\cite{Eidelman:2004wy}).
We provide a confidence level of the fit so that
one can know the consistency of the measurements included in the average,
and attach comments in case some care needs to be taken in the interpretation.
Note that, in general, results obtained from data samples with low statistics
will exhibit some non-Gaussian behaviour.
For measurements where one error is given, 
it represents the total error, 
where statistical and systematic uncertainties have been added in quadrature.
If two errors are given, the first is statistical and the second systematic.
If more than two errors are given,
the origin of the additional uncertainty will be explained in the text.

Averages are computed by maximizing a log-likelihood function 
${\cal L}$ assuming Gaussian statistical and systematic errors.
When observables exhibit significant correlations 
(\eg, sine and cosine coefficients in some time-dependent \CP asymmetries), 
a combined minimization is performed, 
taking into account the correlations. 
Asymmetric errors are treated by defining an asymmetric log-likelihood
function: ${\cal L}_i = (x - x_i)^2/(2\sigma_{i}^2)$,
where $\sigma_i=\sigma_{i,+}$ ($\sigma_i=\sigma_{i,-}$) if $x>x_i$ ($x<x_i$), 
and where $x_i$ is the $i$th measurement of 
the observable $x$ that is averaged.
This example assumes no correlations between observables. 
The correlated case is a straightforward extension to this.

\mysubsection{Time-dependent $\CP$ asymmetries in $b \to c\bar{c}s$ transitions
}
\label{sec:cp_uta:ccs}

\begin{figure}[htb]
  \begin{center}
    \resizebox{0.55\textwidth}{!}{
      \includegraphics{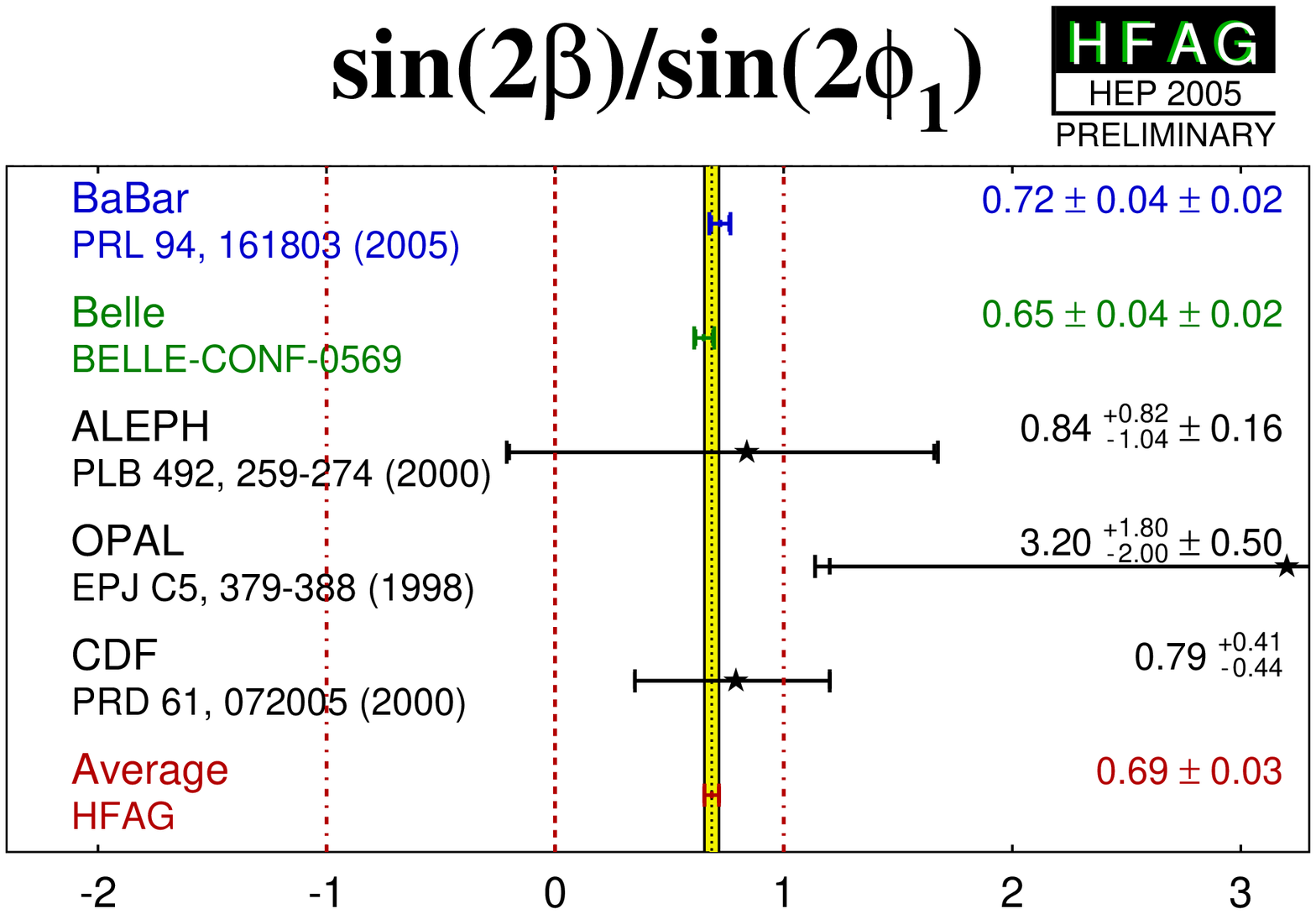}
    }
    \hfill
    \resizebox{0.44\textwidth}{!}{
      \includegraphics{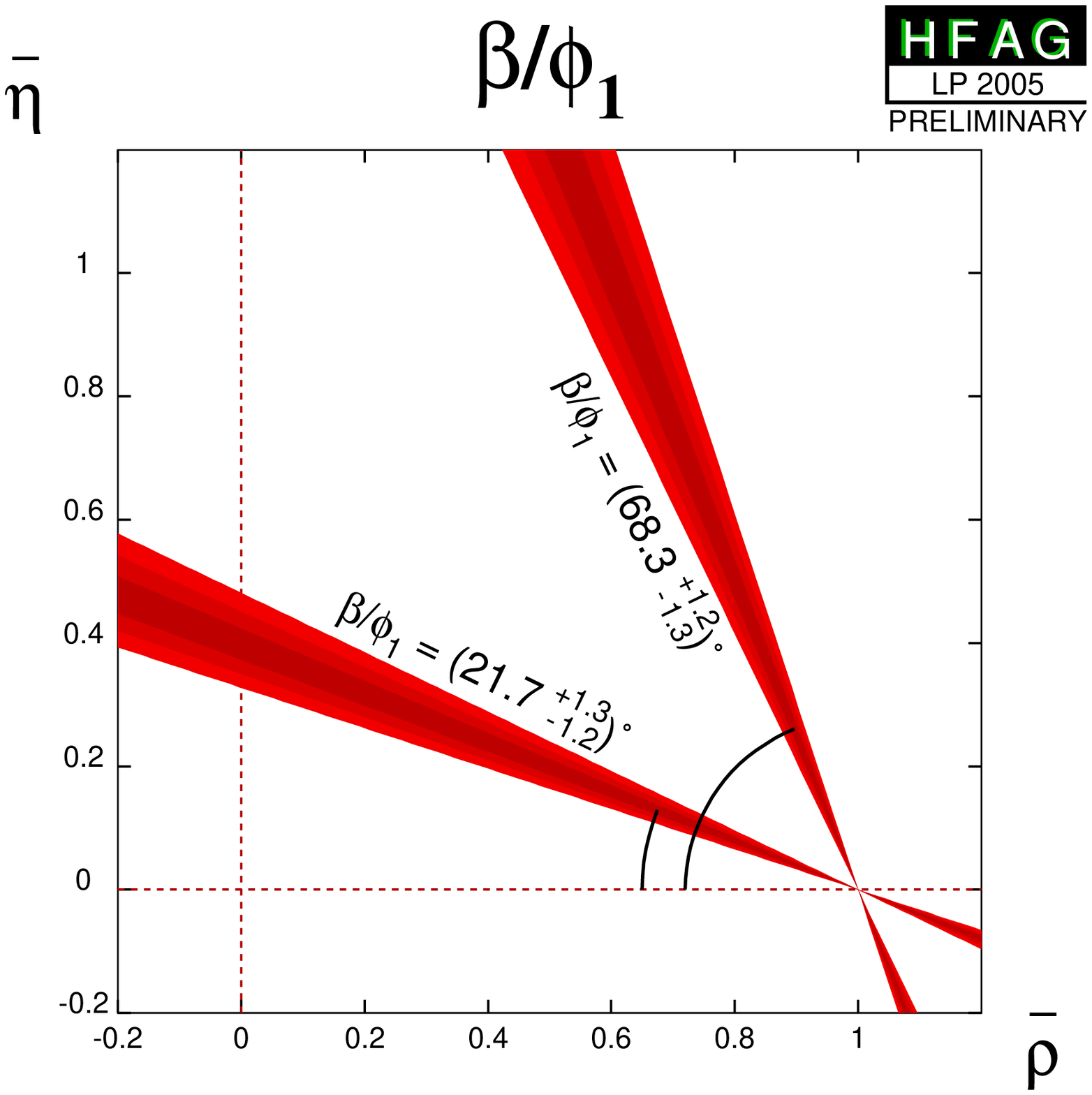}
    }
  \end{center}
  \vspace{-0.5cm}
  \caption{
    (Left) Average of measurements of $S_{b \to c\bar c s}$.
    (Right) Constraints on the $(\rhobar,\etabar)$ plane,
    obtained from the average of $-\etacp S_{b \to c\bar c s}$ 
    and Eq.~\ref{eq:cp_uta:sin2beta}.
  }
  \label{fig:cp_uta:ccs}
\end{figure}

In the Standard Model, the time-dependent parameters for
$b \to c\bar c s$ transitions are predicted to be: 
$S_{b \to c\bar c s} = - \etacp \sin(2\beta)$,
$C_{b \to c\bar c s} = 0$ to very good accuracy.
The averages for $-\etacp S_{b \to c\bar c s}$ and $C_{b \to c\bar c s}$
are provided in Table~\ref{tab:cp_uta:ccs}.
The averages for $-\etacp S_{b \to c\bar c s}$ 
are shown in Fig.~\ref{fig:cp_uta:ccs}.

Both \babar\  and \belle\ have used the $\etacp = -1$ modes
$\jpsi \KS$, $\psi(2S) \KS$, $\chi_{c1} \KS$ and $\eta_c \KS$, 
as well as $\jpsi \KL$, which has $\etacp = +1$
and $\jpsi K^{*0}(892)$, which is found to have $\etacp$ close to $+1$
based on the measurement of $\left| A_\perp \right|$ 
(see Sec.~\ref{sec:cp_uta:common_inputs}).
ALEPH, OPAL and CDF use only the $\jpsi \KS$ final state.
In the latest result from \belle, 
only $\jpsi \KS$ and $\jpsi \KL$ are used.
In future updates, it is hoped to perform separate averages for 
each charmonium-kaon final state.

\begin{table}
  \begin{center}
    \caption{
      $S_{b \to c\bar c s}$ and $C_{b \to c\bar c s}$.
    }
    \vspace{0.2cm}
    \setlength{\tabcolsep}{0.0pc}
    \begin{tabular*}{\textwidth}{@{\extracolsep{\fill}}lrcc} \hline 
      \mc{2}{l}{Experiment} & 
      $- \etacp S_{b \to c\bar c s}$ & $C_{b \to c\bar c s}$ \\
      \hline
      \babar & \cite{ref:cp_uta:ccs:babar} & 
      $0.722 \pm 0.040 \pm 0.023$ & $\ph{-}0.051 \pm 0.033 \pm 0.014$ \\
      \belle & \cite{ref:cp_uta:ccs:belle} & 
      $0.652 \pm 0.039 \pm 0.020$ & $-0.010 \pm 0.026 \pm 0.036$ \\
      \hline
      \mc{2}{l}{\bf \boldmath $\B$ factory average} & 
      $0.685 \pm 0.032$ & $0.026 \pm 0.041$ \\
      \mc{2}{l}{\small Confidence level} & 
      \small $0.27$ & \small $0.31$ \\
      \hline
      ALEPH & \cite{ref:cp_uta:ccs:aleph} & 
      $0.84 \, ^{+0.82}_{-1.04} \pm 0.16$ \\
      OPAL  & \cite{ref:cp_uta:ccs:opal}  & 
      $3.2 \, ^{+1.8}_{-2.0} \pm 0.5$ \\
      CDF   & \cite{ref:cp_uta:ccs:cdf}   & 
      $0.79 \, ^{+0.41}_{-0.44}$ \\
      \hline
      \mc{2}{l}{\bf Average} & 
      $0.687 \pm 0.032$ & $0.026 \pm 0.041$ \\
      \hline
    \end{tabular*}
    \label{tab:cp_uta:ccs}
  \end{center}
\end{table}

It should be noted that, while the uncertainty in the average for 
$-\etacp S_{b \to c\bar c s}$ is still limited by statistics,
that for $C_{b \to c\bar c s}$ is currently dominated by systematics.
This occurs due to the possible effect of tag side interference on the
$C_{b \to c\bar c s}$ measurement, an effect which is correlated between
the different experiments.
Understanding of this effect may improve in future,
allowing the uncertainty to reduce.

\begin{figure}[p]
  \begin{center}
    \resizebox{0.63\textwidth}{!}{
      \includegraphics{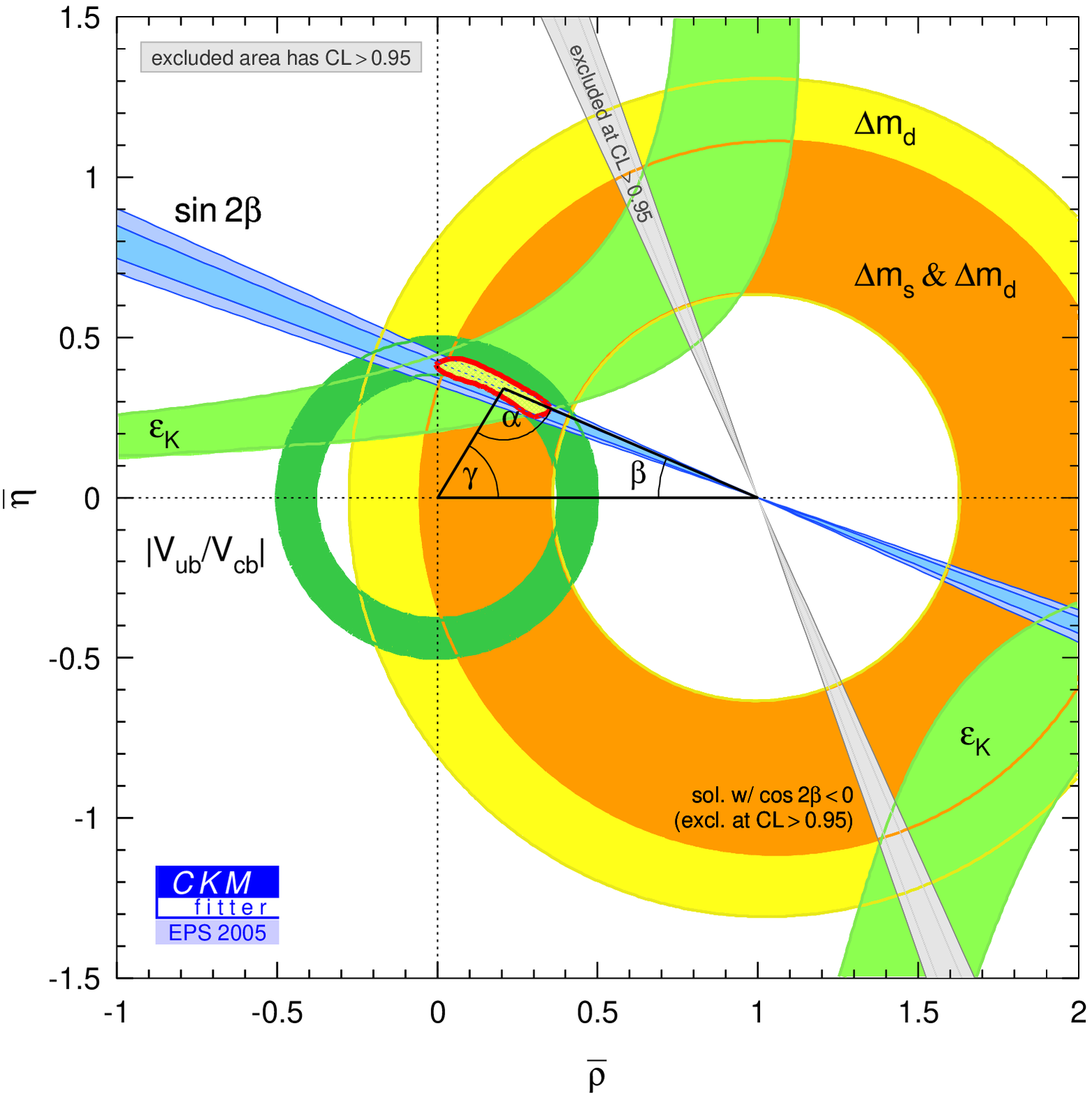}
    }
    \resizebox{0.70\textwidth}{!}{\hspace{0.5cm}
      \includegraphics{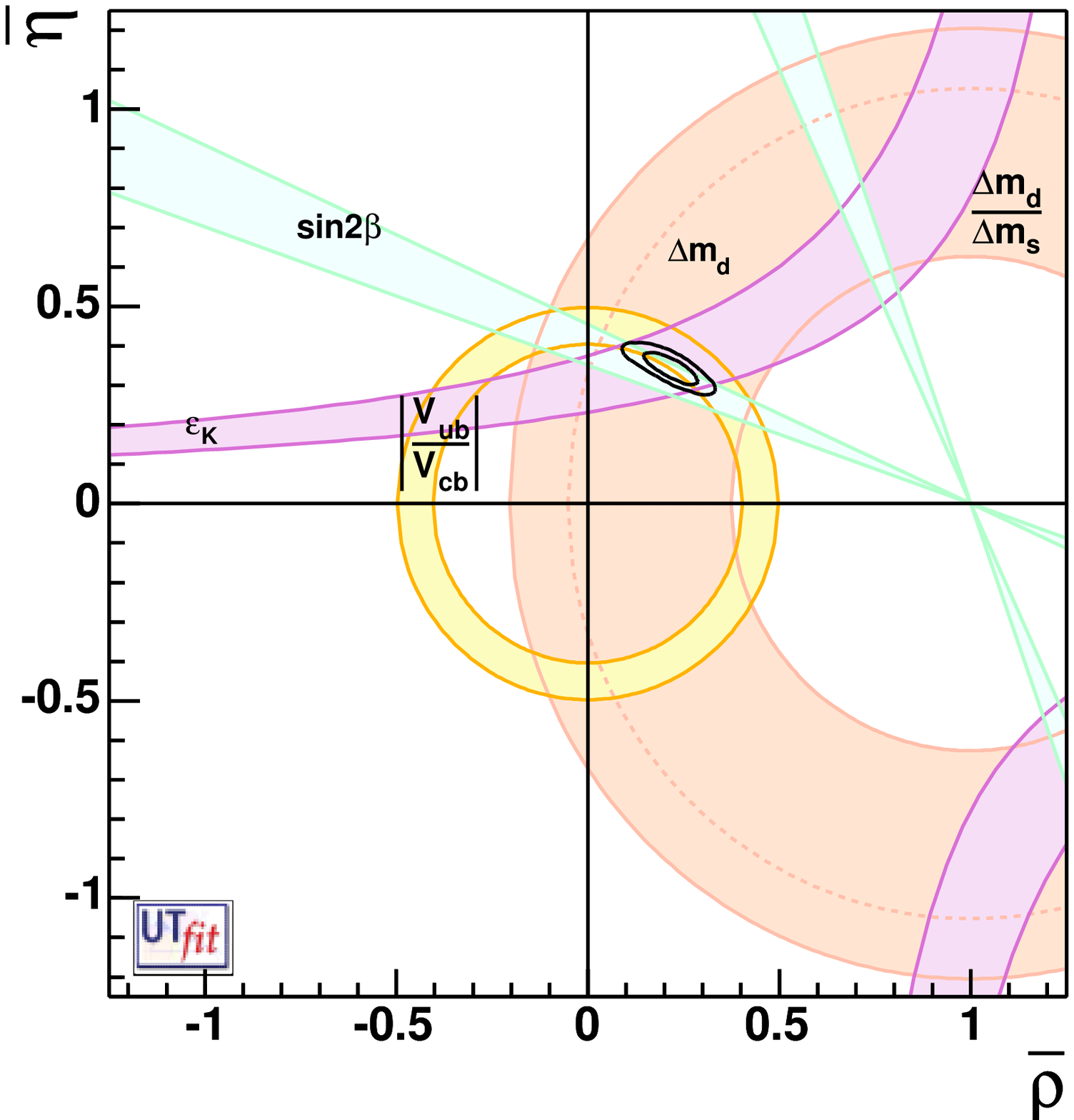}
    }
  \end{center}
  \vspace{-0.5cm}
  \caption{
    Typical Standard Model constraints on the $(\rhobar,\etabar)$ plane,
    from (top)~\cite{ref:cp_uta:ckmfitter} and 
    (bottom)~\cite{ref:cp_uta:utfit}.
  }
  \label{fig:cp_uta:ckmfitter_sin2beta}
\end{figure}

From the average for $-\etacp S_{b \to c\bar c s}$ above, 
we obtain the following solutions for $\beta$
(in $\left[ 0, \pi \right]$):
\begin{equation}
  \beta = \left( 21.7 \, ^{+1.3}_{-1.2} \right)^\circ
  \hspace{5mm}
  {\rm or}
  \hspace{5mm}
  \beta = \left( 68.3 \, ^{+1.2}_{-1.3} \right)^\circ
  \label{eq:cp_uta:sin2beta}
\end{equation}
This result gives a precise constraint on the $(\rhobar,\etabar)$ plane,
as shown in Fig.~\ref{fig:cp_uta:ccs}.
The measurement is in remarkable agreement with other constraints from 
$\CP$ conserving quantities, 
and with $\CP$ violation in the kaon system, in the form of the parameter $\epsilon_K$.
Such comparisons have been performed by various phenomenological groups,
such as CKMfitter~\cite{ref:cp_uta:ckmfitter} 
and UTFit~\cite{ref:cp_uta:utfit}.
Fig.~\ref{fig:cp_uta:ckmfitter_sin2beta} displays the constraints 
obtained from these two groups.

\mysubsection{Time-dependent transversity analysis of $\Bz \to J/\psi K^{*0}$
}
\label{sec:cp_uta:ccs_vv}

$\B$ meson decays to the vector-vector final state $J/\psi K^{*0}$
are also mediated by the $b \to c \bar c s$ transition.
When a final state which is not flavour-specific ($K^{*0} \to \KS \pi^0$) is used,
a time-dependent transversity analysis can be performed 
allowing sensitivity to both 
$\sin(2\beta)$ and $\cos(2\beta)$~\cite{ref:cp_uta:ccs:dqstl}.
Such analyses have been performed by both $\B$ factory experiments.
In principle, the strong phases between the transversity amplitudes
are not uniquely determined by such an analysis, 
leading to a discrete ambiguity in the sign of $\cos(2\beta)$.
The \babar\ collaboration resolves 
this ambiguity using the known variation~\cite{ref:cp_uta:ccs:lass}
of the P-wave phase (fast) relative to the S-wave phase (slow) 
with the invariant mass of the $K\pi$ system 
in the vicinity of the $K^*(892)$ resonance. 
The result is in agreement with the prediction from 
$s$ quark helicity conservation,
and corresponds to Solution II defined by Suzuki~\cite{ref:cp_uta:ccs:suzuki}.
We use this phase convention for the averages given in 
Table~\ref{tab:cp_uta:ccs:psi_kstar}.

\begin{table}
  \begin{center}
    \caption{
      Averages from $\Bz \to J/\psi K^{*0}$ transversity analyses.
    }
    \vspace{0.2cm}
    \setlength{\tabcolsep}{0.0pc}
    \begin{tabular*}{\textwidth}{@{\extracolsep{\fill}}lrccc} \hline 
      \mc{2}{l}{Experiment} & $\sin(2\beta)$ & $\cos(2\beta)$ & Correlation \\
      \hline
      \babar & \cite{ref:cp_uta:ccs:babar:psi_kstar} & 
      $-0.10 \pm 0.57 \pm 0.14$ & $ 3.32 \, ^{+0.76}_{-0.96} \pm 0.27$ & $-0.37$ \\
      \belle & \cite{ref:cp_uta:ccs:belle:psi_kstar} &
      $\ph{-}0.24 \pm 0.31 \pm 0.05$ & $ 0.56 \pm 0.79 \pm 0.11 $ & $+0.22$ \\
      \hline
      \mc{2}{l}{\bf Average} & 
      \mc{2}{c}{\sc In Preparation} \\
      \hline
    \end{tabular*}
    \label{tab:cp_uta:ccs:psi_kstar}
  \end{center}
\end{table}

At present the results are dominated by 
large and non-Gaussian statistical errors.
In addition, there are significant correlations which need to be taken 
into account.
At present, we do not provide averages for the 
results in Table~\ref{tab:cp_uta:ccs:psi_kstar};
nevertheless $\cos(2\beta)>0$ is preferred 
by the experimental data in $J/\psi \Kstar$.

\mysubsection{Time-dependent $\CP$ asymmetries in colour-suppressed $b \to c\bar{u}d$ transitions
}
\label{sec:cp_uta:cud_beta}

Decays of $\B$ mesons to final states such as $D\pi^0$ are 
governed by $b \to c\bar{u}d$ transitions. 
If the final state is a $\CP$ eigenstate, \ie\ $D_{\CP}\pi^0$, 
the usual time-dependence formulae are recovered, 
with the sine coefficient sensitive to $\sin(2\beta)$. 
Since there is no penguin contribution to these decays, 
there is even less associated theoretical uncertainty 
than for $b \to c\bar{c}s$ decays like $\B \to \jpsi \KS$.
Such measurements therefore allow to test the Standard Model prediction
that the $\CP$ violation parameters in $b \to c\bar{u}d$ transitions
are the same as those in $b \to c\bar{c}s$~\cite{ref:cp_uta:cud_beta:gw}.

Note that there is an additional contribution from CKM suppressed
$b \to u \bar{c} d$ decays.
The effect of this contribution is small, and can be taken into 
account in the analysis~\cite{ref:cp_uta:cud_beta:fleischer}.

When multibody $D$ decays, such as $D \to \KS\pi^+\pi^-$ are used, 
a time-dependent analysis of the Dalitz plot of the neutral $D$ decay 
allows a direct determination of the weak phase: $2\beta$. 
(Equivalently, both $\sin(2\beta)$ and $\cos(2\beta)$ can be measured.)
This information allows to resolve the ambiguity in the 
measurement of $2\beta$ from $\sin(2\beta)$~\cite{ref:cp_uta:cud_beta:bgk}.

Results of such an analysis are available from Belle.
The decays $\B \to D\pi^0$, $\B \to D\eta$, $\B \to D\omega$, 
$\B \to D^*\pi^0$ and $\B \to D^*\eta$ are used. 
[This collection of states is denoted by $D^{(*)}h^0$.]
The daughter decays are $D^* \to D\pi^0$ and $D \to \KS\pi^+\pi^-$.
The results are shown in Table~\ref{tab:cp_uta:cud_beta}.

\begin{table}
  \begin{center}
    \caption{
      Averages from $\Bz \to D^{(*)}h^0$ analyses.
    }
    \vspace{0.2cm}
    \setlength{\tabcolsep}{0.0pc}
    \begin{tabular*}{\textwidth}{@{\extracolsep{\fill}}lrc} \hline 
      \mc{2}{l}{Experiment} & $\beta \, (^\circ)$ \\
      \hline
      \belle & \cite{ref:cp_uta:cud_beta:belle} &
      $16 \pm 21 \pm 12$ \\
      \hline
    \end{tabular*}
    \label{tab:cp_uta:cud_beta}
  \end{center}
\end{table}

Again, it is clear that the data prefer $\cos(2\beta)>0$.
Taken in conjunction with the $\jpsi\Kstar$ results,
$\cos(2\beta)<0$ can be considered to be ruled out (at approximately $3\sigma$).

\mysubsection{Time-dependent $\CP$ asymmetries in charmless $b \to q\bar{q}s$ transitions
}
\label{sec:cp_uta:qqs}

The flavour changing neutral current $b \to s$ penguin
can be mediated by any up-type quark in the loop, 
and hence the amplitude can be written as
\begin{equation}
  \label{eq:cp_uta:b_to_s}
  \begin{array}{ccccc}
    A_{b \to s} & = & 
    \mc{3}{l}{F_u V_{ub}V^*_{us} + F_c V_{cb}V^*_{cs} + F_t V_{tb}V^*_{ts}} \\
    & = & (F_u - F_c) V_{ub}V^*_{us} & + & (F_t - F_c) V_{tb}V^*_{ts} \\
    & = & {\cal O}(\lambda^4) & + & {\cal O}(\lambda^2) \\
  \end{array}
\end{equation}
using the unitarity of the CKM matrix.
Therefore, in the Standard Model, 
this amplitude is dominated by $V_{tb}V^*_{ts}$, 
and to within a few degrees ($\delta\beta \lesssim 2^\circ$ for $\beta \approx 20^\circ$) 
the time-dependent parameters can be written\footnote
{
  The presence of a small (${\cal O}(\lambda^2)$) weak phase in 
  the dominant amplitude of the $s$ penguin decays introduces 
  a phase shift given by
  $S_{b \to q\bar q s} = -\eta\sin(2\beta)\cdot(1 + \Delta)$. 
  Using the CKMfitter results for the Wolfenstein 
  parameters~\cite{ref:cp_uta:ckmfitter}, one finds: 
  $\Delta \simeq 0.033$, which corresponds to a shift of 
  $2\beta$ of $+2.1$ degrees. Nonperturbative contributions
  can alter this result.
}
$S_{b \to q\bar q s} \approx - \etacp \sin(2\beta)$,
$C_{b \to q\bar q s} \approx 0$,
assuming $b \to s$ penguin contributions only ($q = u,d,s$).

Due to the large virtual mass scales occurring in the penguin loops,
additional diagrams from physics beyond the Standard Model,
with heavy particles in the loops, may contribute.
In general, these contributions will affect the values of 
$S_{b \to q\bar q s}$ and $C_{b \to q\bar q s}$.
A discrepancy between the values of 
$S_{b \to c\bar c s}$ and $S_{b \to q\bar q s}$
can therefore provide a clean indication of new physics.

However, there is an additional consideration to take into account.
The above argument assumes only the $b \to s$ penguin contributes
to the $b \to q\bar q s$ transition.
For $q = s$ this is a good assumption, 
which neglects only rescattering effects.
However, for $q = u$ there is a colour-suppressed $b \to u$ tree diagram
(of order ${\cal O}(\lambda^4)$), which has a different weak 
(and possibly strong) phase.
In the case $q = d$, any light neutral meson that is formed from
$d \bar{d}$ also has a $u \bar{u}$ component,
and so again there is ``tree pollution''. The \Bz decays to 
$\piz\KS$ and $\omega\KS$ belong to this category.
The mesons $f_0$ and $\etapr$ are expected to have predominant $s\bar s$ parts,
which reduces the possible tree pollution. If the inclusive 
decay $\Bz\to\Kp\Km\Kz$ (excluding $\phi\Kz$) is dominated by
a non-resonant three-body transition, an OZI-rule suppressed 
tree-level diagram can occur through insertion of an 
$s\sbar$ pair. The corresponding penguin-type transition 
proceeds via insertion of a $u\ubar$ pair, which is expected
to be favored over the $s\sbar$ insertion by fragmentation models.
Neglecting rescattering, the final state $\Kz\Kzb\Kz$ 
(reconstructed as $\KS\KS\KS$) has no tree pollution.
Various estimates, using different theoretical approaches,
of the values of $\Delta S = S_{b \to q\bar q s} - S_{b \to c\bar c s}$
exist in the literature~\cite{ref:cp_uta:qqs:theory}.
In general, there is agreement that the modes
$\phi\Kz$, $\etapr\Kz$ and $\Kz\Kzb\Kz$ are the cleanest,
with values of $\left| \Delta S \right|$ at or below the few percent level 
($\Delta S$ is usually positive).

\begin{figure}[htb]
  \begin{center}
    \resizebox{0.45\textwidth}{!}{
      \includegraphics{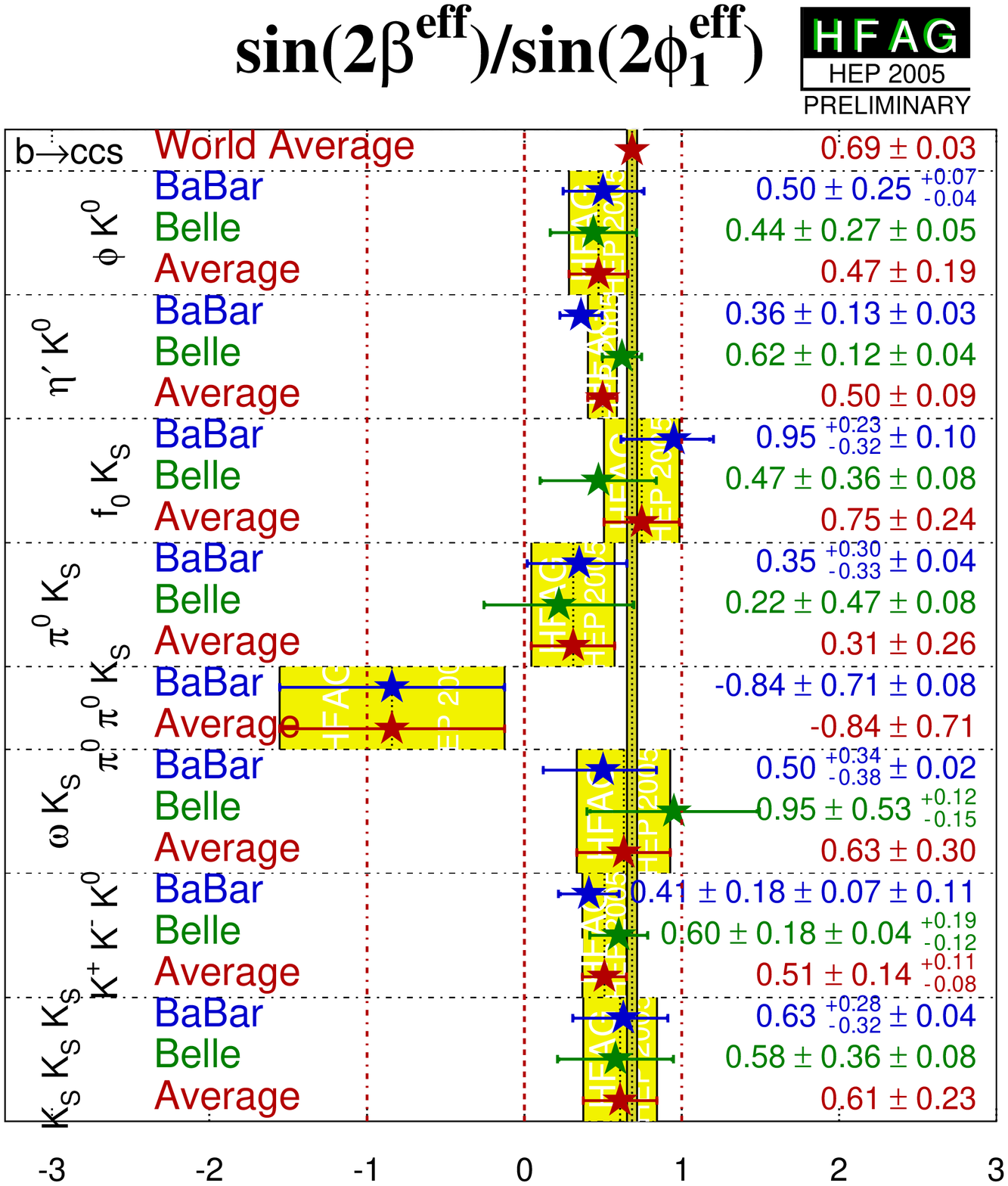}
    }
    \hfill
    \resizebox{0.45\textwidth}{!}{
      \includegraphics{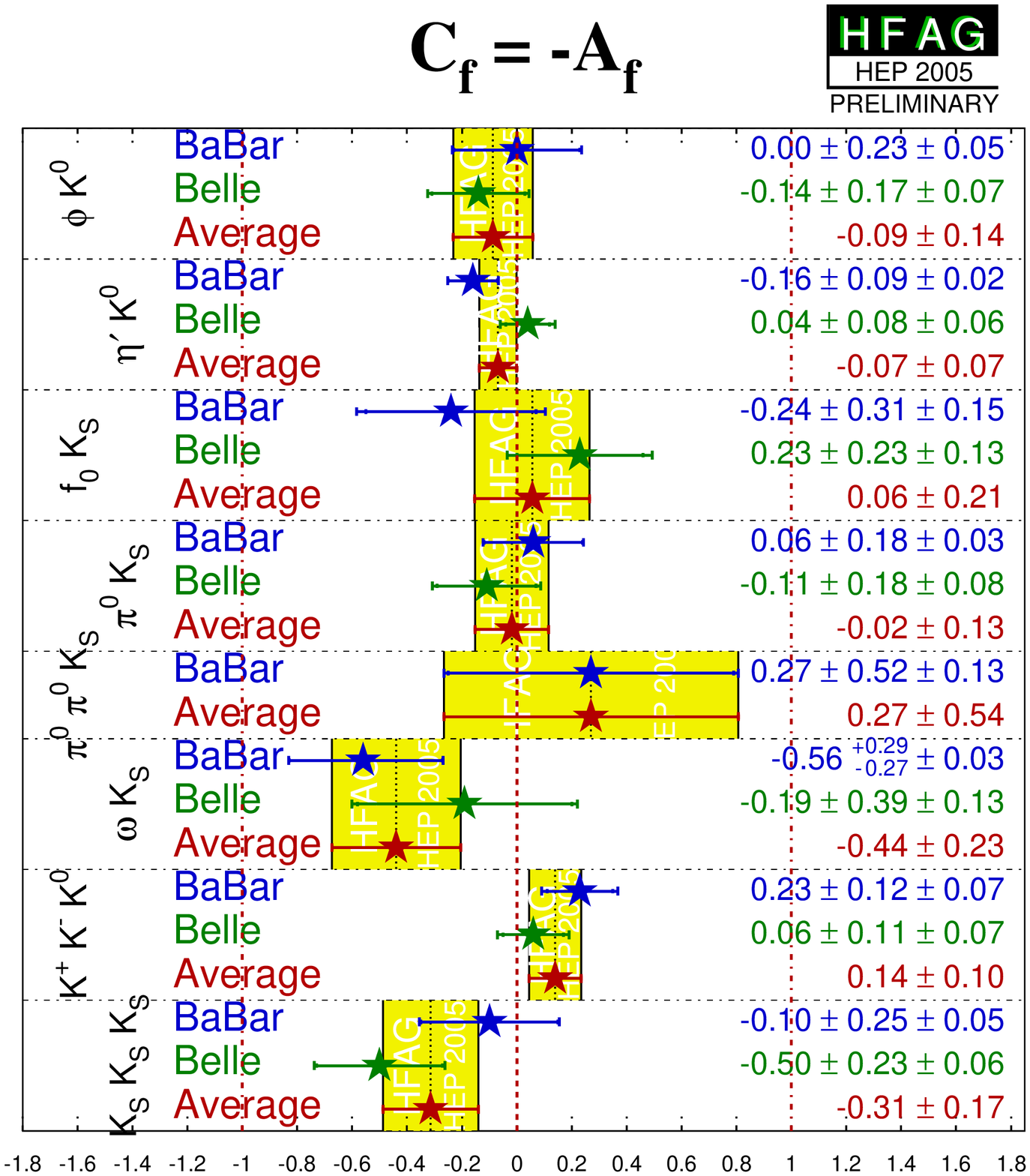}
    }
    \\
    \resizebox{0.45\textwidth}{!}{
      \includegraphics{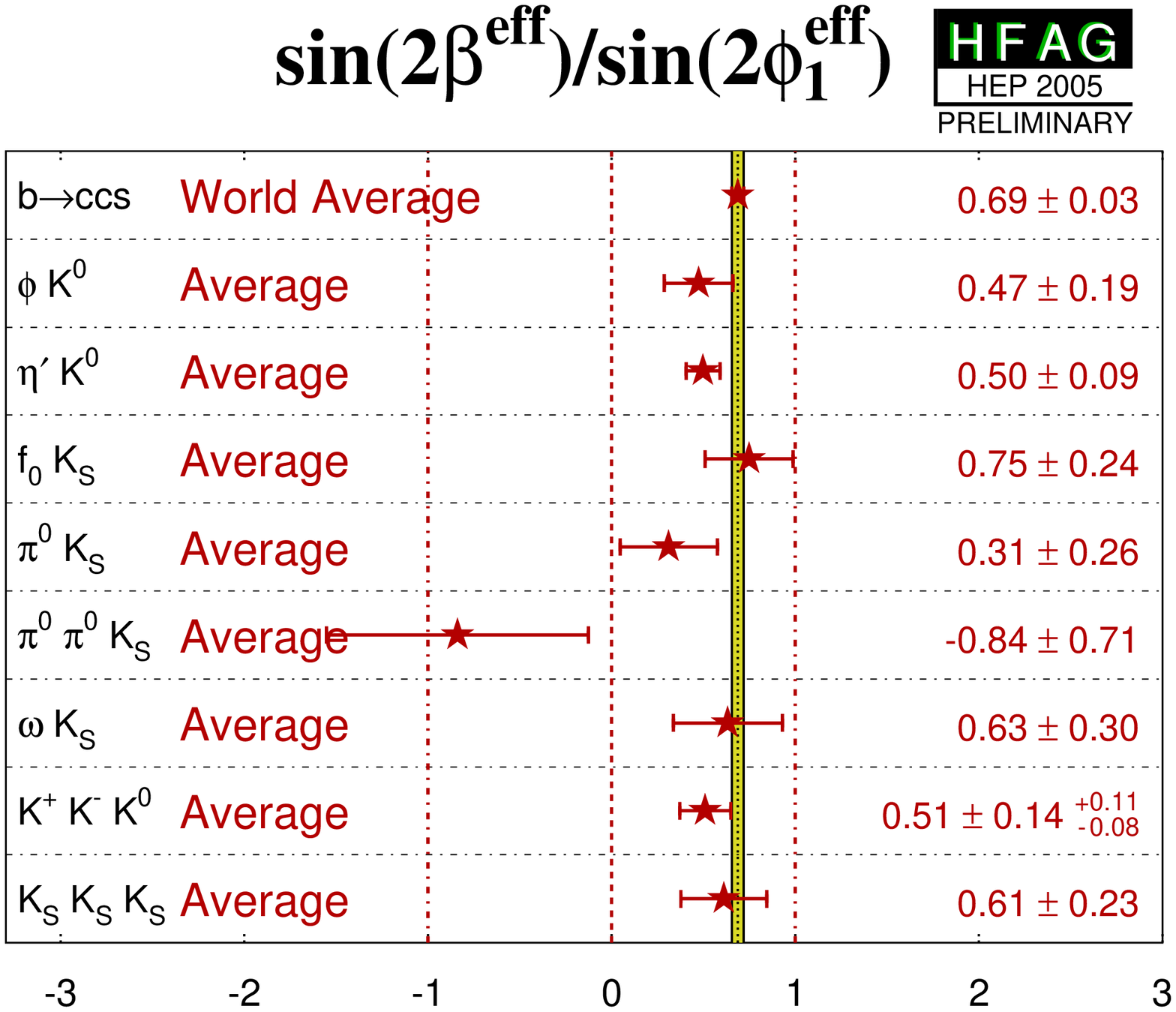}
    }
    \hfill
    \resizebox{0.45\textwidth}{!}{
      \includegraphics{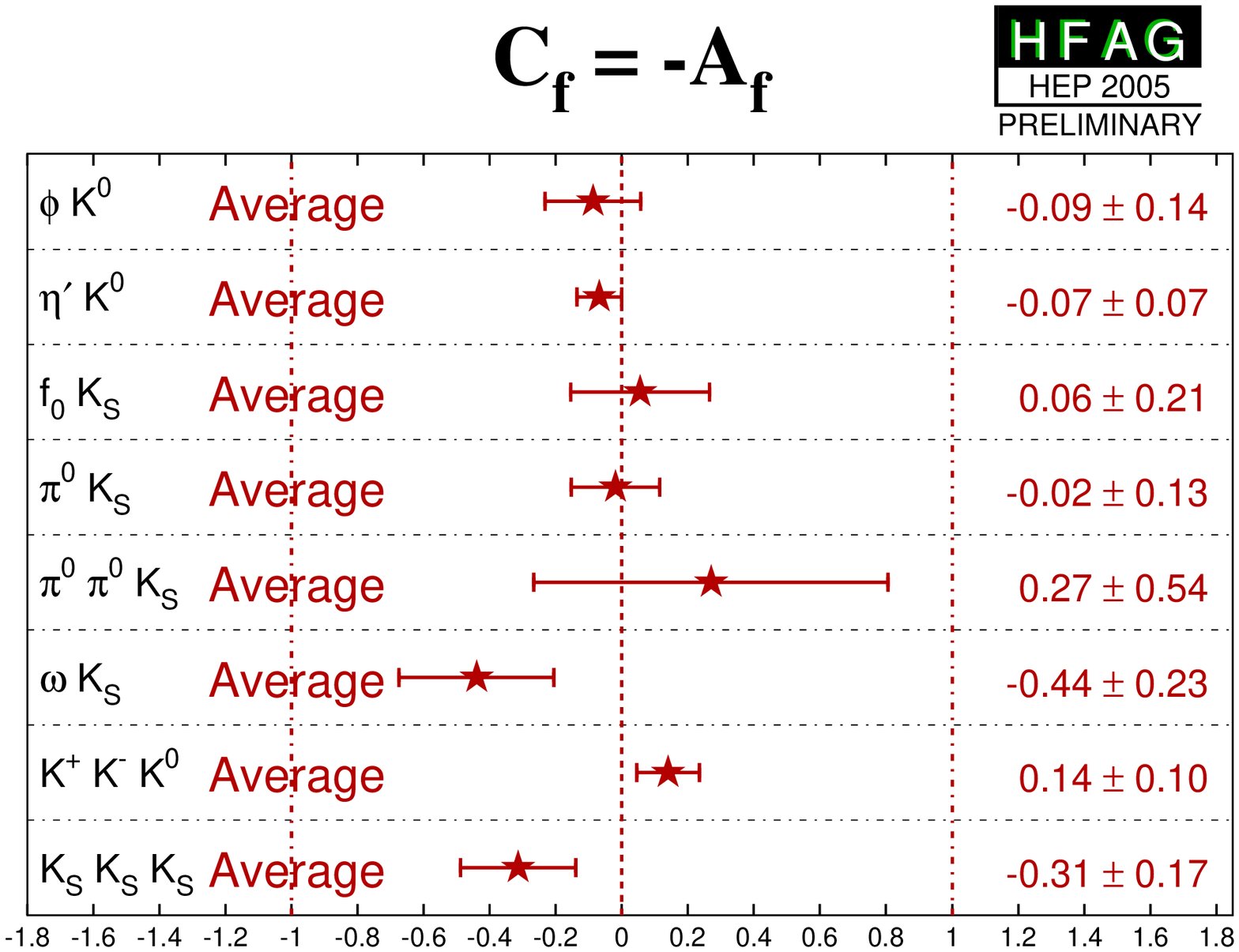}
    }
  \end{center}
  \vspace{-0.8cm}
  \caption{
    (Top)
    Averages of 
    (left) $-\etacp S_{b \to q\bar q s}$ and (right) $C_{b \to q\bar q s}$.
    The $-\etacp S_{b \to q\bar q s}$ figure compares the results to 
    the world average 
    for $-\etacp S_{b \to c\bar c s}$ (see Section~\ref{sec:cp_uta:ccs}).
    (Bottom) Same, but only averages for each mode are shown.
    More figures are available from the HFAG web pages.
  }
  \label{fig:cp_uta:qqs}
\end{figure}

The averages for $-\etacp S_{b \to q\bar q s}$ and $C_{b \to q\bar q s}$
can be found in Table~\ref{tab:cp_uta:qqs},
and are shown in Fig.~\ref{fig:cp_uta:qqs}.
Results from both \babar\  and \belle\ are averaged for the modes
$\phi K^0$, $\etapr K^0$ and $K^+K^- K^0$
($K^0$ indicates that both $\KS$ and $\KL$ are used, 
although \belle\ do not use $K^+K^-\KL$), 
$f_0 \KS$, $\pi^0 \KS$, $\omega\KS$ and $\KS\KS\KS$. 
\babar\ also has results using $\pi^0\pi^0\KS$.
Of these modes,
$\phi\KS$, $\etapr\KS$, $\pi^0 \KS$ and $\omega\KS$ have $\CP$ eigenvalue $\etacp = -1$, 
while $\phi\KL$, $\etapr\KL$, $f_0 \KS$, $\pi^0\pi^0\KS$ and $\KS\KS\KS$ 
have $\etacp = +1$.

The final state $K^+K^- K^0$ 
(contributions from $\phi K^0$ are implicitly excluded) 
is not a $\CP$ eigenstate.
However, the $\CP$ composition can be determined using either an 
isospin argument (used by \belle\ to determine a $\CP$ even fraction of 
$0.93 \pm 0.09 \pm 0.05$~\cite{ref:cp_uta:qqs:belle})
or a moments analysis (used by \babar\ to find $\CP$ even fractions of 
$0.89 \pm 0.08 \pm 0.06$ in $K^+K^-\KS$~\cite{ref:cp_uta:qqs:babar:phik0}
and
$0.92 \pm 0.07 \pm 0.06$ in $K^+K^-\KL$~\cite{ref:cp_uta:qqs:babar:kkk0}).
The uncertainty in the $\CP$ even fraction leads to an 
asymmetric error on $S_{b \to q\bar q s}$, which is taken to be 
correlated among the experiments.
To combine, we rescale the results to the 
average $\CP$ even fraction of $0.91 \pm 0.07$.

\begin{table}
  \begin{center}
    \caption{
      Averages of $-\etacp S_{b \to q\bar q s}$ and $C_{b \to q\bar q s}$.
      Note that the averages are calculated without taking correlations
      into account.
    }
    \vspace{0.2cm}
    \setlength{\tabcolsep}{0.0pc}
    \begin{tabular*}{\textwidth}{@{\extracolsep{\fill}}lrcc} 

      \hline 

      \mc{2}{l}{Experiment} & 
      $- \etacp S_{b \to q\bar q s}$ & $C_{b \to q\bar q s}$ \\

      \hline

      \mc{4}{c}{$\phi K^0$} \\
      \babar & \cite{ref:cp_uta:qqs:babar:phik0} & 
      $ 0.50 \pm 0.25 \, ^{+0.07}_{-0.04}$ & $\ph{-}0.00 \pm 0.23 \pm 0.05$ \\
      \belle & \cite{ref:cp_uta:qqs:belle} & 
      $ 0.44 \pm 0.27 \pm 0.05$ & $ -0.14 \pm 0.17 \pm 0.07$ \\
      \mc{2}{l}{\bf Average} & 
      $ 0.47 \pm 0.19 $ & $ -0.09 \pm 0.14$ \\
      \mc{2}{l}{\small Confidence level} & 
      \small $0.87~(0.2\sigma)$ & \small $0.64~(0.5\sigma)$ \\

      \hline

      \mc{4}{c}{$\etapr K^0$} \\
      \babar & \cite{ref:cp_uta:qqs:babar:etapks} & 
      $ 0.36 \pm 0.13 \pm 0.03 $ & $ -0.16 \pm 0.09 \pm 0.02$ \\
      \belle & \cite{ref:cp_uta:qqs:belle} & 
      $ 0.62 \pm 0.12 \pm 0.04$ & $ \ph{-}0.04 \pm 0.08 \pm 0.06$ \\
      \mc{2}{l}{\bf Average} & 
      $ 0.50 \pm 0.09 $ & $ -0.07 \pm 0.07$ \\
      \mc{2}{l}{\small Confidence level} & 
      \small $0.16~(1.4\sigma)$ & \small $0.14~(1.5\sigma)$ \\

      \hline

      \mc{4}{c}{$f_0 \KS$} \\
      \babar & \cite{ref:cp_uta:qqs:babar:f0ks} &
      $ 0.95 \, ^{+0.23}_{-0.32} \pm 0.10$ & $-0.24 \pm 0.31 \pm 0.15$ \\
      \belle & \cite{ref:cp_uta:qqs:belle} & 
      $ 0.47 \pm 0.36 \pm 0.08$ & $ \ph{-}0.23 \pm 0.23 \pm 0.13$ \\
      \mc{2}{l}{\bf Average} & 
      $ 0.75 \pm 0.24 $ & $ 0.06 \pm 0.21 $ \\
      \mc{2}{l}{\small Confidence level} & 
      \small $0.32~(1.0\sigma)$ & \small $0.28~(1.1\sigma)$ \\

      \hline

      \mc{4}{c}{$\pi^0 \KS$} \\
      \babar & \cite{ref:cp_uta:qqs:babar:pi0ks} &
      $ 0.35 \, ^{+0.30}_{-0.33} \pm 0.04$ & $\ph{-}0.06 \pm 0.18 \pm 0.03$ \\
      \belle & \cite{ref:cp_uta:qqs:belle} & 
      $ 0.22 \pm 0.47 \pm 0.08$ & $ -0.11 \pm 0.18 \pm 0.08$ \\
      \mc{2}{l}{\bf Average} & 
      $ 0.31 \pm 0.26 $ & $ -0.02 \pm 0.13$ \\ 
      \mc{2}{l}{\small Confidence level} & 
      \small $0.82~(0.2\sigma)$ & \small $0.53~(0.6\sigma)$ \\

      \hline

      \mc{4}{c}{$\pi^0 \pi^0 \KS$} \\
      \babar & \cite{ref:cp_uta:qqs:babar:pi0pi0ks} &
      $ -0.84 \pm 0.71 \pm 0.08$ & $ 0.27 \pm 0.52 \pm 0.13$ \\
      \mc{2}{l}{\bf Average} & 
      $ -0.84 \pm 0.71$ & $ 0.27 \pm 0.54$ \\ 

      \hline

      \mc{4}{c}{$\omega \KS$} \\
      \babar & \cite{ref:cp_uta:qqs:babar:omegaks} & 
      $0.50 \, ^{+0.34}_{-0.38} \pm 0.02$ & $-0.56 \, ^{+0.29}_{-0.27} \pm 0.03$ \\
      \belle & \cite{ref:cp_uta:qqs:belle} & 
      $ 0.95 \pm 0.53 \, ^{+0.12}_{-0.15}$ & $ -0.19 \pm 0.39 \pm 0.13 $ \\
      \mc{2}{l}{\bf Average} & 
      $ 0.63 \pm 0.30 $ & $ -0.44 \pm 0.23$ \\
      \mc{2}{l}{\small Confidence level} & 
      \small $0.49~(0.7\sigma)$ & \small $0.46~(0.7\sigma)$ \\

      \hline

      \mc{4}{c}{$K^+K^- K^0$} \\
      \babar & \cite{ref:cp_uta:qqs:babar:kkk0} &
      $ 0.41 \pm 0.18 \pm 0.07 \pm 0.11$ & $ 0.23 \pm 0.13$ \\
      \belle & \cite{ref:cp_uta:qqs:belle} & 
      $ 0.60 \pm 0.18 \pm 0.04 \, ^{+0.19}_{-0.12}$ & $ 0.06 \pm 0.11 \pm 0.07$ \\
      \mc{2}{l}{\bf Average} & 
      $ 0.51 \pm 0.14 \, ^{+0.11}_{-0.08}$ & $ 0.15 \pm 0.09 $ \\ 
      \mc{2}{l}{\small Confidence level} & 
      \small $0.38~(0.9\sigma)$ & \small $0.36~(0.9\sigma)$ \\

      \hline

      \mc{4}{c}{$\KS\KS\KS$} \\
      \babar & \cite{ref:cp_uta:qqs:babar:ksksks} &
      $ 0.63 \, ^{+0.28}_{-0.32} \pm 0.04$ & $ -0.10 \pm 0.25 \pm 0.05$ \\
      \belle & \cite{ref:cp_uta:qqs:belle} & 
      $ 0.58 \pm 0.36 \pm 0.08$ & $ -0.50 \pm 0.23 \pm 0.06$ \\
      \mc{2}{l}{\bf Average} & 
      $ 0.61 \pm 0.23 $ & $ -0.31 \pm 0.17$ \\
      \mc{2}{l}{\small Confidence level} & 
      \small $0.92~(0.1\sigma)$ & $0.25~(1.1\sigma)$ \\

      \hline 





    \end{tabular*}
    \label{tab:cp_uta:qqs}
  \end{center}
\end{table}

As explained above,
each of the modes listed in Table~\ref{tab:cp_uta:qqs} has
different uncertainties within the Standard Model,
and so each may have a different value of $-\etacp S_{b \to q\bar q s}$.
Therefore, there is no strong motivation to make a combined average
over the different modes.
We refer to such an average as a ``na\"\i ve $s$-penguin average.''
It is na\"\i ve not only because of the neglect of the theoretical uncertainty,
but also since possible correlations of systematic effects 
between different modes are neglected.
In spite of these caveats, there remains substantial interest 
in the value of this quantity,
and therefore it is given here:
$\langle -\etacp S_{b \to q\bar q s} \rangle = 0.50 \pm 0.06$,
with confidence level $0.79~(0.3\sigma)$.
Again treating the uncertainties as Gaussian and neglecting correlations,
this value is found to be $2.6\sigma$ below the average $-\etacp S_{b \to c\bar c s}$
given in Sec.~\ref{sec:cp_uta:ccs}.
(The average for $C_{b \to q\bar q s}$ is 
$\langle C_{b \to q\bar q s} \rangle = -0.04 \pm 0.04$
with confidence level $0.30~(1.0\sigma)$).
However, we do not advocate the use of these averages,
and we emphasize that the values should be treated with 
{\it extreme caution}, if at all.
What is unambiguous (although only qualitative) 
is that there is a trend that the values 
of $-\etacp S_{b \to c\bar c s}$ in different modes 
are below the average for $-\etacp S_{b \to c\bar c s}$.

From Table~\ref{tab:cp_uta:qqs} it may be noted 
that the average for $-\etacp S_{b \to c\bar c s}$ in $\etapr K^0$ 
($0.50 \pm 0.09$),
is more than $5\sigma$ away from zero, so that $\CP$ violation in this mode
may now be considered established.
Among other modes,
$\CP$ violation in both $f_0 \KS$ and $K^+K^- K^0$ is near the $3\sigma$ level,
although due to possible non-Gaussian errors in these results
it may be prudent to defer any strong conclusion on these modes.
There is no evidence (above $2\sigma$) for direct $\CP$ violation 
in any $b \to q \bar q s$ mode.

\mysubsection{Time-dependent $\CP$ asymmetries in $b \to c\bar{c}d$ transitions
}
\label{sec:cp_uta:ccd}

The transition $b \to c\bar c d$ can occur via either a $b \to c$ tree
or a $b \to d$ penguin amplitude.  
Similarly to Eq.~(\ref{eq:cp_uta:b_to_s}), the amplitude for 
the $b \to d$ penguin can be written
\begin{equation}
  \label{eq:cp_uta:b_to_d}
  \begin{array}{ccccc}
    A_{b \to d} & = & 
    \mc{3}{l}{F_u V_{ub}V^*_{ud} + F_c V_{cb}V^*_{cd} + F_t V_{tb}V^*_{td}} \\
    & = & (F_u - F_c) V_{ub}V^*_{ud} & + & (F_t - F_c) V_{tb}V^*_{td} \\
    & = & {\cal O}(\lambda^3) & + & {\cal O}(\lambda^3). \\
  \end{array}
\end{equation}
From this it can be seen that the $b \to d$ penguin amplitude 
contains terms with different weak phases at the same order of
CKM suppression.

In the above, we have followed Eq.~(\ref{eq:cp_uta:b_to_s}) 
by eliminating the $F_c$ term using unitarity.
However, we could equally well write
\begin{equation}
  \label{eq:cp_uta:b_to_d_alt}
  \begin{array}{ccccc}
    A_{b \to d} 
    & = & (F_u - F_t) V_{ub}V^*_{ud} & + & (F_c - F_t) V_{cb}V^*_{cd}, \\
    & = & (F_c - F_u) V_{cb}V^*_{cd} & + & (F_t - F_u) V_{tb}V^*_{td}. \\
  \end{array}
\end{equation}
Since the $b \to c\bar{c}d$ tree amplitude 
has the weak phase of $V_{cb}V^*_{cd}$,
either of the above expressions allow the penguin to be decomposed into 
parts with weak phases the same and different to the tree amplitude
(the relative weak phase can be chosen to be either $\beta$ or $\gamma$).
However, if the tree amplitude dominates,
there is little sensitivity to any phase 
other than that from $\Bz$\textendash$\Bzb$ mixing.

\begin{figure}[htb]
  \begin{center}
    \begin{tabular}{cc}
      \resizebox{0.46\textwidth}{!}{
        \includegraphics{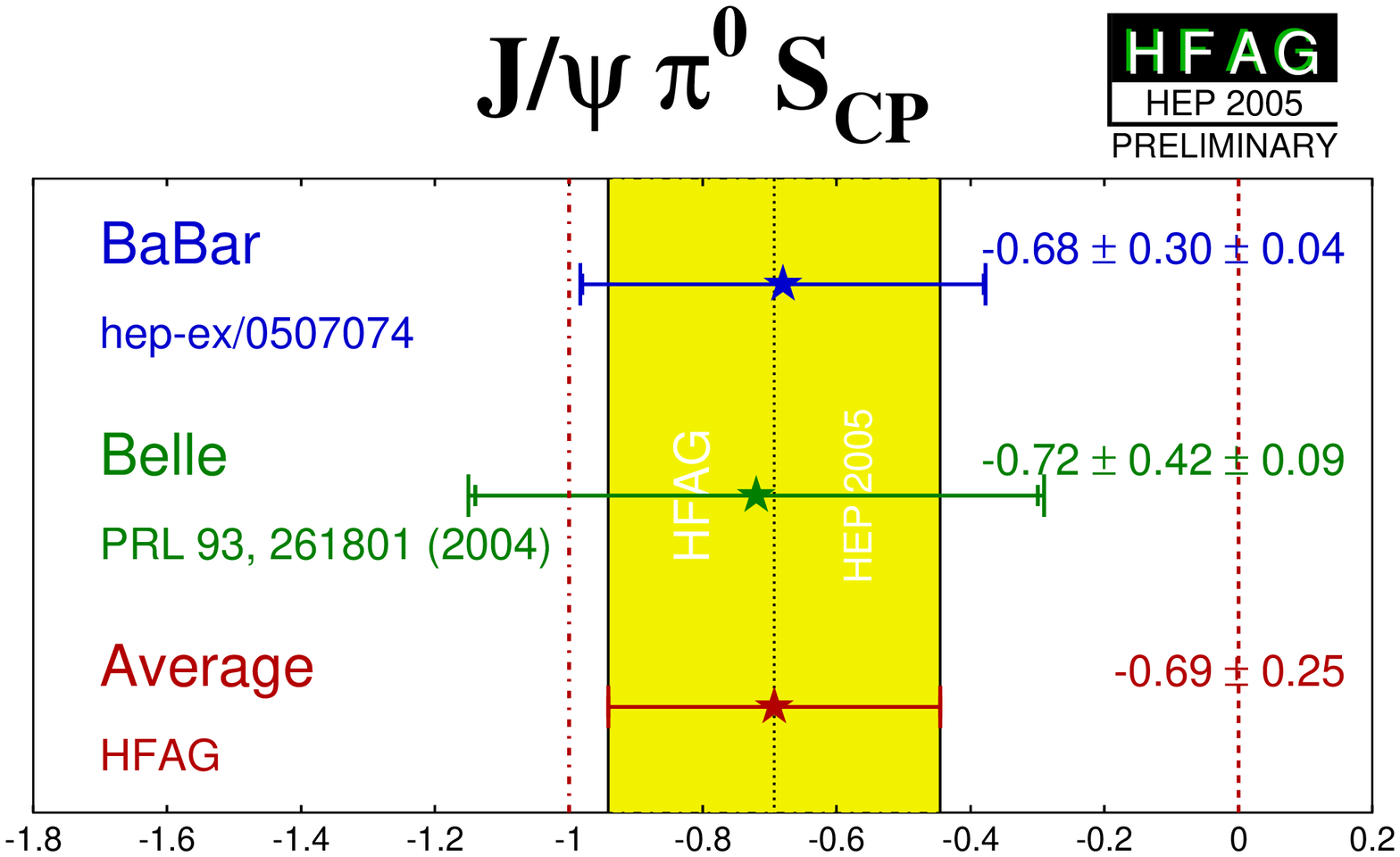}
      }
      &
      \resizebox{0.46\textwidth}{!}{
        \includegraphics{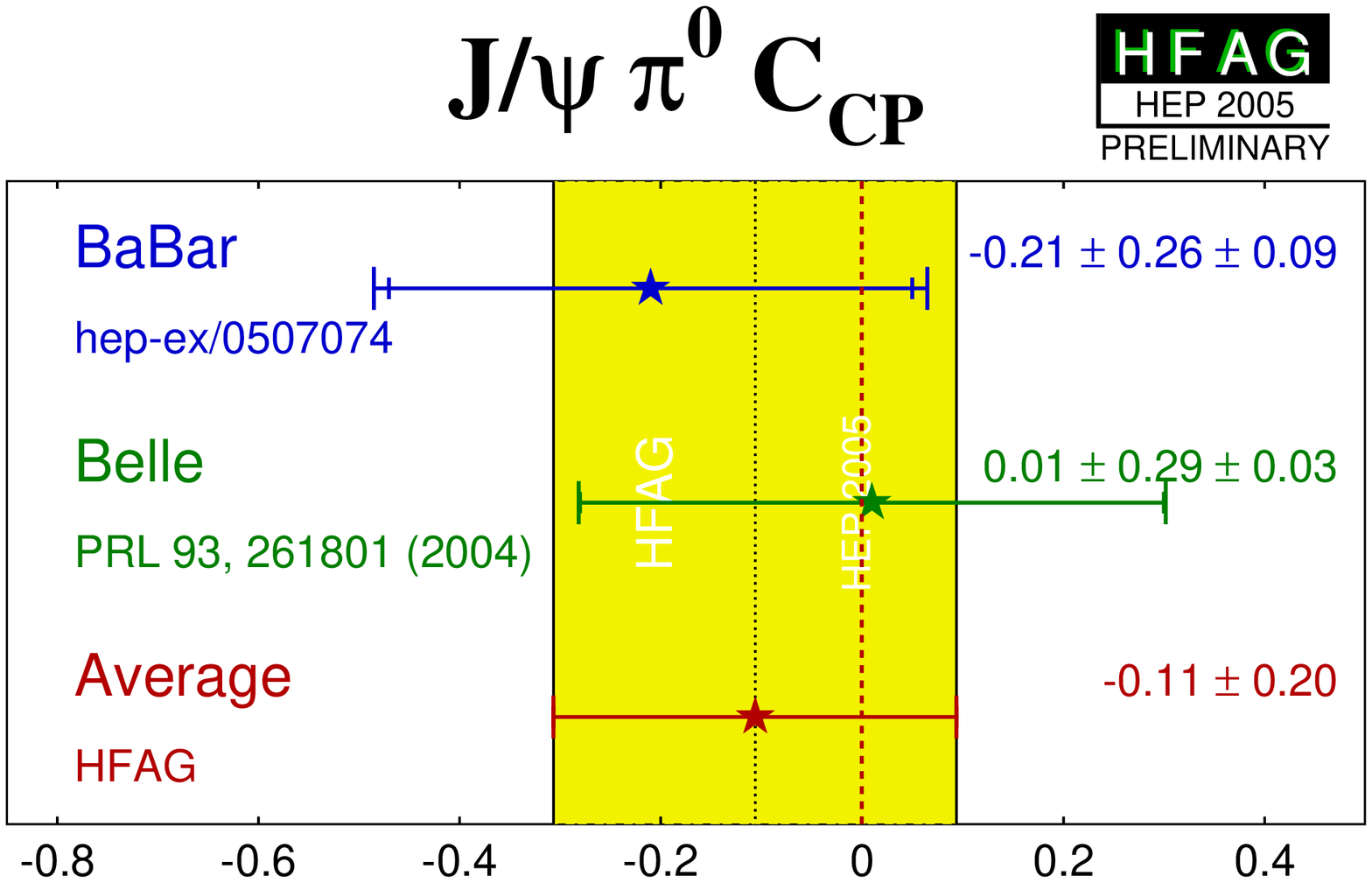}
      }
    \end{tabular}
  \end{center}
  \vspace{-0.8cm}
  \caption{
    Averages of 
    (left) $S_{b \to c\bar c d}$ and (right) $C_{b \to c\bar c d}$ 
    for the mode $\Bz \to J/ \psi \pi^0$.
  }
  \label{fig:cp_uta:ccd:psipi0}
\end{figure}

The $b \to c\bar{c}d$ transitions can be investigated with studies 
of various different final states. 
Results are available from both \babar\  and \belle\ 
using the final states $\jpsi \pi^0$, $D^{*+}D^{*-}$ and $D^{*\pm}D^{\mp}$,
and \babar\ have also used the final state $D^+D^-$;
the averages of these results are given in Table~\ref{tab:cp_uta:ccd}.
The results using the $\CP$ eigenstate ($\etacp = +1$) $\jpsi \pi^0$
are shown in Fig.~\ref{fig:cp_uta:ccd:psipi0}.

\begin{figure}[htb]
  \begin{center}
    \begin{tabular}{cc}
      \resizebox{0.46\textwidth}{!}{
        \includegraphics{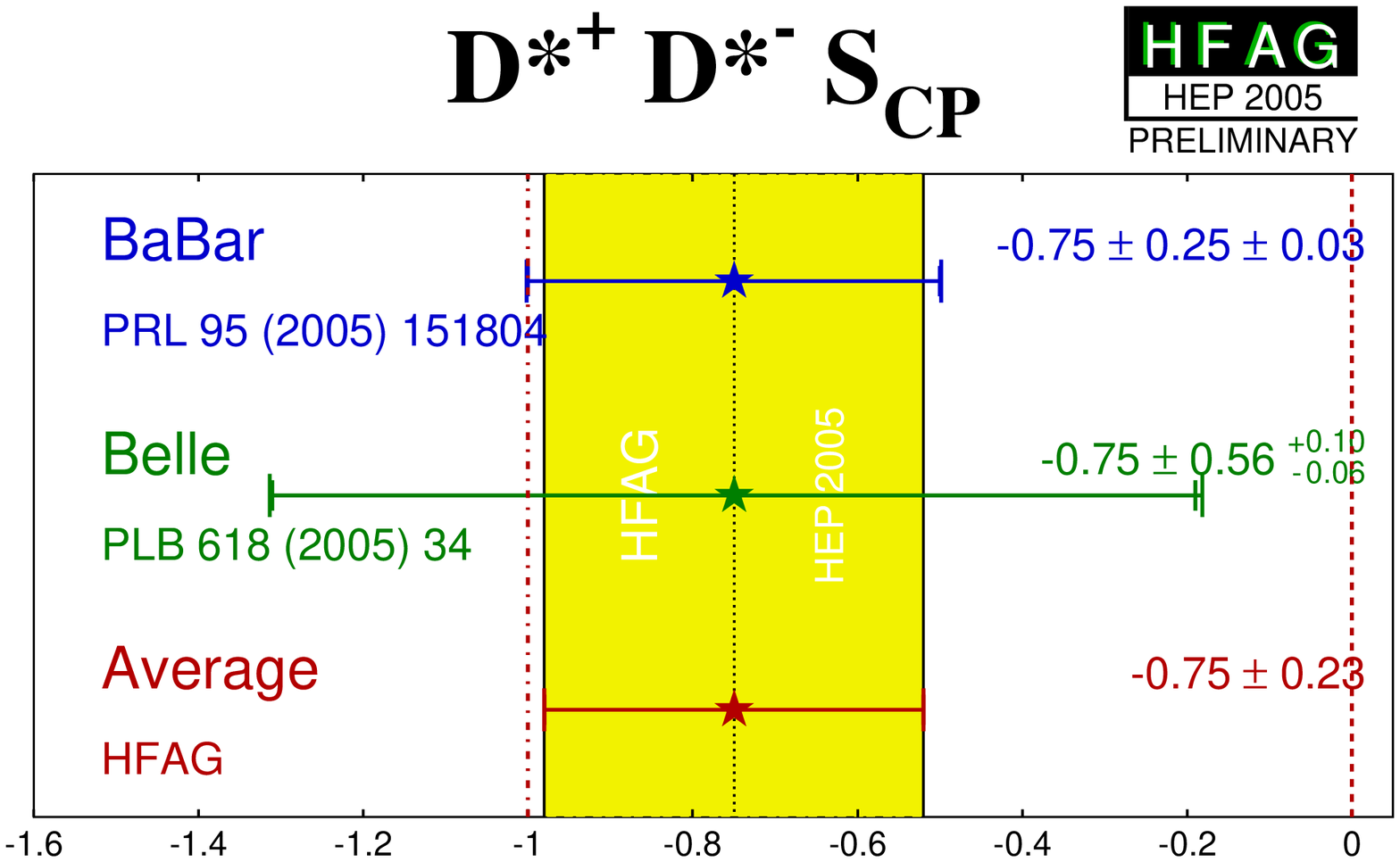}
      }
      &
      \resizebox{0.46\textwidth}{!}{
        \includegraphics{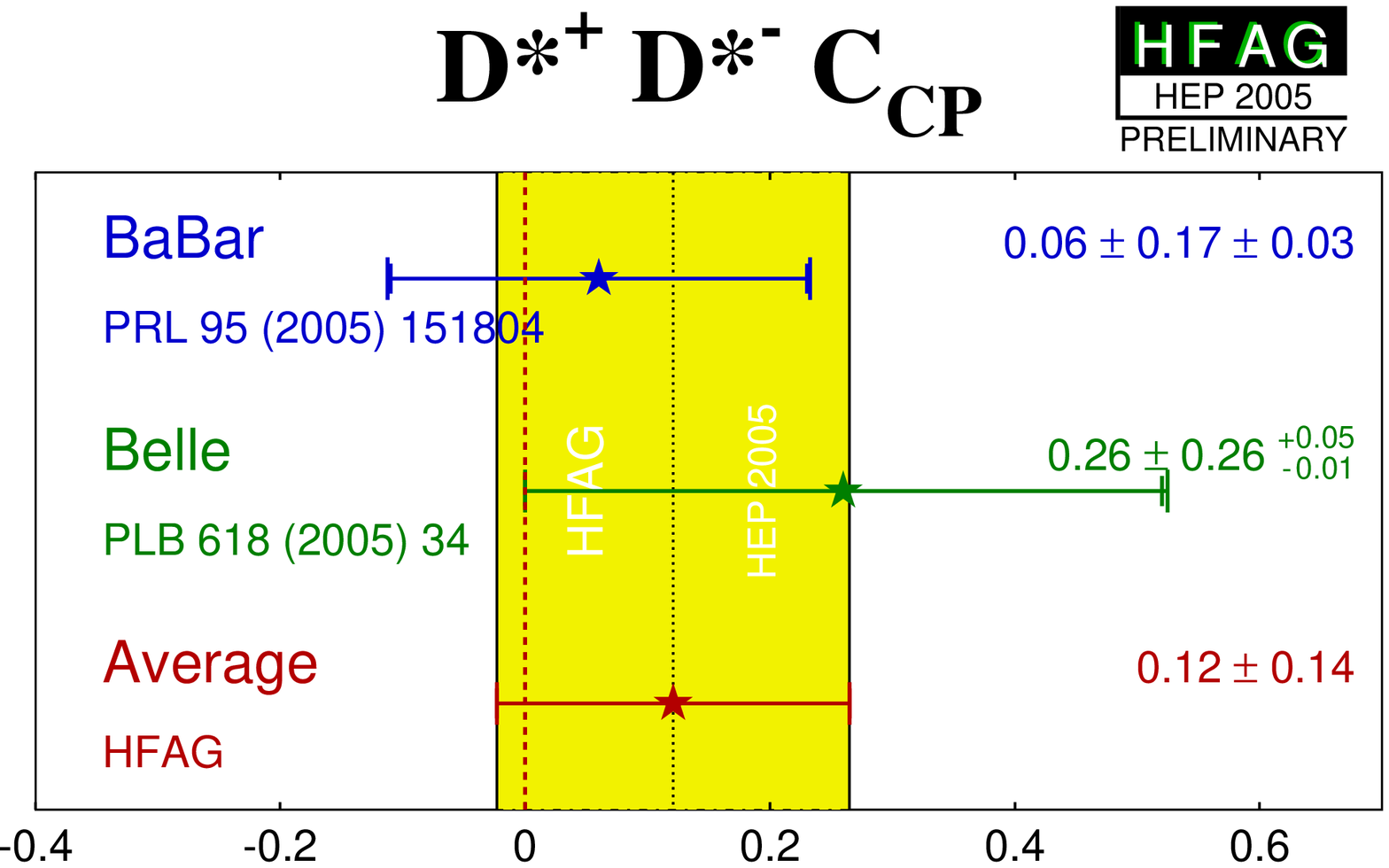}
      }
    \end{tabular}
  \end{center}
  \vspace{-0.8cm}
  \caption{
    Averages of 
    (left) $S_{b \to c\bar c d}$ and (right) $C_{b \to c\bar c d}$ 
    for the mode $\Bz \to D^{*+}D^{*-}$.
  }
  \label{fig:cp_uta:ccd:dstardstar}
\end{figure}

The vector-vector mode $D^{*+}D^{*-}$ 
is found to be dominated by the $\CP$ even longitudinally polarized component;
\babar\ measures a $\CP$ odd fraction of 
$0.125 \pm 0.044 \pm 0.007$~\cite{ref:cp_uta:ccd:babar:dstardstar} while
\belle\ measures a $\CP$ odd fraction of 
$0.19  \pm 0.08  \pm 0.01 $~\cite{ref:cp_uta:ccd:belle:dstardstar}
(here we do not average these fractions and rescale the inputs,
however the average is almost independent of the treatment).
We treat the uncertainty due to the error in the $\CP$-odd fractions
(quoted as a third uncertainty) as a correlated systematic error.
Results using $D^{*+}D^{*-}$ are shown in Fig.~\ref{fig:cp_uta:ccd:dstardstar}.

For the non-$\CP$ eigenstate mode $D^{*\pm}D^{\mp}$
\babar\ uses fully reconstructed events while 
\belle\ combines both fully and partially reconstructed samples.
The most recent results from \babar\ do not include 
a measurement of the overall asymmetry $A$.
At present we perform uncorrelated averages of the parameters in the 
$D^{*\pm}D^{\mp}$ system, using only the information from \belle\ on $A$.

\begin{table}
  \begin{center}
    \caption{
      Averages for $b \to c \bar c d$ modes.
      Note that the averages are calculated without taking correlations
      into account.
    }
    \vspace{0.2cm}
    \setlength{\tabcolsep}{0.0pc}
    \begin{tabular*}{\textwidth}{@{\extracolsep{\fill}}lrcc} \hline 
      \mc{2}{l}{Experiment} & 
      $S_{b \to c\bar c d}$ & $C_{b \to c\bar c d}$ \\
      \hline
      \mc{4}{c}{$\jpsi \pi^0$} \\
      \babar & \cite{ref:cp_uta:ccd:babar:psipi0} & 
      $ -0.68 \pm 0.30 \pm 0.04$ & $ -0.21 \pm 0.26 \pm 0.09$ \\
      \belle & \cite{ref:cp_uta:ccd:belle:psipi0} & 
      $ -0.72 \pm 0.42 \pm 0.09$ & $\ph{-}0.01 \pm 0.29 \pm 0.03$ \\
      \mc{2}{l}{\bf Average} & 
      $ -0.69 \pm 0.25 $ & $ -0.11 \pm 0.20$ \\
      \mc{2}{l}{\small Confidence level} & 
      \small $0.94~(0.1\sigma)$ & \small $0.58~(0.6\sigma)) $ \\
      \hline
      \mc{4}{c}{$D^+D^-$} \\
      \babar & \cite{ref:cp_uta:ccd:babar:dd_dstard} &
      $ -0.29 \pm 0.63 \pm 0.06$ & $ 0.11 \pm 0.35 \pm 0.06$ \\
      \mc{2}{l}{\bf Average} & 
      $ -0.29 \pm 0.63$ & $ 0.11 \pm 0.36$ \\
      \hline
      \mc{4}{c}{$D^{*+}D^{*-}$} \\
      \babar & \cite{ref:cp_uta:ccd:babar:dstardstar} &
      $ -0.75 \pm 0.25 \pm 0.03$ & $ 0.06 \pm 0.17 \pm 0.03$ \\
      \belle & \cite{ref:cp_uta:ccd:belle:dstardstar} &
      $ -0.75 \pm 0.56 \pm 0.10 \pm 0.06$ & $ 0.26 \pm 0.26 \pm 0.05 \pm 0.01$ \\
      \mc{2}{l}{\bf Average} &
      $ -0.75 \pm 0.23 $ & $  0.12 \pm 0.14 $ \\
      \mc{2}{l}{\small Confidence level} & 
      \small $1.00~(0.0\sigma)$ & \small $0.52~(0.6\sigma)$ \\
      \hline
    \end{tabular*}

    \vspace{2ex}

    \resizebox{\textwidth}{!}{
      \setlength{\tabcolsep}{0.0pc}
      \begin{tabular}{@{\extracolsep{2mm}}lrccccc} \hline 
        \mc{2}{l}{Experiment} & 
        $S_{+-}$ & $C_{+-}$ & $S_{-+}$ & $C_{-+}$ & $A$ \\
        \hline
        \mc{7}{c}{$D^{*\pm}D^{\mp}$} \\        
        \babar & \cite{ref:cp_uta:ccd:babar:dd_dstard} &
        $ -0.54 \pm 0.35 \pm 0.07$ & $\ph{-}0.09 \pm 0.25 \pm 0.06$ &
        $ -0.29 \pm 0.33 \pm 0.07$ & $ 0.17 \pm 0.24 \pm 0.04$ \\
        \belle & \cite{ref:cp_uta:ccd:belle:dstard} &
        $ -0.55 \pm 0.39 \pm 0.12$ & $-0.37 \pm 0.22 \pm 0.06$ & 
        $ -0.96 \pm 0.43 \pm 0.12$ & $ 0.23 \pm 0.25 \pm 0.06$ & $\ph{-}0.07 \pm 0.08 \pm 0.04$ \\
        \mc{2}{l}{\bf Average} &
        $-0.54 \pm 0.27$ & $-0.16 \pm 0.17$ &
        $-0.53 \pm 0.27$ & $ 0.20 \pm 0.18$ & $0.07 \pm 0.09 $ \\
        \mc{2}{l}{\small Confidence level} & 
        \small $0.99~(0.0\sigma)$ & \small $0.18~(1.3\sigma)$ & 
        \small $0.23~(1.2\sigma)$ & \small $0.87~(0.2\sigma)$ \\        
        \hline 
      \end{tabular}
    }

    \label{tab:cp_uta:ccd}
  \end{center}
\end{table}

In the absence of the penguin contribution (tree dominance),
the time-dependent parameters would be given by
$S_{b \to c\bar c d} = - \etacp \sin(2\beta)$,
$C_{b \to c\bar c d} = 0$,
$S_{+-} = \sin(2\beta + \delta)$,
$S_{-+} = \sin(2\beta - \delta)$,
$C_{+-} = - C_{-+}$ and 
$A_{+-} = 0$,
where $\delta$ is the strong phase difference between the 
$D^{*+}D^-$ and $D^{*-}D^+$ decay amplitudes.
In the presence of the penguin contribution,
there is no clean interpretation in terms of CKM parameters,
however
direct $\CP$ violation may be observed as any of
$C_{b \to c\bar c d} \neq 0$, $C_{+-} \neq - C_{-+}$ or $A_{+-} \neq 0$.

\begin{figure}[htb]
  \begin{center}
    \begin{tabular}{cc}
      \resizebox{0.46\textwidth}{!}{
        \includegraphics{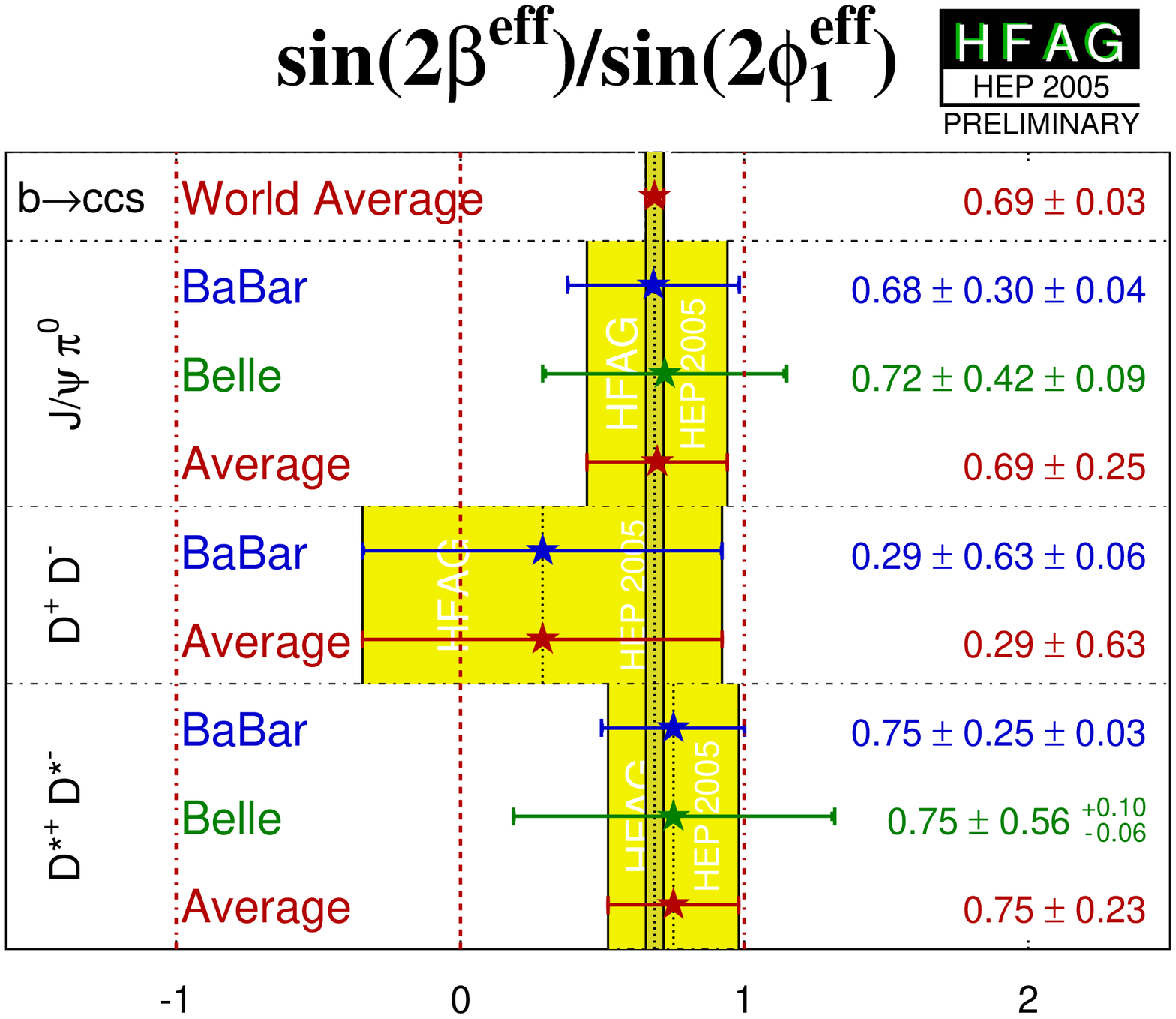}
      }
      &
      \resizebox{0.46\textwidth}{!}{
        \includegraphics{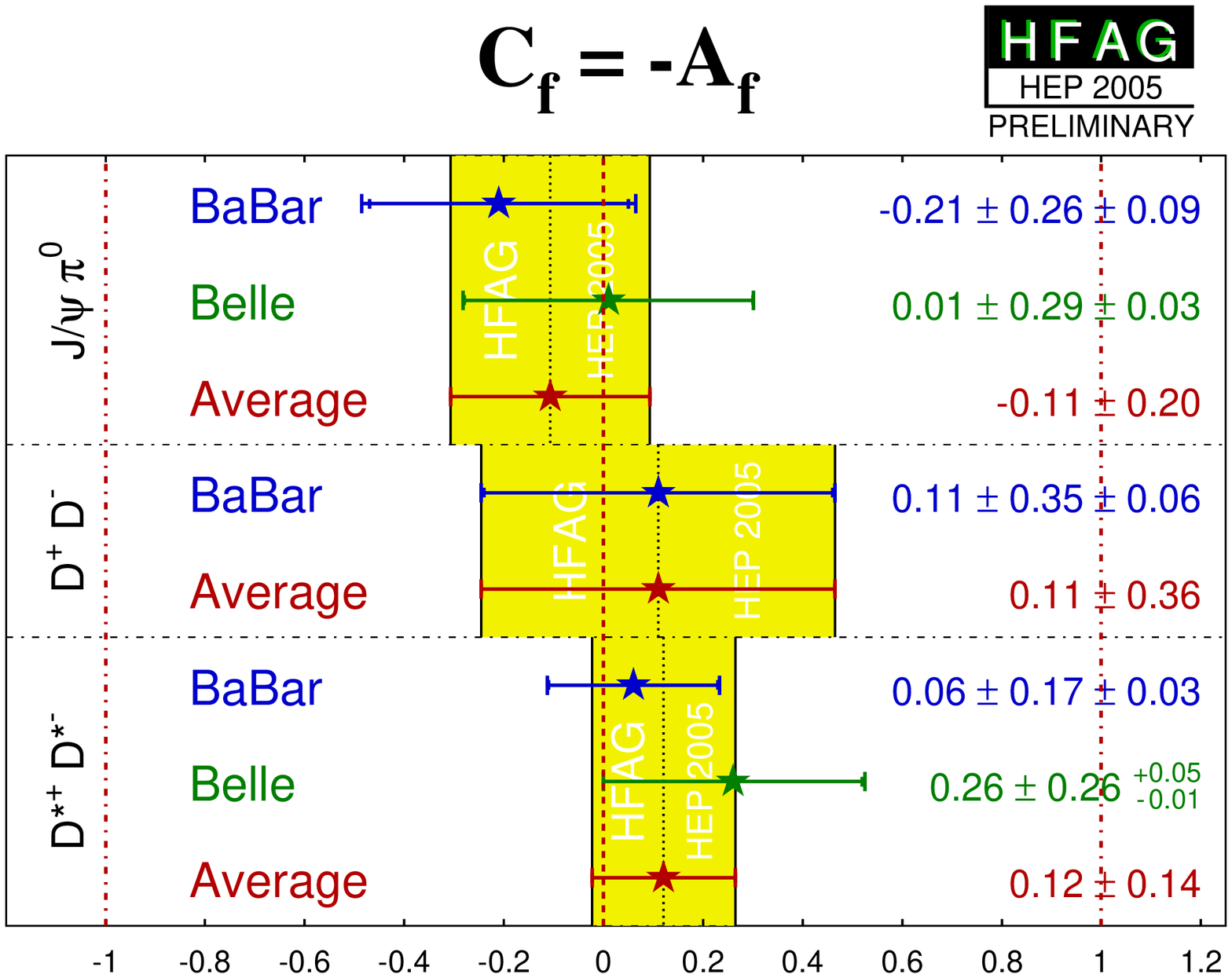}
      }
    \end{tabular}
  \end{center}
  \vspace{-0.8cm}
  \caption{
    Averages of 
    (left) $-\etacp S_{b \to c\bar c d}$ and (right) $C_{b \to c\bar c d}$.
    The $-\etacp S_{b \to q\bar q s}$ figure compares the results to 
    the world average 
    for $-\etacp S_{b \to c\bar c s}$ (see Section~\ref{sec:cp_uta:ccs}).    
  }
  \label{fig:cp_uta:ccd}
\end{figure}

The averages for the $b \to c\bar c d$ modes 
are shown in Fig.~\ref{fig:cp_uta:ccd}.
All results are consistent with tree dominance,
and with the Standard Model.
The average of $S_{b \to c\bar c d}$ in the $D^{*+}D^{*-}$ final state
is about $3\sigma$ from zero;
however, due to the large uncertainty and possible non-Gaussian effects,
any strong conclusion should be deferred.


\mysubsection{Time-dependent asymmetries in $b \to s\gamma$ transitions
}
\label{sec:cp_uta:bsg}

The radiative decays $b \to s\gamma$ produce photons 
which are highly polarized in the Standard Model.
The decays $\Bz \to F \gamma$ and $\Bzb \to F \gamma$ 
produce photons with opposite helicities, 
and since the polarization is, in principle, observable,
these final states cannot interfere.
The finite mass of the $s$ quark introduces small corrections
to the limit of maximum polarization,
but any large mixing induced $\CP$ violation would be a signal for new physics.
Since a single weak phase dominates the $b \to s \gamma$ transition in the 
Standard Model, the cosine term is also expected to be small.

Atwood {\it et al.}~\cite{ref:cp_uta:bsg:aghs} have shown that 
an inclusive analysis with respect to $\KS\pi^0\gamma$ can be performed,
since the properties of the decay amplitudes 
are independent of the angular momentum of the $\KS\pi^0$ system. 
However, if non-dipole operators contribute significantly to the amplitudes, 
then the Standard Model mixing-induced $\CP$ violation could be larger 
than the na\"\i ve expectation $S \simeq -2 (m_s/m_b) \sin \left(2\beta\right)$,
and the $\CP$ parameters may vary over the $\KS\pi^0\gamma$ Dalitz plot, 
for example as a function of the $\KS\pi^0$ invariant mass.

With the above in mind, 
we quote two averages: one for $K^*(892)$ candidates only, 
and the other one for the inclusive $\KS\pi^0\gamma$ decay (including the $K^*(892)$).
If the Standard Model dipole operator is dominant, 
both should give the same quantities 
(the latter naturally with smaller statistical error). 
If not, care needs to be taken in interpretation of the inclusive parameters, 
while the results on the $K^*(892)$ resonance remain relatively clean.
Results from \babar\ and \belle\ are used for both averages;
both experiments use the invariant mass range 
$0.60 \ {\rm GeV}/c^2 < M_{\KS\pi^0} < 1.80 \ {\rm GeV}/c^2$
in the inclusive analysis.

\begin{table}
  \begin{center}
    \caption{
      Averages for $b \to s \gamma$ modes.
      Note that the averages are calculated without taking correlations
      into account.
    }
    \vspace{0.2cm}
    \setlength{\tabcolsep}{0.0pc}
    \begin{tabular*}{\textwidth}{@{\extracolsep{\fill}}lrccc} \hline 
      \mc{2}{l}{Experiment} & 
      $S_{b \to s \gamma}$ & $C_{b \to s \gamma}$ & Correlation \\
      \hline
      \mc{4}{c}{$\Kstar(892)\gamma$} \\
      \babar & \cite{ref:cp_uta:bsg:babar:kspi0gamma} & 
      $-0.21 \pm 0.40 \pm 0.05$ & $-0.40 \pm 0.23 \pm 0.04$ &  -0.064 \\
      \belle & \cite{ref:cp_uta:bsg:belle:kspi0gamma} &  
      $ 0.01 \pm 0.52 \pm 0.11$ & $ -0.11 \pm 0.33 \pm 0.09$ &  0.002 \\
      \mc{2}{l}{\bf Average} & 
      $-0.13 \pm 0.32$ & $-0.31 \pm 0.19 $ \\
      \mc{2}{l}{\small Confidence level} & 
      $0.74~(0.3\sigma)$ & $0.48~(0.7\sigma)$ \\
      \hline       
      \mc{4}{c}{$\KS \pi^0 \gamma$ (including $\Kstar(892)\gamma$)} \\
      \babar & \cite{ref:cp_uta:bsg:babar:kspi0gamma} & 
      $-0.06 \pm 0.37$ & $-0.48 \pm 0.22$ \\
      \belle & \cite{ref:cp_uta:bsg:belle:kspi0gamma} & 
      $ 0.08 \pm 0.41 \pm 0.10$ & $ -0.12 \pm 0.27 \pm 0.10$ & 0.004 \\
      \mc{2}{l}{\bf Average} & 
      $ 0.00 \pm 0.28$ & $ -0.35 \pm 0.17$ \\
      \mc{2}{l}{\small Confidence level} & 
      $0.80~(0.3\sigma)$ & $0.32~(1.0\sigma)$ \\
      \hline 
    \end{tabular*}
      
    \label{tab:cp_uta:bsg}
  \end{center}
\end{table}

\begin{figure}[htb]
  \begin{center}
    \begin{tabular}{cc}
      \resizebox{0.46\textwidth}{!}{
        \includegraphics{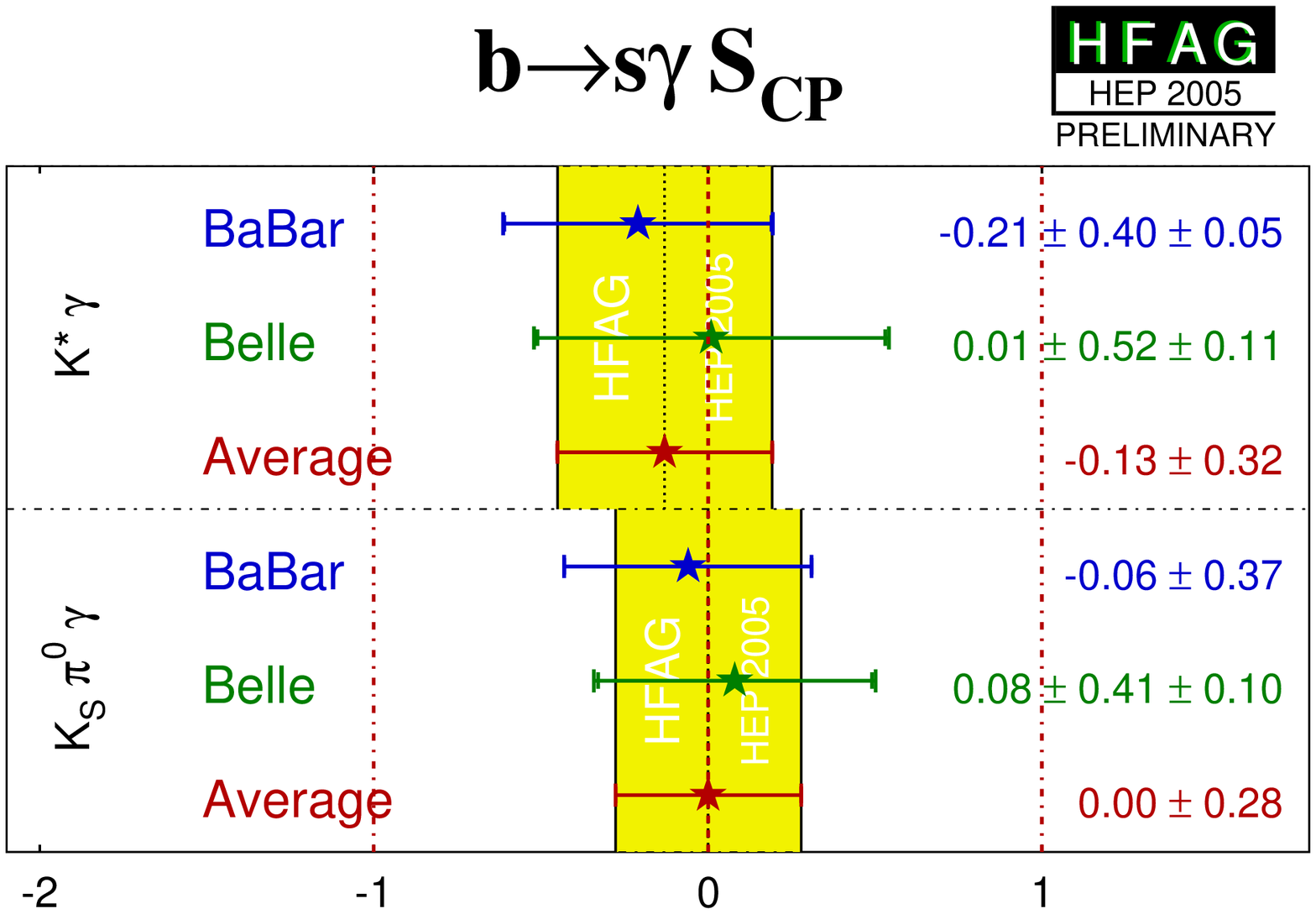}
      }
      &
      \resizebox{0.46\textwidth}{!}{
        \includegraphics{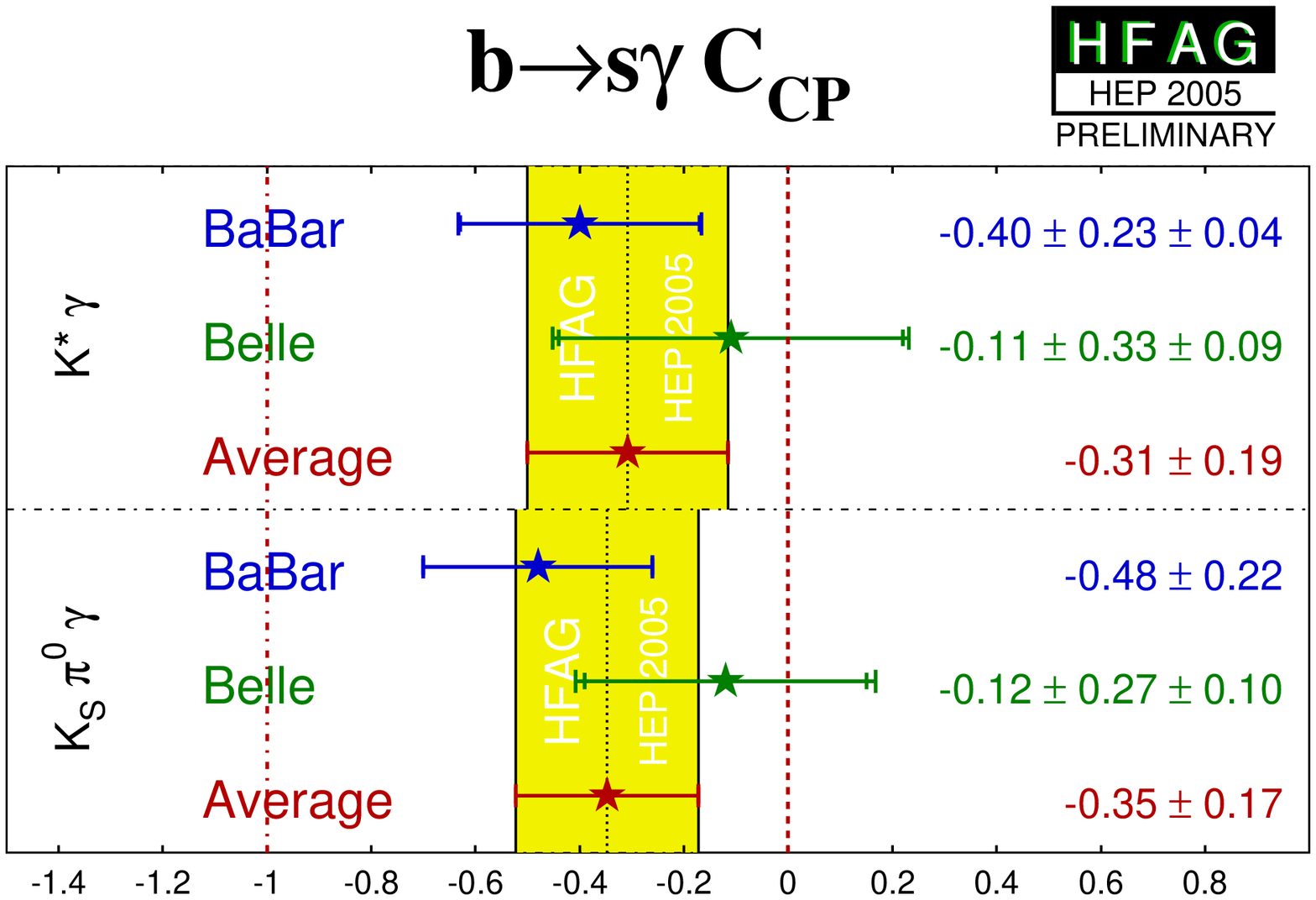}
      }
    \end{tabular}
  \end{center}
  \vspace{-0.8cm}
  \caption{
    Averages of (left) $S_{b \to s \gamma}$ and (right) $C_{b \to s \gamma}$.
    Recall that the data for $K^*\gamma$ is a subset of that for $\KS\pi^0\gamma$.
  }
  \label{fig:cp_uta:bsg}
\end{figure}

The results are shown in Table~\ref{tab:cp_uta:bsg},
and in Fig.~\ref{fig:cp_uta:bsg}.
No significant $\CP$ violation results are seen;
the results are consistent with the Standard Model
and with other measurements in the $b \to s\gamma$ system (see Sec.~\ref{sec:rare}).

\mysubsection{Time-dependent $\CP$ asymmetries in $b \to u\bar{u}d$ transitions
}
\label{sec:cp_uta:uud}

The $b \to u \bar u d$ transition can be mediated by either 
a $b \to u$ tree amplitude or a $b \to d$ penguin amplitude.
These transitions can be investigated using 
the time dependence of $\Bz$ decays to final states containing light mesons.
Results are available from both \babar\ and \belle\ for the 
$\CP$ eigenstate ($\etacp = +1$) $\pi^+\pi^-$ final state
and for the vector-vector final state $\rho^+\rho^-$,
which is found to be dominated by the $\CP$ even
longitudinally polarized component
(\babar\ measure $f_{\rm long} = 
0.978 \pm 0.014 \, ^{+0.021}_{-0.029}$~\cite{ref:cp_uta:uud:babar:rhorho}
while \belle\ measure $f_{\rm long} = 
0.951 \, ^{+0.033}_{-0.039} \, ^{+0.029}_{-0.031}$~\cite{ref:cp_uta:uud:belle:rhorho}).

\begin{figure}[htb]
  \begin{center}
    \begin{tabular}{cc}
      \resizebox{0.46\textwidth}{!}{\includegraphics{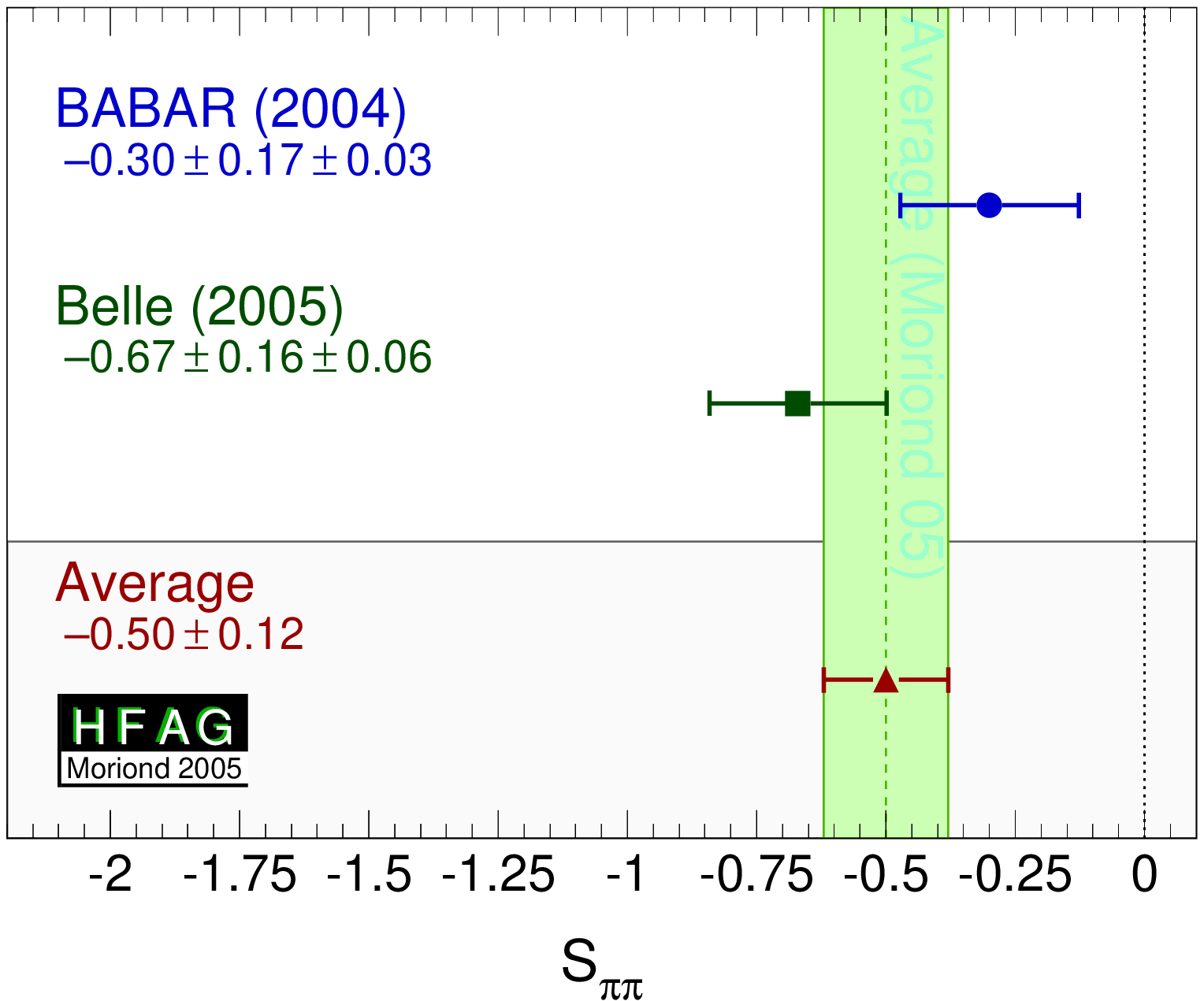}}
      &
      \resizebox{0.46\textwidth}{!}{\includegraphics{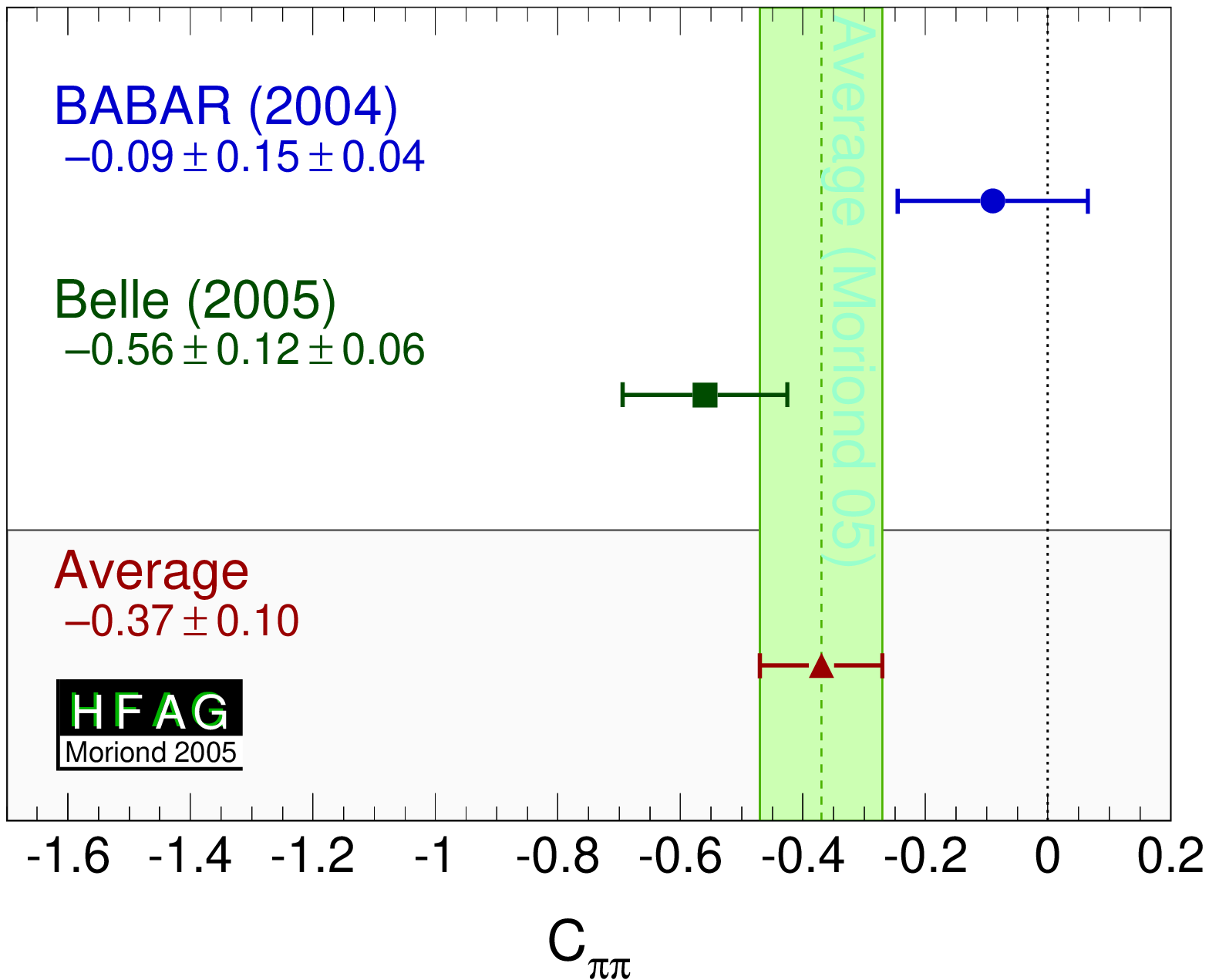}}
    \end{tabular}
  \end{center}
  \vspace{-0.8cm}
  \caption{
    Averages of 
    (left) $S_{b \to u\bar u d}$ and (right) $C_{b \to u\bar u d}$ 
    for the mode $\Bz \to \pi^+\pi^-$.
  }
  \label{fig:cp_uta:uud:pipi}
\end{figure}

For the non-$\CP$ eigenstate $\rho^{\pm}\pi^{\mp}$, 
\belle\ has performed a quasi-two-body analysis,
while  \babar\ performs a time-dependent Dalitz plot (DP) analysis
of the $\pi^+\pi^-\pi^0$ final state~\cite{ref:cp_uta:uud:snyderquinn};
such an analysis allows direct measurements of the phases.
These results, and averages, are listed in Table~\ref{tab:cp_uta:uud}.
The averages for $\pi^+\pi^-$ are shown in Fig.~\ref{fig:cp_uta:uud:pipi}.

\begin{table}
  \begin{center}
    \caption{
      Averages for $b \to u \bar u d$ modes.
    }
    \vspace{0.2cm}
    \setlength{\tabcolsep}{0.0pc}
    \begin{tabular*}{\textwidth}{@{\extracolsep{\fill}}lrccc} \hline 
      \mc{2}{l}{Experiment} & 
      $S_{b \to u\bar u d}$ & $C_{b \to u\bar u d}$ & Correlation \\
      \hline
      & \mc{3}{c}{$\pi^+\pi^-$} \\
      \babar & ~\cite{ref:cp_uta:uud:babar:pipi} & 
      $-0.30 \pm 0.17 \pm 0.03$ & $-0.09 \pm 0.15 \pm 0.04$ & -0.016 \\
      \belle & ~\cite{ref:cp_uta:uud:belle:pipi} & 
      $-0.67 \pm 0.16 \pm 0.06$ & $-0.56 \pm 0.12 \pm 0.06$ & -0.09 \\
      \mc{2}{l}{\bf Average} & 
      $-0.50 \pm 0.12$ & $-0.37 \pm 0.10$ & -0.056 \\
      \mc{2}{l}{\small Confidence level} & 
      \mc{2}{c}{\small combined average: $0.019~(2.3\sigma)$} \\
      \hline
      & \mc{3}{c}{$\rho^+\rho^-$} \\
      \babar & ~\cite{ref:cp_uta:uud:babar:rhorho} &
      $-0.33 \pm 0.24 \, ^{+0.08}_{-0.14}$ & $-0.03 \pm 0.18 \pm 0.09$ & $-0.042$ \\
      \belle & ~\cite{ref:cp_uta:uud:belle:rhorho} &
      $ 0.09 \pm 0.42 \pm 0.08$ & $ 0.00 \pm 0.30 \, ^{+0.09}_{-0.10}$ & 0.06 \\
      \mc{2}{l}{\bf Average} & 
      $-0.21 \pm 0.22$ & $-0.03 \pm 0.17$ & 0.01 \\
       \mc{2}{l}{\small Confidence level} & 
      \mc{2}{c}{\small combined average: xxx } \\    
      \hline 
    \end{tabular*}

    \vspace{2ex}

    \resizebox{\textwidth}{!}{
      \setlength{\tabcolsep}{0.0pc}
      \begin{tabular}{@{\extracolsep{2mm}}lrccccc} \hline 
        & \mc{6}{c}{$\rho^{\pm}\pi^{\mp}$ Q2B/DP analysis} \\
        \mc{2}{l}{Experiment} & 
        $S_{\rho\pi}$ & $C_{\rho\pi}$ & $\Delta S_{\rho\pi}$ & $\Delta C_{\rho\pi}$ & ${\cal A}_{CP}^{\rho\pi}$ \\
        \hline
        \babar & ~\cite{ref:cp_uta:uud:babar:rhopi} &
        $-0.10 \pm 0.14 \pm 0.04$ & $ 0.34 \pm 0.11 \pm 0.05$ &
        $\ph{-}0.22 \pm 0.15 \pm 0.03$ & $ 0.15 \pm 0.11 \pm 0.03$ & $-0.088 \pm 0.049 \pm 0.013$ \\
        \belle & ~\cite{ref:cp_uta:uud:belle:rhopi} & 
        $-0.28 \pm 0.23 \, ^{+0.10}_{-0.08}$ & $ 0.25 \pm 0.17 \, ^{+0.02}_{-0.06}$ &
        $-0.30 \pm 0.24 \pm 0.09$ & $ 0.38 \pm 0.18 \, ^{+0.02}_{-0.04}$ & $-0.16 \pm 0.10 \pm 0.02$ \\
        \mc{2}{l}{\bf Average} & 
        $-0.13 \pm 0.13$ & $ 0.31 \pm 0.10$ &
        $ 0.09 \pm 0.13$ & $ 0.22 \pm 0.10$ & $-0.102 \pm 0.045$ \\
        \hline
        & & & \mc{2}{c}{${\cal A}^{+-}_{\rho\pi}$} & \mc{2}{c}{${\cal A}^{-+}_{\rho\pi}$} \\
        \hline
        \babar & \cite{ref:cp_uta:uud:babar:rhopi} &
        & \mc{2}{c}{$\ph{-}0.25 \pm 0.17 \, ^{+0.02}_{-0.06}$} & \mc{2}{c}{$-0.47 ^{\,+0.14}_{\,-0.15} \pm 0.06$} \\         
        \belle & \cite{ref:cp_uta:uud:belle:rhopi} & 
        & \mc{2}{c}{$-0.02 \pm 0.16^{\,+0.05}_{\,-0.02}$} & \mc{2}{c}{$-0.53 \pm 0.29^{\,+0.09}_{\,-0.04}$} \\
        \mc{2}{l}{\bf Average} & 
        & \mc{2}{c}{$-0.15 \pm 0.09$} & \mc{2}{c}{$-0.47^{\,+0.13}_{\,-0.14}$} \\
        \hline 
      \end{tabular}
    }

    \vspace{2ex}

    \setlength{\tabcolsep}{0.0pc}
    \begin{tabular*}{\textwidth}{@{\extracolsep{\fill}}lrcc} \hline 
      \mc{4}{c}{$\rho^{\pm}\pi^{\mp}$ DP analysis} \\
      \mc{2}{l}{Experiment} & $\alpha \ (^\circ)$ & $\delta_{+-} \ (^\circ)$ \\
      \hline
      \babar & \cite{ref:cp_uta:uud:babar:rhopi} &
      $113 \, ^{+27}_{-17} \pm 6$ & $-67 \, ^{+28}_{-31} \pm 7$ \\ 
      \hline 
    \end{tabular*}
      
    \label{tab:cp_uta:uud}
  \end{center}
\end{table}

\begin{figure}[htb]
  \begin{center}
    \begin{tabular}{cc}
      \resizebox{0.46\textwidth}{!}{\includegraphics{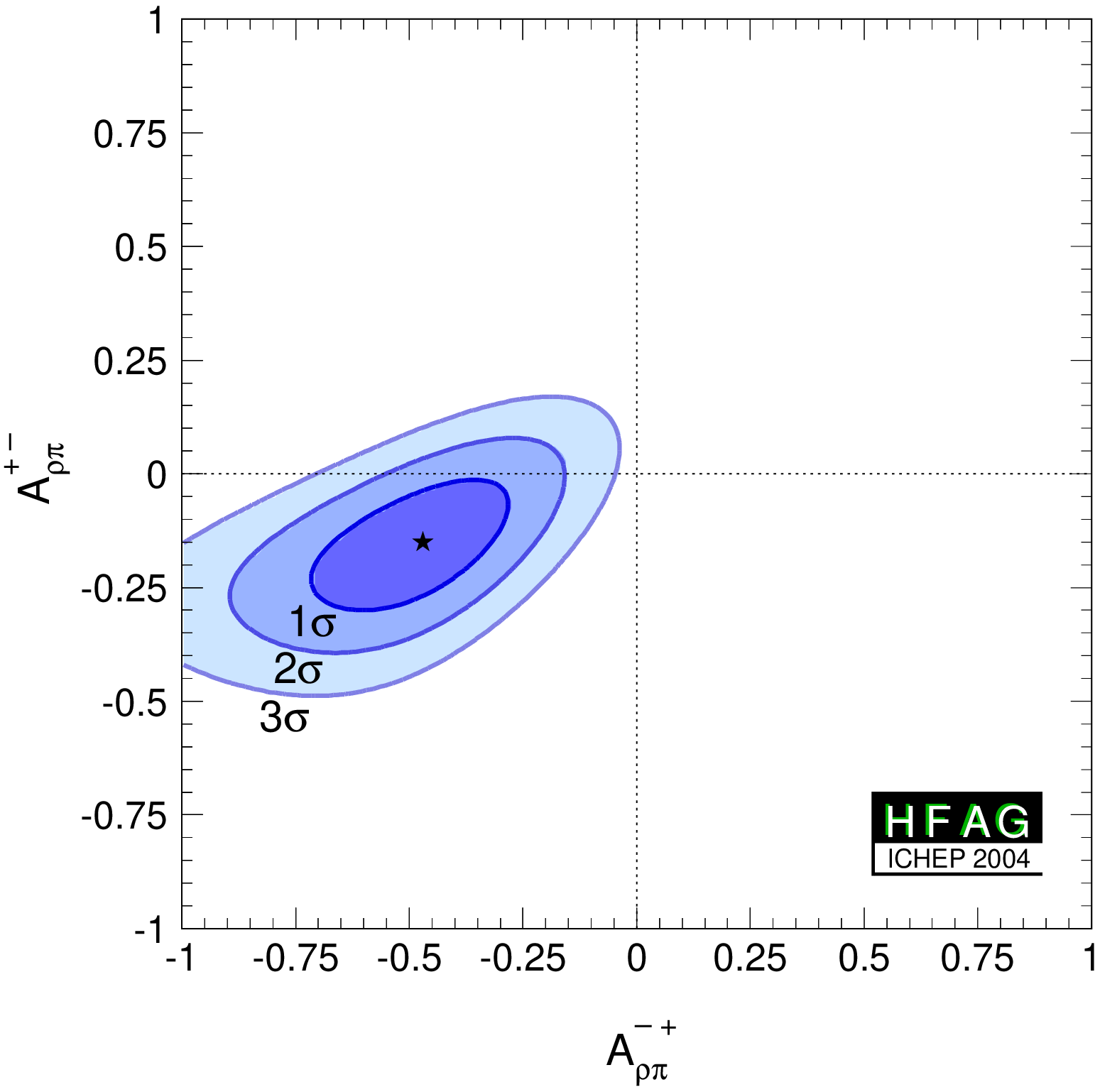}}
    \end{tabular}
  \end{center}
  \vspace{-0.8cm}
  \caption{
    Direct $\CP$ violation in $\Bz\to\rho^\pm\pi^\mp$. The no-\CP violation
    hypothesis is excluded at the $3.4\sigma$ level.
  }
  \label{fig:cp_uta:uud:rhopi-dircp}
\end{figure}

If the penguin contribution is negligible, 
the time-dependent parameters for $\Bz \to \pi^+\pi^-$ and $\Bz \to \rho^+\rho^-$ 
are given by
$S_{b \to u\bar u d} = \etacp \sin(2\alpha)$ and
$C_{b \to u\bar u d} = 0$.
With the notation described in Sec.~\ref{sec:cp_uta:notations}
(Eq.~(\ref{eq:cp_uta:non-cp-s_and_deltas})), 
the time-dependent parameters for the Q2B $\Bz \to \rho^\pm\pi^\mp$ analysis are,
neglecting penguin contributions, given by
$S_{\rho\pi}   = \sqrt{1 - (\frac{\Delta C}{2})^2}\sin(2\alpha)\cos(\delta)$,
$\Delta S_{\rho\pi} = \sqrt{1 - (\frac{\Delta C}{2})^2}\cos(2\alpha)\sin(\delta)$ and
$C_{\rho\pi} = {\cal A}_{\CP}^{\rho\pi} = 0$,
where $\delta=\arg(A_{-+}A^*_{+-})$ is the strong phase difference 
between the $\rho^-\pi^+$ and $\rho^+\pi^-$ decay amplitudes.
In the presence of the penguin contribution, there is no straightforward 
interpretation of the Q2B observables in the $\Bz \to \rho^\pm\pi^\mp$ system
in terms of CKM parameters.
However direct $\CP$ violation may arise,
resulting in either or both of $C_{\rho\pi} \neq 0$ and ${\cal A}_{\CP}^{\rho\pi} \neq 0$.
Equivalently,
direct $\CP$ violation may be seen by either of
the decay-type-specific observables ${\cal A}^{+-}_{\rho\pi}$ 
and ${\cal A}^{-+}_{\rho\pi}$, defined in Eq.~(\ref{eq:cp_uta:non-cp-directcp}), 
deviating from zero.
Results and averages for these parameters
are also given in Table~\ref{tab:cp_uta:uud}.
They exhibit a linear correlation coefficient of $+0.59$.
The significance of observing direct $\CP$ violation 
computed from the difference of the $\chi^2$ obtained in the nominal average, 
compared to setting 
$C_{\rho\pi} = {\cal A}^{\rho\pi}_{\CP} = 0$
is found to be $3.4\sigma$ in this mode. 
The confidence level
contours of ${\cal A}^{+-}_{\rho\pi}$ versus ${\cal A}^{-+}_{\rho\pi}$ 
are shown in Fig.~\ref{fig:cp_uta:uud:rhopi-dircp}.


Some difference is seen between the 
\babar\ and \belle\ measurements in the $\pi^+\pi^-$ system.
The confidence level of the average is $0.019$,
which corresponds to a $2.3\sigma$ discrepancy.  Since there is no
evidence of systematic problems in either analysis,
we do not rescale the errors of the averages.
The average for $S_{b \to u\bar u d}$ in $\Bz \to \pi^+\pi^-$
is more than $4\sigma$ away from zero, 
while that for $C_{b \to u\bar u d}$ is more than $3\sigma$ for zero.
Due to the possible discrepancy mentioned above,
only a cautious interpretation should be made.
Nevertheless, the averages give (at least) a strong indication 
for $\CP$ violation in $\Bz \to \pi^+\pi^-$
(for which \belle\ has already claimed observation).

The precision of the measured $\CP$ violation parameters in
$b \to u\bar{u}d$ transitions allows constraints to be set on the UT angle $\alpha$. 
In addition to the value of $\alpha$ from the \babar\ time-dependent DP analysis,
given in Table~\ref{tab:cp_uta:uud},
constraints have been obtained with various methods:
\begin{itemize}
\item 
  Both \babar~\cite{ref:cp_uta:uud:babar:pipi}
  and  \belle~\cite{ref:cp_uta:uud:belle:pipi} have performed 
  isospin analyses in the $\pi\pi$ system.
  \babar\ exclude $29^\circ < \alpha < 61^\circ$ at the $90\%$ C.L. while
  \belle\ exclude $19^\circ < \alpha < 71^\circ$ at the $95.4\%$ C.L.
  In both cases, only solutions in $0^\circ$\textendash$180^\circ$ are considered.
\item
  Both experiments have also performed isospin analyses in the $\rho\rho$ system.
  \babar~\cite{ref:cp_uta:uud:babar:rhorho} obtain 
  $\alpha = \left( 100 \pm 13 \right)^\circ$,
  while \belle~\cite{ref:cp_uta:uud:belle:rhorho} obtain 
  $\alpha = \left(  87 \pm 17 \right)^\circ$.
  The largest contribution to the uncertainty is due to the 
  possible penguin contribution, limited by the knowledge of the 
  $\Bz \to \rho^0\rho^0$ branching fraction~\cite{ref:cp_uta:uud:babar:rho0rho0},
  and is correlated between the measurements.  
\item 
  Each experiment has obtained a value of $\alpha$ from combining its 
  results in the different $b \to u \bar{u} d$ modes 
  (with some input also from HFAG).
  These values have appeared in talks, but not in publications,
  and are not listed here.
\item The CKMfitter group~\cite{ref:cp_uta:ckmfitter} uses the 
  measurements from \belle\ and \babar\ given in Table~\ref{tab:cp_uta:uud},
  with other branching fractions and \CP asymmetries in 
  $\B\to\pi\pi,~\rho\pi$ and $\rho\rho$ modes, 
  to perform isospin analyses for each system.  They then combine the results
  to obtain $\alpha = (98.6 \, ^{+12.6}_{-8.1})^\circ$.
  A similar analysis is performed by the UTFit group~\cite{ref:cp_uta:utfit}.
\end{itemize}
Note that each method suffers from ambiguities in the solutions.
The model assumption in the $\Bz \to \pi^+\pi^-\pi^0$ analysis 
allows to resolve some of the multiple solutions, 
and results in a single preferred value for $\alpha$ in $\left[ 0, \pi \right]$.
All the above measurements correspond to the choice
that is in agreement with the global CKM fit.

At present we make no attempt to provide an HFAG average for $\alpha$.
More details on procedures to calculate a best fit value for $\alpha$ 
can be found in Refs.~\cite{ref:cp_uta:ckmfitter,ref:cp_uta:utfit}.


\mysubsection{Time-dependent $\CP$ asymmetries in $b \to c\bar{u}d / u\bar{c}d$ transitions
}
\label{sec:cp_uta:cud}

Non-$\CP$ eigenstates such as $D^\pm\pi^\mp$, $D^{*\pm}\pi^\mp$ and $D^\pm\rho^\mp$ can be produced 
in decays of $\Bz$ mesons either via Cabibbo favoured ($b \to c$) or
doubly Cabibbo suppressed ($b \to u$) tree amplitudes. 
Since no penguin contribution is possible,
these modes are theoretically clean.
The ratio of the magnitudes of the suppressed and favoured amplitudes, $R$,
is sufficiently small (predicted to be about $0.02$),
that terms of ${\cal O}(R^2)$ can be neglected, 
and the sine terms give sensitivity to the combination of UT angles $2\beta+\gamma$.

\begin{figure}[htb]
  \begin{center}
    \begin{tabular}{cc}
      \resizebox{0.46\textwidth}{!}{
        \includegraphics{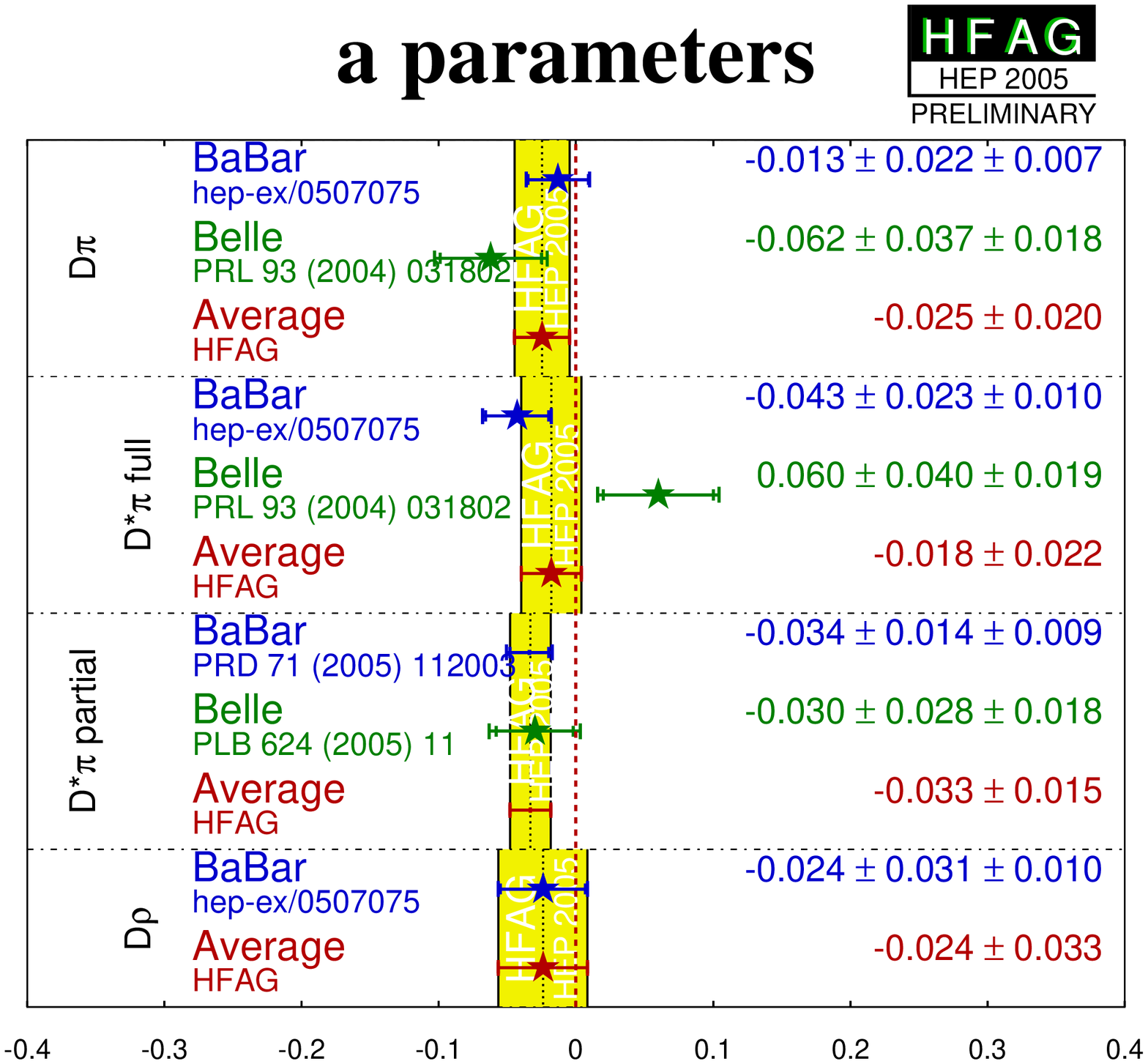}
      }
      &
      \resizebox{0.46\textwidth}{!}{
        \includegraphics{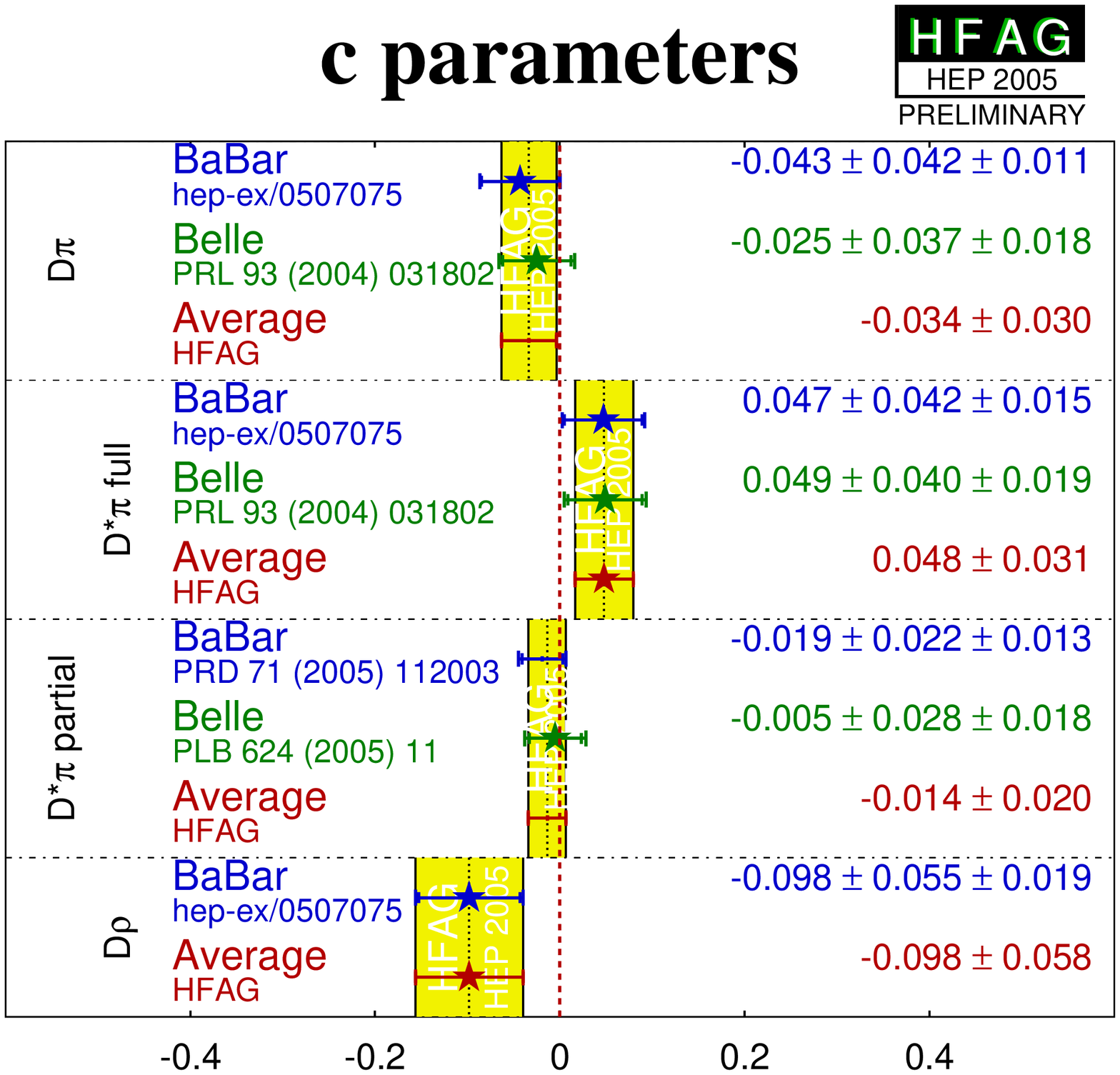}
      }
    \end{tabular}
  \end{center}
  \vspace{-0.8cm}
  \caption{
    Averages for $b \to c\bar{u}d / u\bar{c}d$ modes.
  }
  \label{fig:cp_uta:cud}
\end{figure}

\begin{figure}[htb]
  \begin{center}
    \begin{tabular}{cc}
      \resizebox{0.46\textwidth}{!}{
        \includegraphics{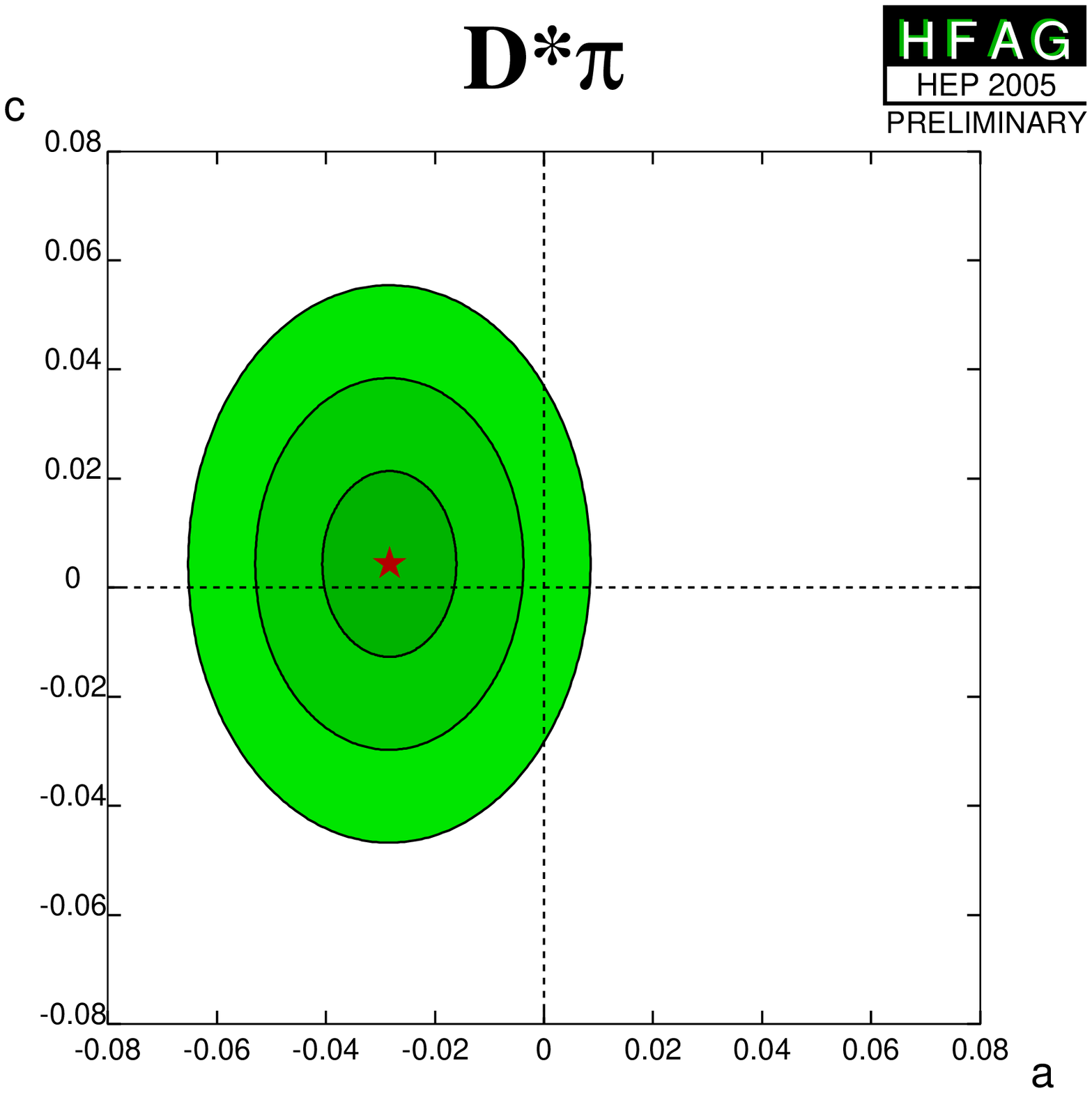}
      }
      &
      \resizebox{0.46\textwidth}{!}{
        \includegraphics{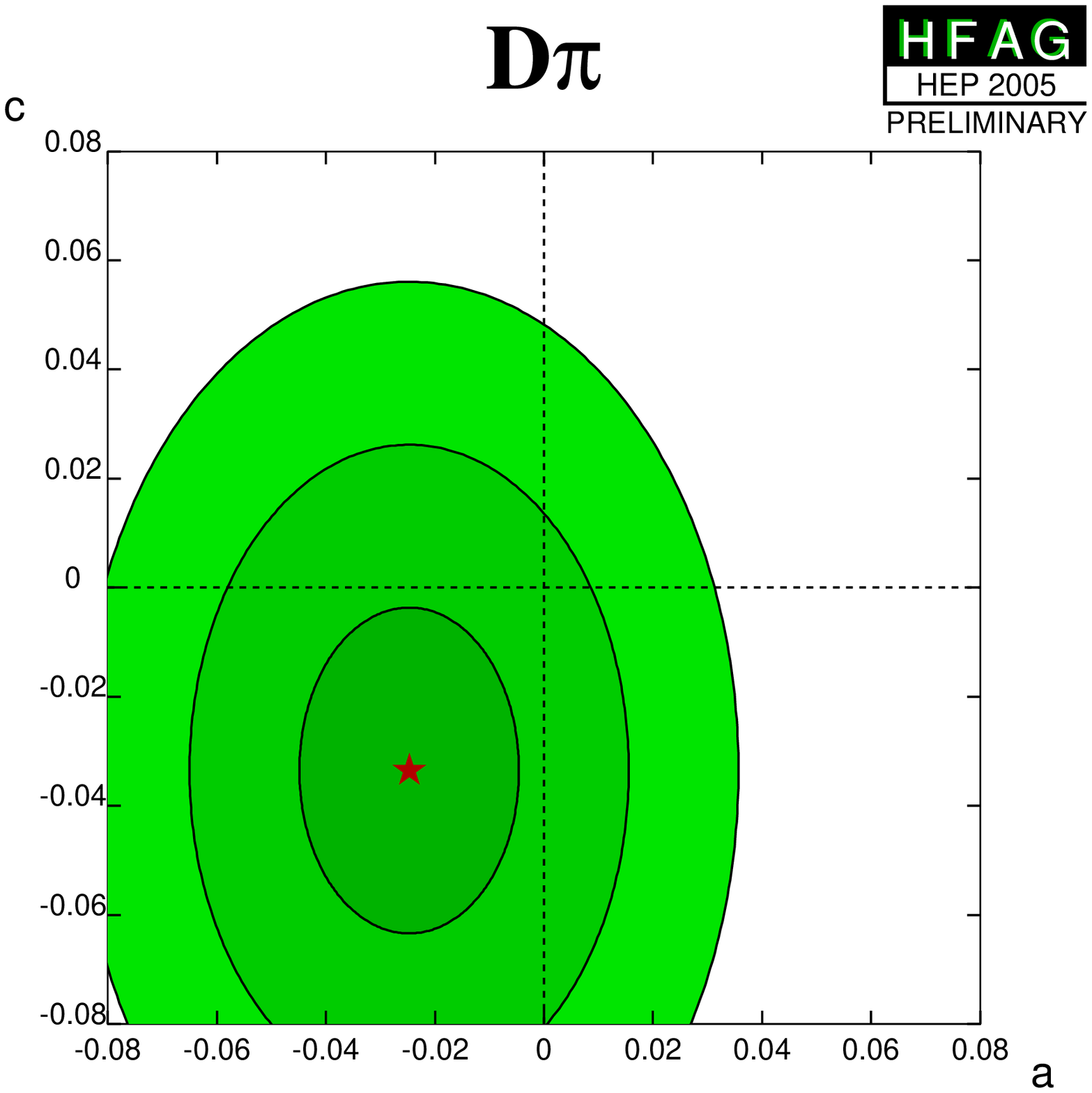}
      } \\
      \resizebox{0.46\textwidth}{!}{
        \includegraphics{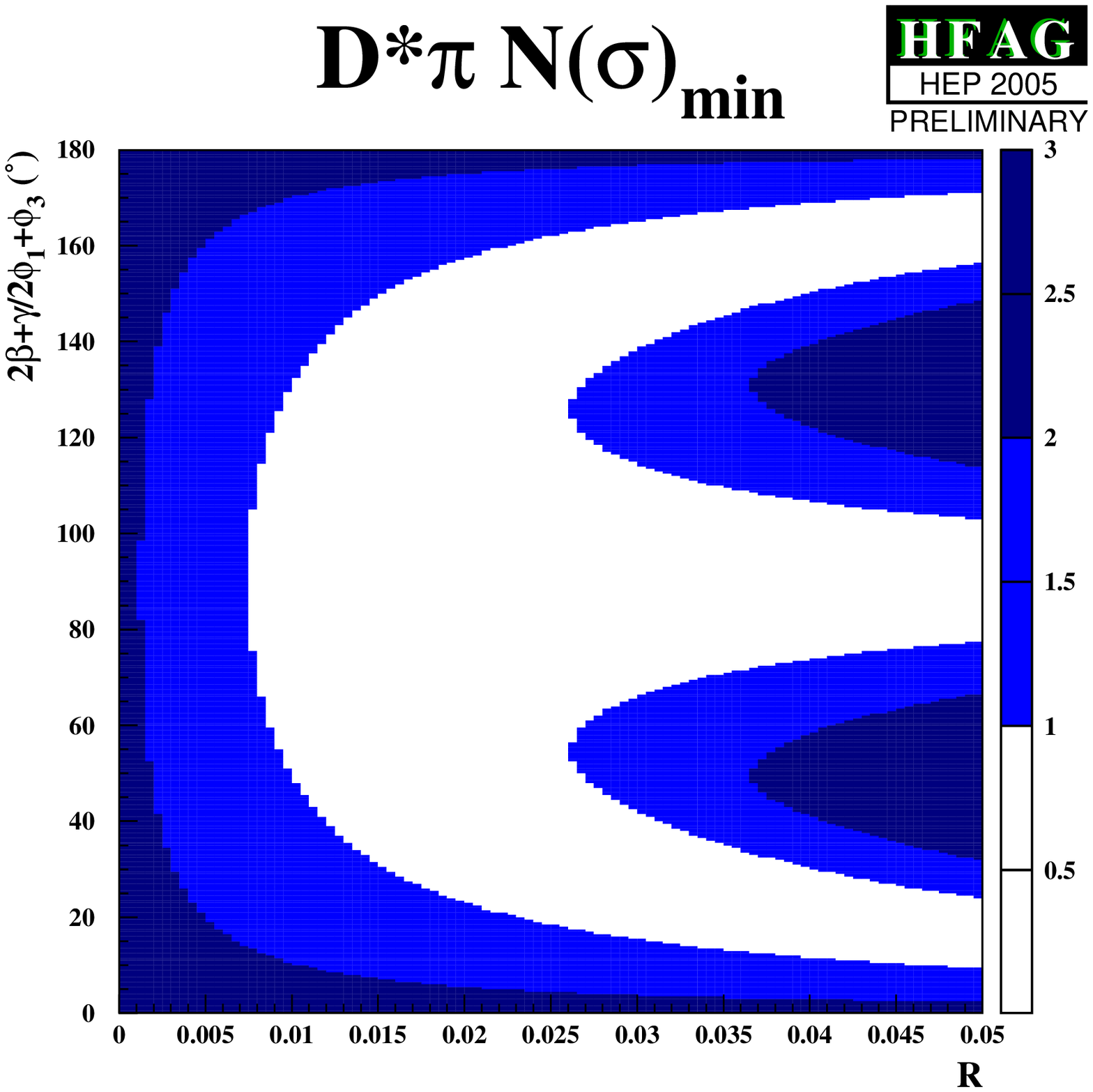}
      }
      &
      \resizebox{0.46\textwidth}{!}{
        \includegraphics{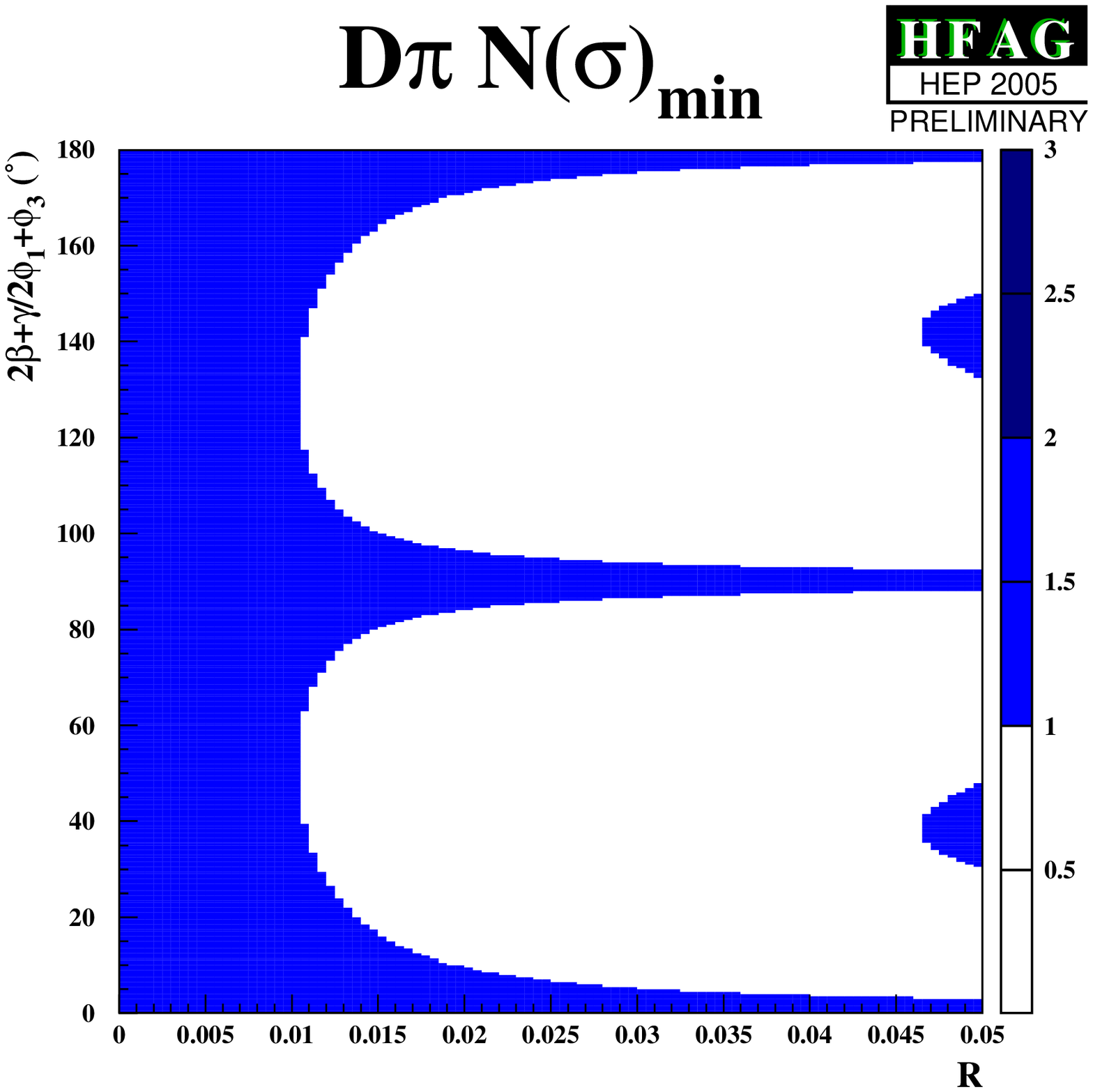}
      }          
    \end{tabular}
  \end{center}
  \vspace{-0.8cm}
  \caption{
    Results from $b \to c\bar{u}d / u\bar{c}d$ modes.
    (Top) Constraints in $c$ {\it vs.} $a$ space.
    (Bottom) Constraints in $2\beta+\gamma$ {\it vs.} $R$ space.
    (Left) $D^*\pi$ and (right) $D\pi$ modes.
  }
  \label{fig:cp_uta:cud_constraints}
\end{figure}

As described in Sec.~\ref{sec:cp_uta:notations:non_cp:dstarpi},
the averages are given in terms of parameters $a$ and $c$.
$\CP$ violation would appear as $a \neq 0$.
Results are available from both \babar\ and \belle\ in the modes
$D^\pm\pi^\mp$ and $D^{*\pm}\pi^\mp$; for the latter mode both experiments 
have used both full and partial reconstruction techniques.
Results are also available from \babar\ using $D^\pm\rho^\mp$.
These results, and their averages, are listed in Table~\ref{tab:cp_uta:cud},
and are shown in Fig.~\ref{fig:cp_uta:cud}.
The constraints in $c$ {\it vs.} $a$ space for the $D\pi$ and $D^*\pi$ modes
are shown in Fig.~\ref{fig:cp_uta:cud_constraints}.

\begin{table}
  \begin{center}
    \caption{
      Averages for $b \to c\bar{u}d / u\bar{c}d$ modes.
    }
    \vspace{0.2cm}
    \setlength{\tabcolsep}{0.0pc}
    \begin{tabular*}{\textwidth}{@{\extracolsep{\fill}}lrcc} \hline 
      \mc{2}{l}{Experiment} & $a$ & $c$ \\
      \hline
      \mc{4}{c}{$D^{*\pm}\pi^{\mp}$} \\
      \babar (full rec.) & \cite{ref:cp_uta:cud:babar:full} &
      $ -0.043 \pm 0.023 \pm 0.010$ & $ 0.047 \pm 0.042 \pm 0.015 $ \\
      \belle (full rec.) & \cite{ref:cp_uta:cud:belle:full} &
      $\ph{-}0.060 \pm 0.040 \pm 0.019$ & $\ph{-}0.049 \pm 0.040 \pm 0.019$ \\
      \babar (partial rec.) & \cite{ref:cp_uta:cud:babar:partial} &
      $ -0.034 \pm 0.014 \pm 0.009$ & $-0.019 \pm 0.022 \pm 0.013$ \\
      \belle (partial rec.) & \cite{ref:cp_uta:cud:belle:partial} &
      $ -0.030 \pm 0.028 \pm 0.018$ & $-0.005 \pm 0.028 \pm 0.018$ \\
      \mc{2}{l}{\bf Average} & 
      $ -0.028 \pm 0.012$ & $ 0.004 \pm 0.017  $ \\
      \mc{2}{l}{\small Confidence level} & 
      \small $0.22~(1.2\sigma)$ & \small $0.42~(0.8\sigma)$ \\
      \hline
      \mc{4}{c}{$D^{\pm}\pi^{\mp}$} \\
      \babar (full rec.) & \cite{ref:cp_uta:cud:babar:full} &
      $ -0.013 \pm 0.022 \pm 0.007$ & $ -0.043 \pm 0.042 \pm 0.011$ \\
      \belle (full rec.) & \cite{ref:cp_uta:cud:belle:full} &
      $ -0.062 \pm 0.037 \pm 0.018$ & $ -0.025 \pm 0.037 \pm 0.018$ \\
      \mc{2}{l}{\bf Average} & $ -0.025 \pm 0.020$ & $ -0.034 \pm 0.030 $ \\
      \mc{2}{l}{\small Confidence level} & 
      \small $0.30~(1.0\sigma)$ & \small $0.76~(0.3\sigma)$ \\
      \hline
      \mc{4}{c}{$D^{\pm}\rho^{\mp}$} \\
      \babar (full rec.) & \cite{ref:cp_uta:cud:babar:full} &
      $ -0.024 \pm 0.031 \pm 0.010$ & $ -0.098 \pm 0.055 \pm 0.019$ \\
      \mc{2}{l}{\bf Average} & $ -0.024 \pm 0.033 $ & $ -0.098 \pm 0.058$ \\
      \hline 
    \end{tabular*}
    \label{tab:cp_uta:cud}
  \end{center}
\end{table}

For each of $D\pi$, $D^*\pi$ and $D\rho$, 
there are two measurements ($a$ and $c$, or $S^+$ and $S^-$) 
which depend on three unknowns ($R$, $\delta$ and $2\beta+\gamma$), 
of which two are different for each decay mode. 
Therefore, there is not enough information to solve directly for $2\beta+\gamma$. 
However, for each choice of $R$ and $2\beta+\gamma$, 
one can find the value of $\delta$ that allows $a$ and $c$ to be closest 
to their measured values, 
and calculate the distance in terms of numbers of standard deviations.
(We currently neglect experimental correlations in this analysis.) 
These values of $N(\sigma)_{\rm min}$ can then be plotted 
as a function of $R$ and $2\beta+\gamma$
(and can trivially be converted to confidence levels). 
These plots are given for the $D\pi$ and $D^*\pi$ modes 
in Figure~\ref{fig:cp_uta:cud_constraints}; 
the uncertainties in the $D\rho$ mode are currently too large 
to give any meaningful constraint.

The constraints can be tightened if one is willing 
to use theoretical input on the values of $R$ and/or $\delta$. 
One popular choice is the use of SU(3) symmetry to obtain 
$R$ by relating the suppressed decay mode to $\B$ decays 
involving $D_s$ mesons. 
More details can be found 
in Refs.~\cite{ref:cp_uta:ckmfitter,ref:cp_uta:utfit}.

\mysubsection{Rates and asymmetries in $\Bmp \to \DorDstar K^{(*)\mp}$ decays
}
\label{sec:cp_uta:cus}

As explained in Sec.~\ref{sec:cp_uta:notations:cus},
rates and asymmetries in $\Bmp \to \DorDstar K^{(*)\mp}$ decays
are sensitive to $\gamma$.
Various methods using different $\DorDstar$ final states exist.

\begin{figure}[htb]
  \begin{center}
    \begin{tabular}{cc}
      \resizebox{0.46\textwidth}{!}{
        \includegraphics{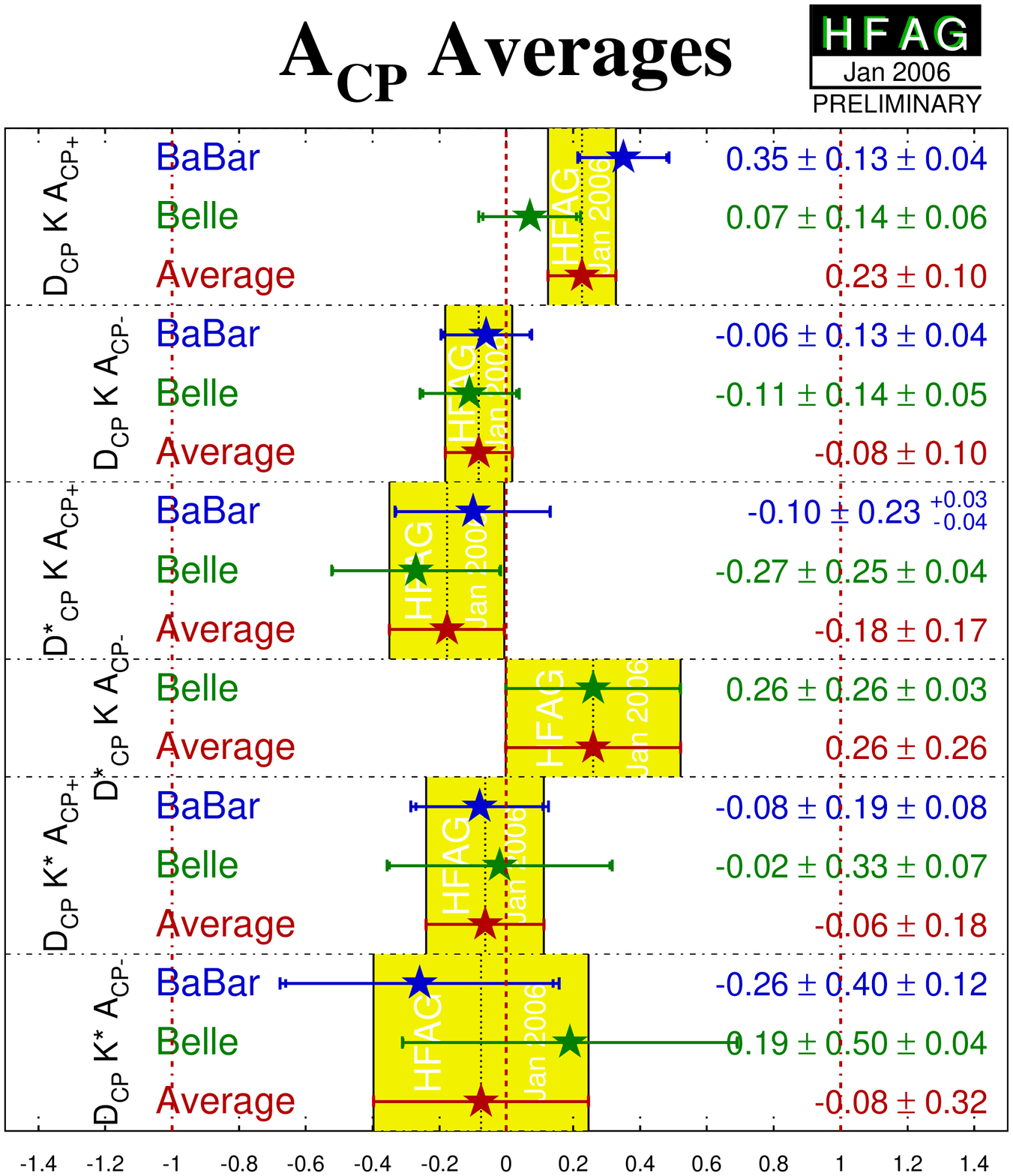}
      }
      &
      \resizebox{0.46\textwidth}{!}{
        \includegraphics{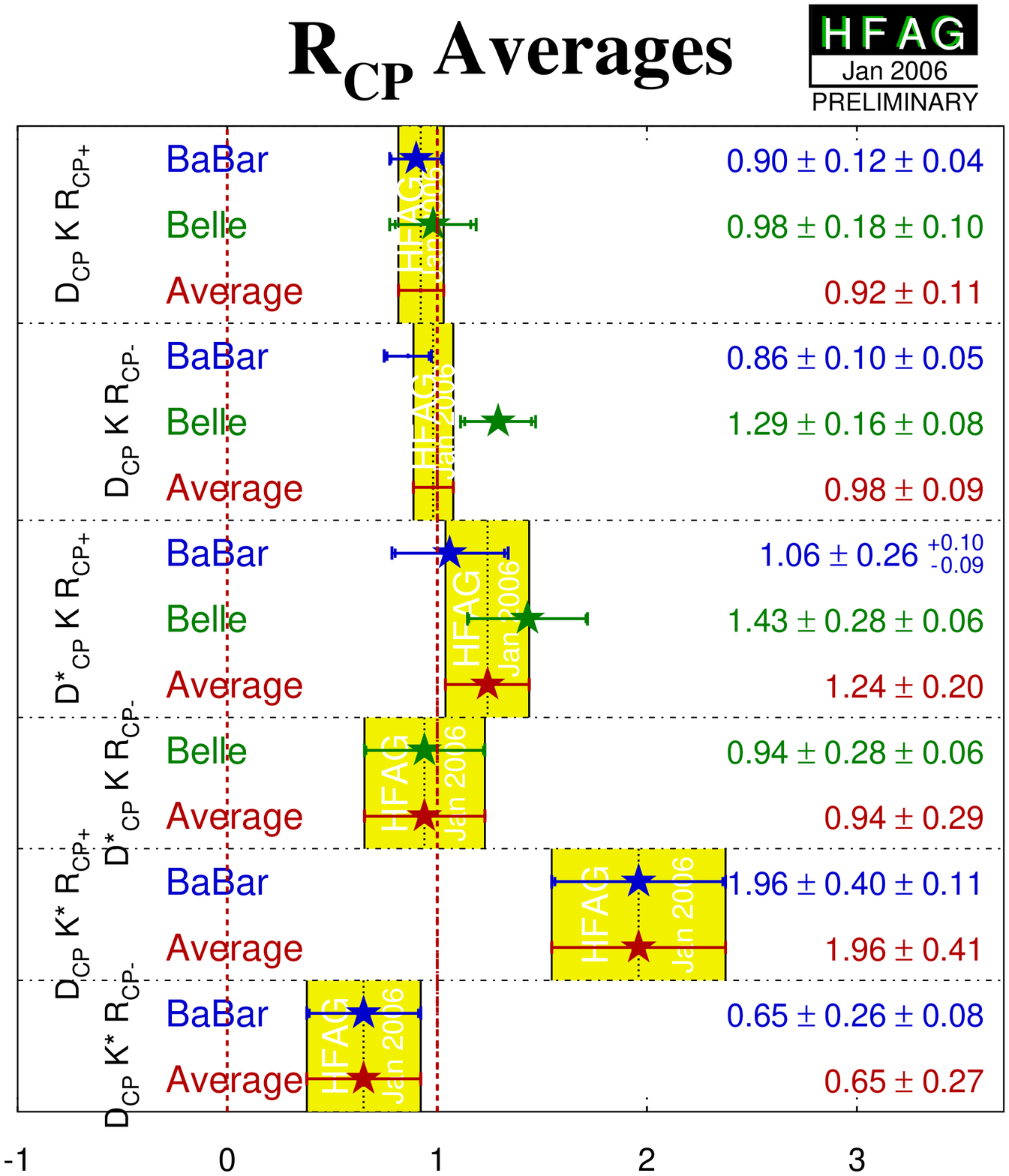}
      }
    \end{tabular}
  \end{center}
  \vspace{-0.8cm}
  \caption{
    Averages of $A_{\CP}$ and $R_{\CP}$ from GLW analyses.
  }
  \label{fig:cp_uta:cus:glw}
\end{figure}

Results are available from both \babar\ and \belle\ on GLW analyses
in the decay modes $\Bmp \to D\Kmp$, $\Bmp \to \Dstar\Kmp$ and $\Bmp \to D\Kstarmp$.
Both experiments use the 
$\CP$ even $D$ decay final states $K^+K^-$ and $\pi^+\pi^-$ in all three modes; 
both experiments also use only the $\Dstar \to D\pi^0$ decay, 
which gives $\CP(\Dstar) = \CP(D)$. 
For $\CP$ odd $D$ decay final states, 
\belle\ uses $\KS\pi^0$, $\KS\eta$ and $\KS\phi$ in all three analyses, 
and also use $\KS\omega$ in $D\Kmp$ and $\Dstar\Kmp$ analyses. 
\babar\ uses $\KS\pi^0$ only for $D\Kmp$ analysis; 
for $D\Kstarmp$ analysis they also use $\KS\phi$ and $\KS\omega$
(and assign an asymmetric systematic error due to $\CP$ even pollution 
in these $\CP$ odd channels~\cite{ref:cp_uta:cus:babar:dstarcpk}).
The results and averages are given in Table~\ref{tab:cp_uta:cus:glw}
and shown in Fig.~\ref{fig:cp_uta:cus:glw}.

\begin{table}
  \begin{center}
    \caption{
      Averages from GLW analyses of $b \to c\bar{u}s / u\bar{c}s$ modes.
    }
    \vspace{0.2cm}
    \resizebox{\textwidth}{!}{
      \setlength{\tabcolsep}{0.0pc}
      \begin{tabular}{@{\extracolsep{2mm}}lrcccc} \hline 
        \mc{2}{l}{Experiment} & 
        $A_{\CP+}$ & $A_{\CP-}$ & $R_{\CP+}$ & $R_{\CP-}$ \\
        \hline
        \mc{6}{c}{$D_{\CP} K^-$} \\
        \babar & \cite{ref:cp_uta:cus:babar:dcpk} &
        $\ph{-}0.35 \pm 0.13 \pm 0.04$ & $\ph{-}0.06 \pm 0.13 \pm 0.04$ & 
        $ 0.90 \pm 0.12 \pm 0.04$ & $ 0.86 \pm 0.10 \pm 0.05$ \\
        \belle & \cite{ref:cp_uta:cus:belle:dcpk} &
        $\ph{-}0.06 \pm 0.14 \pm 0.05$ & $-0.12 \pm 0.14 \pm 0.05$ & 
        $ 1.13 \pm 0.16 \pm 0.08$ & $ 1.17 \pm 0.14 \pm 0.14$ \\
        \mc{2}{l}{\bf Average} & 
        $ 0.22 \pm 0.10$ & $-0.09 \pm 0.10$ & $ 0.98 \pm 0.10$ & $ 0.94 \pm 0.10$ \\
        \hline
        \mc{6}{c}{$\Dstar_{\CP} K^-$} \\
        \babar & \cite{ref:cp_uta:cus:babar:dstarcpk} &
        $-0.10 \pm 0.23 \, ^{+0.03}_{-0.04}$ & & 
        $ 1.06 \pm 0.26 \, ^{+0.10}_{-0.09}$ &  \\
        \belle & \cite{ref:cp_uta:cus:belle:dcpk} &
        $-0.20 \pm 0.22 \pm 0.04$ & $ 0.13 \pm 0.30 \pm 0.08$ & 
        $ 1.41 \pm 0.25 \pm 0.06$ & $ 1.15 \pm 0.31 \pm 0.12$ \\
        \mc{2}{l}{\bf Average} & 
        $-0.15 \pm 0.16$ & $ 0.13 \pm 0.31$ & $ 1.25 \pm 0.19$ & $ 1.15 \pm 0.33$ \\
        \hline
        \mc{6}{c}{$D_{\CP} K^{*-}$} \\
        \babar & \cite{ref:cp_uta:cus:babar:dcpkstar} &
        $-0.08 \pm 0.19 \pm 0.08$ & $-0.26 \pm 0.40 \pm 0.12$ &
        $ 1.96 \pm 0.40 \pm 0.11$ & $ 0.65 \pm 0.26 \pm 0.08$ \\
        \belle & \cite{ref:cp_uta:cus:belle:dcpkstar} &
        $-0.02 \pm 0.33 \pm 0.07$ & $\ph{-}0.19 \pm 0.50 \pm 0.04$ \\
        \mc{2}{l}{\bf Average} & 
        $-0.06 \pm 0.18$ & $-0.08 \pm 0.32$ & $1.96 \pm 0.41$ & $0.65 \pm 0.27$ \\
        \hline
      \end{tabular}
    }
    \label{tab:cp_uta:cus:glw}
  \end{center}
\end{table}

\begin{figure}[htb]
  \begin{center}
    \resizebox{0.46\textwidth}{!}{
      \includegraphics{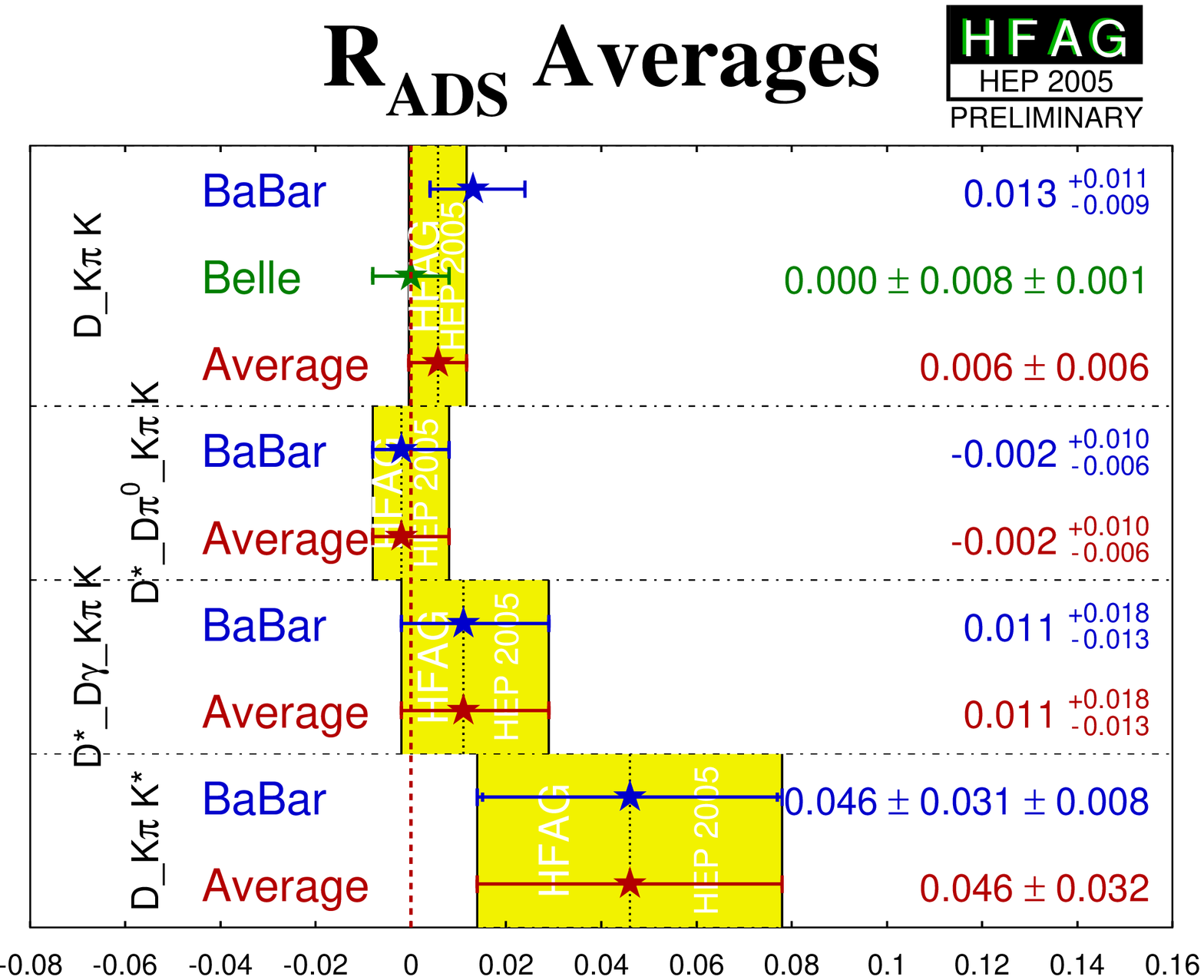}
    }
  \end{center}
  \vspace{-0.8cm}
  \caption{
    Averages of $R_{\rm ADS}$.
  }
  \label{fig:cp_uta:cus:ads}
\end{figure}

For ADS analysis, both \babar\ and \belle\ have studied
the mode $\Bmp \to D\Kmp$;
\belle\ has also studied $\Bmp \to D\pi^\mp$
and \babar\ has also analyzed the $\Bmp \to \Dstar\Kmp$ 
and $\Bmp \to D\Kstarmp$ modes
($\Dstar \to D\pi^0$ and $\Dstar \to D\gamma$ are studied separately;
$\Kstarmp$ is reconstructed as $\KS\pi^\mp$).
In all cases the suppressed decay $D \to K^+\pi^-$ has been used.
The results and averages are given in Table~\ref{tab:cp_uta:cus:ads}
and shown in Fig.~\ref{fig:cp_uta:cus:ads}.
Note that although no clear signals for these modes have yet been seen,
the central values are given.
In $\Bm \to \Dstar\Km$ decays there is an effective shift of $\pi$
in the strong phase difference between the cases that the $\Dstar$ is 
reconstructed as $D\pi^0$ and $D\gamma$~\cite{ref:cp_uta:cus:bg}.
As a consequence, the different $D^*$ decay modes are treated separately.

\begin{table}
  \begin{center} 
    \caption{
      Averages from ADS analyses of $b \to c\bar{u}s / u\bar{c}s$ and 
      $b \to c\bar{u}d / u\bar{c}d$ modes.
    }
    \vspace{0.2cm}
    \setlength{\tabcolsep}{0.0pc}
    \begin{tabular*}{\textwidth}{@{\extracolsep{\fill}}lrcc} \hline 
      \mc{2}{l}{Experiment} & $A_{\rm ADS}$ & $R_{\rm ADS}$ \\
      \hline
      \mc{4}{c}{$D K^-$, $D \to K^+\pi^-$} \\
      \babar & \cite{ref:cp_uta:cus:babar:dk_ads} &
      & $ 0.013 \, ^{+0.011}_{-0.009}$ \\ 
      \belle & \cite{ref:cp_uta:cus:belle:dk_ads} &
      & $ 0.000 \pm 0.008 \pm 0.001$ \\
      \mc{2}{l}{\bf Average} & 
      & $ 0.006 \pm 0.006$ \\
      \hline
      \mc{4}{c}{$\Dstar K^-$, $\Dstar \to D\pi^0$, $D \to K^+\pi^-$} \\
      \babar & \cite{ref:cp_uta:cus:babar:dk_ads} &
      & $-0.002 \, ^{+0.010}_{-0.006}$ \\
      \hline
      \mc{4}{c}{$\Dstar K^-$, $\Dstar \to D\gamma$, $D \to K^+\pi^-$} \\
      \babar & \cite{ref:cp_uta:cus:babar:dk_ads} &
      & $ 0.011 \, ^{+0.018}_{-0.013}$ \\
      \hline
      \mc{4}{c}{$D K^{*-}$, $D \to K^+\pi^-$, $K^{*-} \to \KS \pi^-$} \\
      \babar & \cite{ref:cp_uta:cus:babar:dkstar_ads} &
      $ -0.22 \pm 0.61 \pm 0.17$ & $ 0.046 \pm 0.031 \pm 0.008$ \\
      \hline 
      \mc{4}{c}{$D \pi^-$, $D \to K^+\pi^-$} \\
      \belle & \cite{ref:cp_uta:cus:belle:dk_ads} &
      $ 0.10 \pm 0.22 \pm 0.06$ & $ 0.0035 \, ^{+0.0008}_{-0.0007} \pm 0.0003$ \\
      \hline 
    \end{tabular*}
    \label{tab:cp_uta:cus:ads}
  \end{center}
\end{table}

For the Dalitz plot analysis, both \babar\ and \belle\ have studied
the modes $\Bmp \to D\Kmp$, $\Bmp \to \Dstar\Kmp$ and $\Bmp \to D\Kstarmp$.
For $\Bmp \to \Dstar\Kmp$,
\belle\ has used only $\Dstar \to D\pi^0$,
while \babar\ has used both $\Dstar$ decay modes and 
taken the effective shift in the strong phase difference into account.
In all cases the decay $D \to \KS\pi^+\pi^-$ has been used.
Results are given in Table~\ref{tab:cp_uta:cus:dalitz}.

The parameters measured in the different analyses are explained in
Sec.~\ref{sec:cp_uta:notations:cus}.
\belle\ directly extract $\gamma$, $r_B$ and $\delta_B$ for each decay mode
and perform a frequentist statistical procedure to correct
for bias originating from the positive definite nature of $r_B$.
Results from $D\Kmp$ and $\Dstar\Kmp$ are used 
to obtain a combined value of $\gamma$;
results for $D\Kstarmp$ are not currently included in this procedure.
\babar\ measure the $(x_\pm,y_\pm)$ variables,
and perform a frequentist statistical procedure,
using all three $\B$ decay modes,
to convert these into measurements of $\gamma$, $r_B$ and $\delta_B$.

Both experiments reconstruct $\Kstarmp$ as $\KS\pi^\mp$,
but the treatment of possible nonresonant $\KS\pi^\mp$ differs:
\belle\ assign an additional model uncertainty,
while \babar\ use a reparametrization 
suggested by Gronau~\cite{ref:cp_uta:cus:gronau}.
The parameters $r_B$ and $\delta_B$ are replaced with 
effective parameters $\kappa r_s$ and $\delta_s$;
no attempt is made to extract the true hadronic parameters 
of the $\Bmp \to D\Kstarmp$ decay.

At present, we make no attempt to average the results of the 
Dalitz plot analyses.
Additionally, we have not attempted to combine the results 
of the GLW, ADS and Dalitz analyses in order to obtain the 
most precise determination of $\gamma$ (and associated parameters, such as $r_B$).
Such attempts have been made by the CKMfitter and UTFit groups;
see Refs.~\cite{ref:cp_uta:ckmfitter,ref:cp_uta:utfit}.

\begin{table}
  \begin{center} 
    \caption{
      Averages from Dalitz plot analyses of $b \to c\bar{u}s / u\bar{c}s$ modes.
    }
    \vspace{0.2cm}

      \setlength{\tabcolsep}{0.0pc}
      \begin{tabular*}{\textwidth}{@{\extracolsep{\fill}}lrcccc} \hline 
        \mc{2}{l}{Experiment} & $x_+$ & $y_+$ & $x_-$ & $y_-$ \\
        \hline
        \mc{6}{c}{$D K^-$, $D \to \KS \pi^+\pi^-$} \\
        \babar & \cite{ref:cp_uta:cus:babar:dk_dalitz} & & 
        $\ph{-}0.02 \pm 0.08 \pm 0.02 \pm 0.02$ & & $\ph{-}0.06 \pm 0.09 \pm 0.04 \pm 0.04$ \\
        & & 
        \hspace{-15mm} $-0.13 \pm 0.07 \pm 0.03 \pm 0.03$ \hspace{-15mm} & & 
        \hspace{-15mm} $\ph{-}0.08 \pm 0.07 \pm 0.03 \pm 0.02$ \hspace{-15mm} \\
        \hline
        \mc{6}{c}{$\Dstar K^-$, $\Dstar \to D\pi^0$ \& $D\gamma$, $D \to \KS \pi^+\pi^-$} \\
        \babar & \cite{ref:cp_uta:cus:babar:dk_dalitz} & &
        $\ph{-}0.01 \pm 0.12 \pm 0.04 \pm 0.06$ & & $-0.14 \pm 0.11 \pm 0.02 \pm 0.03$ \\
        & & 
        \hspace{-15mm} $\ph{-}0.14 \pm 0.09 \pm 0.03 \pm 0.03$  \hspace{-15mm} & & 
        \hspace{-15mm} $-0.13 \pm 0.09 \pm 0.03 \pm 0.02$ \hspace{-15mm} \\        
        \hline
        \mc{6}{c}{$D K^{*-}$, $D \to \KS \pi^+\pi^-$, $K^{*-} \to \KS\pi^-$} \\
        \babar & \cite{ref:cp_uta:cus:babar:dk_dalitz} & &
        $-0.01 \pm 0.32 \pm 0.18 \pm 0.05$ & & $\ph{-}0.26 \pm 0.30 \pm 0.16 \pm 0.03$ \\
        & &
        \hspace{-15mm} $-0.07 \pm 0.23 \pm 0.13 \pm 0.03$ \hspace{-15mm} & & 
        \hspace{-15mm} $-0.20 \pm 0.20 \pm 0.11 \pm 0.03$ \hspace{-15mm} \\
        \hline
      \end{tabular*}

    \vspace{2ex}

    \setlength{\tabcolsep}{0.0pc}
    \begin{tabular*}{\textwidth}{@{\extracolsep{\fill}}lrccc} \hline 
      \mc{2}{l}{Experiment} & $\gamma \ (^\circ)$ & $\delta_B \ (^\circ)$ & $r_B$ \\
      \hline
      \mc{5}{c}{$D K^-$, $D \to \KS \pi^+\pi^-$} \\
      \babar & \cite{ref:cp_uta:cus:babar:dk_dalitz} &
      $$ & $104 \pm 45 \, ^{+17}_{-21} \, ^{+16}_{-24}$ & $ 0.12 \pm 0.08 \pm 0.03 \pm 0.04$ \\
      \belle & \cite{ref:cp_uta:cus:belle:dk_dalitz} &
      $ 64 \pm 19 \pm 13 \pm 11$ & $ 157 \pm 19 \pm 11 \pm 21$ & $ 0.21 \pm 0.08 \pm 0.03 \pm 0.04$ \\
      \mc{2}{l}{\bf Average} & 
      \mc{3}{c}{\sc in preparation} \\
      \hline
      \mc{5}{c}{$\Dstar K^-$, $\Dstar \to D\pi^0$ or $D\gamma$, $D \to \KS \pi^+\pi^-$} \\
      \babar & \cite{ref:cp_uta:cus:babar:dk_dalitz} &
      $$ & $ 296 \pm 41 \, ^{+14}_{-12} \pm 15$ & $ 0.17 \pm 0.10 \pm 0.03 \pm 0.03$ \\
      \belle & \cite{ref:cp_uta:cus:belle:dk_dalitz} &
      $ 75 \pm 57 \pm 11 \pm 11$ & $ 321 \pm 57 \pm 11 \pm 21$ & $ 0.12 \, ^{+0.16}_{-0.11} \pm 0.02 \pm 0.04$ \\
      \mc{2}{l}{\bf Average} & 
      \mc{3}{c}{\sc in preparation} \\
      \hline
      \mc{5}{c}{$D K^{*-}$, $D \to \KS \pi^+\pi^-$} \\
      \belle & \cite{ref:cp_uta:cus:belle:dkstar_dalitz} &
      $112 \pm 35 \pm 9 \pm 11 \pm 8$ & $353 \pm 35 \pm 8 \pm 21 \pm 49$ & $ 0.25 \pm 0.18 \pm 0.09 \pm 0.04 \pm 0.08$ \\
      \hline
      \mc{5}{c}{$D K^-$ and $\Dstar K^-$ combined} \\
      \belle & \cite{ref:cp_uta:cus:belle:dk_dalitz} &
      $ 68 \, ^{+14}_{-15} \pm 13 \pm 11$ \\
      \hline
      \mc{5}{c}{$D K^-$, $\Dstar K^-$ and $DK^{*-}$ combined} \\
      \babar & \cite{ref:cp_uta:cus:babar:dk_dalitz} &
      $ 67 \pm 28 \pm 13 \pm 11$ \\
      \hline 
    \end{tabular*}

    \label{tab:cp_uta:cus:dalitz}
  \end{center}
\end{table}


\clearpage


\section{Semileptonic $B$ decays}
\label{slbdecays}

Major updates of \Vub\ in both inclusive and exclusive $B$ decays
have been made since the last HFAG document~\cite{hfag_hepex_winter2005}, 
and are described here in detail.


The determination of \vub\ from inclusive decays involves 
intense ongoing activity in both experiment and theory.
HFAG subgroups have recently determined updated values for the heavy quark
parameters $m_b$ and $\mu_{\pi}^2$ based on moments measured in
inclusive $\Bb\to X_c\ell\nub$ and $\Bb\to X_s\gamma$ decays.  In addition, the
theoretical tools have improved and 
have been incorporated by the experiments.  A comprehensive determination
of \vub\ from inclusive decays based on the results presented at the 2005
summer conferences is given below.

Several new measurements of the exclusive decay
$\Bb\to\pi\ell\nub$ were presented
at the 2005 Summer conferences. Their precision is at a level
that calls for improved calculations of the form factors and in particular their
normalization. An average of these results and the subsequent
determination of \vub\ is discussed below.

In the following, brief descriptions of all parameters
and analyses (published or preliminary) relevant for the
determination of the combined results are given.  The description is
based on the information available on the web page at\\
\medskip
 \centerline{\tt http://www.slac.stanford.edu/xorg/hfag/semi/eps05/eps05.shtml}
\medskip
The values for $\vub$ from inclusive decays have been updated relative
to the web page using the current HFAG value of the average $B$
meson lifetime: $\langle\tau_B\rangle=1.585\pm 0.007$~ps.


\subsection{Methodology}
\label{sec:slmethodology}
The method for extracting averages is described in
section~\ref{sec:method}. In the following, the method has been
extended to take into account the fact that measurement errors often
depend on the measured value, i.e. are relative errors.
Furthermore, an effort has been made to separate statistical, and
different sources of systematic and theoretical errors.

For measurements with Gaussian errors, the usual estimator for the
average of a set of measurements is obtained by minimizing the following
$\chi^2$:
\begin{equation}
\chi^2(t) = \sum_i^N \frac{\left(y_i-t\right)^2}{\sigma^2_i} ,
\label{eq:chi2t}
\end{equation}
where $y_i$ is the measured value for input $i$ and $\sigma_i^2$ is the
variance of the distribution from which $y_i$ was drawn.  The value $\hat{t}$
of $t$ at minimum $\chi^2$ is our estimator for the average.  (This
discussion is given for independent measurements for the sake of
simplicity; the generalization to correlated measurements is
straightforward, and has been used when averaging results.) 
The true $\sigma_i$ are unknown but typically the error as assigned by the
experiment $\sigma_i^{\mathrm{raw}}$ is used as an estimator for it.
Caution is advised,
however, in the case where $\sigma_i^{\mathrm{raw}}$
depends on the value measured for $y_i$. Examples of this include
an uncertainty in any multiplicative factor (like
an acceptance) that enters the determination of $y_i$, i.e. the $\sqrt{N}$
dependence of Poisson statistics, where $y_i \propto N$
and $\sigma_i \propto \sqrt{N}$.
Failing to account for this type of
dependence when averaging leads to a biased average.
Biases in the average can be avoided (or at least reduced)
by minimizing the following
$\chi^2$:
\begin{equation}
\chi^2(t) = \sum_i^N \frac{\left(y_i-t\right)^2}{\sigma^2_i(\hat{t})} .
\label{eq:chi2that}
\end{equation}
In the above $\sigma_i(\hat{t})$ is the uncertainty
assigned to input $i$ that includes the assumed dependence of the
stated error on the value measured.  As an example, consider 
a pure acceptance error, for which
$\sigma_i(\hat{t}) = (\hat{t} / y_i)\times\sigma_i^{\mathrm{raw}}$ .
It is easily verified that solving Eq.~\ref{eq:chi2that} 
leads to the correct behavior, namely
$$ 
\hat{t} = \frac{\sum_i^N y_i^3/(\sigma_i^{\mathrm{raw}})^2}{\sum_i^N y_i^2/(\sigma_i^{\mathrm{raw}})^2},
$$
i.e. weighting by the inverse square of the 
fractional uncertainty, $\sigma_i^{\mathrm{raw}}/y_i$.

It is sometimes difficult to assess the dependence of $\sigma_i^{\mathrm{raw}}$ on
$\hat{t}$ from the errors quoted by experiments.  
As a result, the sensitivity
to different assumptions on these dependences has been
studied for the averages given in this section.

Another issue that needs careful treatment is the question of correlation
among different measurements, e.g. due to using the same theory for
calculating acceptances.  A common practice is to set the correlation
coefficient to unity to indicate full correlation.  However, this is
not a ``conservative'' thing to do, and can in fact lead to a significantly
underestimated uncertainty on the average.  In the absence of
better information, the most conservative choice of correlation coefficient
between two measurements $i$ and $j$
is the one that maximizes the uncertainty on $\hat{t}$
due to that pair of measurements:
\begin{equation}
\sigma_{\hat{t}(i,j)}^2 = \frac{\sigma_i^2\,\sigma_j^2\,(1-\rho_{ij}^2)}
   {\sigma_i^2 + \sigma_j^2 - 2\,\rho_{ij}\,\sigma_i\,\sigma_j} ,
\label{eq:correlij}
\end{equation}
namely
\begin{equation}
\rho_{ij} = \mathrm{min}\left(\frac{\sigma_i}{\sigma_j},\frac{\sigma_j}{\sigma_i}\right) ,
\label{eq:correlrho}
\end{equation}
which corresponds to setting $\sigma_{\hat{t}(i,j)}^2=\mathrm{min}(\sigma_i^2,\sigma_j^2)$.
Setting $\rho_{ij}=1$ when $\sigma_i\ne\sigma_j$ can lead to a significant
underestimate of the uncertainty on $\hat{t}$, as can be seen
from Eq.~\ref{eq:correlij}.

Finally, a note on the breakdown of the error sources
contributing to the overall uncertainty on the average.
The overall covariance matrix is constructed from a number of
individual sources, e.g.
$\mathbf{V} = \mathbf{V_{stat}+V_{sys}+V_{th}}$.
The variance on the average $\hat{t}$ can be written
\begin{eqnarray}
\sigma^2_{\hat{t}} 
 &=& 
\frac{ \sum_{i,j}\left(\mathbf{V^{-1}}\, 
\mathbf{[V_{stat}+V_{sys}+V_{th}]}\, \mathbf{V^{-1}}\right)_{ij}}
{\left(\sum_{i,j} V^{-1}_{ij}\right)^2}
= \sigma^2_{stat} + \sigma^2_{sys} + \sigma^2_{th} .
\end{eqnarray}
Written in this form, one can readily determine the 
contribution of each source of uncertainty to the overall uncertainty
on the average.  This breakdown of the uncertainties is used below.

\subsection{Exclusive Cabibbo-favored decays}
\label{slbdecays_b2cexcl}

There were no major updates in this area; the reader is referred to the
averages given in Ref.~\cite{hfag_hepex_winter2005}.

\subsection{Inclusive Cabibbo-favored decays}
\label{slbdecays_b2cincl}

Aspects of the theory and phenomenology of inclusive Cabibbo-favored
$B$ decays and their use in the determination of \vcb\ in the context
of the Heavy Quark Expansion (HQE), an Operator Product Expansion
based on HQET, are described in many places~\cite{b2xintros}.

\begin{figure}
\begin{center}
\includegraphics[width=4.0in]{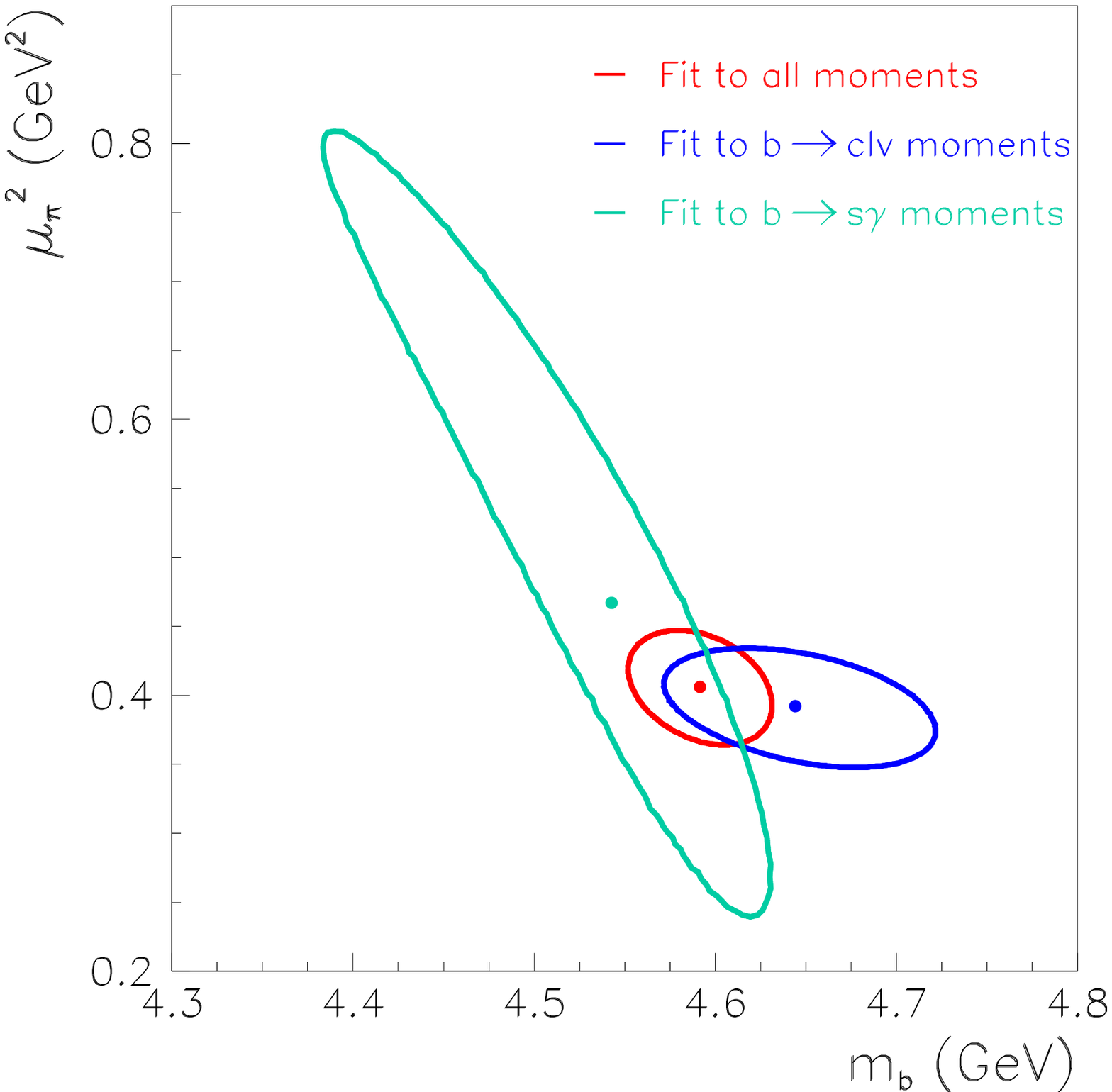}
\end{center}
\caption{Results of HQE fits to moments in $\Bb\to X_s\gamma$
and $\Bb\to X_c\ell\nub$ decays from Ref.~\protect{\cite{hepph-0507253}}.
The quantities shown are in the kinetic mass scheme.}
\label{fig:mupi2mb}
\end{figure}

Updated values for the parameters $m_b$ and $\mu_{\pi}^2$ are
used below in the determination of
\vub\ from inclusive decays.  These are taken from a fit to 
energy and mass moments in
$\Bb\to X_c\ell\nub$ decays~\cite{ref:CLEOmoments,ref:CLEOmoments1,ref:BABARmoments,ref:BABARmoments1,ref:BABARbclnumomentsfits,ref:BELLEmoments,ref:BELLEmoments1,ref:DELPHImoments,ref:CDFmoments}
and to photon energy moments
in $\Bb\to X_s\gamma$ decays~\cite{ref:CLEObsg,ref:BELLEbsg,ref:BABARbsgexcl,ref:BABARbsgincl}
in the ``kinetic'' mass scheme~\cite{hepph-0507253}.
The fit results are shown in Fig.~\ref{fig:mupi2mb}.
These values are translated
into the shape-function mass scheme~\cite{ref:sfmasstranslation,ref:BLNP}
for use in the extraction of $\vub$, giving
$m_b(\mathrm{SF})=4.60\pm 0.04$~GeV and 
$\mu_{\pi}^2(\mathrm{SF})=0.20\pm 0.04$~GeV${}^2$ 
with correlation coefficient -0.26.
Similar fits in
other mass schemes, e.g. the ''1S'' scheme~\cite{Bauer:2004ve}, have
not yet been updated to include the latest measurements; that work is
in progress.  Once it is complete the full set of parameters (including
\vcb\ and the $\cbf(\Bb\to X\ell\nub)$) will be given.
Previously published fits in this area can be found in 
Refs.~\cite{Bauer:2004ve,Battaglia:2002tm,Mahmood:2004kq,Aubert:2004aw}).
For an average of the total semileptonic branching fraction the reader
is referred to Ref.~\cite{hfag_hepex_winter2005}.

\subsection{Exclusive Cabibbo-suppressed decays}
\label{slbdecays_b2uexcl}

\begin{figure}
\begin{center}
\includegraphics[width=6.0in]{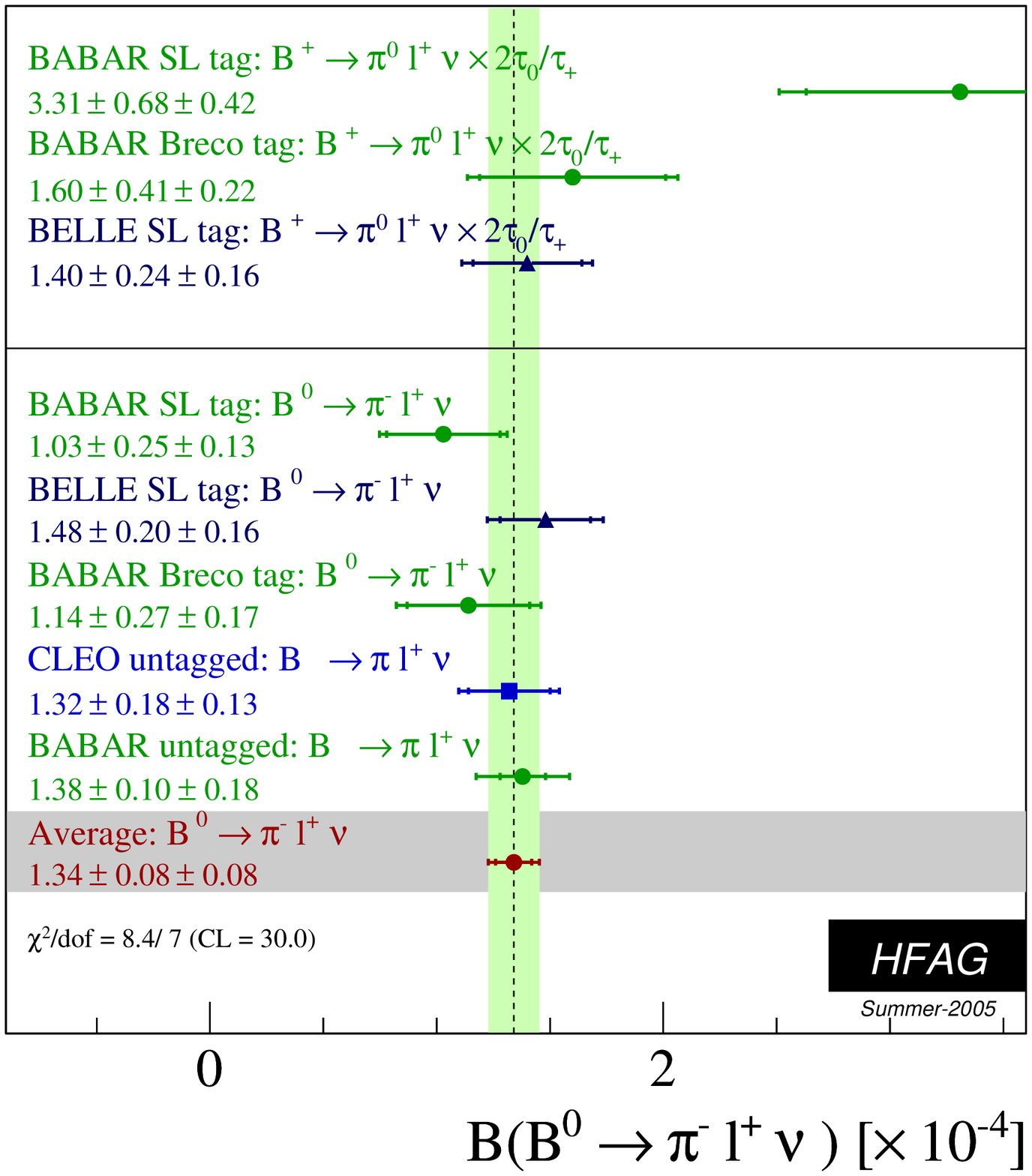}
\end{center}
\caption{Measurements of $\cbf(\Bb\to\pi\ell\nub)$ and their average.}
\label{fig:pilnubf}
\end{figure}

Here we list results on exclusive semileptonic branching fractions
and determinations of $\vub$ based on $\Bb\to\pi\ell\nub$ decays.
The measurements are based on two different event selections: Tagged
events, in which case the second $B$ meson in the event is fully
reconstructed in either a hadronic decay or in a 
Cabibbo-favored semileptonic
decay; and Untagged events, in which case the selection infers the momentum
of the undetected neutrino based on measurements of the total 
momentum sum of detected particles and knowledge of the initial state.
The results for the full and partial branching fraction are given
in Table~\ref{tab:pilnubf} and shown in Fig.~\ref{fig:pilnubf}.  

When averaging these results, systematic uncertainties due to external
inputs, e.g. form factor shapes and background estimates from the
modeling of $\Bb\to X_c\ell\nub$ and $\Bb\to X_u\ell\nub$ decays, are
treated as fully correlated (in the sense of Eq.~\ref{eq:correlrho}).
Uncertainties due to experimental reconstruction effects are treated
as fully correlated among measurements from a given experiment.  Varying
the assumed dependence of the quoted errors on the measured value (see
Eq.~\ref{eq:chi2that}) for error sources where the
dependence was not obvious had no significant impact.


\begin{table}[!htb]
\caption{Summary of exclusive determinations of $\cbf(\Bb\to\pi
\ell\nub)$.  The errors quoted
correspond to statistical and systematic uncertainties, respectively.
Measured branching fractions for $B\rightarrow \pi^0 l \nu$ have been
multiplied by $2\times \tau_{B^0}/\tau_{B^+}$ in accordance with
isospin symmetry.}
\begin{center}
\begin{small}
\begin{tabular}{lcc}
\hline\hline
  & $\cbf [10^{-4}]$   & $\cbf(q^2>16\,\gev^2/c^2) [10^{-4}]$ \\
CLEO $\pi^+,\pi^0$~\cite{ref:CLEOpilnu}  & $1.32\pm 0.18\pm 0.13\ $ 
& $0.25\pm 0.09\pm 0.05$ \\ 
BABAR $\pi^+,\pi^0$~\cite{ref:BABARpilnu} & $1.38\pm 0.10\pm 0.18\ $
& $0.49\pm 0.05\pm 0.06$ \\  \hline
Average of untagged & $1.35 \pm 0.10\pm 0.14\ $ 
& $0.40\pm 0.05\pm 0.05$ \\ \hline\hline
BELLE SL $\pi^+$~\cite{ref:BELLEpilnuSL} & $1.48\pm 0.20\pm 0.16\ $
& $0.40\pm 0.12\pm 0.05$ \\ 
BELLE SL $\pi^0$~\cite{ref:BELLEpilnuSL} & $1.40\pm 0.24\pm 0.16\ $ 
& $0.41\pm 0.15\pm 0.04$ \\ 
BABAR SL $\pi^+$~\cite{ref:BABARpilnuSL} & $1.03\pm 0.25\pm 0.13\ $ 
& $0.21\pm 0.14\pm 0.05$ \\ 
BABAR SL $\pi^0$~\cite{ref:BABARpilnuSLz}& $3.31\pm 0.68\pm 0.42\ $ 
& n/a \\ 
BABAR had $\pi^+$~\cite{ref:BABARpilnuBreco} & $1.14\pm 0.27\pm 0.17\ $ 
& $0.70\pm 0.22\pm 0.11$ \\ 
BABAR had $\pi^0$~\cite{ref:BABARpilnuBreco} & $1.60\pm 0.41 \pm 0.22\ $
& $0.46\pm 0.20\pm 0.04$ \\  \hline
Average of tagged & $1.34 \pm 0.11\pm 0.09\ $
& $0.38\pm 0.07\pm 0.04$ \\  \hline\hline
Average & $1.34 \pm 0.08\pm 0.08\ $ 
& $0.40\pm 0.04\pm 0.04$ \\ 
\hline\hline
\end{tabular}\\
\end{small}
\end{center}
\label{tab:pilnubf}
\end{table}

The determination of \vub\ from the $\Bb\to\pi\ell\nub$ decays is
shown in Table~\ref{tab:pilnuvub} and uses our average for branching
fraction given in Table~\ref{tab:pilnubf}.  Two theoretical approaches are
used: unquenched ($N_f=2+1$) Lattice QCD and 
QCD sum rules.
Lattice calculations of the FF are limited to small hadron momenta, i.e.
large $q^2$, while calculations based on light cone sum rules are restricted
to small $q^2$. More precise calculations of the FF, in particular their
normalization, are needed to reduce the overall uncertainties.


\begin{table}[hbtf]
\begin{center}
\caption{\label{tab:pilnuvub}
Determinations of \vub\ based on the average total and partial
$\Bb\to\pi\ell\nub$ decay branching fraction stated in
Table~\ref{tab:pilnubf}.The first
uncertainty is experimental, the second theoretical.  The
full or partial \cbf\ are used as indicated.
}
\vspace{5mm} {
\begin{tabular}{lc}
\hline\hline
Method & $\Vub [10^{-3}]$ \\
LCSR, full $q^2$~\cite{BallZwicky} & $3.36\pm 0.15 {}^{+0.66}_{-0.41}$ \\ 
LCSR, $q^2<16\,\gev^2/c^2$~\cite{BallZwicky}   & $3.25\pm 0.17 {}^{+0.54}_{-0.36}$ \\ \hline
HPQCD, full $q^2$~\cite{HPQCDpilnu}& $3.92\pm 0.17 {}^{+0.76}_{-0.48}$ \\ 
HPQCD, $q^2>16\,\gev^2/c^2$~\cite{HPQCDpilnu}  & $4.44\pm 0.30 {}^{+0.67}_{-0.46}$ \\ 
FNAL, full $q^2$~\cite{FNALpilnu}  & $3.74\pm 0.16 {}^{+0.86}_{-0.51}$ \\ 
FNAL, $q^2>16\,\gev^2/c^2$~\cite{FNALpilnu}    & $3.76\pm 0.25 {}^{+0.65}_{-0.43}$ \\ 
\hline\hline
\end{tabular}\\
}
\end{center}
\end{table}

Branching fractions for other $\Bb\to X_u\ell\nub$ decays are given in
Table~\ref{tab:xslother}.  At this time the determination of $\Vub$
from these other channels looks less promising than for
$\Bb\to\pi\ell\nub$.

\begin{table}[!htb]
\caption{Summary of branching fractions to $\cbf(\Bb\to X\ell\nub)$ decays
other than $\Bb\to\pi\ell\nub$.  The errors quoted
correspond to statistical and systematic
uncertainties, respectively.  Where a third uncertainty is quoted it
corresponds to uncertainties from form factor shapes.}
\begin{center}
\begin{small}
\begin{tabular}{llc}
\hline\hline
Experiment & Mode & $\cbf [10^{-4}]$  \\
CLEO~\cite{ref:CLEOpilnu} & $\Bz\to\rho^-\ell\nub$ & $2.17\pm 0.34\ {}^{+\ 0.47}_{-\ 0.54}\ \pm 0.41$\\ 
CLEO~\cite{ref:CLEOrholnu} & $\Bz\to\rho^-\ell\nub$ & $2.69\pm 0.41\ {}^{+\ 0.35}_{-\ 0.40}\ \pm 0.50$ \\ 
BABAR~\cite{ref:BABARpilnuBreco} & $\Bz\to\rho^-\ell\nub$ & $2.57\pm 0.52\pm 0.59\ \quad\ \,\,\quad  $ \\ 
BABAR~\cite{ref:BABARrholnu} & $\Bz\to\rho^-\ell\nub$ & $3.29\pm 0.42\pm 0.47\pm 0.60 $ \\ 
BABAR~\cite{ref:BABARpilnu} & $\Bz\to\rho^-\ell\nub$ & $2.14\pm 0.21\pm 0.51\pm 0.28 $ \\ 
BELLE~\cite{ref:BELLEpilnuSL} & $\Bz\to\rho^-\ell\nub$ & $2.07\pm 0.47\pm 0.25\pm 0.14$ \\ \hline
CLEO~\cite{ref:CLEOpilnu} & $\Bp\to\eta\ell\nub$ & $0.84\pm 0.31\pm 0.16 \pm 0.09$ \\ 
BELLE~\cite{ref:BELLEpilnuSL} & $\Bp\to\rho^0\ell\nub$ & $1.39\pm 0.23\pm 0.16\pm 0.02$ \\ 
BELLE~\cite{ref:BELLEomegalnu} & $\Bp\to\omega\ell\nub$ & $1.3\ \,\pm 0.4\ \,\pm 0.2\ \,\pm 0.3\ \,$ \\ 
\hline\hline
\end{tabular}\\
\end{small}
\end{center}
\label{tab:xslother}
\end{table}



\subsection{Inclusive Cabibbo-suppressed decays}
\label{slbdecays_b2uincl}

The large background from $\Bb\to X_c\ell\nub$ decays is the chief
experimental limitation in determinations of $\vub$.  Cuts designed to
reject this background limit the acceptance for $\Bb\to X_u\ell\nub$
decays.  The calculation of partial rates for these restricted
acceptances is more complicated and requires the resummation of an
infinite number of terms into non-perturbative ``shape functions'' at
each order in the $1/m_b$ expansion.  The leading shape function is
the same in semileptonic and radiative $\Bb\to X_s\gamma$ decays.
Subleading shape functions
differ between semileptonic and radiative decays, however.  
Theoretical uncertainties arise from the modeling of these subleading
shape functions and from higher order perturbative and
non-perturbative contributions, including weak annihilation~\cite{ref:WA}.  
The various extractions of \vub\, presented here are based on calculations by
Bosch, Lange, Neubert and Paz
(BLNP)~\cite{ref:BLNP,ref:Neubert-new-1,ref:Neubert-new-2,ref:Neubert-new-3,ref:Neubert-new-4}.  The dominant error remains the uncertainty of the $b$-quark mass,
even though recent HQE fits to moments have significantly reduced this
uncertainty.   The results of such fits are shown in
Fig.~\ref{fig:mupi2mb}. The
relative large uncertainty in $\mu_\pi^2$ has a much smaller impact on the shape
function error on \vub. 

\begin{figure}
\begin{center}
\includegraphics[width=6.0in]{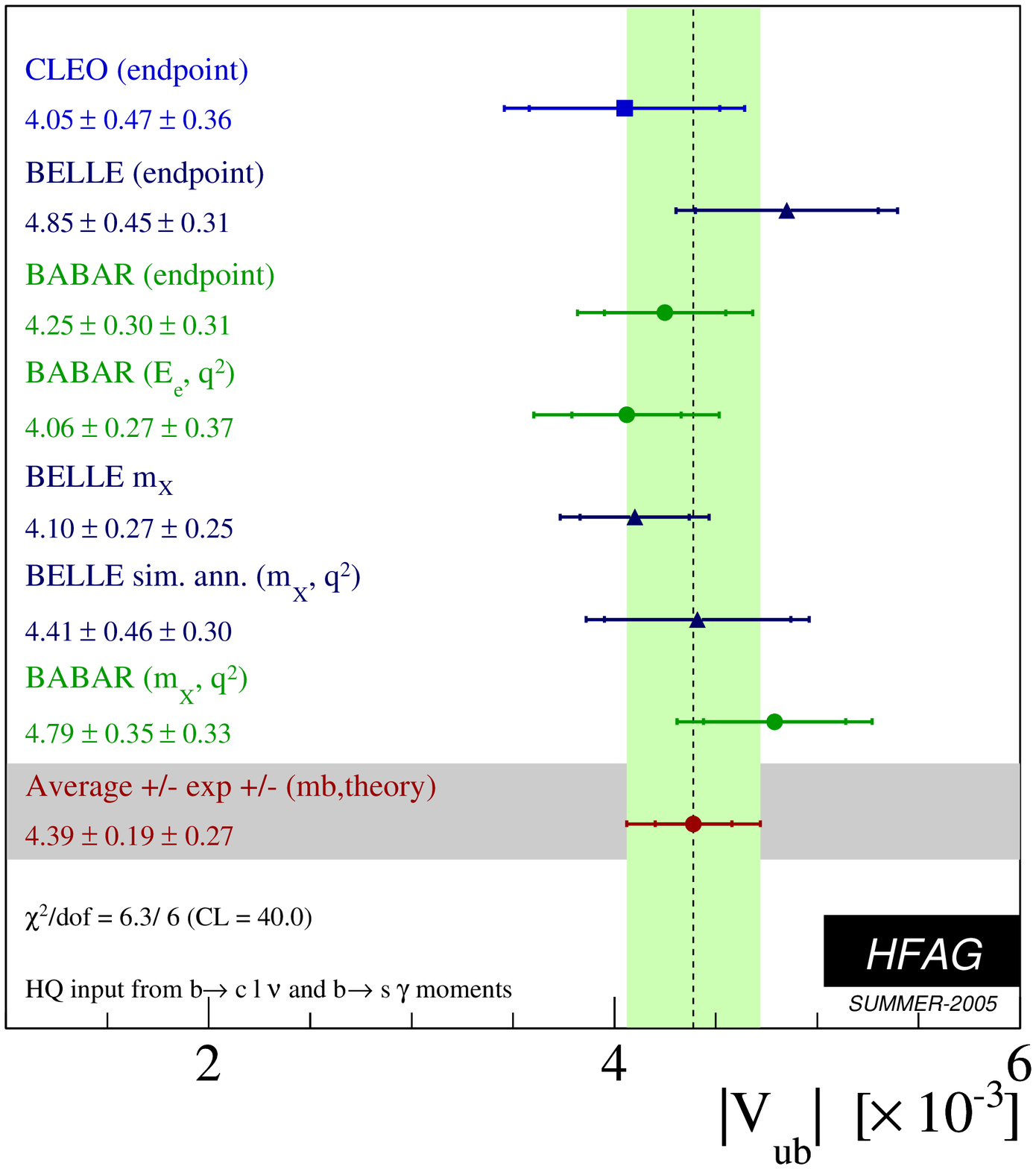}
\end{center}
\caption{Measurements of $\vub$ from inclusive semileptonic decays 
and their average.}
\label{fig:inclusivevub}
\end{figure}

Measurements of partial decay rates for $\Bb\to X_u\ell\nub$
transitions from $\Upsilon(4S)$ decays are given in
Table~\ref{tab:bulnu}, along with extracted values for $\vub$, which
are also shown in Fig.~\ref{fig:inclusivevub}.
Earlier measurements from
LEP~\cite{ref:ALEPHvub,ref:L3vub,ref:OPALvub,ref:DELPHIvub} are less
precise and cannot readily be used in a consistent framework with the
$\Upsilon(4S)$ results.  The recent measurements
tend to include a larger fraction $f_u$ of the total
phase space for $\Bb\to X_u\ell\nub$ decay than did earlier measurements.

The systematic errors associated with the modeling of $\Bb\to
X_c\ell\nub$ and $\Bb\to X_u\ell\nub$ decays and the theoretical
uncertainties are taken as fully correlated among all measurements
in the sense of Eq.~\ref{eq:correlrho}.  Reconstruction-related
uncertainties are taken as fully correlated within a given experiment.
From the three results quoted in Ref.~\cite{ref:belle-mx}, only one is
used in the average, as they are based on the same dataset and are
highly correlated.  The other experimental results have negligible
statistical correlation.  The assumed dependence of the quoted error
on the measured value was input for each source of error, as
discussed in section~\ref{sec:slmethodology}.  The average
is given in Table~\ref{tab:bulnu}.  The breakdown of the uncertainties
on the average (in percent) is $\pm 2.1$ (statistical), $\pm 2.4$ (experimental), $\pm
1.9$ ($b\to c\ell\nub$ model), $\pm 2.4$ ($b\to u\ell\nub$ model),
$\pm 4.7$ ($m_b$ and $\mu_{\pi}^2$), $\pm 3.5$ (subleading shape functions), $\pm 1.9$
(weak annihilation). No uncertainty is assigned to the assumption of
quark-hadron duality. The average \vub\, corresponds to an
inclusive charmless semileptonic $B$ decay average branching fraction
$\mathcal{B}(B\rightarrow X_u l \nu_l)=(2.16 \pm 0.33) \times
10^{-3}$. 

\begin{table}[!htb]
\caption{Summary of inclusive determinations of partial branching
  fractions for $B\rightarrow X_u l \nu$ decays and $\vub$.
  The errors quoted on \vub\ correspond to
experimental and theoretical uncertainties, respectively.  The
$s_\mathrm{h}^{\mathrm{max}}$ variable is described in Ref.~\cite{ref:shmax} }
\begin{center}
\begin{small}
\begin{tabular}{llccc}
\hline\hline
& accepted region & $f_u$
& $\Delta\cbf [10^{-4}]$ & $\Vub [10^{-3}]$\\
\hline
$*$CLEO~\cite{ref:cleo-endpoint}
& $E_e>2.1\,\gev$ & 0.19
& $3.3\pm 0.2\pm 0.7$           & $4.05\pm 0.47\pm 0.36$ \\ 
$*$BABAR~\cite{ref:babar-elq2}
& $E_e>2.0\,\gev$, $s_\mathrm{h}^{\mathrm{max}}<3.5\,\mathrm{GeV^2}$ & 0.19
& $3.5\pm 0.3\pm 0.3$           & $4.06\pm 0.27\pm 0.36$ \\
$*$BABAR~\cite{ref:babar-endpoint}
& $E_e>2.0\,\gev$  & 0.26
& $5.3\pm 0.3\pm 0.5$           & $4.25\pm 0.30\pm 0.31$ \\ 
$*$BELLE~\cite{ref:belle-endpoint}
& $E_e>1.9\,\gev$  & 0.34
& $8.5\pm 0.4\pm 1.5$           & $4.85\pm 0.45\pm 0.31$ \\ 
$*$BABAR~\cite{ref:babar-q2mx}
& $M_X<1.7\,\gev/c^2, q^2>8\,\gev^2/c^2$ & 0.34
& $8.7\pm 0.9\pm 0.9$           & $4.79 \pm 0.35 \pm 0.33$ \\ 
$*$BELLE~\cite{ref:belle-mxq2Anneal}
& $M_X<1.7\,\gev/c^2, q^2>8\,\gev^2/c^2$ & 0.34
& $7.4\pm 0.9\pm 1.3$           & $4.41 \pm 0.46 \pm 0.30$ \\ 
BELLE~\cite{ref:belle-mx}
& $M_X<1.7\,\gev/c^2, q^2>8\,\gev^2/c^2$ & 0.34
& $8.4\pm 0.8\pm 1.0$           & $4.68 \pm 0.37 \pm 0.32$ \\
BELLE~\cite{ref:belle-mx}
& $P_+<0.66\,\gev$  & 0.57
& $11.0\pm 1.0\pm 1.6$          & $4.14 \pm 0.35 \pm 0.29 $ \\
$*$BELLE~\cite{ref:belle-mx}
& $M_X<1.7\,\gev/c^2$   & 0.66
& $12.4\pm 1.1\pm 1.2$          & $4.10 \pm 0.27 \pm 0.25$ \\ \hline\hline
Average of $*$ 
& $\chi^2=6.3/6$, CL=0.40 &
&                    & $4.39 \pm 0.19 \pm 0.27$ \\

\hline\hline
\end{tabular}\\
\end{small}
\end{center}
\label{tab:bulnu}
\end{table}

\clearpage 
\mysection{Charmless \B-decay branching fractions and
               their asymmetries }
\label{sec:rare}

The aim of this section is to provide the branching fractions and
the partial rate asymmetries ($A_{CP}$) of charmless 
$B$ decays. The asymmetry is defined as 
$A_{CP} = \frac{N_{\Bbar} -N_B}{N_{\Bbar} +N_B}$, where $N_{\Bbar}$ 
and $N_B$ are respectively the number of $\Bzb/\Bm$ and $\Bz/\Bp$ decaying
into a specific final state. 
Four different $B$ decay categories are considered: 
charmless mesonic, baryonic, radiative and leptonic. Measurements supported 
with  written documents are accepted in  
the averages; written documents could be journal papers, 
conference contributed papers, preprints or conference proceedings.  
Results from  $A_{CP}$ measurements  obtained from time dependent analyses 
are listed and described in Sec.~\ref{sec:cp_uta}. Measurements of  charmful 
baryonic $B$ decays, which were included in our previous averages
\cite{hfag_hepex_2004,hfag_hepex_winter2005}, are now shown    
in Section 7, which deals with $B$ decays to charm.  

So far all branching fractions assume equal production of charged and
neutral $B$ pairs.  The best measurements to date show that this is
still a good approximation (see Sec.~\ref{sec:bfraction}).
For branching fractions, we provide either averages or the most stringent
90\% confidence level upper limits.  If one or more experiments have
measurements with $>$4$\sigma$ significance for a decay channel, all
available central values 
for that channel are used in the averaging.  We also give central values
and errors for cases where the significance of the average value is at
least $3 \sigma$, even if no single measurement is above $4 \sigma$. 
Since some decay modes are sensitive to the contribution of
 new physics and the current experimental upper limits are not far from the 
Standard Model expectation, it's better to provide the combined upper limits or
averages rather than to list the most stringent upper limits. For instance, 
$B^+\to \tau^+ \nu$ is one of these decays. In our update of Summer 2005,
the combined averages are given for the decays $B^+\to \tau^+ \nu$ and 
$B^0 \to K^+K^-$ although no significant signals are observed. 
Their upper limits can be estimated assuming that the errors are 
Gaussian.  For $A_{CP}$ we provide averages in all cases.  

Our averaging is performed by maximizing the likelihood,
   $\displaystyle {\mathcal L} = \prod_i {\mathcal P}_i(x),$  
where ${\mathcal P_i}$ is the probability density function (PDF) of the
$i$th  measurement, and $x$ is the branching fraction or $A_{CP}$.
The PDF is modeled by an asymmetric Gaussian function with the measured
central value as its mean and the quadratic sum of the statistical
and systematic errors as the standard deviations. The experimental
uncertainties are considered to be uncorrelated with each other when the 
averaging is performed. No error scaling is applied when the fit $\chi^2$ is 
greater than 1 since we believe that tends to overestimate the errors
except in cases of extreme disagreement (we have no such cases).
One exception to consider the correlated systematic errors is the inclusive
$b\to s\gamma$ mode, which is sensitive to physics beyond the Standard Model.
We tried to include as many measurements as possible and take the common
systematic errors into account when performing the average. The details are 
described in section~\ref{sec:btosg}.

At present, we have measurements of more than 250  decay modes, reported in
more than 150 papers. Because the number of references is so large, we do
not include them with the tables shown here but the full set of
references is available quickly from active gifs at the 
``Summer 2005'' link on 
the rare web page: {\tt http://www.slac.stanford.edu/xorg/hfag/rare/index.html}

\mysubsection{Mesonic charmless decays}

\begin{table}[!htbp]
\vspace{-1.5cm}
\begin{center}
\caption{
Branching fractions (BF) of charmless mesonic $B^+$ decays
(in units of $10^{-6}$). Upper limits are at 90\% CL.
Values in {\red red} ({\blue blue}) are new {\red published}
({\blue preliminary}) result since PDG2004  [as of July 15, 2005].
}
\scriptsize
\begin{tabular}{|lccccccc|} 
\sgline
RPP\#   & Mode & PDG2004 Avg. & BABAR & Belle & CLEO & CDF & New Avg. \\
\sglinespb
117                                               & 
$K^0\pi^+$                                        & 
$18.8\pm 2.1$                                     & 
\red{$\err{26.0}{1.3}{1.0}$}                      & 
\red{$\err{22.0}{1.9}{1.1}$}                      & 
$\aerrsy{18.8}{3.7}{3.3}{2.1}{1.8}$               & 
\nodata                                           & 
$24.1 \pm 1.3$                                    \\

118                                               & 
$K^+\pi^0$                                        & 
$12.9\pm 1.2$                                     & 
\red{$\err{12.0}{0.7}{0.6}$}                      & 
\red{$\berr{12.0}{1.3}{1.3}{0.9}$}                & 
$\aerrsy{12.9}{2.4}{2.2}{1.2}{1.1}$               & 
\nodata                                           & 
$12.1 \pm 0.8$                                    \\

119                                               & 
$\etapr K^+$                                      & 
$78\pm 5$                                         & 
\red{$\err{68.9}{2.0}{3.2}$}                      & 
\blue{$\err{68.6}{2.1}{3.6}$}                     & 
$\aerr{80}{10}{9}{7}$                             & 
\nodata                                           & 
$69.4 \pm 2.7$                                    \\
\end{tabular}
\end{center}

\vspace{-0.4cm} 
$\dag$Product BF - daughter BF taken to be 100\%; 
~$\ddag$Larger of two solutions taken; 
~\S $M_{\phi\phi}<2.85$ GeV/$c^2$
\end{table}
\clearpage


\begin{table}
\begin{center}
\caption{
Branching fractions of charmless mesonic $B^0$ decays 
(in units of $10^{-6}$).
Upper limits are at 90\% CL.
Values in {\red red} ({\blue blue}) are new {\red published}
({\blue preliminary}) result since PDG2004  [as of July 15, 2005].
}
\scriptsize
\begin{tabular}{|lccccccc|} 
\sgline
RPP\#   & Mode & PDG2004 Avg. & BABAR & Belle & CLEO & CDF & New Avg. \\
\hline
123                                               & 
$K^+\pi^-$                                        & 
$18.5\pm 1.1$                                     & 
\blue{$\err{19.2}{0.6}{0.6}$}                     & 
\red{$\err{18.5}{1.0}{0.7}$}                      & 
$\aerrsy{18.0}{2.3}{2.1}{1.2}{0.9}$               & 
\nodata                                           & 
$18.9 \pm 0.7$                                    \\

124                                               & 
$K^0\pi^0$                                        & 
$\cerr{9.5}{2.1}{1.9}$                            & 
\blue{$\err{11.4}{0.9}{0.6}$}                     & 
\red{$\berr{11.7}{2.3}{1.2}{1.3}$}                & 
$\aerrsy{12.8}{4.0}{3.3}{1.7}{1.4}$               & 
\nodata                                           & 
$11.5 \pm 1.0$                                    \\

125                                               & 
$\etapr K^0$                                      & 
$63\pm 7$                                         & 
\red{$\err{67.4}{3.3}{3.2}$}                      & 
\blue{$\err{56.6}{3.6}{3.3}$}                     & 
$\aerr{89}{18}{16}{9}$                            & 
\nodata                                           & 
$63.2 \pm 3.3$                                    \\
\end{tabular}
\end{center}
\vspace{-0.4cm}
 $\dag$Product BF - daughter BF taken to be 100\%, 
 $\ddag$Relative BF converted to absolute BF
\end{table}

\clearpage

\begin{table}
\begin{center}
\caption{
 Relative branching fractions of $B^0\to K^+K^-, K^+\pi^-, \pi^+\pi^-$.
Values in {\red red} ({\blue blue}) are new {\red published}
({\blue preliminary}) result since PDG2004  [as of July 15, 2005].
}
\small
\begin{tabular}{|lccccc|} \hline
RPP\# &Mode & PDG2004 Avg.  &  CDF & D0
 & New Avg.  \\ \sglinespb
$132$                                             & 
$\mathcal{B}(B^0\rightarrow K^+K^-)/\mathcal{B}(B^0\rightarrow K^+\pi^-)$& 
\nodata                                           & 
\blue{$<0.17$}                                    & 
\nodata                                           & 
{$<0.17$}                                         \\

$176$                                             & 
$\mathcal{B}(B^0\rightarrow\pi^+\pi^-)/\mathcal{B}(B^0\rightarrow K^+\pi^-)$& 
\nodata                                           & 
\blue{$0.24\pm0.06\pm0.05$}                       & 
\nodata                                           & 
$0.24 \pm 0.08$                                   \\

\hline
\end{tabular}
\end{center}
\end{table}

\mysubsection{Radiative and leptonic decays}
\begin{table}[!hb]
\begin{center}
\caption{
Branchign fractions of semileptonic and radiative $B^+$ decays 
(in units of $10^{-6}$). Upper limits are at 90\% CL.
Values in {\red red} ({\blue blue}) are new {\red published}
({\blue preliminary}) result since PDG2004  [as of July 15, 2005].
}
\end{center}
\vspace{-0.3cm}
\scriptsize
\begin{tabular}{|lcccccc|} \hline
RPP\# &Mode & PDG2004 Avg. & BABAR  & Belle & CLEO
 & New Avg.  \\
 \sglinespb


160                                               & 
$K^*(892)^+\gamma$                                & 
$38\pm 5$                                         & 
{\red $38.7\pm 2.8 \pm 2.6$}                      & 
{\red $42.5\pm 3.1\pm 2.4$}                       & 
$\aerr{37.6}{8.9}{8.3}{2.8} $                     & 
$40.3 \pm 2.6$                                    \\

161                                               & 
$K_1(1270)^+\gamma$                               & 
$<99$                                             & 
\nodata                                           & 
{\red $\err{43}{9}{9}$}                           & 
\nodata                                           & 
$43 \pm 12$                                       \\

162                                               & 
$K^+\phi\gamma$                                   & 
$3.4\pm 1.0$                                      & 
\nodata                                           & 
$\err{3.4}{0.9}{0.4}$                             & 
\nodata                                           & 
$3.4 \pm 1.0$                                     \\
\hline
\end{tabular}
~\S~$M_{K\pi\pi} < 2.4$ GeV/$c^2$;
~\ddag~Central values are not significant.
\end{table}

\begin{table}[!htbp]
\begin{center}
\caption{
Branching fractions of semileptonic and radiative $B^0$ decays
(in units of $10^{-6}$).  Upper limits are at 90\% CL.
Values in {\red red} ({\blue blue}) are new {\red published}
({\blue preliminary}) result since PDG2004  [as of July 15, 2005].
}
\end{center}
\hspace*{-0.0cm}
\scriptsize
\begin{tabular}{|lcccccc|} \hline
RPP\# &Mode & PDG2004 Avg. & BABAR  & Belle & CLEO
 & New Avg.  \\ \sglinespb
162                                               & 
$K^*(892)^0\gamma$                                & 
$43\pm 4$                                         & 
{\red $39.2\pm 2.0\pm 2.4$}                       & 
{\red $40.1\pm 2.1\pm 1.7$}                       & 
$\aerr{45.5}{7.2}{6.8}{3.4}$                      & 
$40.1 \pm 2.0$                                    \\

163                                               & 
$K^0\phi\gamma$                                   & 
$<8.3$                                            & 
\nodata                                           & 
$<8.3$                                            & 
\nodata                                           & 
$<8.3$                                            \\

164                                               & 
$K^+\pi^-\gamma$~\dag                             & 
$4.6\pm 1.4$                                      & 
\nodata                                           & 
$\aerrsy{4.6}{1.3}{1.2}{0.5}{0.7}$                & 
\nodata                                           & 
$4.6 \pm 1.4$                                     \\

$~-$                                              & 
$K^0\pi^+\pi^-\gamma$                             & 
New                                               & 
{\blue $\err{18.5}{2.1}{1.2}$}                    & 
{\red $\err{24}{4}{3}$}                           & 
\nodata                                           & 
$19.5 \pm 2.2$                                    \\
\hline
\end{tabular}
\hspace{+0.6cm}
~\dag~$1.25$ GeV/$c^2 < M_{K\pi} < 1.6$ GeV/$c^2$;
~\ddag~ Central values are not significant.
\end{table}

\begin{table}
\begin{center}
\caption{
Branching fractions of  semileptonic and radiative $B$ decays
(in units of $10^{-6}$).  Upper limits are at 90\% CL.
Values in {\red red} ({\blue blue}) are new {\red published}
({\blue preliminary}) result since PDG2004  [as of July 15, 2005].
}
\end{center}
\hspace*{-0.0cm}
\scriptsize
\begin{tabular}{|lcccccc|} \hline
RPP\# &Mode & PDG2004 Avg. & BABAR  & Belle & CLEO
 & New Avg.  \\ \sglinespb
~60                                               & 
$K_3^*(1780)\gamma $                              & 
$<3000$                                           & 
\nodata                                           & 
{\red $<2.8$}                                    & 
\nodata                                           & 
{ $<2.8$}                                         \\


$~-$                                              & 
$s\gamma$ with baryons                            & 
New                                               & 
\nodata                                           & 
\nodata                                           & 
\red $<38$~\dag                                   & 
 $<38$~\dag                                       \\
\hline
\end{tabular}

 ~\dag $E_\gamma > 2.0$ GeV; ~\ddag $M(\ell^+\ell^-)>0.2$ GeV/$c^2$
\end{table}
\clearpage

\begin{table}[!htbp]
\begin{center}
\caption{
 Branching fractions of leptonic  $B$ decays
(in units of $10^{-6}$). Upper limits are at 90\% CL.
Values in {\red red} ({\blue blue}) are new {\red published}
({\blue preliminary}) result since PDG2004  [as of July 15, 2005].
}
\vspace{0.3cm}
\begin{tabular}{|lcccccccc|} \hline
RPP\# &Mode & PDG2004 Avg. & BABAR  & Belle & CLEO &CDF & D0
 & New Avg.  \\ \sglinespb
~12                                               & 
$e^+\nu$                                          & 
$<15$                                             & 
\nodata                                           & 
{\blue $<5.4$}                                    & 
$<15$                                             & 
\nodata                                           & 
\nodata                                           & 
{ $<5.4$}                                         \\

~13                                               & 
$\mu^+\nu$                                        & 
$<21$                                             & 
{\red $<6.6$}                                     & 
{\blue $<2.0$}                                    & 
$<21$                                             & 
\nodata                                           & 
\nodata                                           & 
{ $<2.0$}                                         \\

~14                                               & 
$\tau^+\nu$                                       & 
$<570$                                            & 
{\blue $\cerr{130}{100}{90}$}                     & 
{\blue $\cerr{81}{58}{45}$}                       & 
$<840$                                            & 
\nodata                                           & 
\nodata                                           & 
$\cerr{92}{51}{41}$                               \\
\hline
\end{tabular}
\end{center}
\end{table}

\mysubsection{$B\to s\gamma$}
\label{sec:btosg}

The decay $b \to s\gamma$ proceeds through a process of
flavor changing neutral current. Since the charged Higgs or SUSY particles may
contribute in the penguin loop, the branching fraction is sensitive to physics
beyond the Standard Model. Experimentally, the branching fraction is measured
using either a semi-inclusive or an inclusive approach. A minimum
photon energy requirement is applied in the analysis and the branching fraction
is corrected based on the theoretical model for the photon energy spectrum
(shape function). Although there are several experimental results
available, only one measurement each for \babar, Belle and CLEO
is used in the HFAG average \cite{hfag_hepex_2004, hfag_hepex_winter2005} to avoid dealing with correlated
 errors for results reported from the same experiment. Furthermore, the
 model uncertainties from the shape function should be highly
correlated but no proper action was made in our previous averages.
To perform the average with better precision and good accuracy, it is
important to use as many experimental
results as possible and to handle the shape function issue in a proper
way. In this note, we report the updated average of $b\to s\gamma$ branching
fraction by implementing a common  shape function.

Several shape function schemes are commonly used.
Usually one is chosen to obtain the extrapolation factor,
defined as the ratio of the $b\to s\gamma$ branching fractions
with minimum photon energies above and at 1.6 GeV,
and the difference between various schemes are treated as the
model uncertainty. Recently O. Buchm\"uller and H. Fl\"acher have calculated the
extrapolation factors \cite{ref:bf}.
Table \ref{tab:factor} lists the extrapolation factors with various photon
energy cuts for three different schemes and the average. The appropriate
approach to average the experimental results is to first convert them
according to the average extrapolation factors and then perform the average,
assuming that the errors of the extrapolation factors are 100\% correlated.

\begin{table}[h]
\caption{Extrapolation factor in various scheme with various minimum
   photon energy requirement (in GeV).}
{\small
\begin{tabular}{|lccccc|} \hline
Scheme & $E_\gamma > 1.7$ & $E_\gamma > 1.8$ & $E_\gamma > 1.9$ &
$E_\gamma > 2.0$ & $E_\gamma > 2.242$ \\ \hline
Kinetic \cite{kine} & $0.986\pm 0.001$ & $0.968\pm 0.002$ & $0.939\pm 0.005$ & $0.903\pm0.009$ & $0.656\pm 0.031$ \\
Neubert SF \cite{neusf}& $0.982\pm 0.002$ & $0.962\pm 0.004$ & $0.930\pm 0.008$ & $0.888\pm 0.014$ & $0.665\pm 0.035$ \\
Kagan-Neubert \cite{kagan} & $0.988\pm 0.002$ & $0.970\pm 0.005$ & $0.940\pm 0.009$ & $0.892\pm 0.014$ & $0.643\pm 0.033$\\ \hline
Average & $0.985\pm 0.004$ & $0.967\pm 0.006$ & $0.936\pm 0.010$ & $0.894\pm 0.016$ &  $0.655\pm 0.037$ \\ \hline
\end{tabular}
}
\label{tab:factor}
\end{table}

After surveying all available experimental results, we choose the most updated
ones from each experiment for the average. Since the $b\to s\gamma$
branching fraction from $\Upsilon(4S)$ and $Z$ pole decays are not equal
footing, we drop the ALEPH measurement in this average \cite{aleph}.
Finally the five shown in Table \ref{tab:measurement} are selected.  They
have provided in their papers either the $b\to s\gamma$ branching fraction at
a certain photon energy cut or the extrapolation factor used.
Therefore we are able to convert them to the values at $E_{\rm min}= 1.6$
GeV using the information in Table \ref{tab:factor}.  The errors are,
in order, statistical, systematic and shape-function systematic,
except for the \babar\ inclusive where there is a second systematic
error (third quoted error) due to theoretical uncertainties.
Moreover, in the three inclusive analyses a possible $b\to d\gamma$
contamination has been considered according to the theoretical expectation
of $(4.0\pm 1.6)$\%. The uncertainty from the $b\to d\gamma$
fraction in the three inclusive measurements should not be considered
independently.  For those three measurements, a fourth uncertainty for
the $b\to d\gamma$ fraction is included.
We perform the average assuming that the systematic errors of the shape
function and the $d\gamma$ fraction are correlated, and the other systematic
errors and the statistical errors are Gaussian and uncorrelated.
The obtained average is
${\cal B}(b\to s\gamma) = (355\pm 24^{+9}_{-10}\pm3)\times 10^{-6}$ with
a $\chi^2$/DOF$= 0.74/4$, where the
errors are combined statistical and systematic, systematic due to the shape
function, and the $d\gamma$ fraction. The last two errors are estimated to
be the difference of the average after simultaneously varying the central
value of each experimental result by $\pm 1\sigma$. Although  a small fraction
of events was used in both the semi-inclusive and inclusive analyses in the
same experiment, we neglect their statistical correlations. Some
other correlated systematic errors, such as photon detection and the background
suppression, are not considered in our new average.
In the future it would be better if each collaboration would provide a single
combined result so that the average can be performed more accurately and easily.

\begin{table}[h]
\caption{Reported branching fraction, minimum photon energy, branching fraction
at minimum photon energy  and converted branching fraction for the
decay $b\to s\gamma$. All the branching fractions are in units of $10^{-6}$.
See text for an explanation of the errors.
}
\begin{tabular}{|lcccc|} \hline
Mode & Reported $\cal{B}$ & $E_{\rm min}$ & $\cal{B}$ at $E_{\rm min}$ & Modified ${\cal{B}}~(E_{\rm min}=1.6)$ \\ \hline
CLEO Inc. \cite{cleo}& $321 \pm 43\pm 27^{+18}_{-10}$ & 2.0 & $306\pm 41\pm 26$ & $329\pm 44\pm 28 \pm 6\pm 6$ \\
Belle Semi.\cite{belle1} & $336\pm 53 \pm 42^{+50}_{-54}$ & 2.24 & $-$ &
$369\pm 58 \pm 46^{+56}_{-60}$\\
Belle Inc.\cite{belle2} & $355\pm 32^{+30+11}_{-31-7}$ & 1.8 &$351\pm32\pm29$&
$350\pm 32^{+30}_{-31} \pm 2\pm 2$ \\
\babar\ Semi.\cite{babar1} & $335\pm 19^{+56+4}_{-41-9}$ & 1.9 &
$327\pm 18^{+55+4}_{-43-9}$ & $349\pm 20^{+59+4}_{-46-3}$ \\
\babar\ Inc.\cite{babar2} & $-$ & 1.9 & $367\pm 29\pm 34\pm 29$ &
$392\pm 31\pm 36\pm 30 \pm 4\pm 6$ \\\hline
\end{tabular}
\label{tab:measurement}
\end{table}

\mysubsection{Baryonic decays}

\begin{table}[!htbp]
\begin{center}
\caption{
 Branching fractions of  baryonic $B^+$ decays (in units of $10^{-6}$).
Upper limits are at 90\% CL.
values in {\red red} ({\blue blue}) are new {\red published}
({\blue preliminary}) result since PDG2004  [as of July 15, 2005].
}
\end{center}
\begin{center}
\footnotesize
\begin{tabular}{|lcccccc|}
\sgline
RPP\#   & Mode & PDG2004 Avg. & BABAR & Belle & CLEO & New Avg. \\
\sgline
201                                               & 
$p\overline p \pi^+$                              & 
$<3.7$                                            & 
\nodata                                           & 
{\red $\aerr{3.06}{0.73}{0.62}{0.37} \ddag$ }     & 
$<160$                                            & 
$\cerr{3.06}{0.82}{0.72}$                         \\

204                                               & 
$p\overline p K^+$                                & 
$\cerr{4.3}{1.2}{1.0}$                            & 
{\blue$\err{6.7}{0.5}{0.4} \dag$}                 & 
{\red $\aerr{5.30}{0.45}{0.39}{0.58} \ddag$}      & 
\nodata                                           & 
$6.10 \pm 0.48$                                   \\

$~$                                               & 
$\Theta^{++} \overline p$ $^*$                    & 
New                                               & 
{\blue$<0.09$}                                    & 
{\red $<0.091$}                                   & 
\nodata                                           & 
{$<0.09$}                                         \\

$~$                                               & 
${\mathcal G} K^+$ $^*$                           & 
New                                               & 
\nodata                                           & 
{\red $<0.41$}                                    & 
\nodata                                           & 
{$<0.41$}                                         \\

$~-$                                              & 
$p\overline p K^{*+}$                             & 
New                                               & 
\nodata                                           & 
{\red$\aerrsy{10.31}{3.62}{2.77}{1.34}{1.65} \ddag$}& 
\nodata                                           & 
$\cerr{10.31}{3.86}{3.22}$                        \\

206                                               & 
$p \overline\Lambda$                              & 
$< 1.5$                                           & 
\nodata                                           & 
{\red $< 0.49$}                                   & 
$< 1.5$                                           & 
{ $< 0.49$}                                       \\

$~-$                                              & 
$p \overline\Lambda(1520)$                        & 
New                                               & 
{\blue $< 1.5$}                                   & 
\nodata                                           & 
\nodata                                           & 
{ $< 1.5$}                                        \\

$-$                                               & 
$\Lambda \overline{\Lambda} K^+$                  & 
New                                               & 
\nodata                                           & 
{\red $\aerr{2.91}{0.90}{0.70}{0.38}\ddag$}       & 
\nodata                                           & 
$\cerr{2.91}{0.98}{0.80}$                         \\

$-$                                               & 
$\Lambda \overline{\Lambda} \pi^+$                & 
New                                               & 
\nodata                                           & 
{\red $<2.8 \ddag$ }                              & 
\nodata                                           & 
{ $<2.8 \ddag$ }                                  \\


\hline
\end{tabular}
\end{center}

\hspace{1.cm}$\dag$ Charmonium decays to $p\bar p$ have been
statistically subtracted.

\hspace{1.cm}$\ddag$ The charmonium mass region has been vetoed.

\hspace{1.cm}$^*$ Product BF - daughter BF taken to be 100\%:

\hspace{1.cm}$~~~~ \Theta(1540)^{++}\to K^+p$ (pentaquark candidate);

\hspace{1.cm}$~~~~ {\mathcal G}(2220)\to p\overline p$ (glueball candidate).\\
\end{table}

\begin{table}[!htbp]
\begin{center}
\caption{
Branching fractions of  baryonic $B^0$ decays
(in units of $10^{-6}$). Upper limits are at 90\% CL.
values in {\red red} ({\blue blue}) are new {\red published}
({\blue preliminary}) result since PDG2004  [as of July 15, 2005].
}
\vskip 0.25cm
\end{center}
\begin{center}
\footnotesize
\begin{tabular}{|lcccccc|}
\sgline
RPP\#   & Mode & PDG2004 Avg. & BABAR & Belle & CLEO & New Avg. \\
\sgline
212                                               & 
$p \overline{p}$                                  & 
$< 1.2$                                           & 
{\red $<0.27$}                                    & 
{\red $<0.41$}                                    & 
$<1.4$                                            & 
{ $<0.27$}                                        \\

214                                               & 
$p \overline{p} K^0$                              & 
$<7.2$                                            & 
\nodata                                           & 
{\red  $\aerr{1.20}{0.32}{0.22}{0.14} \ddag$}     & 
\nodata                                           & 
$\cerr{1.20}{0.35}{0.26}$                         \\

$~$                                               & 
$\Theta^+ \bar{p}$ $\ddag$                            & 
New                                               & 
\nodata                                           & 
{\red  $<0.23$}                                   & 
\nodata                                           & 
{ $<0.23$}                                        \\

$~-$                                              & 
$p \overline{p} K^{*0}$                           & 
New                                               & 
\nodata                                           & 
{\red $<7.6 \ddag$}                               & 
\nodata                                           & 
{ $<7.6 \ddag$}                                   \\

215                                               & 
$p \overline\Lambda \pi^-$                        & 
$\cerr{4.0}{1.1}{1.0}$                            & 
\nodata                                           & 
\red {$\aerr{3.27}{0.62}{0.51}{0.39}$}            & 
$<13$                                             & 
$\cerr{3.27}{0.73}{0.64}$                         \\

216                                               & 
$p \overline\Lambda K^-$                          & 
$<0.82$                                           & 
\nodata                                           & 
$< 0.82$                                          & 
\nodata                                           & 
$< 0.82$                                          \\

217                                               & 
$p \overline\Sigma^0 \pi^-$                       & 
$<3.8$                                            & 
\nodata                                           & 
$< 3.8$                                           & 
\nodata                                           & 
$< 3.8$                                           \\

218                                               & 
$\Lambda \overline\Lambda$                        & 
$<1.0$                                            & 
\nodata                                           & 
{\red $<0.69$}                                    & 
$<1.2$                                            & 
{ $<0.69$}                                        \\


\hline
\end{tabular}
\end{center}
\hspace{1.cm}$\dag$ Product BF - daughter BF taken to be 100\%:
$\ddag$ The charmonium mass region has been

 \hspace{1.2cm} vetoed. $\Theta(1540)^+\to p K_S^0$ (pentaquark candidate).
\end{table}

\clearpage
\mysubsection{$B_s$ decays}

\begin{table}[!htbp]
\caption{
 $B_s$   
branching fractions (in units of $10^{-6}$). Upper limits are at 90\% CL.
Values in {\red red} ({\blue blue}) are new {\red published}
({\blue preliminary}) result since PDG2004  [as of July 15, 2005].
}
\vskip 0.25cm

\begin{center}
\begin{tabular}{|lccccc|} \hline
RPP\# &Mode & PDG2004 Avg. &  CDF & D0
 & New Avg.  \\ \sglinespb
$15$                                              & 
$\phi\phi$                                        & 
$<1183 $                                          & 
\blue $\aerr{14}{6}{5}{6}$ $\dag$                 & 
\nodata                                           & 
$\cerr{14}{8}{7}$                                 \\

$24$                                              & 
$\mu^+\mu^-$                                      & 
$<2.0 $                                           & 
\blue$<0.15$                                      & 
\blue$<0.30$                                      & 
$<0.16$                                           \\

$26$                                              & 
$e^{\pm}\mu^{\mp}$                                & 
$<6.1 $                                           & 
$<6.1 $                                           & 
\nodata                                           & 
$<6.1 $                                           \\

$27$                                              & 
$\mu^+\mu^-\phi$                                  & 
$<47 $                                            & 
$<47 $                                            & 
\blue$<3.2$ $\dag$                                & 
$<3.2$ $\dag$                                     \\

\hline
\end{tabular}
\end{center}
\hspace{1.cm} $\dag$Relative BF converted to absolute BF
\end{table}

\begin{table}[h]
\begin{center}
\caption{
 $B_s$ rare relative  
branching fractions. 
Values in {\red red} ({\blue blue}) are new {\red published}
({\blue preliminary}) result since PDG2004  [as of July 15, 2005].
}
\vskip 0.25cm

{\scriptsize
\begin{tabular}{|lccccc|} \hline
RPP\# &Mode & PDG2004  &  CDF & D0
 & New Avg.  \\ 
 & & Avg. & & & \\ \sglinespb
$~9$                                              & 
$\mathcal{B}(B^0_s\rightarrow\pi^+\pi^-)/\mathcal{B}(B^0_s\rightarrow K^+K^-)$& 
\nodata                                           & 
\blue$<0.10 $                                     & 
\nodata                                           & 
$<0.10 $                                          \\

$15$                                              & 
$\mathcal{B}(B^0_s\rightarrow\phi\phi)/\mathcal{B}(B^0_s\rightarrow J/\psi\phi)$& 
\nodata                                           & 
\blue$(10^{+5}_{-4}\pm1)\times 10^{-3} $          & 
\nodata                                           & 
$\cerr{10}{7}{6}$                                 \\

$16$                                              & 
$\it{f}_s\mathcal{B}(B^0_s\rightarrow K^+\pi^-)/\it{f}_d\mathcal{B}(B^0_d\rightarrow K^+\pi^-)$& 
\nodata                                           & 
\blue$<0.11 $                                     & 
\nodata                                           & 
$<0.11 $                                          \\

$17$                                              & 
$\it{f}_s\mathcal{B}(B^0_s\rightarrow K^+K^-)/\it{f}_d\mathcal{B}(B^0_d\rightarrow K^+\pi^-)$& 
\nodata                                           & 
\blue$0.50\pm0.08\pm0.07 $                        & 
\nodata                                           & 
$0.50 \pm 0.11$                                   \\
\hline
\end{tabular}
}
\end{center}
\end{table}

\mysubsection{Charge asymmetries}

\begin{sidewaystable}[!htbp]
\parbox{8.2in}{\caption{
 $CP$ asymmetries for charmless hadronic charged $B$ decays.
Values in {\red red} ({\blue blue}) are 
new {\red published}
({\blue preliminary}) result since PDG2004  [as of July 15, 2005].
}}

\vspace*{ 0.5cm}
\scriptsize
\begin{tabular}{|lccccccc|} 
\hline
RPP\#   & Mode & PDG2004 Avg. & BABAR & Belle & CLEO & CDF & New Avg. \\
\sglinespb

117                                               & 
$K^0\pi^+$                                        & 
$0.03\pm 0.08$                                    & 
{\blue$\err{-0.09}{0.05}{0.01}$}                  & 
{\red$\err{0.05}{0.05}{0.01}$}                    & 
$\err{0.18}{0.24}{0.02}$                          & 
\nodata                                           & 
$-0.02 \pm 0.04$                                  \\

118                                               & 
$K^+\pi^0$                                        & 
$-0.10\pm 0.08$                                   & 
{\red $\err{0.06}{0.06}{0.01}$}                   & 
{\blue $\err{0.04}{0.04}{0.02}$}                  & 
$\err{-0.29}{0.23}{0.02}$                         & 
\nodata                                           & 
$0.04 \pm 0.04$                                   \\
\end{tabular}
\end{sidewaystable}

\begin{sidewaystable}[!htbp]
\begin{center}
\parbox{7.5in}{\caption{
Charmless hadronic $CP$ asymmetries for $B^\pm/B^0$ admixtures.
Values in {\red red} ({\blue blue}) are new {\red published}
({\blue preliminary}) result since PDG2004  [as of July 15, 2005].
}}

\vspace*{0.5cm} 
\scriptsize
\begin{tabular}{|lccccccc|}
\hline
RPP\#    & Mode & PDG2004 Avg. & BABAR & Belle & CLEO & CDF & New Avg. \\
\sglinespb
~56                                               & 
$K^*\gamma$                                       & 
$-0.01\pm 0.07$                                   & 
{\red $-0.013\pm 0.036\pm 0.010$}                 & 
{\red $-0.015\pm 0.044\pm 0.012$}                 & 
$0.08\pm 0.13\pm 0.03$                            & 
\nodata                                           & 
$-0.010 \pm 0.028$                                \\


%

~67                                               & 
$s\gamma$                                         & 
$\err{-0.079}{0.108}{0.022}$                      & 
\red$\err{0.025}{0.050}{0.015}$                   & 
\red$\err{0.002}{0.050}{0.026}$                   & 
$\err{-0.079}{0.108}{0.022}$                      & 
\nodata                                           & 
$0.004 \pm 0.036$                                 \\


103                                               & 
$s\ell\ell$                                       & 
New                                               & 
\red$\err{-0.22}{0.26}{0.02}$                     & 
\nodata                                           & 
\nodata                                           & 
\nodata                                           & 
$-0.22 \pm 0.26$                                  \\
\hline
\end{tabular}
\end{center}

\vskip -1cm
\begin{center}
\parbox{8.5in}{\caption{
 $CP$ asymmetries for charmless hadronic neutral $B$ decays.
Values in {\red red} ({\blue blue}) are new {\red published}
({\blue preliminary}) result since PDG2004  [as of July 15, 2005].
}}

\vspace*{ 0.5cm}
\scriptsize
\begin{tabular}{|lccccccc|} 
\sgline
RPP\#   & Mode & PDG2004 Avg. & BABAR & Belle & CLEO & CDF & New Avg. \\
\sglinespb

123                                               & 
$K^+\pi^-$                                        & 
$-0.09\pm 0.04$                                   & 
{\red $\err{-0.133}{0.030}{0.009}$}               & 
{\blue $\err{-0.113}{0.022}{0.008}$}              & 
$\err{-0.04}{0.16}{0.02}$                         & 
{\blue $\err{-0.04}{0.08}{0.01}$}                 & 
$-0.115 \pm 0.018$                                \\

124                                               & 
$K^0\pi^0$ $\dag$                                 & 
\nodata                                           & 
\nodata                                           & 
\nodata                                           & 
\nodata                                           & 
\nodata                                           & 
\nodata                                           \\
\end{tabular}
\vskip 0.5cm\end{center}

\large
\hspace{-1.9cm}
\dag~Measurements of time-dependent $CP$ asymmetries are listed on 
the Unitarity Triangle home page. (http://www.slac.stanford.edu/xorg/hfag/triangle/index.html) 

\end{sidewaystable}

 \clearpage
\mysubsection{Polarization measurements}
%

%

\begin{table}[!htbp]
\caption{
 Longitudinal polarization fraction $f_L$ for $B^+$ decays.
Values in {\red red} ({\blue blue}) are new {\red published}
({\blue preliminary}) result since PDG2004  [as of July 15, 2005].
\vspace{0.3cm}
}
 
\begin{center}
\begin{tabular}{|lccccc|} 
\sgline
RPP\#   & Mode & PDG2004 Avg. & BABAR & Belle & New Avg. \\
\sglinespb

%

$~-$                                              & 
$K^{*0}\rho^+$                                    & 
New                                               & 
\blue{$\err{0.79}{0.08}{0.04}$}                   & 
\blue{$\berr{0.43}{0.11}{0.05}{0.02}$}            & 
$0.66 \pm 0.07$                                   \\

138                                               & 
$K^{*+}\rho^0$                                    & 
$\aerr{0.96}{0.04}{0.15}{0.04}$                   & 
$\aerr{0.96}{0.04}{0.15}{0.04}$                   & 
\nodata                                           & 
$\cerr{0.96}{0.06}{0.15}$                         \\

%

156                                               & 
$\phi K^{*+}$                                     & 
$0.46\pm0.12\pm0.03$                              & 
$\err{0.46}{0.12}{0.03}$                          & 
\red{$\err{0.52}{0.08}{0.03}$}                    & 
$0.50 \pm 0.07$                                   \\

182                                               & 
$\rho^+\rho^0$                                    & 
$0.96\pm0.06$                                     & 
$\aerr{0.97}{0.03}{0.07}{0.04}$                   & 
$\err{0.95}{0.11}{0.02}$                          & 
$\cerr{0.97}{0.05}{0.07}$                         \\

186                                               & 
$\omega\rho^+$                                    & 
New                                               & 
\red{$\aerr{0.88}{0.12}{0.15}{0.03}$}             & 
\nodata                                           & 
$\cerr{0.88}{0.12}{0.15}$                         \\

\sglinespt
\end{tabular}
\end{center}
\end{table}

\begin{table}
\caption{
Full angular analysis of $B^+\to\phi K^{*+}$.
Values in {\red red} ({\blue blue}) are new {\red published}
({\blue preliminary}) result since PDG2004  [as of July 15, 2005].
\vspace{0.3cm}
}

\begin{center}
\begin{tabular}{|lccccc|} 
\sgline
\RPP & Parameter & PDG2004 Avg. & BABAR & Belle & New Avg. \\
\sglinespb
\nodata                                           & 
$f_\perp$                                         & 
New                                               & 
\nodata                                           & 
\red{$\err{0.19}{0.08}{0.02}$}                    & 
$0.19 \pm 0.08$                                   \\

\nodata                                           & 
$\phi_\parallel$                                  & 
New                                               & 
\nodata                                           & 
\red{$\err{2.10}{0.28}{0.04}$}                    & 
$2.10 \pm 0.28$                                   \\

\nodata                                           & 
$\phi_\perp$                                      & 
New                                               & 
\nodata                                           & 
\red{$\err{2.31}{0.30}{0.07}$}                    & 
$2.31 \pm 0.31$                                   \\

\sglinespt
\end{tabular}
\end{center}
\hspace*{2.0cm}BR, $f_L$ and $A_{CP}$ are tabulated separately.
\end{table}

\large

\begin{table}[!htbp]
\caption{
Longitudinal polarization fraction $f_L$ for $B^0$ decays.
Values in {\red red} ({\blue blue}) are new {\red published}
({\blue preliminary}) result since PDG2004  [as of July 15, 2005].
\vspace{0.3cm}
}

\begin{center}
\begin{tabular}{|lccccc|} 
\sgline
RPP\#   & Mode & PDG2004 Avg. & BABAR & Belle & New Avg. \\
\hline




154                                               & 
$\phi K^{*0}$                                     & 
$0.57\pm0.11$                                     & 
\red{$\err{0.52}{0.05}{0.02}$}                    & 
\red{$\err{0.45}{0.05}{0.02}$}                    & 
$0.48 \pm 0.04$                                   \\

%

%

%

%

203                                               & 
$\rho^+\rho^-$                                    & 
New                                               & 
\red{$\berr{0.99}{0.03}{0.04}{0.03}$}             & 
\blue{$\aerr{0.941}{0.034}{0.040}{0.030}$}        & 
$\cerr{0.971}{0.031}{0.030}$                      \\

\sglinespt
\end{tabular}
\end{center}
\end{table}

\begin{table}[!htbp]
\caption{
Full angular analysis of $B^0 \to \phi K^{*0}$.
Values in {\red red} ({\blue blue}) are new {\red published}
({\blue preliminary}) result since PDG2004  [as of July 15, 2005].
\vspace{0.3cm}
}

\begin{center}
\begin{tabular}{|lccccc|} 
\sgline
\RPP & Parameter & PDG2004 Avg. & BABAR & Belle & New Avg. \\
\sglinespb
\nodata                                           & 
$f_\perp = \Lambda_{\perp\perp}$                  & 
New                                               & 
\red{$\err{0.22}{0.05}{0.02}$}                    & 
\red{$\aerr{0.31}{0.06}{0.05}{0.02}$}             & 
$0.26 \pm 0.04$                                   \\

\nodata                                           & 
$\phi_\parallel$                                  & 
New                                               & 
\red{$\aerr{2.34}{0.23}{0.20}{0.05}$}             & 
\red{$\aerr{2.40}{0.28}{0.24}{0.07}$}             & 
$\cerr{2.36}{0.18}{0.16}$                         \\

\nodata                                           & 
$\phi_\perp$                                      & 
New                                               & 
\red{$\err{2.47}{0.25}{0.05}$}                    & 
\red{$\err{2.51}{0.25}{0.06}$}                    & 
$2.49 \pm 0.18$                                   \\

\nodata                                           & 
$A_{CP}^0$                                        & 
New                                               & 
\red{$\err{-0.06}{0.10}{0.01}$}                   & 
\red{$\err{0.13}{0.12}{0.04}$}                    & 
$0.01 \pm 0.08$                                   \\

\nodata                                           & 
$A_{CP}^\perp$                                    & 
New                                               & 
\red{$\err{-0.10}{0.24}{0.05}$}                   & 
\red{$\err{-0.20}{0.18}{0.04}$}                   & 
$-0.16 \pm 0.15$                                  \\

\nodata                                           & 
$\Delta\phi_\parallel$                            & 
New                                               & 
\red{$\aerr{0.27}{0.20}{0.23}{0.05}$}             & 
\red{$\err{-0.32}{0.27}{0.07}$}                   & 
$0.03 \pm 0.18$                                   \\

\nodata                                           & 
$\Delta\phi_\perp$                                & 
New                                               & 
\red{$\err{0.36}{0.25}{0.05}$}                    & 
\red{$\err{-0.30}{0.25}{0.06}$}                   & 
$0.03 \pm 0.18$                                   \\

\hline                                            & 
$f_\parallel = \Lambda_{\parallel\,\parallel}$    & 
New                                               & 
\red{$\err{0.26}{0.05}{0.02}$}                    & 
\red{$\err{0.24}{0.06}{0.02}$}                    & 
$0.25 \pm 0.04$                                   \\

\nodata                                           & 
$\mathcal{A}_T^0 = -0.5\Lambda_{\perp 0}$         & 
New                                               & 
\red{$\err{0.11}{0.05}{0.01}$}                    & 
\red{$\err{-0.08}{0.08}{0.02}$}                   & 
$0.06 \pm 0.04$                                   \\

\nodata                                           & 
$\mathcal{A}_T^\parallel=-0.5\Lambda_{\perp\parallel}$& 
New                                               & 
\red{$\err{-0.02}{0.04}{0.01}$}                   & 
\red{$\err{-0.01}{0.05}{0.01}$}                   & 
$-0.02 \pm 0.03$                                  \\

\nodata                                           & 
$\Lambda_{\parallel 0}$                           & 
New                                               & 
\red{$\err{-0.50}{0.12}{0.03}$}                   & 
\red{$\err{-0.45}{0.11}{0.02}$}                   & 
$-0.47 \pm 0.08$                                  \\

\nodata                                           & 
$\Sigma_{00}$                                     & 
New                                               & 
\red{$\err{0.03}{0.05}{0.01}$}                    & 
\red{$\err{-0.06}{0.05}{0.01}$}                   & 
$-0.02 \pm 0.04$                                  \\

\nodata                                           & 
$\Sigma_{\parallel\parallel}$                     & 
New                                               & 
\red{$\err{-0.05}{0.06}{0.01}$}                   & 
\red{$\err{-0.01}{0.06}{0.01}$}                   & 
$-0.03 \pm 0.04$                                  \\

\nodata                                           & 
$\Sigma_{\perp\perp}$                             & 
New                                               & 
\red{$\aerr{0.02}{0.06}{0.05}{0.01}$}             & 
\red{$\err{0.06}{0.06}{0.01}$}                    & 
$0.04 \pm 0.04$                                   \\

\nodata                                           & 
$\Sigma_{\perp 0}$                                & 
New                                               & 
\red{$\err{-0.41}{0.14}{0.03}$}                   & 
\red{$\aerr{-0.41}{0.16}{0.14}{0.04}$}            & 
$\cerr{-0.41}{0.11}{0.10}$                        \\

\nodata                                           & 
$\Sigma_{\perp\parallel}$                         & 
New                                               & 
\red{$\aerr{-0.06}{0.09}{0.08}{0.02}$}            & 
\red{$\err{-0.06}{0.10}{0.01}$}                   & 
$\cerr{-0.06}{0.07}{0.06}$                        \\

\nodata                                           & 
$\Sigma_{\parallel 0}$                            & 
New                                               & 
\red{$\aerr{0.18}{0.11}{0.13}{0.03}$}             & 
\red{$\err{-0.11}{0.11}{0.02}$}                   & 
$0.01 \pm 0.09$                                   \\

\sglinespt
\end{tabular}
\end{center}
Results below the line have been derived from the primary results.
BR, $f_L$ and $A_{CP}$ are tabulated separately.
\end{table}

\clearpage

\newpage\normalsize
\mysection{$B$ Decays to open charm and charmonium final states }\label{sec:BtoCharm}

This section is the first contribution to the HFAG report from
the ``$B \to$ charm " group\footnote{The HFAG/Charm group was formed in the spring of 2005; it performs its work using an XML database backed web application.}.
The mandate of the group
is to compile measurements and perform averages of all available quantities related to 
$B$ decays to charmed particles, excluding CP related quantities. To date the 
group has analyzed a total of 330 measurements reported in 124 papers, principally branching fractions.
The group aims to organize and present the copious information on $B$ decays to charmed
particles obtained from a combined sample of more than one billion $B$ mesons
from the BABAR, Belle and CDF Collaborations. 

Branching fractions for rare $B$-meson decays or decay chains of a few $10^{-7}$ are now being measured with 
statistical uncertainties typically below $30\%$. 
Results for more common decays, with branching fractions around $10^{-4}$, are becoming precision measurements, 
with uncertainties typically at the $3\%$ level. 
Some decays have been observed for the first time, for example $\bar{B}^0\to D_s^-\pi^+$ 
or $B^-\to \psi(3770)K^-$, with a branching fraction of $(2.74^{+0.67}_{-0.74})\times 10^{-5}$ and $(4.4\pm1.1)\times 10^{-4}$, respectively.

Among the many results, we highlight the great improvements that have been attained towards a deeper understanding 
of recently discovered new states with either hidden or open charm content. 
The new average for the branching fraction of the decay chain $B^- \to X(3870) K^-$, where 
$ X(3870)\to J/\psi \pi^+\pi^-$ is $(1.16\pm0.19)\times 10^{-5}$; 
many more $X(3870)$ decay modes have been searched for and several measurements or upper limits are reported. 
With an inclusive approach, an upper limit on the branching fraction for $B^- \to X(3870) K^-$ has been derived 
of $3.2\times 10^{-4}$, $90\%$ C.L.. 
In addition, a new state at about 3940 MeV, $Y(3940)$, has been observed in $B$ decays and the branching fraction for 
$B\to Y(3940) K, Y(3940) \to \omega J/\psi$ has been measured to be $(7.1\pm3.4)\times10^{-5}$. 
Several $B$ decays to $D_{sJ}^{*-}(2317)$ and $D_{sJ}^-(2460)$ have been
observed for the first time and the branching fractions measured. 
The abundance of measurements with many different final states is of the
greatest importance for quantum number assignments, and already some of the proposed theoretical interpretations have been ruled out.

The measurements are classified according to the decaying particle : Charged B, Neutral B or Miscellaneous; the decay products 
and the type of quantity : branching fraction, product of branching fractions, ratio of branching fractions or other quantities. 
For the decay product classification the below precedence order is used to ensure that each measurement appears in only one category. 
\begin{itemize}\addtolength{\itemsep}{-0.4\baselineskip}
\item new particles
\item strange $D$ mesons
\item baryons
\item $J/\psi$
\item charmonium other than $J/\psi$
\item multiple $D$, $D^{*}$ or $D^{**}$ mesons
\item a single $D^{*}$ or $D^{**}$ meson
\item a single $D$ meson
\end{itemize}
  
Within each table the measurements are color coded according to the
publication status and age. Table~\ref{tab:hfc99999} provides a key to the
color scheme and categories used. When viewing the tables with most pdf
viewers every number, label and average provides hyperlinks to the corresponding 
reference and individual quantity web pages on the HFAG/Charm group website
\hfhref{}{http://hfag.phys.ntu.edu.tw}.
The links provided in the captions of the table lead to the corresponding compilation
pages.  Both the individual and compilation webpages provide a graphical view
of the results, in a variety of formats.

Tables \ref{tab:hfc01101} to \ref{tab:hfc03300} provide either limits at 90\%
confidence level or measurements with statistical and systematic uncertainties 
and in some cases a third error corresponding to correlated systematics. 
For details on the meanings of the uncertainties and access to the references 
click on the numbers to visit the corresponding web pages.  Where there are
multiple determinations of the same quantity by one experiment the table
footnotes act to distinguish the methods or datasets used; such cases are
visually highlighted in the table by presenting the measurements on the lines
beneath the quantity label.
Where both limits and measured values of a quantity are available the limits 
are presented in the tables but are not used in the determination of the
average. Where only limits are available the most stringent is presented in
the Average column of the tables.
Where available the PDG 2004 result is also presented.

\clearpage

\begin{\hftabletype}[\hftableposn]

\begin{center}
\caption{Key to the colors used to classify the results presented in tables \ref{tab:hfc01101} to \ref{tab:hfc03300}. When viewing these tables in a pdf viewer each number, label and average provides a hyperlink to the corresponding online version provided by the charm subgroup website \hfurl{}. Where an experiment has multiple determinations of a single quantity they are distinguished by the table footnotes.}
 \label{tab:hfc99999} 
 
\hfmetadata{  }

\begin{tabular}{ll}  
Class
&  Definition
\\\hline 

{ \hfwaitingtext{waiting} } \hftabletlcell 
 & { \hfdeftext{Results without a preprint available } } 
\\%

{ \hfpubhottext{pubhot} }  
 & { \hfdeftext{Results published in 2005} } 
\\%

{ \hfprehottext{prehot} }  
 & { \hfdeftext{Preprint released in 2005} } 
\\%

{ \hfpubtext{pub} }  
 & { \hfdeftext{Results published after or during the last PDG year} } 
\\%

{ \hfpretext{pre} }  
 & { \hfdeftext{Preprint released after or during the last PDG year} } 
\\%

{ \hfpuboldtext{pubold} }  
 & { \hfdeftext{Results published before the last PDG year} } 
\\%

{ \hfpreoldtext{preold} }  
 & { \hfdeftext{Preprint released before the last PDG year} } 
\\%

{ \hferrortext{error} }  
 & { \hfdeftext{Incomplete information to classify } } 
\\%

{ \hfsuperceededtext{superceeded} }  
 & { \hfdeftext{Results superceeded by more recent measurements from the same experiment } } 
\\%

{ \hfinactivetext{inactive} }  
 & { \hfdeftext{Results in the process of being entered into the database} } 
\\%

{ \hfnoquotext{noquo} }  
 & { \hfdeftext{Results without quotes} } 
\\\hline 

\end{tabular}

\end{center}


\begin{center}
\hfnewcaption{Branching fractions}{charged B}{\hfnewp}{in units of $10^{-4}$}{upper limits are at 90\% CL}{/hfagc/00101.html}{tab:hfc01101}
\hfmetadata{ [ .tml created 2006-02-08T20:01:29.5+08:00] }

\begin{tabular}{lccccc}  
\hflmode
&  \hflpdg2004
&  \hflbelle
&  \hflbabar
&  \hflcdf
&  \hflavg
\\\hline 

 \hfhref{/hfagc/BR_-521_-321+81.html}{ \hflabel{X(3870) K^-} } \hftabletlcell\hftableblcell 
 &  { \,}  
 &  { \,}  
 &  \hfhref{/hfagc/0505003.html}{ \hfprehot{<3.2} }  
 &  { \,}  
 &  \hfhref{/hfagc/BR_-521_-321+81.html}{ \hfavg{<3.2} } 
\\\hline 

\end{tabular}

\end{center}

\end{\hftabletype}
\clearpage  
\begin{\hftabletype}[\hftableposn]

\begin{center}
\hfnewcaption{Product branching fractions}{charged B}{\hfnewp}{in units of $10^{-4}$}{upper limits are at 90\% CL}{/hfagc/00101.html}{tab:hfc02101}
\hfmetadata{ [ .tml created 2006-02-08T20:01:30.89+08:00] }

\begin{tabular}{lccccc}  
\hflmode
&  \hflpdg2004
&  \hflbelle
&  \hflbabar
&  \hflcdf
&  \hflavg
\\\hline 

 \hfhref{/hfagc/BR_-521_-321+81xBR_81_22+443.html}{ \hflabel{K^- X(3870) [ \gamma J/\psi(1S) ]} } \hftabletlcell 
 &  { \,}  
 &  \hfhref{/hfagc/0506060.html}{ \hfprehot{0.018\pm0.006\pm0.001} }  
 &  { \,}  
 &  { \,}  
 &  \hfhref{/hfagc/BR_-521_-321+81xBR_81_22+443.html}{ \hfavg{0.018\pm0.006} } 
\\%

 \hfhref{/hfagc/BR_-521_-321+81xBR_81_221+443.html}{ \hflabel{K^- X(3870) [ J/\psi(1S) \eta ]} }  
 &  { \,}  
 &  { \,}  
 &  \hfhref{/hfagc/0505002.html}{ \hfpub{<0.077} }  
 &  { \,}  
 &  \hfhref{/hfagc/BR_-521_-321+81xBR_81_221+443.html}{ \hfavg{<0.077} } 
\\%

 \hfhref{/hfagc/BR_-521_-321+81xBR_81_-211+211+443.html}{ \hflabel{K^- X(3870) [ \pi^+ \pi^- J/\psi(1S) ]} }  
 &  \hfhref{/hfagc/BR_-521_-321+81xBR_81_-211+211+443.html}{ \hfpdg{0.136\pm0.031} }  
 &  \hfhref{/hfagc/0506061.html}{ \hfprehot{0.131\pm0.024\pm0.013} }  
 &  \hfhref{/hfagc/0507002.html}{ \hfpub{0.101\pm0.025\pm0.010} }  
 &  { \,}  
 &  \hfhref{/hfagc/BR_-521_-321+81xBR_81_-211+211+443.html}{ \hfavg{0.116\pm0.019} } 
\\%

 \hfhref{/hfagc/BR_-521_-321+86xBR_86_-211+211+443.html}{ \hflabel{K^- Y(4260) [ J/\psi(1S) \pi^+ \pi^- ]} }  
 &  { \,}  
 &  { \,}  
 &  \hfhref{/hfagc/0507002.html}{ \hfpub{0.200\pm0.070\pm0.020} }  
 &  { \,}  
 &  \hfhref{/hfagc/BR_-521_-321+86xBR_86_-211+211+443.html}{ \hfavg{0.200\pm0.073} } 
\\%

 \hfhref{/hfagc/BR_-521_-311+-82xBR_-82_-211+111+443.html}{ \hflabel{\bar{K}^0 X^{-}(3870) [ J/\psi(1S) \pi^- \pi^0 ]} }  
 &  { \,}  
 &  { \,}  
 &  \hfhref{/hfagc/0506009.html}{ \hfpubhot{<0.22} }  
 &  { \,}  
 &  \hfhref{/hfagc/BR_-521_-311+-82xBR_-82_-211+111+443.html}{ \hfavg{<0.22} } 
\\%

 \hfhref{/hfagc/BR_-521_-321+81xBR_81_-411+411.html}{ \hflabel{K^- X(3870) [ D^+ D^- ]} }  
 &  { \,}  
 &  \hfhref{/hfagc/0506075.html}{ \hfpub{<0.40} }  
 &  { \,}  
 &  { \,}  
 &  \hfhref{/hfagc/BR_-521_-321+81xBR_81_-411+411.html}{ \hfavg{<0.40} } 
\\%

 \hfhref{/hfagc/BR_-521_-321+81xBR_81_-421+111+421.html}{ \hflabel{K^- X(3870) [ D^0 \bar{D}^0 \pi^0 ]} }  
 &  { \,}  
 &  \hfhref{/hfagc/0506075.html}{ \hfpub{<0.60} }  
 &  { \,}  
 &  { \,}  
 &  \hfhref{/hfagc/BR_-521_-321+81xBR_81_-421+111+421.html}{ \hfavg{<0.60} } 
\\%

 \hfhref{/hfagc/BR_-521_-321+81xBR_81_-421+421.html}{ \hflabel{K^- X(3870) [ D^0 \bar{D}^0 ]} }  
 &  { \,}  
 &  \hfhref{/hfagc/0506075.html}{ \hfpub{<0.60} }  
 &  { \,}  
 &  { \,}  
 &  \hfhref{/hfagc/BR_-521_-321+81xBR_81_-421+421.html}{ \hfavg{<0.60} } 
\\%

 \hfhref{/hfagc/BR_-521_-85+421xBR_-85_-431+-211+211.html}{ \hflabel{D^0 D_{sJ}^{-}(2460) [ D_s^- \pi^+ \pi^- ]} }  
 &  { \,}  
 &  \hfhref{/hfagc/0506040.html}{ \hfpubold{<2.2} }  
 &  { \,}  
 &  { \,}  
 &  \hfhref{/hfagc/BR_-521_-85+421xBR_-85_-431+-211+211.html}{ \hfavg{<2.2} } 
\\%

 \hfhref{/hfagc/BR_-521_-85+421xBR_-85_-431+111.html}{ \hflabel{D^0 D_{sJ}^{-}(2460) [ D_s^- \pi^0 ]} }  
 &  { \,}  
 &  \hfhref{/hfagc/0506040.html}{ \hfpubold{<2.7} }  
 &  { \,}  
 &  { \,}  
 &  \hfhref{/hfagc/BR_-521_-85+421xBR_-85_-431+111.html}{ \hfavg{<2.7} } 
\\%

 \hfhref{/hfagc/BR_-521_-85+421xBR_-85_-431+22.html}{ \hflabel{D^0 D_{sJ}^{-}(2460) [ D_s^- \gamma ]} }  
 &  \hfhref{/hfagc/BR_-521_-85+421xBR_-85_-431+22.html}{ \hfpdg{5.6\pm2.3} }  
 &  \hfhref{/hfagc/0506040.html}{ \hfpubold{5.6\pm^{1.6}_{1.5}\pm1.7} }  
 &  \hfhref{/hfagc/0506025.html}{ \hfpub{6.0\pm2.0\pm1.0\pm^{2.0}_{1.0}} }  
 &  { \,}  
 &  \hfhref{/hfagc/BR_-521_-85+421xBR_-85_-431+22.html}{ \hfavg{5.8\pm^{1.7}_{1.9}} } 
\\%

 \hfhref{/hfagc/BR_-521_-84+421xBR_-84_-433+22.html}{ \hflabel{D^0 D_{sJ}^{*}(2317)^{-} [ D_s^{*-} \gamma ]} }  
 &  { \,}  
 &  \hfhref{/hfagc/0506040.html}{ \hfpubold{<7.6} }  
 &  { \,}  
 &  { \,}  
 &  \hfhref{/hfagc/BR_-521_-84+421xBR_-84_-433+22.html}{ \hfavg{<7.6} } 
\\%

 \hfhref{/hfagc/BR_-521_-84+423xBR_-84_-431+111.html}{ \hflabel{D^{*0}(2007) D_{sJ}^{*}(2317)^{-} [ D_s^- \pi^0 ]} }  
 &  { \,}  
 &  { \,}  
 &  \hfhref{/hfagc/0506025.html}{ \hfpub{9.0\pm6.0\pm2.0\pm^{3.0}_{2.0}} }  
 &  { \,}  
 &  \hfhref{/hfagc/BR_-521_-84+423xBR_-84_-431+111.html}{ \hfavg{9.0\pm^{7.0}_{6.6}} } 
\\%

 \hfhref{/hfagc/BR_-521_-84+421xBR_-84_-431+111.html}{ \hflabel{D^0 D_{sJ}^{*}(2317)^{-} [ D_s^- \pi^0 ]} }  
 &  \hfhref{/hfagc/BR_-521_-84+421xBR_-84_-431+111.html}{ \hfpdg{8.1\pm3.7} }  
 &  \hfhref{/hfagc/0506040.html}{ \hfpubold{8.1\pm^{3.0}_{2.7}\pm2.4} }  
 &  \hfhref{/hfagc/0506025.html}{ \hfpub{10.0\pm3.0\pm1.0\pm^{4.0}_{2.0}} }  
 &  { \,}  
 &  \hfhref{/hfagc/BR_-521_-84+421xBR_-84_-431+111.html}{ \hfavg{8.9\pm^{2.7}_{3.2}} } 
\\%

 \hfhref{/hfagc/BR_-521_-85+421xBR_-85_-433+22.html}{ \hflabel{D^0 D_{sJ}^{-}(2460) [ D_s^{*-} \gamma ]} }  
 &  { \,}  
 &  \hfhref{/hfagc/0506040.html}{ \hfpubold{<9.8} }  
 &  { \,}  
 &  { \,}  
 &  \hfhref{/hfagc/BR_-521_-85+421xBR_-85_-433+22.html}{ \hfavg{<9.8} } 
\\%

 \hfhref{/hfagc/BR_-521_-85+423xBR_-85_-431+22.html}{ \hflabel{D^{*0}(2007) D_{sJ}^{-}(2460) [ D_s^- \gamma ]} }  
 &  { \,}  
 &  { \,}  
 &  \hfhref{/hfagc/0506025.html}{ \hfpub{14.0\pm4.0\pm3.0\pm^{5.0}_{3.0}} }  
 &  { \,}  
 &  \hfhref{/hfagc/BR_-521_-85+423xBR_-85_-431+22.html}{ \hfavg{14.0\pm^{7.1}_{5.8}} } 
\\%

 \hfhref{/hfagc/BR_-521_-85+421xBR_-85_-433+111.html}{ \hflabel{D^0 D_{sJ}^{-}(2460) [ D_s^{*-} \pi^0 ]} }  
 &  \hfhref{/hfagc/BR_-521_-85+421xBR_-85_-433+111.html}{ \hfpdg{11.9\pm6.5} }  
 &  \hfhref{/hfagc/0506040.html}{ \hfpubold{11.9\pm^{6.1}_{4.9}\pm3.6} }  
 &  \hfhref{/hfagc/0506025.html}{ \hfpub{27\pm7\pm5\pm^{9}_{6}} }  
 &  { \,}  
 &  \hfhref{/hfagc/BR_-521_-85+421xBR_-85_-433+111.html}{ \hfavg{15.0\pm^{5.3}_{5.8}} } 
\\%

 \hfhref{/hfagc/BR_-521_-85+423xBR_-85_-433+111.html}{ \hflabel{D^{*0}(2007) D_{sJ}^{-}(2460) [ D_s^{*-} \pi^0 ]} } \hftableblcell 
 &  { \,}  
 &  { \,}  
 &  \hfhref{/hfagc/0506025.html}{ \hfpub{76\pm17\pm18\pm^{26}_{16}} }  
 &  { \,}  
 &  \hfhref{/hfagc/BR_-521_-85+423xBR_-85_-433+111.html}{ \hfavg{76\pm^{36}_{29}} } 
\\\hline 

\end{tabular}

\end{center}


\begin{center}
\hfnewcaption{Branching fractions}{charged B}{\hfdstr}{in units of $10^{-5}$}{upper limits are at 90\% CL}{/hfagc/00102.html}{tab:hfc01102}
\hfmetadata{ [ .tml created 2006-02-08T20:01:35.91+08:00] }

\begin{tabular}{lccccc}  
\hflmode
&  \hflpdg2004
&  \hflbelle
&  \hflbabar
&  \hflcdf
&  \hflavg
\\\hline 

 \hfhref{/hfagc/BR_-521_-431+333.html}{ \hflabel{D_s^- \phi(1020)} } \hftabletlcell 
 &  \hfhref{/hfagc/BR_-521_-431+333.html}{ \hfpdg{<32} }  
 &  { \,}  
 &  \hfhref{/hfagc/0506020.html}{ \hfprehot{<0.190} }  
 &  { \,}  
 &  \hfhref{/hfagc/BR_-521_-431+333.html}{ \hfavg{<0.190} } 
\\%

 \hfhref{/hfagc/BR_-521_-433+333.html}{ \hflabel{D_s^{*-} \phi(1020)} }  
 &  \hfhref{/hfagc/BR_-521_-433+333.html}{ \hfpdg{<40} }  
 &  { \,}  
 &  \hfhref{/hfagc/0506020.html}{ \hfprehot{<1.20} }  
 &  { \,}  
 &  \hfhref{/hfagc/BR_-521_-433+333.html}{ \hfavg{<1.20} } 
\\%

 \hfhref{/hfagc/BR_-521_-431+111.html}{ \hflabel{D_s^- \pi^0} } \hftableblcell 
 &  \hfhref{/hfagc/BR_-521_-431+111.html}{ \hfpdg{<20} }  
 &  { \,}  
 &  \hfhref{/hfagc/0506015.html}{ \hfwaiting{<2.8} }  
 &  { \,}  
 &  \hfhref{/hfagc/BR_-521_-431+111.html}{ \hfavg{<2.8} } 
\\\hline 

\end{tabular}

\end{center}

\end{\hftabletype}
\clearpage  
\begin{\hftabletype}[\hftableposn]

\begin{center}
\hfnewcaption{Branching fractions}{charged B}{\hfbary}{in units of $10^{-5}$}{upper limits are at 90\% CL}{/hfagc/00103.html}{tab:hfc01103}
\hfmetadata{ [ .tml created 2006-02-08T20:01:40.912+08:00] }

\begin{tabular}{lccccc}  
\hflmode
&  \hflpdg2004
&  \hflbelle
&  \hflbabar
&  \hflcdf
&  \hflavg
\\\hline 

 \hfhref{/hfagc/BR_-521_-2212+3122+443.html}{ \hflabel{J/\psi(1S) \Lambda \bar{p}} } \hftabletlcell 
 &  \hfhref{/hfagc/BR_-521_-2212+3122+443.html}{ \hfpdg{1.20\pm0.80} }  
 &  { \,}  
 &  \hfhref{/hfagc/0506010.html}{ \hfpubold{1.16\pm^{0.74}_{0.53}\pm^{0.42}_{0.18}} }  
 &  { \,}  
 &  \hfhref{/hfagc/BR_-521_-2212+3122+443.html}{ \hfavg{1.16\pm^{0.85}_{0.56}} } 
\\%

 \hfhref{/hfagc/BR_-521_-2212+2212+413.html}{ \hflabel{D^{*+}(2010) p \bar{p}} }  
 &  { \,}  
 &  \hfhref{/hfagc/0506083.html}{ \hfpubold{<1.50} }  
 &  { \,}  
 &  { \,}  
 &  \hfhref{/hfagc/BR_-521_-2212+2212+413.html}{ \hfavg{<1.50} } 
\\%

 \hfhref{/hfagc/BR_-521_-2212+2212+411.html}{ \hflabel{D^+ p \bar{p}} }  
 &  { \,}  
 &  \hfhref{/hfagc/0506083.html}{ \hfpubold{<1.50} }  
 &  { \,}  
 &  { \,}  
 &  \hfhref{/hfagc/BR_-521_-2212+2212+411.html}{ \hfavg{<1.50} } 
\\%

 \hfhref{/hfagc/BR_-521_-2212+4114.html}{ \hflabel{\Sigma_c^{*0} \bar{p}} }  
 &  \hfhref{/hfagc/BR_-521_-2212+4114.html}{ \hfpdg{<4.6} }  
 &  \hfhref{/hfagc/0506086.html}{ \hfpubold{<4.6} }  
 &  { \,}  
 &  { \,}  
 &  \hfhref{/hfagc/BR_-521_-2212+4114.html}{ \hfavg{<4.6} } 
\\%

 \hfhref{/hfagc/BR_-521_-2212+4112.html}{ \hflabel{\Sigma_c^0 \bar{p}} }  
 &  \hfhref{/hfagc/BR_-521_-2212+4112.html}{ \hfpdg{<8.0} }  
 &  \hfhref{/hfagc/0506086.html}{ \hfpubold{<9.3} }  
 &  { \,}  
 &  { \,}  
 &  \hfhref{/hfagc/BR_-521_-2212+4112.html}{ \hfavg{<9.3} } 
\\%

 \hfhref{/hfagc/BR_-521_-211+-2212+4122.html}{ \hflabel{\Lambda_c^+ \bar{p} \pi^-} } \hftableblcell 
 &  \hfhref{/hfagc/BR_-521_-211+-2212+4122.html}{ \hfpdg{21\pm7} }  
 &  \hfhref{/hfagc/0506086.html}{ \hfpubold{18.7\pm^{4.3}_{4.0}\pm2.8\pm4.9} }  
 &  { \,}  
 &  { \,}  
 &  \hfhref{/hfagc/BR_-521_-211+-2212+4122.html}{ \hfavg{18.7\pm^{7.1}_{6.9}} } 
\\\hline 

\end{tabular}

\end{center}


\begin{center}
\hfnewcaption{Branching fractions}{charged B}{\hfjpsi}{in units of $10^{-4}$}{upper limits are at 90\% CL}{/hfagc/00104.html}{tab:hfc01104}
\hfmetadata{ [ .tml created 2006-02-08T20:01:48.709+08:00] }

\begin{tabular}{lccccc}  
\hflmode
&  \hflpdg2004
&  \hflbelle
&  \hflbabar
&  \hflcdf
&  \hflavg
\\\hline 

 \hfhref{/hfagc/BR_-521_-211+421+443.html}{ \hflabel{J/\psi(1S) D^0 \pi^-} } \hftabletlcell 
 &  { \,}  
 &  \hfhref{/hfagc/0506038.html}{ \hfpubhot{<0.25} }  
 &  \hfhref{/hfagc/0506005.html}{ \hfpubhot{<0.52} }  
 &  { \,}  
 &  \hfhref{/hfagc/BR_-521_-211+421+443.html}{ \hfavg{<0.25} } 
\\%

 \hfhref{/hfagc/BR_-521_-321+333+443.html}{ \hflabel{J/\psi(1S) \phi(1020) K^-} }  
 &  \hfhref{/hfagc/BR_-521_-321+333+443.html}{ \hfpdg{0.52\pm0.17} }  
 &  { \,}  
 &  \hfhref{/hfagc/0506012.html}{ \hfpubold{0.44\pm0.14\pm0.05\pm0.01} }  
 &  { \,}  
 &  \hfhref{/hfagc/BR_-521_-321+333+443.html}{ \hfavg{0.44\pm0.15} } 
\\%

 \hfhref{/hfagc/BR_-521_-211+443.html}{ \hflabel{J/\psi(1S) \pi^-} }  
 &  \hfhref{/hfagc/BR_-521_-211+443.html}{ \hfpdg{0.40\pm0.05} }  
 &  \hfhref{/hfagc/0506076.html}{ \hfpubold{0.38\pm0.06\pm0.03} }  
 &  \hfhref{/hfagc/0506004.html}{ \hfpub{0.54\pm0.04\pm0.02} }  
 &  { \,}  
 &  \hfhref{/hfagc/BR_-521_-211+443.html}{ \hfavg{0.48\pm0.04} } 
\\%

 \hfhref{/hfagc/BR_-521_-321+221+443.html}{ \hflabel{J/\psi(1S) \eta K^-} }  
 &  { \,}  
 &  { \,}  
 &  \hfhref{/hfagc/0505002.html}{ \hfpub{1.08\pm0.23\pm0.24\pm0.03} }  
 &  { \,}  
 &  \hfhref{/hfagc/BR_-521_-321+221+443.html}{ \hfavg{1.08\pm0.33} } 
\\%

 \hfhref{/hfagc/BR_-521_-411+443.html}{ \hflabel{J/\psi(1S) D^-} }  
 &  { \,}  
 &  { \,}  
 &  \hfhref{/hfagc/0506003.html}{ \hfpubhot{<1.20} }  
 &  { \,}  
 &  \hfhref{/hfagc/BR_-521_-411+443.html}{ \hfavg{<1.20} } 
\\%

 \hfhref{/hfagc/BR_-521_-321+443.html}{ \hflabel{J/\psi(1S) K^-} }  
 &  \hfhref{/hfagc/BR_-521_-321+443.html}{ \hfpdg{10.0\pm0.4} }  
 &  { \,}  
 &  { \,}  
 &  { \,}  
 &  \hfhref{/hfagc/BR_-521_-321+443.html}{ \hfavg{10.3\pm0.4} } 
\\%

{ \hflabel{\,} }  
 &  { \,}  
 &  \hfhref{/hfagc/0506076.html}{ \hfpubold{10.1\pm0.2\pm0.7\pm0.2} }  
 &  \hfhref{/hfagc/0506109.html}{ \hfpubhot{10.6\pm0.2\pm0.4\pm0.2} } 
\hffootnotemark{1}
 
 &  { \,}  
 &  { \,} 
\\%

{ \hflabel{\,} }  
 &  { \,}  
 &  { \,}  
 &  \hfhref{/hfagc/0507001.html}{ \hfpubhot{10.1\pm0.9\pm0.6} } 
\hffootnotemark{2}
 
 &  { \,}  
 &  { \,} 
\\%

{ \hflabel{\,} }  
 &  { \,}  
 &  { \,}  
 &  \hfhref{/hfagc/0505003.html}{ \hfprehot{8.1\pm1.3\pm0.7} } 
\hffootnotemark{3}
 
 &  { \,}  
 &  { \,} 
\\%

 \hfhref{/hfagc/BR_-521_-211+-321+211+443.html}{ \hflabel{J/\psi(1S) K^- \pi^+ \pi^-} }  
 &  \hfhref{/hfagc/BR_-521_-211+-321+211+443.html}{ \hfpdg{7.7\pm2.0} }  
 &  { \,}  
 &  \hfhref{/hfagc/0506005.html}{ \hfpubhot{11.6\pm0.7\pm0.9} }  
 &  \hfhref{/hfagc/0506091.html}{ \hfpubold{6.9\pm1.8\pm1.2} }  
 &  \hfhref{/hfagc/BR_-521_-211+-321+211+443.html}{ \hfavg{10.6\pm1.0} } 
\\%

 \hfhref{/hfagc/BR_-521_-323+443.html}{ \hflabel{J/\psi(1S) K^{*-}(892)} }  
 &  \hfhref{/hfagc/BR_-521_-323+443.html}{ \hfpdg{13.5\pm1.0} }  
 &  \hfhref{/hfagc/0506073.html}{ \hfpubold{12.8\pm0.7\pm1.4\pm0.2} }  
 &  \hfhref{/hfagc/0506109.html}{ \hfpubhot{14.5\pm0.5\pm0.9\pm0.2} }  
 &  \hfhref{/hfagc/0506093.html}{ \hfpubold{15.8\pm4.7\pm2.7} }  
 &  \hfhref{/hfagc/BR_-521_-323+443.html}{ \hfavg{14.0\pm0.9} } 
\\%

 \hfhref{/hfagc/BR_-521_-10323+443.html}{ \hflabel{J/\psi(1S) K_1^-(1270)} } \hftableblcell 
 &  \hfhref{/hfagc/BR_-521_-10323+443.html}{ \hfpdg{18.0\pm5.2} }  
 &  \hfhref{/hfagc/0506072.html}{ \hfpubold{18.0\pm3.4\pm3.0\pm2.5} }  
 &  { \,}  
 &  { \,}  
 &  \hfhref{/hfagc/BR_-521_-10323+443.html}{ \hfavg{18.0\pm5.2} } 
\\\hline 

\end{tabular}

\begin{description}\addtolength{\itemsep}{\hffootitemsep\baselineskip}

	  \item[ \hffootnotemark{1} ] \hffootnotetext{ MEASUREMENT OF BRANCHING FRACTIONS AND CHARGE ASYMMETRIES FOR EXCLUSIVE B DECAYS TO CHARMONIUM (\hfbb{124}) } 
     ;  \hffootnotetext{ $B^- \rightarrow J/\psi K^-$ with $J/\psi$ to leptons } 
	  \item[ \hffootnotemark{2} ] \hffootnotetext{ MEASUREMENT OF THE $B^+ \rightarrow p \overline{p} K^+$ BRANCHING FRACTION AND STUDY OF THE DECAY DYNAMICS (\hfbb{232}) } 
     ;  \hffootnotetext{ $B^- \rightarrow J/\psi K^-$ with $J/\psi \rightarrow p\overline{p}$ } 
	  \item[ \hffootnotemark{3} ] \hffootnotetext{ Measurements of the absolute branching fractions of $B^\pm \rightarrow K^\pm X_{c\overline{c}}$ (\hfbb{231.8}) } 
     ;  \hffootnotetext{ $B^- \rightarrow J/\psi K^-$ (inclusive) } 
\end{description}

\end{center}

\end{\hftabletype}
\clearpage  
\begin{\hftabletype}[\hftableposn]

\begin{center}
\hfnewcaption{Product branching fractions}{charged B}{\hfjpsi}{in units of $10^{-4}$}{upper limits are at 90\% CL}{/hfagc/00104.html}{tab:hfc02104}
\hfmetadata{ [ .tml created 2006-02-08T20:01:52.142+08:00] }

\begin{tabular}{lccccc}  
\hflmode
&  \hflpdg2004
&  \hflbelle
&  \hflbabar
&  \hflcdf
&  \hflavg
\\\hline 

 \hfhref{/hfagc/BR_-521_-321+10443xBR_10443_-211+211+443.html}{ \hflabel{K^- h_c(1P) [ J/\psi(1S) \pi^+ \pi^- ]} } \hftabletlcell\hftableblcell 
 &  { \,}  
 &  { \,}  
 &  \hfhref{/hfagc/0506005.html}{ \hfpubhot{<0.034} }  
 &  { \,}  
 &  \hfhref{/hfagc/BR_-521_-321+10443xBR_10443_-211+211+443.html}{ \hfavg{<0.034} } 
\\\hline 

\end{tabular}

\end{center}


\begin{center}
\hfnewcaption{Ratios of branching fractions}{charged B}{\hfjpsi}{in units of $10^{0}$}{upper limits are at 90\% CL}{/hfagc/00104.html}{tab:hfc03104}
\hfmetadata{ [ .tml created 2006-02-08T20:01:53.023+08:00] }

\begin{tabular}{lccccc}  
\hflmode
&  \hflpdg2004
&  \hflbelle
&  \hflbabar
&  \hflcdf
&  \hflavg
\\\hline 

 \hfhref{/hfagc/BR_-521_-211+443xoBR_-521_-321+443.html}{ \hflabel{\frac{{\cal{B}} ( B^- \to J/\psi(1S) \pi^- )}{{\cal{B}} ( B^- \to J/\psi(1S) K^- )}} } \hftabletlcell 
 &  \hfhref{/hfagc/BR_-521_-211+443xoBR_-521_-321+443.html}{ \hfpdg{0.040\pm0.005} }  
 &  { \,}  
 &  \hfhref{/hfagc/0506004.html}{ \hfpub{0.054\pm0.004\pm0.001} }  
 &  \hfhref{/hfagc/0506095.html}{ \hfpubold{0.050\pm^{0.019}_{0.017}\pm0.001} }  
 &  \hfhref{/hfagc/BR_-521_-211+443xoBR_-521_-321+443.html}{ \hfavg{0.053\pm0.004} } 
\\%

 \hfhref{/hfagc/BR_-521_-20323+443xoBR_-521_-10323+443.html}{ \hflabel{\frac{{\cal{B}} ( B^- \to J/\psi(1S) K_1^-(1400) )}{{\cal{B}} ( B^- \to J/\psi(1S) K_1^-(1270) )}} }  
 &  \hfhref{/hfagc/BR_-521_-20323+443xoBR_-521_-10323+443.html}{ \hfpdg{<0.30} }  
 &  \hfhref{/hfagc/0506072.html}{ \hfpubold{<0.30} }  
 &  { \,}  
 &  { \,}  
 &  \hfhref{/hfagc/BR_-521_-20323+443xoBR_-521_-10323+443.html}{ \hfavg{<0.30} } 
\\%

 \hfhref{/hfagc/BR_-521_-321+10441xoBR_-521_-321+443.html}{ \hflabel{\frac{{\cal{B}} ( B^- \to \chi_{c0}(1P) K^- )}{{\cal{B}} ( B^- \to J/\psi(1S) K^- )}} }  
 &  \hfhref{/hfagc/BR_-521_-321+10441xoBR_-521_-321+443.html}{ \hfpdg{0.60\pm0.20} }  
 &  \hfhref{/hfagc/0506080.html}{ \hfpubold{0.60\pm^{0.21}_{0.18}\pm0.05\pm0.08} }  
 &  { \,}  
 &  { \,}  
 &  \hfhref{/hfagc/BR_-521_-321+10441xoBR_-521_-321+443.html}{ \hfavg{0.60\pm^{0.23}_{0.20}} } 
\\%

 \hfhref{/hfagc/BR_-521_-321+441xoBR_-521_-321+443.html}{ \hflabel{\frac{{\cal{B}} ( B^- \to \eta_c(1S) K^- )}{{\cal{B}} ( B^- \to J/\psi(1S) K^- )}} }  
 &  { \,}  
 &  { \,}  
 &  { \,}  
 &  { \,}  
 &  \hfhref{/hfagc/BR_-521_-321+441xoBR_-521_-321+443.html}{ \hfavg{1.12\pm0.20} } 
\\%

{ \hflabel{\,} }  
 &  { \,}  
 &  { \,}  
 &  \hfhref{/hfagc/0506001.html}{ \hfpub{1.28\pm0.10\pm0.38} } 
\hffootnotemark{1}
 
 &  { \,}  
 &  { \,} 
\\%

{ \hflabel{\,} }  
 &  { \,}  
 &  { \,}  
 &  \hfhref{/hfagc/0505003.html}{ \hfprehot{1.06\pm0.23\pm0.04} } 
\hffootnotemark{2}
 
 &  { \,}  
 &  { \,} 
\\%

 \hfhref{/hfagc/BR_-521_-323+443xoBR_-521_-321+443.html}{ \hflabel{\frac{{\cal{B}} ( B^- \to J/\psi(1S) K^{*-}(892) )}{{\cal{B}} ( B^- \to J/\psi(1S) K^- )}} }  
 &  \hfhref{/hfagc/BR_-521_-323+443xoBR_-521_-321+443.html}{ \hfpdg{1.40\pm0.11} }  
 &  { \,}  
 &  \hfhref{/hfagc/0506109.html}{ \hfpubhot{1.37\pm0.05\pm0.08} }  
 &  \hfhref{/hfagc/0506094.html}{ \hfpubold{1.92\pm0.60\pm0.17} }  
 &  \hfhref{/hfagc/BR_-521_-323+443xoBR_-521_-321+443.html}{ \hfavg{1.38\pm0.09} } 
\\%

 \hfhref{/hfagc/BR_-521_-10323+443xoBR_-521_-321+443.html}{ \hflabel{\frac{{\cal{B}} ( B^- \to J/\psi(1S) K_1^-(1270) )}{{\cal{B}} ( B^- \to J/\psi(1S) K^- )}} } \hftableblcell 
 &  { \,}  
 &  \hfhref{/hfagc/0506072.html}{ \hfpubold{1.80\pm0.34\pm0.34} }  
 &  { \,}  
 &  { \,}  
 &  \hfhref{/hfagc/BR_-521_-10323+443xoBR_-521_-321+443.html}{ \hfavg{1.80\pm0.48} } 
\\\hline 

\end{tabular}

\begin{description}\addtolength{\itemsep}{\hffootitemsep\baselineskip}

	  \item[ \hffootnotemark{1} ] \hffootnotetext{ Branching Fraction Measurements of $B \rightarrow \eta_c K$ Decays (\hfbb{86.1}) } 
     ;  \hffootnotetext{ Ratio $B^- \rightarrow \eta_{c} K^-$ to $B^- \rightarrow J/\psi K^-$ with $\eta_c \rightarrow K\overline{K}\pi$ } 
	  \item[ \hffootnotemark{2} ] \hffootnotetext{ Measurements of the absolute branching fractions of $B^\pm \rightarrow K^\pm X_{c\overline{c}}$ (\hfbb{231.8}) } 
     ;  \hffootnotetext{ Ratio $B^- \rightarrow \eta_c K^-$ to $B^- \rightarrow J/\psi K^-$  (inclusive analysis) } 
\end{description}

\end{center}

\end{\hftabletype}
\clearpage  
\begin{\hftabletype}[\hftableposn]

\begin{center}
\hfnewcaption{Branching fractions}{charged B}{\hfochm}{in units of $10^{-3}$}{upper limits are at 90\% CL}{/hfagc/00105.html}{tab:hfc01105}
\hfmetadata{ [ .tml created 2006-02-08T20:01:57.771+08:00] }

\begin{tabular}{lccccc}  
\hflmode
&  \hflpdg2004
&  \hflbelle
&  \hflbabar
&  \hflcdf
&  \hflavg
\\\hline 

 \hfhref{/hfagc/BR_-521_-323+445.html}{ \hflabel{\chi_{c2}(1P) K^{*-}(892)} } \hftabletlcell 
 &  { \,}  
 &  { \,}  
 &  \hfhref{/hfagc/0506006.html}{ \hfpubhot{<0.012} }  
 &  { \,}  
 &  \hfhref{/hfagc/BR_-521_-323+445.html}{ \hfavg{<0.012} } 
\\%

 \hfhref{/hfagc/BR_-521_-321+445.html}{ \hflabel{\chi_{c2}(1P) K^-} }  
 &  { \,}  
 &  { \,}  
 &  { \,}  
 &  { \,}  
 &  \hfhref{/hfagc/BR_-521_-321+445.html}{ \hfavg{<0.030} } 
\\%

{ \hflabel{\,} }  
 &  { \,}  
 &  { \,}  
 &  \hfhref{/hfagc/0506006.html}{ \hfpubhot{<0.030} } 
\hffootnotemark{1}
 
 &  { \,}  
 &  { \,} 
\\%

{ \hflabel{\,} }  
 &  { \,}  
 &  { \,}  
 &  \hfhref{/hfagc/0505003.html}{ \hfsuperceeded{<0.200} } 
\hffootnotemark{2a}
 
 &  { \,}  
 &  { \,} 
\\%

 \hfhref{/hfagc/BR_-521_-211+10441.html}{ \hflabel{\chi_{c0}(1P) \pi^-} }  
 &  { \,}  
 &  { \,}  
 &  \hfhref{/hfagc/0506014.html}{ \hfpubhot{<0.061} }  
 &  { \,}  
 &  \hfhref{/hfagc/BR_-521_-211+10441.html}{ \hfavg{<0.061} } 
\\%

 \hfhref{/hfagc/BR_-521_-321+10441.html}{ \hflabel{\chi_{c0}(1P) K^-} }  
 &  \hfhref{/hfagc/BR_-521_-321+10441.html}{ \hfpdg{0.60\pm0.20} }  
 &  { \,}  
 &  { \,}  
 &  { \,}  
 &  \hfhref{/hfagc/BR_-521_-321+10441.html}{ \hfavg{0.191\pm0.040} } 
\\%

{ \hflabel{\,} }  
 &  { \,}  
 &  \hfhref{/hfagc/0506080.html}{ \hfpubold{0.60\pm^{0.21}_{0.18}\pm0.07\pm0.09} }  
 &  \hfhref{/hfagc/0506008.html}{ \hfpub{0.27\pm0.07} } 
\hffootnotemark{3}
 
 &  { \,}  
 &  { \,} 
\\%

{ \hflabel{\,} }  
 &  { \,}  
 &  { \,}  
 &  \hfhref{/hfagc/0506013.html}{ \hfpubhot{0.134\pm0.045\pm0.015\pm0.014} } 
\hffootnotemark{4}
 
 &  { \,}  
 &  { \,} 
\\%

{ \hflabel{\,} }  
 &  { \,}  
 &  { \,}  
 &  \hfhref{/hfagc/0505003.html}{ \hfprehot{<0.180} } 
\hffootnotemark{2c}
 
 &  { \,}  
 &  { \,} 
\\%

 \hfhref{/hfagc/BR_-521_-323+20443.html}{ \hflabel{\chi_{c1}(1P) K^{*-}(892)} }  
 &  \hfhref{/hfagc/BR_-521_-323+20443.html}{ \hfpdg{<2.1} }  
 &  { \,}  
 &  \hfhref{/hfagc/0506109.html}{ \hfpubhot{0.29\pm0.10\pm0.09\pm0.03} }  
 &  { \,}  
 &  \hfhref{/hfagc/BR_-521_-323+20443.html}{ \hfavg{0.29\pm0.14} } 
\\%

 \hfhref{/hfagc/BR_-521_-321+100441.html}{ \hflabel{\eta_c(2S) K^-} }  
 &  { \,}  
 &  { \,}  
 &  \hfhref{/hfagc/0505003.html}{ \hfprehot{0.34\pm0.18\pm0.03} }  
 &  { \,}  
 &  \hfhref{/hfagc/BR_-521_-321+100441.html}{ \hfavg{0.34\pm0.18} } 
\\%

 \hfhref{/hfagc/BR_-521_-321+30443.html}{ \hflabel{\psi(3770) K^-} }  
 &  { \,}  
 &  \hfhref{/hfagc/0506075.html}{ \hfpub{0.48\pm0.11\pm0.07} }  
 &  \hfhref{/hfagc/0505003.html}{ \hfprehot{0.35\pm0.25\pm0.03} }  
 &  { \,}  
 &  \hfhref{/hfagc/BR_-521_-321+30443.html}{ \hfavg{0.45\pm0.12} } 
\\%

 \hfhref{/hfagc/BR_-521_-321+20443.html}{ \hflabel{\chi_{c1}(1P) K^-} }  
 &  \hfhref{/hfagc/BR_-521_-321+20443.html}{ \hfpdg{0.68\pm0.12} }  
 &  { \,}  
 &  { \,}  
 &  { \,}  
 &  \hfhref{/hfagc/BR_-521_-321+20443.html}{ \hfavg{0.63\pm0.06} } 
\\%

{ \hflabel{\,} }  
 &  { \,}  
 &  { \,}  
 &  \hfhref{/hfagc/0506109.html}{ \hfpubhot{0.58\pm0.03\pm0.06} } 
\hffootnotemark{5a}
 
 &  \hfhref{/hfagc/0506091.html}{ \hfpubold{1.55\pm0.54\pm0.20} }  
 &  { \,} 
\\%

{ \hflabel{\,} }  
 &  { \,}  
 &  { \,}  
 &  \hfhref{/hfagc/0505003.html}{ \hfprehot{0.80\pm0.14\pm0.07} } 
\hffootnotemark{2b}
 
 &  { \,}  
 &  { \,} 
\\%

 \hfhref{/hfagc/BR_-521_-321+100443.html}{ \hflabel{\psi(2S) K^-} }  
 &  \hfhref{/hfagc/BR_-521_-321+100443.html}{ \hfpdg{0.68\pm0.04} }  
 &  { \,}  
 &  { \,}  
 &  { \,}  
 &  \hfhref{/hfagc/BR_-521_-321+100443.html}{ \hfavg{0.63\pm0.04} } 
\\%

{ \hflabel{\,} }  
 &  { \,}  
 &  \hfhref{/hfagc/0506076.html}{ \hfpubold{0.69\pm0.06} }  
 &  \hfhref{/hfagc/0506109.html}{ \hfpubhot{0.62\pm0.03\pm0.04\pm0.02} } 
\hffootnotemark{5b}
 
 &  \hfhref{/hfagc/0506092.html}{ \hfpubold{0.55\pm0.10\pm0.06} }  
 &  { \,} 
\\%

{ \hflabel{\,} }  
 &  { \,}  
 &  { \,}  
 &  \hfhref{/hfagc/0505003.html}{ \hfprehot{0.49\pm0.16\pm0.04} } 
\hffootnotemark{2e}
 
 &  { \,}  
 &  { \,} 
\\%

 \hfhref{/hfagc/BR_-521_-323+100443.html}{ \hflabel{\psi(2S) K^{*-}(892)} }  
 &  \hfhref{/hfagc/BR_-521_-323+100443.html}{ \hfpdg{0.92\pm0.22} }  
 &  \hfhref{/hfagc/0506077.html}{ \hfpreold{0.81\pm0.08\pm0.09} }  
 &  \hfhref{/hfagc/0506109.html}{ \hfpubhot{0.59\pm0.08\pm0.09\pm0.02} }  
 &  { \,}  
 &  \hfhref{/hfagc/BR_-521_-323+100443.html}{ \hfavg{0.71\pm0.09} } 
\\%

 \hfhref{/hfagc/BR_-521_-321+441.html}{ \hflabel{\eta_c(1S) K^-} }  
 &  \hfhref{/hfagc/BR_-521_-321+441.html}{ \hfpdg{0.90\pm0.27} }  
 &  { \,}  
 &  { \,}  
 &  { \,}  
 &  \hfhref{/hfagc/BR_-521_-321+441.html}{ \hfavg{0.98\pm0.13} } 
\\%

{ \hflabel{\,} }  
 &  { \,}  
 &  \hfhref{/hfagc/0506074.html}{ \hfpubold{1.25\pm0.14\pm^{0.10}_{0.12}\pm0.38} }  
 &  \hfhref{/hfagc/0506001.html}{ \hfpub{1.29\pm0.09\pm0.13\pm0.36} } 
\hffootnotemark{6}
 
 &  { \,}  
 &  { \,} 
\\%

{ \hflabel{\,} }  
 &  { \,}  
 &  { \,}  
 &  \hfhref{/hfagc/0507001.html}{ \hfpubhot{1.38\pm^{0.23}_{0.15}\pm0.15\pm0.42} } 
\hffootnotemark{7}
 
 &  { \,}  
 &  { \,} 
\\%

{ \hflabel{\,} }  
 &  { \,}  
 &  { \,}  
 &  \hfhref{/hfagc/0505003.html}{ \hfprehot{0.87\pm0.15} } 
\hffootnotemark{2d}
 
 &  { \,}  
 &  { \,} 
\\%

 \hfhref{/hfagc/BR_-521_-323+10441.html}{ \hflabel{\chi_{c0}(1P) K^{*-}(892)} } \hftableblcell 
 &  { \,}  
 &  { \,}  
 &  \hfhref{/hfagc/0506006.html}{ \hfpubhot{<2.9} }  
 &  { \,}  
 &  \hfhref{/hfagc/BR_-521_-323+10441.html}{ \hfavg{<2.9} } 
\\\hline 

\end{tabular}

\begin{description}\addtolength{\itemsep}{\hffootitemsep\baselineskip}

	  \item[ \hffootnotemark{1} ] \hffootnotetext{ SEARCH FOR FACTORIZATION-SUPPRESSED $B \rightarrow \chi_c K^{(*)}$ DECAYS (\hfbb{124}) } 
     ;  \hffootnotetext{ $B^- \rightarrow \chi_{c2} K^-$ with $\chi_{c2} \rightarrow J/\psi \gamma$ } 
	  \item[ \hffootnotemark{2} ] \hffootnotetext{ Measurements of the absolute branching fractions of $B^\pm \rightarrow K^\pm X_{c\overline{c}}$ (\hfbb{231.8}) } 
     ;  \hffootnotemark{2a}  \hffootnotetext{ $B^- \rightarrow \chi_{c2} K^-$ (inclusive) }  ;  \hffootnotemark{2b}  \hffootnotetext{ $B^- \rightarrow \chi_{c1} K^-$ (inclusive) }  ;  \hffootnotemark{2c}  \hffootnotetext{ $B^- \rightarrow \chi_{c0} K^-$ (inclusive) }  ;  \hffootnotemark{2d}  \hffootnotetext{ $B^- \rightarrow \eta_c K^-$ (inclusive) }  ;  \hffootnotemark{2e}  \hffootnotetext{ $B^- \rightarrow \psi(2S) K^-$ (inclusive) } 
	  \item[ \hffootnotemark{3} ] \hffootnotetext{ MEASUREMENT OF THE BRANCHING FRACTION FOR $B^\pm \rightarrow \chi_{c0} K^\pm$. (\hfbb{88.9}) } 
     ;  \hffootnotetext{ $B^- \rightarrow \chi_{c0} K^-$ with $\chi_{c0}\rightarrow K^+ K^-, \pi^+ \pi^-$ } 
	  \item[ \hffootnotemark{4} ] \hffootnotetext{ Dalitz-plot analysis of the decays $B^\pm \rightarrow K^\pm \pi^\mp \pi^\pm$ (\hfbb{226}) } 
     ;  \hffootnotetext{ $B^- \rightarrow \chi_{c0} K^-$ with $\chi_{c0} \rightarrow \pi^+ \pi^-$ (Dalitz analysis) } 
	  \item[ \hffootnotemark{5} ] \hffootnotetext{ MEASUREMENT OF BRANCHING FRACTIONS AND CHARGE ASYMMETRIES FOR EXCLUSIVE B DECAYS TO CHARMONIUM (\hfbb{124}) } 
     ;  \hffootnotemark{5a}  \hffootnotetext{ $B^- \rightarrow \chi_{c1} K^-$ with $\chi_{c1}$ to $J/\psi \gamma$ }  ;  \hffootnotemark{5b}  \hffootnotetext{ $B^- \rightarrow \psi(2S) K^-$ with $\psi(2S)$ to leptons } 
	  \item[ \hffootnotemark{6} ] \hffootnotetext{ Branching Fraction Measurements of $B \rightarrow \eta_c K$ Decays (\hfbb{86.1}) } 
     ;  \hffootnotetext{ $B^- \rightarrow \eta_{c} K^-$ with $\eta_c \rightarrow K\overline{K}\pi$ } 
	  \item[ \hffootnotemark{7} ] \hffootnotetext{ MEASUREMENT OF THE $B^+ \rightarrow p \overline{p} K^+$ BRANCHING FRACTION AND STUDY OF THE DECAY DYNAMICS (\hfbb{232}) } 
     ;  \hffootnotetext{ $B^- \rightarrow \eta_c K^-$ with $\eta_c \rightarrow p\overline{p}$ } 
\end{description}

\end{center}

\end{\hftabletype}
\clearpage  
\begin{\hftabletype}[\hftableposn]

\begin{center}
\hfnewcaption{Ratios of branching fractions}{charged B}{\hfochm}{in units of $10^{0}$}{upper limits are at 90\% CL}{/hfagc/00105.html}{tab:hfc03105}
\hfmetadata{ [ .tml created 2006-02-08T20:02:01.375+08:00] }

\begin{tabular}{lccccc}  
\hflmode
&  \hflpdg2004
&  \hflbelle
&  \hflbabar
&  \hflcdf
&  \hflavg
\\\hline 

 \hfhref{/hfagc/BR_-521_-323+20443xoBR_-521_-321+20443.html}{ \hflabel{\frac{{\cal{B}} ( B^- \to \chi_{c1}(1P) K^{*-}(892) )}{{\cal{B}} ( B^- \to \chi_{c1}(1P) K^- )}} } \hftabletlcell 
 &  { \,}  
 &  { \,}  
 &  \hfhref{/hfagc/0506109.html}{ \hfpubhot{0.51\pm0.17\pm0.16} }  
 &  { \,}  
 &  \hfhref{/hfagc/BR_-521_-323+20443xoBR_-521_-321+20443.html}{ \hfavg{0.51\pm0.23} } 
\\%

 \hfhref{/hfagc/BR_-521_-323+100443xoBR_-521_-321+100443.html}{ \hflabel{\frac{{\cal{B}} ( B^- \to \psi(2S) K^{*-}(892) )}{{\cal{B}} ( B^- \to \psi(2S) K^- )}} } \hftableblcell 
 &  { \,}  
 &  { \,}  
 &  \hfhref{/hfagc/0506109.html}{ \hfpubhot{0.96\pm0.15\pm0.09} }  
 &  { \,}  
 &  \hfhref{/hfagc/BR_-521_-323+100443xoBR_-521_-321+100443.html}{ \hfavg{0.96\pm0.17} } 
\\\hline 

\end{tabular}

\end{center}


\begin{center}
\hfnewcaption{Branching fractions}{charged B}{\hfmuld}{in units of $10^{-3}$}{upper limits are at 90\% CL}{/hfagc/00106.html}{tab:hfc01106}
\hfmetadata{ [ .tml created 2006-02-08T20:02:04.241+08:00] }

\begin{tabular}{lccccc}  
\hflmode
&  \hflpdg2004
&  \hflbelle
&  \hflbabar
&  \hflcdf
&  \hflavg
\\\hline 

 \hfhref{/hfagc/BR_-521_-413+421.html}{ \hflabel{D^0 D^{*-}(2010)} } \hftabletlcell 
 &  \hfhref{/hfagc/BR_-521_-413+421.html}{ \hfpdg{<13.0} }  
 &  \hfhref{/hfagc/0506037.html}{ \hfprehot{0.46\pm0.07\pm0.06} }  
 &  { \,}  
 &  { \,}  
 &  \hfhref{/hfagc/BR_-521_-413+421.html}{ \hfavg{0.46\pm0.09} } 
\\%

 \hfhref{/hfagc/BR_-521_-411+421.html}{ \hflabel{D^0 D^-} }  
 &  \hfhref{/hfagc/BR_-521_-411+421.html}{ \hfpdg{<6.7} }  
 &  \hfhref{/hfagc/0506037.html}{ \hfprehot{0.56\pm0.08\pm0.06} }  
 &  { \,}  
 &  { \,}  
 &  \hfhref{/hfagc/BR_-521_-411+421.html}{ \hfavg{0.56\pm0.10} } 
\\%

 \hfhref{/hfagc/BR_-521_-321+-411+411.html}{ \hflabel{D^+ D^- K^-} }  
 &  \hfhref{/hfagc/BR_-521_-321+-411+411.html}{ \hfpdg{<0.40} }  
 &  \hfhref{/hfagc/0506075.html}{ \hfpub{<0.90} }  
 &  \hfhref{/hfagc/0506112.html}{ \hfpubold{<0.40} }  
 &  { \,}  
 &  \hfhref{/hfagc/BR_-521_-321+-411+411.html}{ \hfavg{<0.40} } 
\\%

 \hfhref{/hfagc/BR_-521_-321+-413+411.html}{ \hflabel{D^+ D^{*-}(2010) K^-} }  
 &  { \,}  
 &  { \,}  
 &  \hfhref{/hfagc/0506112.html}{ \hfpubold{<0.70} }  
 &  { \,}  
 &  \hfhref{/hfagc/BR_-521_-321+-413+411.html}{ \hfavg{<0.70} } 
\\%

 \hfhref{/hfagc/BR_-521_-321+-421+421.html}{ \hflabel{D^0 \bar{D}^0 K^-} }  
 &  \hfhref{/hfagc/BR_-521_-321+-421+421.html}{ \hfpdg{1.90\pm0.40} }  
 &  \hfhref{/hfagc/0506075.html}{ \hfpub{1.17\pm0.21\pm0.15} }  
 &  \hfhref{/hfagc/0506112.html}{ \hfpubold{1.90\pm0.30\pm0.30} }  
 &  { \,}  
 &  \hfhref{/hfagc/BR_-521_-321+-421+421.html}{ \hfavg{1.37\pm0.22} } 
\\%

 \hfhref{/hfagc/BR_-521_-321+-411+413.html}{ \hflabel{D^{*+}(2010) D^- K^-} }  
 &  { \,}  
 &  { \,}  
 &  \hfhref{/hfagc/0506112.html}{ \hfpubold{1.50\pm0.30\pm0.20} }  
 &  { \,}  
 &  \hfhref{/hfagc/BR_-521_-321+-411+413.html}{ \hfavg{1.50\pm0.36} } 
\\
\hline 

\end{tabular}

\end{center}

\end{\hftabletype}
\clearpage  
\begin{\hftabletype}[\hftableposn]

\begin{center}
\hfnewcaption{Product branching fractions}{charged B}{\hfmuld}{in units of $10^{-4}$}{upper limits are at 90\% CL}{/hfagc/00106.html}{tab:hfc02106}
\hfmetadata{ [ .tml created 2006-02-08T20:02:07.887+08:00] }

\begin{tabular}{lccccc}  
\hflmode
&  \hflpdg2004
&  \hflbelle
&  \hflbabar
&  \hflcdf
&  \hflavg
\\\hline 

 \hfhref{/hfagc/BR_-521_-211+10423xBR_10423_-211+211+423.html}{ \hflabel{\pi^- D_1^0(2420) [ D^{*0}(2007) \pi^- \pi^+ ]} } \hftabletlcell 
 &  { \,}  
 &  \hfhref{/hfagc/0506030.html}{ \hfpubhot{<0.060} }  
 &  { \,}  
 &  { \,}  
 &  \hfhref{/hfagc/BR_-521_-211+10423xBR_10423_-211+211+423.html}{ \hfavg{<0.060} } 
\\%

 \hfhref{/hfagc/BR_-521_-211+425xBR_425_-211+211+423.html}{ \hflabel{\pi^- D_2^{*0}(2460) [ D^{*0}(2007) \pi^- \pi^+ ]} }  
 &  { \,}  
 &  \hfhref{/hfagc/0506030.html}{ \hfpubhot{<0.22} }  
 &  { \,}  
 &  { \,}  
 &  \hfhref{/hfagc/BR_-521_-211+425xBR_425_-211+211+423.html}{ \hfavg{<0.22} } 
\\%

 \hfhref{/hfagc/BR_-521_-211+425xBR_425_-211+413.html}{ \hflabel{\pi^- D_2^{*0}(2460) [ D^{*+}(2010) \pi^- ]} }  
 &  { \,}  
 &  \hfhref{/hfagc/0506071.html}{ \hfpub{1.80\pm0.30\pm0.30\pm0.20} }  
 &  \hfhref{/hfagc/0506113.html}{ \hfpreold{1.80\pm0.30\pm0.50} }  
 &  { \,}  
 &  \hfhref{/hfagc/BR_-521_-211+425xBR_425_-211+413.html}{ \hfavg{1.80\pm0.36} } 
\\%

 \hfhref{/hfagc/BR_-521_-211+10423xBR_10423_-211+211+421.html}{ \hflabel{\pi^- D_1^0(2420) [ D^0 \pi^- \pi^+ ]} }  
 &  { \,}  
 &  \hfhref{/hfagc/0506030.html}{ \hfpubhot{1.85\pm0.29\pm0.35\pm^{0.00}_{0.46}} }  
 &  { \,}  
 &  { \,}  
 &  \hfhref{/hfagc/BR_-521_-211+10423xBR_10423_-211+211+421.html}{ \hfavg{1.85\pm^{0.45}_{0.65}} } 
\\%

 \hfhref{/hfagc/BR_-521_-211+425xBR_425_-211+411.html}{ \hflabel{\pi^- D_2^{*0}(2460) [ D^+ \pi^- ]} }  
 &  { \,}  
 &  \hfhref{/hfagc/0506071.html}{ \hfpub{3.4\pm0.3\pm0.6\pm0.4} }  
 &  \hfhref{/hfagc/0506113.html}{ \hfpreold{2.9\pm0.2\pm0.5} }  
 &  { \,}  
 &  \hfhref{/hfagc/BR_-521_-211+425xBR_425_-211+411.html}{ \hfavg{3.1\pm0.4} } 
\\
\hline 

\end{tabular}

\end{center}


\begin{center}
\hfnewcaption{Ratios of branching fractions}{charged B}{\hfmuld}{in units of $10^{0}$}{upper limits are at 90\% CL}{/hfagc/00106.html}{tab:hfc03106}
\hfmetadata{ [ .tml created 2006-02-08T20:02:10.064+08:00] }

\begin{tabular}{lccccc}  
\hflmode
&  \hflpdg2004
&  \hflbelle
&  \hflbabar
&  \hflcdf
&  \hflavg
\\\hline 

 \hfhref{/hfagc/BR_-521_-321+423xoBR_-521_-211+423.html}{ \hflabel{\frac{{\cal{B}} ( B^- \to D^{*0}(2007) K^- )}{{\cal{B}} ( B^- \to D^{*0}(2007) \pi^- )}} } \hftabletlcell 
 &  \hfhref{/hfagc/BR_-521_-321+423xoBR_-521_-211+423.html}{ \hfpdg{0.078\pm0.021} }  
 &  \hfhref{/hfagc/0506081.html}{ \hfpubold{0.078\pm0.019\pm0.009} }  
 &  \hfhref{/hfagc/0601003.html}{ \hfpubhot{0.081\pm0.004\pm^{0.004}_{0.003}} }  
 &  { \,}  
 &  \hfhref{/hfagc/BR_-521_-321+423xoBR_-521_-211+423.html}{ \hfavg{0.081\pm0.005} } 
\\%

 \hfhref{/hfagc/BR_-521_-321+421xoBR_-521_-211+421.html}{ \hflabel{\frac{{\cal{B}} ( B^- \to D^0 K^- )}{{\cal{B}} ( B^- \to D^0 \pi^- )}} }  
 &  \hfhref{/hfagc/BR_-521_-321+421xoBR_-521_-211+421.html}{ \hfpdg{0.083\pm0.010} }  
 &  \hfhref{/hfagc/0506082.html}{ \hfpubold{0.077\pm0.005\pm0.006} }  
 &  \hfhref{/hfagc/0601002.html}{ \hfpub{0.083\pm0.003\pm0.002} }  
 &  { \,}  
 &  \hfhref{/hfagc/BR_-521_-321+421xoBR_-521_-211+421.html}{ \hfavg{0.082\pm0.004} } 
\\%

 \hfhref{/hfagc/BR_-521_-211+425xoBR_-521_-211+10423.html}{ \hflabel{\frac{{\cal{B}} ( B^- \to D_2^{*0}(2460) \pi^- )}{{\cal{B}} ( B^- \to D_1^0(2420) \pi^- )}} }  
 &  { \,}  
 &  { \,}  
 &  \hfhref{/hfagc/0506113.html}{ \hfpreold{0.80\pm0.07\pm0.16} }  
 &  { \,}  
 &  \hfhref{/hfagc/BR_-521_-211+425xoBR_-521_-211+10423.html}{ \hfavg{0.80\pm0.17} } 
\\%

 \hfhref{/hfagc/BR_-521_-211+421xoBR_-511_-211+411.html}{ \hflabel{\frac{{\cal{B}} ( B^- \to D^0 \pi^- )}{{\cal{B}} ( \bar{B}^0 \to D^+ \pi^- )}} } \hftableblcell 
 &  { \,}  
 &  { \,}  
 &  { \,}  
 &  \hfhref{/hfagc/0506099.html}{ \hfprehot{1.97\pm0.10\pm0.21} }  
 &  \hfhref{/hfagc/BR_-521_-211+421xoBR_-511_-211+411.html}{ \hfavg{1.97\pm0.23} } 
\\\hline 

\end{tabular}

\end{center}

\end{\hftabletype}
\clearpage  
\begin{\hftabletype}[\hftableposn]

\begin{center}
\hfnewcaption{Branching fractions}{charged B}{\hfsgdx}{in units of $10^{-4}$}{upper limits are at 90\% CL}{/hfagc/00107.html}{tab:hfc01107}
\hfmetadata{ [ .tml created 2006-02-08T20:02:13.184+08:00] }

\begin{tabular}{lccccc}  
\hflmode
&  \hflpdg2004
&  \hflbelle
&  \hflbabar
&  \hflcdf
&  \hflavg
\\\hline 

 \hfhref{/hfagc/BR_-521_-311+-413.html}{ \hflabel{D^{*-}(2010) \bar{K}^0} } \hftabletlcell 
 &  \hfhref{/hfagc/BR_-521_-311+-413.html}{ \hfpdg{<0.95} }  
 &  { \,}  
 &  \hfhref{/hfagc/0506016.html}{ \hfpubhot{<0.090} }  
 &  { \,}  
 &  \hfhref{/hfagc/BR_-521_-311+-413.html}{ \hfavg{<0.090} } 
\\%

 \hfhref{/hfagc/BR_-521_-321+423.html}{ \hflabel{D^{*0}(2007) K^-} }  
 &  \hfhref{/hfagc/BR_-521_-321+423.html}{ \hfpdg{3.6\pm1.0} }  
 &  \hfhref{/hfagc/0506081.html}{ \hfpubold{3.6\pm0.9\pm0.4\pm0.3} }  
 &  { \,}  
 &  { \,}  
 &  \hfhref{/hfagc/BR_-521_-321+423.html}{ \hfavg{3.6\pm1.0} } 
\\%

 \hfhref{/hfagc/BR_-521_-323+423.html}{ \hflabel{D^{*0}(2007) K^{*-}(892)} }  
 &  \hfhref{/hfagc/BR_-521_-323+423.html}{ \hfpdg{7.2\pm3.4} }  
 &  { \,}  
 &  \hfhref{/hfagc/0506105.html}{ \hfpub{8.3\pm1.1\pm1.0\pm0.3} }  
 &  { \,}  
 &  \hfhref{/hfagc/BR_-521_-323+423.html}{ \hfavg{8.3\pm1.5} } 
\\%

 \hfhref{/hfagc/BR_-521_-321+311+423.html}{ \hflabel{D^{*0}(2007) K^- K^0} }  
 &  \hfhref{/hfagc/BR_-521_-321+311+423.html}{ \hfpdg{<10.6} }  
 &  \hfhref{/hfagc/0506090.html}{ \hfpubold{<10.6} }  
 &  { \,}  
 &  { \,}  
 &  \hfhref{/hfagc/BR_-521_-321+311+423.html}{ \hfavg{<10.6} } 
\\%

 \hfhref{/hfagc/BR_-521_-211+-211+413.html}{ \hflabel{D^{*+}(2010) \pi^- \pi^-} }  
 &  \hfhref{/hfagc/BR_-521_-211+-211+413.html}{ \hfpdg{21\pm6} }  
 &  \hfhref{/hfagc/0506071.html}{ \hfpub{12.5\pm0.8\pm2.2} }  
 &  \hfhref{/hfagc/0506113.html}{ \hfpreold{12.2\pm0.5\pm1.8} }  
 &  { \,}  
 &  \hfhref{/hfagc/BR_-521_-211+-211+413.html}{ \hfavg{12.3\pm1.5} } 
\\%

 \hfhref{/hfagc/BR_-521_-321+313+423.html}{ \hflabel{D^{*0}(2007) K^- K^{*0}(892)} }  
 &  \hfhref{/hfagc/BR_-521_-321+313+423.html}{ \hfpdg{15.0\pm4.0} }  
 &  \hfhref{/hfagc/0506090.html}{ \hfpubold{15.3\pm3.1\pm2.9} }  
 &  { \,}  
 &  { \,}  
 &  \hfhref{/hfagc/BR_-521_-321+313+423.html}{ \hfavg{15.3\pm4.2} } 
\\
\hline 

\end{tabular}

\end{center}


\begin{center}
\hfnewcaption{Branching fractions}{charged B}{\hfsgld}{in units of $10^{-4}$}{upper limits are at 90\% CL}{/hfagc/00108.html}{tab:hfc01108}
\hfmetadata{ [ .tml created 2006-02-08T20:02:20.148+08:00] }

\begin{tabular}{lccccc}  
\hflmode
&  \hflpdg2004
&  \hflbelle
&  \hflbabar
&  \hflcdf
&  \hflavg
\\\hline 

 \hfhref{/hfagc/BR_-521_-311+-411.html}{ \hflabel{D^- \bar{K}^0} } \hftabletlcell 
 &  { \,}  
 &  { \,}  
 &  \hfhref{/hfagc/0506016.html}{ \hfpubhot{<0.050} }  
 &  { \,}  
 &  \hfhref{/hfagc/BR_-521_-311+-411.html}{ \hfavg{<0.050} } 
\\%

 \hfhref{/hfagc/BR_-521_-321+421.html}{ \hflabel{D^0 K^-} }  
 &  \hfhref{/hfagc/BR_-521_-321+421.html}{ \hfpdg{3.7\pm0.6} }  
 &  \hfhref{/hfagc/0506082.html}{ \hfpubold{3.8\pm0.2\pm0.3\pm0.2} }  
 &  { \,}  
 &  { \,}  
 &  \hfhref{/hfagc/BR_-521_-321+421.html}{ \hfavg{3.8\pm0.4} } 
\\%

 \hfhref{/hfagc/BR_-521_-321+311+421.html}{ \hflabel{D^0 K^- K^0} }  
 &  \hfhref{/hfagc/BR_-521_-321+311+421.html}{ \hfpdg{5.5\pm1.6} }  
 &  \hfhref{/hfagc/0506090.html}{ \hfpubold{5.5\pm1.4\pm0.8} }  
 &  { \,}  
 &  { \,}  
 &  \hfhref{/hfagc/BR_-521_-321+311+421.html}{ \hfavg{5.5\pm1.6} } 
\\%

 \hfhref{/hfagc/BR_-521_-323+421.html}{ \hflabel{D^0 K^{*-}(892)} }  
 &  \hfhref{/hfagc/BR_-521_-323+421.html}{ \hfpdg{6.1\pm2.3} }  
 &  { \,}  
 &  \hfhref{/hfagc/0506104.html}{ \hfpub{6.3\pm0.7\pm0.5} }  
 &  { \,}  
 &  \hfhref{/hfagc/BR_-521_-323+421.html}{ \hfavg{6.3\pm0.9} } 
\\%

 \hfhref{/hfagc/BR_-521_-321+313+421.html}{ \hflabel{D^0 K^- K^{*0}(892)} }  
 &  \hfhref{/hfagc/BR_-521_-321+313+421.html}{ \hfpdg{7.5\pm1.7} }  
 &  \hfhref{/hfagc/0506090.html}{ \hfpubold{7.5\pm1.3\pm1.1} }  
 &  { \,}  
 &  { \,}  
 &  \hfhref{/hfagc/BR_-521_-321+313+421.html}{ \hfavg{7.5\pm1.7} } 
\\%

 \hfhref{/hfagc/BR_-521_-211+-211+411.html}{ \hflabel{D^+ \pi^- \pi^-} } \hftableblcell 
 &  \hfhref{/hfagc/BR_-521_-211+-211+411.html}{ \hfpdg{<14.0} }  
 &  \hfhref{/hfagc/0506071.html}{ \hfpub{10.2\pm0.4\pm1.5} }  
 &  \hfhref{/hfagc/0506113.html}{ \hfpreold{8.7\pm0.4\pm1.3} }  
 &  { \,}  
 &  \hfhref{/hfagc/BR_-521_-211+-211+411.html}{ \hfavg{9.4\pm1.0} } 
\\\hline 

\end{tabular}

\end{center}

\end{\hftabletype}
\clearpage  
\begin{\hftabletype}[\hftableposn]

\begin{center}
\hfnewcaption{Branching fractions}{neutral B}{\hfnewp}{in units of $10^{-4}$}{upper limits are at 90\% CL}{/hfagc/00201.html}{tab:hfc01201}
\hfmetadata{ [ .tml created 2006-02-08T20:02:31.059+08:00] }

\begin{tabular}{lccccc}  
\hflmode
&  \hflpdg2004
&  \hflbelle
&  \hflbabar
&  \hflcdf
&  \hflavg
\\\hline 

 \hfhref{/hfagc/BR_-511_-321+82.html}{ \hflabel{X^{+}(3870) K^-} } \hftabletlcell\hftableblcell 
 &  { \,}  
 &  { \,}  
 &  \hfhref{/hfagc/0505003.html}{ \hfprehot{<5.0} }  
 &  { \,}  
 &  \hfhref{/hfagc/BR_-511_-321+82.html}{ \hfavg{<5.0} } 
\\\hline 

\end{tabular}

\end{center}


\begin{center}
\hfnewcaption{Product branching fractions}{neutral B}{\hfnewp}{in units of $10^{-4}$}{upper limits are at 90\% CL}{/hfagc/00201.html}{tab:hfc02201}
\hfmetadata{ [ .tml created 2006-02-08T20:02:32.237+08:00] }

\begin{tabular}{lccccc}  
\hflmode
&  \hflpdg2004
&  \hflbelle
&  \hflbabar
&  \hflcdf
&  \hflavg
\\\hline 

 \hfhref{/hfagc/BR_-511_-85+211xBR_-85_-431+22.html}{ \hflabel{\pi^+ D_{sJ}^{-}(2460) [ D_s^- \gamma ]} } \hftabletlcell 
 &  { \,}  
 &  \hfhref{/hfagc/0506032.html}{ \hfpubhot{<0.040} }  
 &  { \,}  
 &  { \,}  
 &  \hfhref{/hfagc/BR_-511_-85+211xBR_-85_-431+22.html}{ \hfavg{<0.040} } 
\\%

 \hfhref{/hfagc/BR_-511_-311+81xBR_81_-211+211+443.html}{ \hflabel{\bar{K}^0 X(3870) [ J/\psi(1S) \pi^+ \pi^- ]} }  
 &  { \,}  
 &  { \,}  
 &  \hfhref{/hfagc/0507002.html}{ \hfpub{0.051\pm0.028\pm0.007} }  
 &  { \,}  
 &  \hfhref{/hfagc/BR_-511_-311+81xBR_81_-211+211+443.html}{ \hfavg{0.051\pm0.029} } 
\\%

 \hfhref{/hfagc/BR_-511_-321+82xBR_82_111+211+443.html}{ \hflabel{K^- X^{+}(3870) [ J/\psi(1S) \pi^+ \pi^0 ]} }  
 &  { \,}  
 &  { \,}  
 &  \hfhref{/hfagc/0506009.html}{ \hfpubhot{<0.054} }  
 &  { \,}  
 &  \hfhref{/hfagc/BR_-511_-321+82xBR_82_111+211+443.html}{ \hfavg{<0.054} } 
\\%

 \hfhref{/hfagc/BR_-511_-321+85xBR_85_22+431.html}{ \hflabel{K^- D_{sJ}^{+}(2460) [ D_s^+ \gamma ]} }  
 &  { \,}  
 &  { \,}  
 &  { \,}  
 &  { \,}  
 &  \hfhref{/hfagc/BR_-511_-321+85xBR_85_22+431.html}{ \hfavg{<0.086} } 
\\%

{ \hflabel{\,} }  
 &  { \,}  
 &  \hfhref{/hfagc/0506051.html}{ \hfprehot{<0.086} } 
\hffootnotemark{1a}
 
 &  { \,}  
 &  { \,}  
 &  { \,} 
\\%

{ \hflabel{\,} }  
 &  { \,}  
 &  \hfhref{/hfagc/0506032.html}{ \hfsuperceeded{<0.094} } 
\hffootnotemark{2a}
 
 &  { \,}  
 &  { \,}  
 &  { \,} 
\\%

 \hfhref{/hfagc/BR_-511_-84+211xBR_-84_-431+111.html}{ \hflabel{\pi^+ D_{sJ}^{*}(2317)^{-} [ D_s^- \pi^0 ]} }  
 &  { \,}  
 &  \hfhref{/hfagc/0506032.html}{ \hfpubhot{<0.25} }  
 &  { \,}  
 &  { \,}  
 &  \hfhref{/hfagc/BR_-511_-84+211xBR_-84_-431+111.html}{ \hfavg{<0.25} } 
\\
\hline 

\end{tabular}

\begin{description}\addtolength{\itemsep}{\hffootitemsep\baselineskip}

	  \item[ \hffootnotemark{1} ] \hffootnotetext{ Improved Measurements of $\bar{B}^0 \rightarrow D_{sJ}^+ K^-$ decays (\hfbb{386}) } 
     ;  \hffootnotemark{1a}  \hffootnotetext{ improved limit }  ;  \hffootnotemark{1b}  \hffootnotetext{ improved measurement } 
	  \item[ \hffootnotemark{2} ] \hffootnotetext{ Observation of $\bar{B}^0 \to D_{sJ}^{*}(2317)^{+} K^- $
decay (\hfbb{152}) } 
     ;  \hffootnotemark{2a}  \hffootnotetext{ first measurement }  ;  \hffootnotemark{2b}  \hffootnotetext{ first measurement } 
\end{description}

\end{center}

\end{\hftabletype}
\clearpage  
\begin{\hftabletype}[\hftableposn]

\begin{center}
\hfnewcaption{Ratios of branching fractions}{neutral B}{\hfnewp}{in units of $10^{0}$}{upper limits are at 90\% CL}{/hfagc/00201.html}{tab:hfc03201}
\hfmetadata{ [ .tml created 2006-02-08T20:02:37.247+08:00] }

\begin{tabular}{lccccc}  
\hflmode
&  \hflpdg2004
&  \hflbelle
&  \hflbabar
&  \hflcdf
&  \hflavg
\\\hline 

 \hfhref{/hfagc/BR_-511_-311+81xoBR_-521_-321+81.html}{ \hflabel{\frac{{\cal{B}} ( \bar{B}^0 \to X(3870) \bar{K}^0 )}{{\cal{B}} ( B^- \to X(3870) K^- )}} } \hftabletlcell\hftableblcell 
 &  { \,}  
 &  { \,}  
 &  \hfhref{/hfagc/0507002.html}{ \hfpub{0.50\pm0.30\pm0.05} }  
 &  { \,}  
 &  \hfhref{/hfagc/BR_-511_-311+81xoBR_-521_-321+81.html}{ \hfavg{0.50\pm0.30} } 
\\\hline 

\end{tabular}

\end{center}


\begin{center}
\hfnewcaption{Branching fractions}{neutral B}{\hfdstr}{in units of $10^{-3}$}{upper limits are at 90\% CL}{/hfagc/00202.html}{tab:hfc01202}
\hfmetadata{ [ .tml created 2006-02-08T20:02:42.756+08:00] }

\begin{tabular}{lccccc}  
\hflmode
&  \hflpdg2004
&  \hflbelle
&  \hflbabar
&  \hflcdf
&  \hflavg
\\\hline 

 \hfhref{/hfagc/BR_-511_-431+10211.html}{ \hflabel{D_s^- a_0^+(1450)} } \hftabletlcell 
 &  { \,}  
 &  { \,}  
 &  \hfhref{/hfagc/0506017.html}{ \hfprehot{<0.019} }  
 &  { \,}  
 &  \hfhref{/hfagc/BR_-511_-431+10211.html}{ \hfavg{<0.019} } 
\\%

 \hfhref{/hfagc/BR_-511_-431+213.html}{ \hflabel{D_s^- \rho^+(770)} }  
 &  \hfhref{/hfagc/BR_-511_-431+213.html}{ \hfpdg{<0.70} }  
 &  { \,}  
 &  \hfhref{/hfagc/0506022.html}{ \hfpre{<0.019} }  
 &  { \,}  
 &  \hfhref{/hfagc/BR_-511_-431+213.html}{ \hfavg{<0.019} } 
\\%

 \hfhref{/hfagc/BR_-511_-321+433.html}{ \hflabel{D_s^{*+} K^-} }  
 &  \hfhref{/hfagc/BR_-511_-321+433.html}{ \hfpdg{<0.025} }  
 &  { \,}  
 &  \hfhref{/hfagc/0505001.html}{ \hfpubold{<0.025} }  
 &  { \,}  
 &  \hfhref{/hfagc/BR_-511_-321+433.html}{ \hfavg{<0.025} } 
\\%

 \hfhref{/hfagc/BR_-511_-431+211.html}{ \hflabel{D_s^- \pi^+} }  
 &  \hfhref{/hfagc/BR_-511_-431+211.html}{ \hfpdg{0.027\pm0.010} }  
 &  \hfhref{/hfagc/0506089.html}{ \hfpubold{0.024\pm^{0.010}_{0.008}\pm0.004\pm0.006} }  
 &  \hfhref{/hfagc/0505001.html}{ \hfpubold{0.032\pm0.009\pm0.007\pm0.008} }  
 &  { \,}  
 &  \hfhref{/hfagc/BR_-511_-431+211.html}{ \hfavg{0.027\pm0.009} } 
\\%

 \hfhref{/hfagc/BR_-511_-433+10211.html}{ \hflabel{D_s^{*-} a_0^+(1450)} }  
 &  { \,}  
 &  { \,}  
 &  \hfhref{/hfagc/0506017.html}{ \hfprehot{<0.036} }  
 &  { \,}  
 &  \hfhref{/hfagc/BR_-511_-433+10211.html}{ \hfavg{<0.036} } 
\\%

 \hfhref{/hfagc/BR_-511_-321+431.html}{ \hflabel{D_s^+ K^-} }  
 &  \hfhref{/hfagc/BR_-511_-321+431.html}{ \hfpdg{0.038\pm0.013} }  
 &  \hfhref{/hfagc/0506089.html}{ \hfpubold{0.046\pm^{0.012}_{0.011}\pm0.006\pm0.012} }  
 &  \hfhref{/hfagc/0505001.html}{ \hfpubold{0.032\pm0.010\pm0.007\pm0.008} }  
 &  { \,}  
 &  \hfhref{/hfagc/BR_-511_-321+431.html}{ \hfavg{0.037\pm0.011} } 
\\%

 \hfhref{/hfagc/BR_-511_-433+211.html}{ \hflabel{D_s^{*-} \pi^+} }  
 &  \hfhref{/hfagc/BR_-511_-433+211.html}{ \hfpdg{<0.041} }  
 &  { \,}  
 &  \hfhref{/hfagc/0505001.html}{ \hfpubold{<0.041} }  
 &  { \,}  
 &  \hfhref{/hfagc/BR_-511_-433+211.html}{ \hfavg{<0.041} } 
\\%

 \hfhref{/hfagc/BR_-511_-431+433.html}{ \hflabel{D_s^- D_s^{*+}} }  
 &  { \,}  
 &  { \,}  
 &  \hfhref{/hfagc/0506021.html}{ \hfprehot{<0.130} }  
 &  { \,}  
 &  \hfhref{/hfagc/BR_-511_-431+433.html}{ \hfavg{<0.130} } 
\\%

 \hfhref{/hfagc/BR_-511_-431+431.html}{ \hflabel{D_s^- D_s^+} }  
 &  { \,}  
 &  \hfhref{/hfagc/0506066.html}{ \hfprehot{<0.200} }  
 &  \hfhref{/hfagc/0506021.html}{ \hfprehot{<0.100} }  
 &  { \,}  
 &  \hfhref{/hfagc/BR_-511_-431+431.html}{ \hfavg{<0.100} } 
\\%

 \hfhref{/hfagc/BR_-511_-431+215.html}{ \hflabel{D_s^- a_2^+(1320)} }  
 &  { \,}  
 &  { \,}  
 &  \hfhref{/hfagc/0506017.html}{ \hfprehot{<0.190} }  
 &  { \,}  
 &  \hfhref{/hfagc/BR_-511_-431+215.html}{ \hfavg{<0.190} } 
\\%

 \hfhref{/hfagc/BR_-511_-433+215.html}{ \hflabel{D_s^{*-} a_2^+(1320)} }  
 &  { \,}  
 &  { \,}  
 &  \hfhref{/hfagc/0506017.html}{ \hfprehot{<0.200} }  
 &  { \,}  
 &  \hfhref{/hfagc/BR_-511_-433+215.html}{ \hfavg{<0.200} } 
\\%

 \hfhref{/hfagc/BR_-511_-433+433.html}{ \hflabel{D_s^{*+} D_s^{*-}} }  
 &  { \,}  
 &  { \,}  
 &  \hfhref{/hfagc/0506021.html}{ \hfprehot{<0.24} }  
 &  { \,}  
 &  \hfhref{/hfagc/BR_-511_-433+433.html}{ \hfavg{<0.24} } 
\\%

 \hfhref{/hfagc/BR_-511_-431+411.html}{ \hflabel{D_s^- D^+} }  
 &  \hfhref{/hfagc/BR_-511_-431+411.html}{ \hfpdg{8.0\pm3.0} }  
 &  \hfhref{/hfagc/0506066.html}{ \hfprehot{7.4\pm0.2\pm1.4} }  
 &  { \,}  
 &  { \,}  
 &  \hfhref{/hfagc/BR_-511_-431+411.html}{ \hfavg{7.4\pm1.4} } 
\\%

 \hfhref{/hfagc/BR_-511_-431+413.html}{ \hflabel{D_s^- D^{*+}(2010)} }  
 &  \hfhref{/hfagc/BR_-511_-431+413.html}{ \hfpdg{10.7\pm2.9} }  
 &  { \,}  
 &  \hfhref{/hfagc/0506110.html}{ \hfpubold{10.3\pm1.4\pm1.3\pm2.6} }  
 &  { \,}  
 &  \hfhref{/hfagc/BR_-511_-431+413.html}{ \hfavg{10.3\pm3.2} } 
\\%

 \hfhref{/hfagc/BR_-511_-433+413.html}{ \hflabel{D_s^{*-} D^{*+}(2010)} }  
 &  \hfhref{/hfagc/BR_-511_-433+413.html}{ \hfpdg{19.0\pm5.0} }  
 &  { \,}  
 &  { \,}  
 &  { \,}  
 &  \hfhref{/hfagc/BR_-511_-433+413.html}{ \hfavg{18.9\pm1.8} } 
\\%

{ \hflabel{\,} }  
 &  { \,}  
 &  { \,}  
 &  \hfhref{/hfagc/0506019.html}{ \hfpubhot{18.8\pm0.9\pm1.6\pm0.6} } 
\hffootnotemark{1}
 
 &  { \,}  
 &  { \,} 
\\%

{ \hflabel{\,} } \hftableblcell 
 &  { \,}  
 &  { \,}  
 &  \hfhref{/hfagc/0506110.html}{ \hfpubold{19.7\pm1.5\pm3.0\pm4.9} } 
\hffootnotemark{2}
 
 &  { \,}  
 &  { \,} 
\\\hline 

\end{tabular}

\begin{description}\addtolength{\itemsep}{\hffootitemsep\baselineskip}

	  \item[ \hffootnotemark{1} ] \hffootnotetext{ Measurement of the $\bar{B}^0 \rightarrow D_s^{*-} D^+$ and $D_s^+ \rightarrow \phi \pi^+$ branching
fractions (\hfbb{123}) } 
     ;  \hffootnotetext{ $\bar{B}^0 \rightarrow D_s^{*-} D^{*+}$ } 
	  \item[ \hffootnotemark{2} ] \hffootnotetext{ Measurement of $\bar{B}^0 \rightarrow D_s^{(*)}D^*$ Branching Fractions and $D_s^*D^*$
 Polarization with a Partial Reconstruction technique (\hfbb{22.7}) } 
     ;  \hffootnotetext{ $\bar{B}^0 \rightarrow D_s^{*-} D^{*+}$ } 
\end{description}

\end{center}


\begin{center}
\hfnewcaption{Product branching fractions}{neutral B}{\hfdstr}{in units of $10^{-4}$}{upper limits are at 90\% CL}{/hfagc/00202.html}{tab:hfc02202}
\hfmetadata{ [ .tml created 2006-02-08T20:02:46.413+08:00] }

\begin{tabular}{lccccc}  
\hflmode
&  \hflpdg2004
&  \hflbelle
&  \hflbabar
&  \hflcdf
&  \hflavg
\\\hline 

 \hfhref{/hfagc/BR_-511_-431+411xBR_-431_-211+333xBR_333_-321+321.html}{ \hflabel{D^+ D_s^- [ \pi^- \phi(1020) [ K^+ K^- ] ]} } \hftabletlcell\hftableblcell 
 &  \hfhref{/hfagc/BR_-511_-431+411xBR_-431_-211+333xBR_333_-321+321.html}{ \hfpdg{1.42\pm0.40} }  
 &  \hfhref{/hfagc/0506066.html}{ \hfprehot{1.47\pm0.05\pm0.21} }  
 &  { \,}  
 &  { \,}  
 &  \hfhref{/hfagc/BR_-511_-431+411xBR_-431_-211+333xBR_333_-321+321.html}{ \hfavg{1.47\pm0.22} } 
\\\hline 

\end{tabular}

\end{center}

\end{\hftabletype}
\clearpage  
\begin{\hftabletype}[\hftableposn]

\begin{center}
\hfnewcaption{Ratios of branching fractions}{neutral B}{\hfdstr}{in units of $10^{0}$}{upper limits are at 90\% CL}{/hfagc/00202.html}{tab:hfc03202}
\hfmetadata{ [ .tml created 2006-02-08T20:02:48.06+08:00] }

\begin{tabular}{lccccc}  
\hflmode
&  \hflpdg2004
&  \hflbelle
&  \hflbabar
&  \hflcdf
&  \hflavg
\\\hline 

 \hfhref{/hfagc/BR_-511_-433+411xoBR_-511_-431+411.html}{ \hflabel{\frac{{\cal{B}} ( \bar{B}^0 \to D_s^{*-} D^+ )}{{\cal{B}} ( \bar{B}^0 \to D_s^- D^+ )}} } \hftabletlcell 
 &  { \,}  
 &  { \,}  
 &  { \,}  
 &  \hfhref{/hfagc/0601004.html}{ \hfwaiting{0.90\pm0.20\pm0.10} }  
 &  \hfhref{/hfagc/BR_-511_-433+411xoBR_-511_-431+411.html}{ \hfavg{0.90\pm0.22} } 
\\%

 \hfhref{/hfagc/BR_-511_-431+413xoBR_-511_-431+411.html}{ \hflabel{\frac{{\cal{B}} ( \bar{B}^0 \to D_s^- D^{*+}(2010) )}{{\cal{B}} ( \bar{B}^0 \to D_s^- D^+ )}} }  
 &  { \,}  
 &  { \,}  
 &  { \,}  
 &  \hfhref{/hfagc/0601004.html}{ \hfwaiting{1.50\pm0.50\pm0.10} }  
 &  \hfhref{/hfagc/BR_-511_-431+413xoBR_-511_-431+411.html}{ \hfavg{1.50\pm0.51} } 
\\%

 \hfhref{/hfagc/BR_-511_-431+411xoBR_-511_-211+-211+211+411.html}{ \hflabel{\frac{{\cal{B}} ( \bar{B}^0 \to D_s^- D^+ )}{{\cal{B}} ( \bar{B}^0 \to D^+ \pi^+ \pi^- \pi^- )}} }  
 &  { \,}  
 &  { \,}  
 &  { \,}  
 &  \hfhref{/hfagc/0601004.html}{ \hfwaiting{1.99\pm0.13\pm0.11\pm0.45} }  
 &  \hfhref{/hfagc/BR_-511_-431+411xoBR_-511_-211+-211+211+411.html}{ \hfavg{1.99\pm0.48} } 
\\%

 \hfhref{/hfagc/BR_-511_-433+413xoBR_-511_-431+411.html}{ \hflabel{\frac{{\cal{B}} ( \bar{B}^0 \to D_s^{*-} D^{*+}(2010) )}{{\cal{B}} ( \bar{B}^0 \to D_s^- D^+ )}} } \hftableblcell 
 &  { \,}  
 &  { \,}  
 &  { \,}  
 &  \hfhref{/hfagc/0601004.html}{ \hfwaiting{2.6\pm0.5\pm0.2} }  
 &  \hfhref{/hfagc/BR_-511_-433+413xoBR_-511_-431+411.html}{ \hfavg{2.6\pm0.5} } 
\\\hline 

\end{tabular}

\end{center}


\begin{center}
\hfnewcaption{Branching fractions}{neutral B}{\hfbary}{in units of $10^{-5}$}{upper limits are at 90\% CL}{/hfagc/00203.html}{tab:hfc01203}
\hfmetadata{ [ .tml created 2006-02-08T20:02:53.476+08:00] }

\begin{tabular}{lccccc}  
\hflmode
&  \hflpdg2004
&  \hflbelle
&  \hflbabar
&  \hflcdf
&  \hflavg
\\\hline 

 \hfhref{/hfagc/BR_-511_-2212+2212+443.html}{ \hflabel{J/\psi(1S) \bar{p} p} } \hftabletlcell 
 &  \hfhref{/hfagc/BR_-511_-2212+2212+443.html}{ \hfpdg{<0.190} }  
 &  { \,}  
 &  \hfhref{/hfagc/0506010.html}{ \hfpubold{<0.190} }  
 &  { \,}  
 &  \hfhref{/hfagc/BR_-511_-2212+2212+443.html}{ \hfavg{<0.190} } 
\\%

 \hfhref{/hfagc/BR_-511_-2212+4122.html}{ \hflabel{\Lambda_c^+ \bar{p}} }  
 &  \hfhref{/hfagc/BR_-511_-2212+4122.html}{ \hfpdg{2.2\pm0.8} }  
 &  \hfhref{/hfagc/0506087.html}{ \hfpubold{2.2\pm^{0.6}_{0.5}\pm0.3\pm0.6} }  
 &  { \,}  
 &  { \,}  
 &  \hfhref{/hfagc/BR_-511_-2212+4122.html}{ \hfavg{2.2\pm^{0.9}_{0.8}} } 
\\%

 \hfhref{/hfagc/BR_-511_-2212+2212+423.html}{ \hflabel{D^{*0}(2007) p \bar{p}} }  
 &  { \,}  
 &  \hfhref{/hfagc/0506083.html}{ \hfpubold{12.0\pm^{3.3}_{2.9}\pm2.1} }  
 &  \hfhref{/hfagc/0506024.html}{ \hfpre{6.7\pm2.1\pm0.8\pm0.4} }  
 &  { \,}  
 &  \hfhref{/hfagc/BR_-511_-2212+2212+423.html}{ \hfavg{8.0\pm2.0} } 
\\%

 \hfhref{/hfagc/BR_-511_-2212+211+4114.html}{ \hflabel{\Sigma_c^{*0} \bar{p} \pi^+} }  
 &  \hfhref{/hfagc/BR_-511_-2212+211+4114.html}{ \hfpdg{<12.1} }  
 &  \hfhref{/hfagc/0506086.html}{ \hfpubold{<12.1} }  
 &  { \,}  
 &  { \,}  
 &  \hfhref{/hfagc/BR_-511_-2212+211+4114.html}{ \hfavg{<12.1} } 
\\%

 \hfhref{/hfagc/BR_-511_-2212+2212+421.html}{ \hflabel{D^0 p \bar{p}} }  
 &  { \,}  
 &  \hfhref{/hfagc/0506083.html}{ \hfpubold{11.8\pm1.5\pm1.6} }  
 &  \hfhref{/hfagc/0506024.html}{ \hfpre{12.4\pm1.4\pm1.2\pm0.3} }  
 &  { \,}  
 &  \hfhref{/hfagc/BR_-511_-2212+2212+421.html}{ \hfavg{12.2\pm1.4} } 
\\%

 \hfhref{/hfagc/BR_-511_-2212+211+4112.html}{ \hflabel{\Sigma_c^0 \bar{p} \pi^+} }  
 &  \hfhref{/hfagc/BR_-511_-2212+211+4112.html}{ \hfpdg{10.0\pm8.0} }  
 &  \hfhref{/hfagc/0506086.html}{ \hfpubold{<15.9} }  
 &  { \,}  
 &  { \,}  
 &  \hfhref{/hfagc/BR_-511_-2212+211+4112.html}{ \hfavg{<15.9} } 
\\
\hline 

\end{tabular}

\end{center}

\end{\hftabletype}
\clearpage  
\begin{\hftabletype}[\hftableposn]

\begin{center}
\hfnewcaption{Branching fractions}{neutral B}{\hfjpsi}{in units of $10^{-4}$}{upper limits are at 90\% CL}{/hfagc/00204.html}{tab:hfc01204}
\hfmetadata{ [ .tml created 2006-02-08T20:03:02.54+08:00] }

\begin{tabular}{lccccc}  
\hflmode
&  \hflpdg2004
&  \hflbelle
&  \hflbabar
&  \hflcdf
&  \hflavg
\\\hline 

 \hfhref{/hfagc/BR_-511_22+443.html}{ \hflabel{J/\psi(1S) \gamma} } \hftabletlcell 
 &  { \,}  
 &  { \,}  
 &  \hfhref{/hfagc/0506002.html}{ \hfpub{<0.016} }  
 &  { \,}  
 &  \hfhref{/hfagc/BR_-511_22+443.html}{ \hfavg{<0.016} } 
\\%

 \hfhref{/hfagc/BR_-511_333+443.html}{ \hflabel{J/\psi(1S) \phi(1020)} }  
 &  \hfhref{/hfagc/BR_-511_333+443.html}{ \hfpdg{<0.092} }  
 &  { \,}  
 &  \hfhref{/hfagc/0506012.html}{ \hfpubold{<0.090} }  
 &  { \,}  
 &  \hfhref{/hfagc/BR_-511_333+443.html}{ \hfavg{<0.090} } 
\\%

 \hfhref{/hfagc/BR_-511_225+443.html}{ \hflabel{J/\psi(1S) f_2(1270)} }  
 &  { \,}  
 &  \hfhref{/hfagc/0506078.html}{ \hfpre{0.098\pm0.039\pm0.020} }  
 &  { \,}  
 &  { \,}  
 &  \hfhref{/hfagc/BR_-511_225+443.html}{ \hfavg{0.098\pm0.044} } 
\\%

 \hfhref{/hfagc/BR_-511_421+443.html}{ \hflabel{J/\psi(1S) D^0} }  
 &  { \,}  
 &  \hfhref{/hfagc/0506038.html}{ \hfpubhot{<0.200} }  
 &  \hfhref{/hfagc/0506003.html}{ \hfpubhot{<0.130} }  
 &  { \,}  
 &  \hfhref{/hfagc/BR_-511_421+443.html}{ \hfavg{<0.130} } 
\\%

 \hfhref{/hfagc/BR_-511_111+443.html}{ \hflabel{J/\psi(1S) \pi^0} }  
 &  \hfhref{/hfagc/BR_-511_111+443.html}{ \hfpdg{0.22\pm0.04} }  
 &  \hfhref{/hfagc/0506076.html}{ \hfpubold{0.23\pm0.05\pm0.02} }  
 &  \hfhref{/hfagc/0507003.html}{ \hfprehot{0.194\pm0.022\pm0.017} }  
 &  { \,}  
 &  \hfhref{/hfagc/BR_-511_111+443.html}{ \hfavg{0.20\pm0.02} } 
\\%

 \hfhref{/hfagc/BR_-511_113+443.html}{ \hflabel{J/\psi(1S) \rho^0(770)} }  
 &  \hfhref{/hfagc/BR_-511_113+443.html}{ \hfpdg{0.160\pm0.070} }  
 &  \hfhref{/hfagc/0506078.html}{ \hfpre{0.28\pm0.03\pm0.03} }  
 &  \hfhref{/hfagc/0506011.html}{ \hfpubold{0.160\pm0.060\pm0.040} }  
 &  { \,}  
 &  \hfhref{/hfagc/BR_-511_113+443.html}{ \hfavg{0.25\pm0.04} } 
\\
\hline 

\end{tabular}

\end{center}


\begin{center}
\hfnewcaption{Ratios of branching fractions}{neutral B}{\hfjpsi}{in units of $10^{0}$}{upper limits are at 90\% CL}{/hfagc/00204.html}{tab:hfc03204}
\hfmetadata{ [ .tml created 2006-02-08T20:03:09.222+08:00] }

\begin{tabular}{lccccc}  
\hflmode
&  \hflpdg2004
&  \hflbelle
&  \hflbabar
&  \hflcdf
&  \hflavg
\\\hline 

 \hfhref{/hfagc/BR_-511_-10313+443xoBR_-521_-321+443.html}{ \hflabel{\frac{{\cal{B}} ( \bar{B}^0 \to J/\psi(1S) \bar{K}_1^0(1270) )}{{\cal{B}} ( B^- \to J/\psi(1S) K^- )}} } \hftabletlcell 
 &  { \,}  
 &  \hfhref{/hfagc/0506072.html}{ \hfpubold{1.30\pm0.34\pm0.28} }  
 &  { \,}  
 &  { \,}  
 &  \hfhref{/hfagc/BR_-511_-10313+443xoBR_-521_-321+443.html}{ \hfavg{1.30\pm0.44} } 
\\%

 \hfhref{/hfagc/BR_-511_-311+441xoBR_-511_-311+443.html}{ \hflabel{\frac{{\cal{B}} ( \bar{B}^0 \to \eta_c(1S) \bar{K}^0 )}{{\cal{B}} ( \bar{B}^0 \to J/\psi(1S) \bar{K}^0 )}} }  
 &  { \,}  
 &  { \,}  
 &  \hfhref{/hfagc/0506001.html}{ \hfpub{1.34\pm0.19\pm0.13\pm0.38} }  
 &  { \,}  
 &  \hfhref{/hfagc/BR_-511_-311+441xoBR_-511_-311+443.html}{ \hfavg{1.34\pm0.44} } 
\\%

 \hfhref{/hfagc/BR_-511_-313+443xoBR_-511_-311+443.html}{ \hflabel{\frac{{\cal{B}} ( \bar{B}^0 \to J/\psi(1S) \bar{K}^{*0}(892) )}{{\cal{B}} ( \bar{B}^0 \to J/\psi(1S) \bar{K}^0 )}} } \hftableblcell 
 &  \hfhref{/hfagc/BR_-511_-313+443xoBR_-511_-311+443.html}{ \hfpdg{1.48\pm0.12} }  
 &  { \,}  
 &  \hfhref{/hfagc/0506109.html}{ \hfpubhot{1.51\pm0.05\pm0.08} }  
 &  \hfhref{/hfagc/0506094.html}{ \hfpubold{1.39\pm0.36\pm0.10} }  
 &  \hfhref{/hfagc/BR_-511_-313+443xoBR_-511_-311+443.html}{ \hfavg{1.50\pm0.09} } 
\\\hline 

\end{tabular}

\end{center}

\end{\hftabletype}
\clearpage  
\begin{\hftabletype}[\hftableposn]

\begin{center}
\hfnewcaption{Miscellaneous quantities}{neutral B}{\hfjpsi}{in units of $10^{0}$}{upper limits are at 90\% CL}{/hfagc/00204.html}{tab:hfc04204}
\hfmetadata{ [ .tml created 2006-02-08T20:03:10.899+08:00] }

\begin{tabular}{lccccc}  
\hflmode
&  \hflpdg2004
&  \hflbelle
&  \hflbabar
&  \hflcdf
&  \hflavg
\\\hline 

 \hfhref{/hfagc/A0_-511_313+443xoA0_511_313+443.html}{ \hflabel{\frac{{\cal{A}} ( \bar{B}^0 \to J/\psi(1S) K^{*0}(892) )}{{\cal{A}} ( B^0 \to J/\psi(1S) K^{*0}(892) )}} } \hftabletlcell 
 &  { \,}  
 &  { \,}  
 &  \hfhref{/hfagc/0506007.html}{ \hfpub{<0.26} }  
 &  { \,}  
 &  \hfhref{/hfagc/A0_-511_313+443xoA0_511_313+443.html}{ \hfavg{<0.26} } 
\\%

 \hfhref{/hfagc/A0_511_-313+443xoA0_-511_-313+443.html}{ \hflabel{\frac{{\cal{A}} ( B^0 \to J/\psi(1S) \bar{K}^{*0}(892) )}{{\cal{A}} ( \bar{B}^0 \to J/\psi(1S) \bar{K}^{*0}(892) )}} } \hftableblcell 
 &  { \,}  
 &  { \,}  
 &  \hfhref{/hfagc/0506007.html}{ \hfpub{<0.32} }  
 &  { \,}  
 &  \hfhref{/hfagc/A0_511_-313+443xoA0_-511_-313+443.html}{ \hfavg{<0.32} } 
\\\hline 

\end{tabular}

\end{center}

\end{\hftabletype}
\clearpage  
\begin{\hftabletype}[\hftableposn]

\begin{center}
\hfnewcaption{Branching fractions}{neutral B}{\hfochm}{in units of $10^{-3}$}{upper limits are at 90\% CL}{/hfagc/00205.html}{tab:hfc01205}
\hfmetadata{ [ .tml created 2006-02-08T20:03:15.978+08:00] }

\begin{tabular}{lccccc}  
\hflmode
&  \hflpdg2004
&  \hflbelle
&  \hflbabar
&  \hflcdf
&  \hflavg
\\\hline 

 \hfhref{/hfagc/BR_-511_-313+445.html}{ \hflabel{\chi_{c2}(1P) \bar{K}^{*0}(892)} } \hftabletlcell 
 &  { \,}  
 &  { \,}  
 &  \hfhref{/hfagc/0506006.html}{ \hfpubhot{<0.036} }  
 &  { \,}  
 &  \hfhref{/hfagc/BR_-511_-313+445.html}{ \hfavg{<0.036} } 
\\%

 \hfhref{/hfagc/BR_-511_-311+445.html}{ \hflabel{\chi_{c2}(1P) \bar{K}^0} }  
 &  { \,}  
 &  { \,}  
 &  \hfhref{/hfagc/0506006.html}{ \hfpubhot{<0.041} }  
 &  { \,}  
 &  \hfhref{/hfagc/BR_-511_-311+445.html}{ \hfavg{<0.041} } 
\\%

 \hfhref{/hfagc/BR_-511_-313+20443.html}{ \hflabel{\chi_{c1}(1P) \bar{K}^{*0}(892)} }  
 &  \hfhref{/hfagc/BR_-511_-313+20443.html}{ \hfpdg{0.41\pm0.15} }  
 &  { \,}  
 &  \hfhref{/hfagc/0506109.html}{ \hfpubhot{0.33\pm0.04\pm0.05\pm0.03} }  
 &  { \,}  
 &  \hfhref{/hfagc/BR_-511_-313+20443.html}{ \hfavg{0.33\pm0.08} } 
\\%

 \hfhref{/hfagc/BR_-511_-311+20443.html}{ \hflabel{\chi_{c1}(1P) \bar{K}^0} }  
 &  \hfhref{/hfagc/BR_-511_-311+20443.html}{ \hfpdg{0.40\pm0.11} }  
 &  { \,}  
 &  \hfhref{/hfagc/0506109.html}{ \hfpubhot{0.45\pm0.04\pm0.02\pm0.05} }  
 &  { \,}  
 &  \hfhref{/hfagc/BR_-511_-311+20443.html}{ \hfavg{0.45\pm0.07} } 
\\%

 \hfhref{/hfagc/BR_-511_-311+100443.html}{ \hflabel{\psi(2S) \bar{K}^0} }  
 &  \hfhref{/hfagc/BR_-511_-311+100443.html}{ \hfpdg{0.62\pm0.07} }  
 &  \hfhref{/hfagc/0506076.html}{ \hfpubold{0.67\pm0.11} }  
 &  \hfhref{/hfagc/0506109.html}{ \hfpubhot{0.65\pm0.06\pm0.04\pm0.02} }  
 &  { \,}  
 &  \hfhref{/hfagc/BR_-511_-311+100443.html}{ \hfavg{0.65\pm0.07} } 
\\
\hline 

\end{tabular}

\end{center}


\begin{center}
\hfnewcaption{Ratios of branching fractions}{neutral B}{\hfochm}{in units of $10^{0}$}{upper limits are at 90\% CL}{/hfagc/00205.html}{tab:hfc03205}
\hfmetadata{ [ .tml created 2006-02-08T20:03:19.928+08:00] }

\begin{tabular}{lccccc}  
\hflmode
&  \hflpdg2004
&  \hflbelle
&  \hflbabar
&  \hflcdf
&  \hflavg
\\\hline 

 \hfhref{/hfagc/BR_-511_-313+20443xoBR_-511_-311+20443.html}{ \hflabel{\frac{{\cal{B}} ( \bar{B}^0 \to \chi_{c1}(1P) \bar{K}^{*0}(892) )}{{\cal{B}} ( \bar{B}^0 \to \chi_{c1}(1P) \bar{K}^0 )}} } \hftabletlcell 
 &  \hfhref{/hfagc/BR_-511_-313+20443xoBR_-511_-311+20443.html}{ \hfpdg{0.89\pm0.38} }  
 &  { \,}  
 &  \hfhref{/hfagc/0506109.html}{ \hfpubhot{0.72\pm0.11\pm0.12} }  
 &  { \,}  
 &  \hfhref{/hfagc/BR_-511_-313+20443xoBR_-511_-311+20443.html}{ \hfavg{0.72\pm0.16} } 
\\%

 \hfhref{/hfagc/BR_-511_-311+441xoBR_-521_-321+441.html}{ \hflabel{\frac{{\cal{B}} ( \bar{B}^0 \to \eta_c(1S) \bar{K}^0 )}{{\cal{B}} ( B^- \to \eta_c(1S) K^- )}} }  
 &  { \,}  
 &  { \,}  
 &  \hfhref{/hfagc/0506001.html}{ \hfpub{0.87\pm0.13\pm0.07} }  
 &  { \,}  
 &  \hfhref{/hfagc/BR_-511_-311+441xoBR_-521_-321+441.html}{ \hfavg{0.87\pm0.15} } 
\\%

 \hfhref{/hfagc/BR_-511_-313+100443xoBR_-511_-311+100443.html}{ \hflabel{\frac{{\cal{B}} ( \bar{B}^0 \to \psi(2S) \bar{K}^{*0}(892) )}{{\cal{B}} ( \bar{B}^0 \to \psi(2S) \bar{K}^0 )}} }  
 &  { \,}  
 &  { \,}  
 &  \hfhref{/hfagc/0506109.html}{ \hfpubhot{1.00\pm0.14\pm0.09} }  
 &  { \,}  
 &  \hfhref{/hfagc/BR_-511_-313+100443xoBR_-511_-311+100443.html}{ \hfavg{1.00\pm0.17} } 
\\%

 \hfhref{/hfagc/BR_-511_-313+441xoBR_-511_-311+441.html}{ \hflabel{\frac{{\cal{B}} ( \bar{B}^0 \to \eta_c(1S) \bar{K}^{*0}(892) )}{{\cal{B}} ( \bar{B}^0 \to \eta_c(1S) \bar{K}^0 )}} } \hftableblcell 
 &  \hfhref{/hfagc/BR_-511_-313+441xoBR_-511_-311+441.html}{ \hfpdg{1.30\pm0.40} }  
 &  \hfhref{/hfagc/0506074.html}{ \hfpubold{1.33\pm0.36\pm^{0.24}_{0.33}} }  
 &  { \,}  
 &  { \,}  
 &  \hfhref{/hfagc/BR_-511_-313+441xoBR_-511_-311+441.html}{ \hfavg{1.33\pm^{0.43}_{0.49}} } 
\\\hline 

\end{tabular}

\end{center}

\end{\hftabletype}
\clearpage  
\begin{\hftabletype}[\hftableposn]

\begin{center}
\hfnewcaption{Branching fractions}{neutral B}{\hfmuld}{in units of $10^{-3}$}{upper limits are at 90\% CL}{/hfagc/00206.html}{tab:hfc01206}
\hfmetadata{ [ .tml created 2006-02-08T20:03:24.589+08:00] }

\begin{tabular}{lccccc}  
\hflmode
&  \hflpdg2004
&  \hflbelle
&  \hflbabar
&  \hflcdf
&  \hflavg
\\\hline 

 \hfhref{/hfagc/BR_-511_-411+411.html}{ \hflabel{D^- D^+} } \hftabletlcell 
 &  \hfhref{/hfagc/BR_-511_-411+411.html}{ \hfpdg{<0.94} }  
 &  \hfhref{/hfagc/0506037.html}{ \hfprehot{0.32\pm0.06\pm0.05} }  
 &  { \,}  
 &  { \,}  
 &  \hfhref{/hfagc/BR_-511_-411+411.html}{ \hfavg{0.32\pm0.07} } 
\\%

 \hfhref{/hfagc/BR_-511_-413+413.html}{ \hflabel{D^{*+}(2010) D^{*-}(2010)} }  
 &  \hfhref{/hfagc/BR_-511_-413+413.html}{ \hfpdg{0.87\pm0.18} }  
 &  \hfhref{/hfagc/0506036.html}{ \hfpubhot{0.81\pm0.08\pm0.11} }  
 &  \hfhref{/hfagc/0506111.html}{ \hfpubold{0.83\pm0.16\pm0.12} }  
 &  { \,}  
 &  \hfhref{/hfagc/BR_-511_-413+413.html}{ \hfavg{0.82\pm0.11} } 
\\%

 \hfhref{/hfagc/BR_-511_-411+413.html}{ \hflabel{D^{*+}(2010) D^-} }  
 &  \hfhref{/hfagc/BR_-511_-411+413.html}{ \hfpdg{<0.63} }  
 &  { \,}  
 &  \hfhref{/hfagc/0506107.html}{ \hfpubold{0.88\pm0.10\pm0.11\pm0.06} }  
 &  { \,}  
 &  \hfhref{/hfagc/BR_-511_-411+413.html}{ \hfavg{0.88\pm0.16} } 
\\%

 \hfhref{/hfagc/BR_-511_-413+411.html}{ \hflabel{D^{*-}(2010) D^+} }  
 &  \hfhref{/hfagc/BR_-511_-413+411.html}{ \hfpdg{<0.63} }  
 &  \hfhref{/hfagc/0506084.html}{ \hfpubold{1.17\pm0.26\pm^{0.20}_{0.24}\pm0.08} }  
 &  { \,}  
 &  { \,}  
 &  \hfhref{/hfagc/BR_-511_-413+411.html}{ \hfavg{1.17\pm^{0.34}_{0.36}} } 
\\%

 \hfhref{/hfagc/BR_-511_-311+-421+421.html}{ \hflabel{D^0 \bar{D}^0 \bar{K}^0} }  
 &  \hfhref{/hfagc/BR_-511_-311+-421+421.html}{ \hfpdg{<1.40} }  
 &  { \,}  
 &  \hfhref{/hfagc/0506112.html}{ \hfpubold{<1.40} }  
 &  { \,}  
 &  \hfhref{/hfagc/BR_-511_-311+-421+421.html}{ \hfavg{<1.40} } 
\\%

 \hfhref{/hfagc/BR_-511_-321+-421+411.html}{ \hflabel{D^+ \bar{D}^0 K^-} }  
 &  \hfhref{/hfagc/BR_-511_-321+-421+411.html}{ \hfpdg{1.70\pm0.40} }  
 &  { \,}  
 &  \hfhref{/hfagc/0506112.html}{ \hfpubold{1.70\pm0.30\pm0.30} }  
 &  { \,}  
 &  \hfhref{/hfagc/BR_-511_-321+-421+411.html}{ \hfavg{1.70\pm0.42} } 
\\
\hline 

\end{tabular}

\end{center}


\begin{center}
\hfnewcaption{Product branching fractions}{neutral B}{\hfmuld}{in units of $10^{-4}$}{upper limits are at 90\% CL}{/hfagc/00206.html}{tab:hfc02206}
\hfmetadata{ [ .tml created 2006-02-08T20:03:28.557+08:00] }

\begin{tabular}{lccccc}  
\hflmode
&  \hflpdg2004
&  \hflbelle
&  \hflbabar
&  \hflcdf
&  \hflavg
\\\hline 

 \hfhref{/hfagc/BR_-511_-321+415xBR_415_211+421.html}{ \hflabel{K^- D_2^{*+}(2460) [ D^0 \pi^+ ]} } \hftabletlcell 
 &  { \,}  
 &  { \,}  
 &  \hfhref{/hfagc/0506018.html}{ \hfpub{0.183\pm0.040\pm0.031} }  
 &  { \,}  
 &  \hfhref{/hfagc/BR_-511_-321+415xBR_415_211+421.html}{ \hfavg{0.183\pm0.051} } 
\\%

 \hfhref{/hfagc/BR_-511_-211+415xBR_415_-211+211+413.html}{ \hflabel{\pi^- D_2^{*+}(2460) [ D^{*+}(2010) \pi^- \pi^+ ]} }  
 &  { \,}  
 &  \hfhref{/hfagc/0506030.html}{ \hfpubhot{<0.24} }  
 &  { \,}  
 &  { \,}  
 &  \hfhref{/hfagc/BR_-511_-211+415xBR_415_-211+211+413.html}{ \hfavg{<0.24} } 
\\%

 \hfhref{/hfagc/BR_-511_-211+10413xBR_10413_-211+211+413.html}{ \hflabel{\pi^- D_1^+(2420) [ D^{*+}(2010) \pi^- \pi^+ ]} }  
 &  { \,}  
 &  \hfhref{/hfagc/0506030.html}{ \hfpubhot{<0.33} }  
 &  { \,}  
 &  { \,}  
 &  \hfhref{/hfagc/BR_-511_-211+10413xBR_10413_-211+211+413.html}{ \hfavg{<0.33} } 
\\%

 \hfhref{/hfagc/BR_-511_-211+20413xBR_20413_211+423.html}{ \hflabel{\pi^- D_1^+(H) [ D^{*0}(2007) \pi^+ ]} }  
 &  { \,}  
 &  \hfhref{/hfagc/0506070.html}{ \hfpre{<0.70} }  
 &  { \,}  
 &  { \,}  
 &  \hfhref{/hfagc/BR_-511_-211+20413xBR_20413_211+423.html}{ \hfavg{<0.70} } 
\\%

 \hfhref{/hfagc/BR_-511_-211+10413xBR_10413_-211+211+411.html}{ \hflabel{\pi^- D_1^+(2420) [ D^+ \pi^- \pi^+ ]} }  
 &  { \,}  
 &  \hfhref{/hfagc/0506030.html}{ \hfpubhot{0.89\pm0.15\pm0.17\pm^{0.00}_{0.26}} }  
 &  { \,}  
 &  { \,}  
 &  \hfhref{/hfagc/BR_-511_-211+10413xBR_10413_-211+211+411.html}{ \hfavg{0.89\pm^{0.23}_{0.34}} } 
\\%

 \hfhref{/hfagc/BR_-511_-211+10411xBR_10411_211+421.html}{ \hflabel{\pi^- D_0^{*+} [ D^0 \pi^+ ]} }  
 &  { \,}  
 &  \hfhref{/hfagc/0506070.html}{ \hfpre{<1.20} }  
 &  { \,}  
 &  { \,}  
 &  \hfhref{/hfagc/BR_-511_-211+10411xBR_10411_211+421.html}{ \hfavg{<1.20} } 
\\
\hline 

\end{tabular}

\end{center}

\end{\hftabletype}
\clearpage  
\begin{\hftabletype}[\hftableposn]

\begin{center}
\hfnewcaption{Ratios of branching fractions}{neutral B}{\hfmuld}{in units of $10^{0}$}{upper limits are at 90\% CL}{/hfagc/00206.html}{tab:hfc03206}
\hfmetadata{ [ .tml created 2006-02-08T20:03:39.47+08:00] }

%

\end{center}

\end{\hftabletype}
\clearpage  
\begin{\hftabletype}[\hftableposn]

\begin{center}
\hfnewcaption{Branching fractions}{neutral B}{\hfsgdx}{in units of $10^{-3}$}{upper limits are at 90\% CL}{/hfagc/00207.html}{tab:hfc01207}
\hfmetadata{ [ .tml created 2006-02-08T20:03:43.058+08:00] }

%

\begin{description}\addtolength{\itemsep}{\hffootitemsep\baselineskip}

	  \item[ \hffootnotemark{1} ] \hffootnotetext{ Study of $\bar{B}^{0} \to  D^{(*)0} \pi^{+} \pi^{-}$ decays ; Dalitz fit analysis (\hfbb{152}) } 
    
	  \item[ \hffootnotemark{2} ] \hffootnotetext{ Study of $\bar{B^{0}} \to D^{(*) 0} \pi^+ \pi^-$ Decays (\hfbb{31.3}) } 
    
\end{description}

\end{center}

\end{\hftabletype}
\clearpage  

\clearpage 
\begin{\hftabletype}[\hftableposn]

\begin{center}
\hfnewcaption{Branching fractions}{neutral B}{\hfsgld}{in units of $10^{-3}$}{upper limits are at 90\% CL}{/hfagc/00208.html}{tab:hfc01208}
\hfmetadata{ [ .tml created 2006-02-08T20:03:49.744+08:00] }

\begin{tabular}{lccccc}  
\hflmode
&  \hflpdg2004
&  \hflbelle
&  \hflbabar
&  \hflcdf
&  \hflavg
\\\hline 

 \hfhref{/hfagc/BR_-511_-321+-421+211.html}{ \hflabel{\bar{D}^0 K^- \pi^+} } \hftabletlcell 
 &  { \,}  
 &  { \,}  
 &  \hfhref{/hfagc/0506018.html}{ \hfpub{<0.019} }  
 &  { \,}  
 &  \hfhref{/hfagc/BR_-511_-321+-421+211.html}{ \hfavg{<0.019} } 
\\%

 \hfhref{/hfagc/BR_-511_-313+-421.html}{ \hflabel{\bar{D}^0 \bar{K}^{*0}(892)} }  
 &  \hfhref{/hfagc/BR_-511_-313+-421.html}{ \hfpdg{<0.018} }  
 &  \hfhref{/hfagc/0506085.html}{ \hfpubold{<0.018} }  
 &  \hfhref{/hfagc/0506023.html}{ \hfpre{<0.041} }  
 &  { \,}  
 &  \hfhref{/hfagc/BR_-511_-313+-421.html}{ \hfavg{<0.018} } 
\\%

 \hfhref{/hfagc/BR_-511_-313+421.html}{ \hflabel{D^0 \bar{K}^{*0}(892)} }  
 &  \hfhref{/hfagc/BR_-511_-313+421.html}{ \hfpdg{0.048\pm0.012} }  
 &  \hfhref{/hfagc/0506085.html}{ \hfpubold{0.048\pm^{0.011}_{0.010}\pm0.005} }  
 &  \hfhref{/hfagc/0506023.html}{ \hfpre{0.062\pm0.014\pm0.006} }  
 &  { \,}  
 &  \hfhref{/hfagc/BR_-511_-313+421.html}{ \hfavg{0.053\pm0.009} } 
\\%

 \hfhref{/hfagc/BR_-511_-311+421.html}{ \hflabel{D^0 \bar{K}^0} }  
 &  \hfhref{/hfagc/BR_-511_-311+421.html}{ \hfpdg{0.050\pm0.014} }  
 &  \hfhref{/hfagc/0506085.html}{ \hfpubold{0.050\pm^{0.013}_{0.012}\pm0.006} }  
 &  \hfhref{/hfagc/0506023.html}{ \hfpre{0.062\pm0.012\pm0.004} }  
 &  { \,}  
 &  \hfhref{/hfagc/BR_-511_-311+421.html}{ \hfavg{0.056\pm0.009} } 
\\%

 \hfhref{/hfagc/BR_-511_-321+211+421.html}{ \hflabel{D^0 K^- \pi^+} }  
 &  { \,}  
 &  { \,}  
 &  \hfhref{/hfagc/0506018.html}{ \hfpub{0.088\pm0.015\pm0.009} }  
 &  { \,}  
 &  \hfhref{/hfagc/BR_-511_-321+211+421.html}{ \hfavg{0.088\pm0.017} } 
\\%

 \hfhref{/hfagc/BR_-511_331+421.html}{ \hflabel{D^0 \eta^\prime(958)} }  
 &  \hfhref{/hfagc/BR_-511_331+421.html}{ \hfpdg{0.170\pm0.040} }  
 &  \hfhref{/hfagc/0506035.html}{ \hfprehot{0.114\pm0.020\pm^{0.010}_{0.013}} }  
 &  \hfhref{/hfagc/0506106.html}{ \hfpub{0.170\pm0.040\pm0.018\pm0.010} }  
 &  { \,}  
 &  \hfhref{/hfagc/BR_-511_331+421.html}{ \hfavg{0.126\pm0.021} } 
\\%

 \hfhref{/hfagc/BR_-511_225+421.html}{ \hflabel{f_2(1270) D^0} }  
 &  { \,}  
 &  \hfhref{/hfagc/0506070.html}{ \hfpre{0.195\pm0.034\pm0.038\pm^{0.032}_{0.002}} }  
 &  { \,}  
 &  { \,}  
 &  \hfhref{/hfagc/BR_-511_225+421.html}{ \hfavg{0.195\pm0.056} } 
\\%

 \hfhref{/hfagc/BR_-511_-321+411.html}{ \hflabel{D^+ K^-} }  
 &  \hfhref{/hfagc/BR_-511_-321+411.html}{ \hfpdg{0.200\pm0.060} }  
 &  \hfhref{/hfagc/0506081.html}{ \hfpubold{0.20\pm0.05\pm0.02\pm0.03} }  
 &  { \,}  
 &  { \,}  
 &  \hfhref{/hfagc/BR_-511_-321+411.html}{ \hfavg{0.20\pm0.06} } 
\\%

 \hfhref{/hfagc/BR_-511_221+421.html}{ \hflabel{D^0 \eta} }  
 &  \hfhref{/hfagc/BR_-511_221+421.html}{ \hfpdg{0.22\pm0.05} }  
 &  { \,}  
 &  \hfhref{/hfagc/0506106.html}{ \hfpub{0.25\pm0.02\pm0.03\pm0.01} }  
 &  { \,}  
 &  \hfhref{/hfagc/BR_-511_221+421.html}{ \hfavg{0.25\pm0.04} } 
\\%

 \hfhref{/hfagc/BR_-511_111+421.html}{ \hflabel{D^0 \pi^0} }  
 &  \hfhref{/hfagc/BR_-511_111+421.html}{ \hfpdg{0.29\pm0.03} }  
 &  { \,}  
 &  \hfhref{/hfagc/0506106.html}{ \hfpub{0.29\pm0.02\pm0.03\pm0.01} }  
 &  { \,}  
 &  \hfhref{/hfagc/BR_-511_111+421.html}{ \hfavg{0.29\pm0.04} } 
\\%

 \hfhref{/hfagc/BR_-511_113+421.html}{ \hflabel{D^0 \rho^0(770)} }  
 &  \hfhref{/hfagc/BR_-511_113+421.html}{ \hfpdg{0.29\pm0.11} }  
 &  { \,}  
 &  { \,}  
 &  { \,}  
 &  \hfhref{/hfagc/BR_-511_113+421.html}{ \hfavg{0.29\pm0.05} } 
\\%

{ \hflabel{\,} }  
 &  { \,}  
 &  \hfhref{/hfagc/0506088.html}{ \hfpubold{0.29\pm0.10\pm0.04} } 
\hffootnotemark{2}
 
 &  { \,}  
 &  { \,}  
 &  { \,} 
\\%

{ \hflabel{\,} }  
 &  { \,}  
 &  \hfhref{/hfagc/0506070.html}{ \hfpre{0.29\pm0.03\pm0.03\pm^{0.01}_{0.05}} } 
\hffootnotemark{1}
 
 &  { \,}  
 &  { \,}  
 &  { \,} 
\\%

 \hfhref{/hfagc/BR_-511_223+421.html}{ \hflabel{D^0 \omega(782)} }  
 &  \hfhref{/hfagc/BR_-511_223+421.html}{ \hfpdg{0.25\pm0.06} }  
 &  { \,}  
 &  \hfhref{/hfagc/0506106.html}{ \hfpub{0.30\pm0.03\pm0.04\pm0.01} }  
 &  { \,}  
 &  \hfhref{/hfagc/BR_-511_223+421.html}{ \hfavg{0.30\pm0.05} } 
\\%

 \hfhref{/hfagc/BR_-511_-321+311+411.html}{ \hflabel{D^+ K^- K^0} }  
 &  \hfhref{/hfagc/BR_-511_-321+311+411.html}{ \hfpdg{<0.31} }  
 &  \hfhref{/hfagc/0506090.html}{ \hfpubold{<0.31} }  
 &  { \,}  
 &  { \,}  
 &  \hfhref{/hfagc/BR_-511_-321+311+411.html}{ \hfavg{<0.31} } 
\\%

 \hfhref{/hfagc/BR_-511_-323+411.html}{ \hflabel{D^+ K^{*-}(892)} }  
 &  \hfhref{/hfagc/BR_-511_-323+411.html}{ \hfpdg{0.37\pm0.18} }  
 &  { \,}  
 &  \hfhref{/hfagc/0506026.html}{ \hfpubhot{0.46\pm0.06\pm0.05\pm0.02} }  
 &  { \,}  
 &  \hfhref{/hfagc/BR_-511_-323+411.html}{ \hfavg{0.46\pm0.08} } 
\\%

 \hfhref{/hfagc/BR_-511_-211+311+411.html}{ \hflabel{D^+ K^0 \pi^-} }  
 &  { \,}  
 &  { \,}  
 &  \hfhref{/hfagc/0506026.html}{ \hfpubhot{0.49\pm0.07\pm0.04\pm0.03} }  
 &  { \,}  
 &  \hfhref{/hfagc/BR_-511_-211+311+411.html}{ \hfavg{0.49\pm0.09} } 
\\%

 \hfhref{/hfagc/BR_-511_-321+313+411.html}{ \hflabel{D^+ K^- K^{*0}(892)} }  
 &  \hfhref{/hfagc/BR_-511_-321+313+411.html}{ \hfpdg{0.88\pm0.19} }  
 &  \hfhref{/hfagc/0506090.html}{ \hfpubold{0.88\pm0.11\pm0.15} }  
 &  { \,}  
 &  { \,}  
 &  \hfhref{/hfagc/BR_-511_-321+313+411.html}{ \hfavg{0.88\pm0.19} } 
\\%

 \hfhref{/hfagc/BR_-511_-211+211+421.html}{ \hflabel{D^0 \pi^+ \pi^-} }  
 &  \hfhref{/hfagc/BR_-511_-211+211+421.html}{ \hfpdg{0.80\pm0.16} }  
 &  { \,}  
 &  { \,}  
 &  { \,}  
 &  \hfhref{/hfagc/BR_-511_-211+211+421.html}{ \hfavg{0.98\pm0.09} } 
\\%

{ \hflabel{\,} }  
 &  { \,}  
 &  \hfhref{/hfagc/0506088.html}{ \hfpubold{0.80\pm0.06\pm0.15} } 
\hffootnotemark{2}
 
 &  { \,}  
 &  { \,}  
 &  { \,} 
\\%

{ \hflabel{\,} } \hftableblcell 
 &  { \,}  
 &  \hfhref{/hfagc/0506070.html}{ \hfpre{1.07\pm0.06\pm0.10} } 
\hffootnotemark{1}
 
 &  { \,}  
 &  { \,}  
 &  { \,} 
\\\hline 

\end{tabular}

\begin{description}\addtolength{\itemsep}{\hffootitemsep\baselineskip}

	  \item[ \hffootnotemark{1} ] \hffootnotetext{ Study of $\bar{B}^{0} \to  D^{(*)0} \pi^{+} \pi^{-}$ decays ; Dalitz fit analysis (\hfbb{152}) } 
    
	  \item[ \hffootnotemark{2} ] \hffootnotetext{ Study of $\bar{B^{0}} \to D^{(*) 0} \pi^+ \pi^-$ Decays (\hfbb{31.3}) } 
    
\end{description}

\end{center}


\begin{center}
\hfnewcaption{Product branching fractions}{neutral B}{\hfsgld}{in units of $10^{-5}$}{upper limits are at 90\% CL}{/hfagc/00208.html}{tab:hfc02208}
\hfmetadata{ [ .tml created 2006-02-08T20:03:53.581+08:00] }

\begin{tabular}{lccccc}  
\hflmode
&  \hflpdg2004
&  \hflbelle
&  \hflbabar
&  \hflcdf
&  \hflavg
\\\hline 

 \hfhref{/hfagc/BR_-511_-313+421xBR_-323_-321+211.html}{ \hflabel{D^0 \bar{K}^{*0}(892)} } \hftabletlcell\hftableblcell 
 &  { \,}  
 &  { \,}  
 &  \hfhref{/hfagc/0506018.html}{ \hfpub{3.8\pm0.6\pm0.4} }  
 &  { \,}  
 &  \hfhref{/hfagc/BR_-511_-313+421xBR_-323_-321+211.html}{ \hfavg{3.8\pm0.7} } 
\\\hline 

\end{tabular}

\end{center}

\end{\hftabletype}
\clearpage  
\begin{\hftabletype}[\hftableposn]

\begin{center}
\hfnewcaption{Branching fractions}{miscellaneous}{charmed particles }{in units of $10^{-3}$}{upper limits are at 90\% CL}{/hfagc/00300.html}{tab:hfc01300}
\hfmetadata{ [ .tml created 2006-02-08T20:04:03.385+08:00] }

\begin{tabular}{lccccc}  
\hflmode
&  \hflpdg2004
&  \hflbelle
&  \hflbabar
&  \hflcdf
&  \hflavg
\\\hline 

 \hfhref{/hfagc/BR_-5122_-3122+443.html}{ \hflabel{{\cal{B}} ( \bar{\Lambda}_b^0 \to J/\psi(1S) \bar{\Lambda} )} } \hftabletlcell 
 &  \hfhref{/hfagc/BR_-5122_-3122+443.html}{ \hfpdg{0.47\pm0.28} }  
 &  { \,}  
 &  { \,}  
 &  \hfhref{/hfagc/0506097.html}{ \hfpubold{0.47\pm0.21\pm0.19} }  
 &  \hfhref{/hfagc/BR_-5122_-3122+443.html}{ \hfavg{0.47\pm0.28} } 
\\%

 \hfhref{/hfagc/BR_-531_333+443.html}{ \hflabel{{\cal{B}} ( \bar{B}_s^0 \to J/\psi(1S) \phi(1020) )} } \hftableblcell 
 &  \hfhref{/hfagc/BR_-531_333+443.html}{ \hfpdg{0.93\pm0.33} }  
 &  { \,}  
 &  { \,}  
 &  \hfhref{/hfagc/0506094.html}{ \hfpubold{0.93\pm0.28\pm0.17} }  
 &  \hfhref{/hfagc/BR_-531_333+443.html}{ \hfavg{0.93\pm0.33} } 
\\\hline 

\end{tabular}

\end{center}


\begin{center}
\hfnewcaption{Product branching fractions}{miscellaneous}{charmed particles }{in units of $10^{-5}$}{upper limits are at 90\% CL}{/hfagc/00300.html}{tab:hfc02300}
\hfmetadata{ [ .tml created 2006-02-08T20:04:04.608+08:00] }

\begin{tabular}{lccccc}  
\hflmode
&  \hflpdg2004
&  \hflbelle
&  \hflbabar
&  \hflcdf
&  \hflavg
\\\hline 

 \hfhref{/hfagc/BR_97_83+96xBR_83_223+443.html}{ \hflabel{{\cal{B}} ( B \to K Y(3940) [ \omega(782) J/\psi(1S) ] )} } \hftabletlcell\hftableblcell 
 &  { \,}  
 &  \hfhref{/hfagc/0506033.html}{ \hfpubhot{7.1\pm1.3\pm3.1} }  
 &  { \,}  
 &  { \,}  
 &  \hfhref{/hfagc/BR_97_83+96xBR_83_223+443.html}{ \hfavg{7.1\pm3.4} } 
\\\hline 

\end{tabular}

\end{center}


\begin{center}
\hfnewcaption{Ratios of branching fractions}{miscellaneous}{charmed particles }{in units of $10^{0}$}{upper limits are at 90\% CL}{/hfagc/00300.html}{tab:hfc03300}
\hfmetadata{ [ .tml created 2006-02-08T20:04:05.945+08:00] }

\begin{tabular}{lccccc}  
\hflmode
&  \hflpdg2004
&  \hflbelle
&  \hflbabar
&  \hflcdf
&  \hflavg
\\\hline 

 \hfhref{/hfagc/BR_-531_100443+333xoBR_-531_333+443.html}{ \hflabel{\frac{{\cal{B}} ( \bar{B}_s^0 \to \psi(2S) \phi(1020) )}{{\cal{B}} ( \bar{B}_s^0 \to J/\psi(1S) \phi(1020) )}} } \hftabletlcell 
 &  { \,}  
 &  { \,}  
 &  { \,}  
 &  \hfhref{/hfagc/0506101.html}{ \hfwaiting{0.52\pm0.13\pm0.07} }  
 &  \hfhref{/hfagc/BR_-531_100443+333xoBR_-531_333+443.html}{ \hfavg{0.52\pm0.15} } 
\\%

 \hfhref{/hfagc/BR_-531_-211+431xoBR_-511_-211+411.html}{ \hflabel{\frac{{\cal{B}} ( \bar{B}_s^0 \to D_s^+ \pi^- )}{{\cal{B}} ( \bar{B}^0 \to D^+ \pi^- )}} }  
 &  { \,}  
 &  { \,}  
 &  { \,}  
 &  \hfhref{/hfagc/0601001.html}{ \hfwaiting{1.02\pm0.07\pm0.03\pm0.15} }  
 &  \hfhref{/hfagc/BR_-531_-211+431xoBR_-511_-211+411.html}{ \hfavg{1.02\pm0.17} } 
\\%

 \hfhref{/hfagc/BR_-531_-211+-211+211+431xoBR_-511_-211+-211+211+411.html}{ \hflabel{\frac{{\cal{B}} ( \bar{B}_s^0 \to D_s^+ \pi^+ \pi^- \pi^- )}{{\cal{B}} ( \bar{B}^0 \to D^+ \pi^+ \pi^- \pi^- )}} }  
 &  { \,}  
 &  { \,}  
 &  { \,}  
 &  \hfhref{/hfagc/0506098.html}{ \hfwaiting{1.17\pm0.18\pm0.34} }  
 &  \hfhref{/hfagc/BR_-531_-211+-211+211+431xoBR_-511_-211+-211+211+411.html}{ \hfavg{1.17\pm0.38} } 
\\%

 \hfhref{/hfagc/BR_-5122_-4122+211xoBR_-511_-211+411.html}{ \hflabel{\frac{{\cal{B}} ( \bar{\Lambda}_b^0 \to \Lambda_c^- \pi^+ )}{{\cal{B}} ( \bar{B}^0 \to D^+ \pi^- )}} }  
 &  { \,}  
 &  { \,}  
 &  { \,}  
 &  \hfhref{/hfagc/0506103.html}{ \hfpre{3.3\pm0.3\pm0.4\pm1.1} }  
 &  \hfhref{/hfagc/BR_-5122_-4122+211xoBR_-511_-211+411.html}{ \hfavg{3.3\pm1.2} } 
\\%

 \hfhref{/hfagc/BR_-5122_-13+-4122+14xoBR_-5122_-4122+211.html}{ \hflabel{\frac{{\cal{B}} ( \bar{\Lambda}_b^0 \to \Lambda_c^- \mu^+ \nu_\mu )}{{\cal{B}} ( \bar{\Lambda}_b^0 \to \Lambda_c^- \pi^+ )}} } \hftableblcell 
 &  { \,}  
 &  { \,}  
 &  { \,}  
 &  \hfhref{/hfagc/0506102.html}{ \hfwaiting{20.0\pm3.0\pm1.2\pm^{0.9}_{2.2}} }  
 &  \hfhref{/hfagc/BR_-5122_-13+-4122+14xoBR_-5122_-4122+211.html}{ \hfavg{20.0\pm^{3.4}_{3.9}} } 
\\\hline 

\end{tabular}

\end{center}

\end{\hftabletype}
\clearpage  

\clearpage


\clearpage

\section{Summary }
\labs{summary}

This article provides the updated world averages for 
$b$-hadron properties as of at the end of 2005.
A small selection of highlights of the results described in sections
\ref{sec:life_mix}-\ref{sec:BtoCharm} is given in 
Table~\ref{tab_summary}.

\begin{table}
\caption{ Brief summary of the world averages as of at the end of 2005.}
\label{tab_summary}
\begin{center}
\begin{tabular}{|l|c|}
\hline
 {\bf\boldmath \b-hadron lifetimes} &   \\
 ~~$\tau(\Bd)$  & \hfagTAUBD \\
 ~~$\tau(\Bu)$  & \hfagTAUBU \\
 ~~$\tau(\Bs\to~\mbox{flavour specific})$  & \hfagTAUBSSL \\
 ~~$\bar{\tau}(\Bs) = 1/\Gs$  & \hfagTAUBSMEANCON \\
 ~~$\tau(\Bc)$  & \hfagTAUBC \\
 ~~$\tau(\Lb)$  & \hfagTAULB \\
\hline
 {\bf\boldmath \b-hadron fractions} &   \\
 ~~$f^{+-}/f^{00}$ in \Ups\ decays  & \hfagFF \\ 
 ~~$\fBd=\fBu$ at high energy & \hfagFBD \\
 ~~\fBs\ at high energy  & \hfagFBS \\
 ~~\fbb\ at high energy  & \hfagFBB \\
\hline
 {\bf\boldmath \Bd\ and \Bs\ mixing parameters} &   \\
 ~~\dmd &  \hfagDMDWU \\
 ~~$|q/p|_{\particle{d}}$ & \hfagQP  \\
 ~~\dms  &  $> \rm \hfagDMSWLIM~at~\CL{95}$ \\
 ~~$\DGGs = (\Gamma_{\rm L} - \Gamma_{\rm H})/\Gs$ & \hfagDGSGSCON \\
 \hline
 {\bf\boldmath Semileptonic \B decay parameters} &   \\
 ~~${\cal B}(\BzbDstarlnu)$  & $( 5.34 \pm 0.20)\%$ \\
 ~~${\cal B}(\BzbDplnu)$      & $( 2.12 \pm 0.20)\%$ \\
 ~~${\cal B}(\Bb\to  X\ell\nub)$   & $(10.95 \pm 0.15)\%$ \\
 ~~$|V_{\particle{cb}}|F(1)\ (\BzbDstarlnu)$   
       & $ (37.7 \pm 0.9) \times 10^{-3}$ \\
 ~~$|V_{\particle{cb}}|G(1) \ (\BzbDplnu)$   
       & $(42.2 \pm 3.7) \times 10^{-3}$ \\
 ~~$|V_{\particle{ub}}|$ (inclusive)    & $ (4.39 \pm 0.33 ) \times 10^{-3}$ \\
 ~~${\cal B}(\Bb\to  \pi\ell\nub)$   & $(1.34 \pm 0.11) \times 10^{-4}$ \\
\hline
 {\bf\boldmath Rare \B decays} &   \\
~~${\cal B}(b \to s \gamma)$ & $ (355 \pm 24^{+9}_{-10}\pm 3)\times
 10^{-6}$ \\
\hline
\end{tabular}
\end{center}
\end{table}

Concerning the lifetime and mixing averages, the most significant changes
since winter 2005~\cite{hfag_hepex_winter2005}
are due to new measurements from the Tevatron experiments
in the areas of heavy \b-baryon lifetimes and \Bs mixing
parameters (in particular, it is worth noting that the 
\CL{95} lower limit on \dms has now increased to \hfagDMSWLIM, after 
having stayed for several years between 14 and 15\invps). 
In the near future, CDF and \dzero are expected to contribute
with increasing significance in these areas, while the \Bu and \Bd lifetime
and mixing properties, dominated by the \B factory results, have probably
already reached close to their asymptotic precisions.

In the field of semileptonic $B$-meson decays, the advances reported
in this update center on the CKM matrix element $|V_{ub}|$. 
For the first time the extraction of $|V_{ub}|$ has been performed on
a consistent basis (using BLNP~\cite{ref:BLNP}) for all available
measurements of the partial
branching fraction to inclusive charmless semileptonic $B$-meson decays.
The leading uncertainty ($\sim 5\%$)
depends on the precision on the determination of the $b$-quark mass. The overall 
uncertainty of $\sim 8\%$ is approaching the lower bound of $5\%$,
which is the irreducible theoretical uncertainty typically quoted by theorists.
However it will be important in future to not rely on just one
theoretical framework. In exclusive decays, in contrast, 
our average of the branching fraction to $B\to\pi l\nu$
is an input into several approaches of extracting $|V_{ub}|$. They 
rely on light cone sum rule and unquenched lattice QCD calculations.
The uncertainty in $|V_{ub}|$ is dominated by theoretical
uncertainties that are at the level of 15-20\%. In future the
uncertainties will be reduced with use of the measured differential
decay rate in $q^2$.      

Measurements by \babar\ and \belle 
of the time-dependent $\CP$ violation parameter
$S_{b \to c\bar c s}$ in \B decays to charmonium and a neutral kaon
have established $\CP$ violation in $\B$ decays,
and allow a precise extraction of 
the Unitarity Triangle parameter $\stwob / \sin\! 2\phi_1$.
Recent studies of $\B \to \jpsi \Kstar$ (Sec.~\ref{sec:cp_uta:ccs_vv})
and $\B \to D^{(*)}h^0$, where $h^0 = \pi^0$ \etc. (Sec.~\ref{sec:cp_uta:cud_beta})
allow to resolve the ambiguity in the solutions for $\beta / \phi_1$ from
the measurement of $\stwob / \sin\! 2\phi_1$.
Measurements of
time-dependent $\CP$ asymmetries in hadronic $b \to s$ penguin decays
continue to provide insight into possible new physics.
In this update, results from \belle\ have been updated with more statistics,
and both \babar\ and \belle\ have added more decay modes,
including the mode $\KS\pi^0\pi^0$ from \babar\ and charmless modes with $\KL$
in the final state.
Compared to the previous round of averages,
the consistency with the Standard Model expectation is improved.
An intriguing discrepancy still exists, 
although it is not presently a significant effect.
Results from time-dependent analyses 
with the decays $\Bz \to \pi^+\pi^-, \rho^\pm\pi^\mp$ and $\rho^+\rho^-$ provide constraints 
on the Unitarity Triangle angle $\alpha/\phi_2$ (Sec.~\ref{sec:cp_uta:uud}).
The most constraining measurements are 
the results from \babar\ and \belle\ in the $\rho\rho$ system.
Constraints on the third Unitarity Triangle angle $\gamma/\phi_3$
have been obtained by \babar\ and \belle,
using $\Bm \to \DorDstar \Km$ decays (Sec.~\ref{sec:cp_uta:cus}).
At present, the most constraining results 
arise from the Dalitz plot analysis of
the subsequent $D \to \KS\pi^+\pi^-$ decay.

For the rare \B\ decays, many new measurements on branching fractions 
and $CP$ violating asymmetries are included and summarized in
section~\ref{sec:rare}. A special effort has been made to handle the
correlated model uncertainties of shape function in averaging the  $b\to
s\gamma$ results.  The obtained average is highlighted in
Table~\ref{tab_summary}. Results of ratios of branching fractions for
charmless two-body $B$ decays reported by CDF are also added for the
first time.  

Quantities related to $B$ decays to charmed particles, other than
$CP$-asymmetry parameters, have been dealt with by HFAG for the first
time. The huge sample of $B$ mesons available has enabled measurements of 
branching fractions for rare decays or decay chains of a few $10^{-7}$
with statistical uncertainties typically below $30\%$, while results for
more common decays, with branching fractions around $10^{-4}$, are
becoming precision measurements, with uncertainties typically at the
$3\%$ level. We highlight the great improvements that have been attained
toward a deeper understanding of recently discovered new states with
either hidden or open charm content, such as $X(3870)$,
$D_{sJ}^{*-}(2317)$ and $D_{sJ}^{-}(2460)$.

\section*{ Acknowledgments }

We would like to thank collaborators of \babar, Belle, CDF, CLEO, \dzero,
LEP, and SLD experiments who provided fruitful results on \b-hadron
properties and cooperated with the HFAG for averaging.
These results are thanks to the excellent operations of the 
accelerators and collaborations with experimental groups by the 
accelerator groups of PEP-II, KEKB, CESR, Tevatron, LEP, and SLC.


\end{document}